\documentclass[11pt]{article}
\usepackage[utf8]{inputenc}

\pdfoutput=1
\usepackage{bbold}
\usepackage{amssymb}
\usepackage{amsmath}
\usepackage{cancel}
\usepackage[dvips]{graphicx}
\usepackage{setspace}
\usepackage{slashed}
\usepackage{mathtools}
\usepackage{amsfonts}
\usepackage{fancyhdr}
\usepackage{dsfont}
\usepackage{xcolor}
\usepackage{graphicx}
\usepackage{rotating}
\usepackage{float}
\usepackage{comment}
\usepackage{color} 
\usepackage{subcaption}
\usepackage[skip=6pt]{caption}
\usepackage[percent]{overpic}
\usepackage{cite}
\usepackage{braket}
\usepackage{ulem,lipsum}
\definecolor{darkgreen}{rgb}{0,0.5,0}
\definecolor{darkblue}{rgb}{0,0,0.6}
\definecolor{purple}{rgb}{0.4,.2,0.7}

\newcommand{\be}{\begin{equation}}
\newcommand{\ee}{\end{equation}}

\usepackage[colorlinks=true,citecolor=darkgreen,linkcolor=black,urlcolor=purple]{hyperref}
\usepackage{eqparbox}

\usepackage{mcite}
\usepackage{pdfsync}

\makeatletter
\newcommand*{\defeq}{\mathrel{\rlap{%
                     \raisebox{0.3ex}{$\m@th\cdot$}}%
                     \raisebox{-0.3ex}{$\m@th\cdot$}}%
                     =} 
\makeatother

\def\be{\begin{eqnarray}}
\def\ee{\end{eqnarray}}

\newcommand{\tr}{\textrm{Tr}\,}

\newcommand{\bea}{\begin{eqnarray}}
\newcommand{\eea}{\end{eqnarray}}
\def\ben{\begin{equation}}
\def\een{\end{equation}}

     \let\r=v

\def\be{\begin{equation}}
\def\ee{\end{equation}}
\def\ba{\begin{eqnarray}}
\def\ea{\end{eqnarray}}

\def\bal#1\eal{\begin{align}#1\end{align}}
\def\bs#1\es{\begin{split}#1\end{split}}

\usepackage{enumitem}

\allowdisplaybreaks  

\numberwithin{equation}{section}

\def\be{\begin{equation}}
\def\ee{\end{equation}}
\def\ba{\begin{eqnarray}}
\def\ea{\end{eqnarray}}
\def\bal#1\eal{\begin{align}#1\end{align}}

\def\r{\rightarrow}

\def\r{\right}

\usepackage{tikz}
\usetikzlibrary{positioning,arrows}
\usetikzlibrary{decorations.pathmorphing}
\usetikzlibrary{decorations.markings}
\tikzset{
particle/.style={postaction={decorate}},
graviton/.style={decorate, decoration={snake, amplitude=0.8 mm, segment length=1.5 mm, pre length=0.8 mm, post length=0.8 mm}},
photon/.style={
        decoration={complete sines, amplitude=0.15cm, segment length=0.2cm},
        decorate    
    },
gluon/.style={
        decoration={coil, aspect=0.75, mirror, segment length=1.5mm},
        decorate
    }
}

\def \be {\begin{equation}}
\def \ee {\end{equation}}

\usepackage{framed}

\begin{document}
\onehalfspacing

\begin{center}

~
\vskip5mm

{\LARGE  {
QG from SymQRG: AdS$_3$/CFT$_2$ Correspondence as Topological Symmetry-Preserving Quantum RG Flow 
\\
\ \\
}}

\vskip10mm

Ning Bao${}^{1}$, Ling-Yan Hung${}^{2,3}$, Yikun Jiang${}^{1}$, Zhihan Liu${}^{4}$

\vskip5mm
\it{${}^1$ Department of Physics, Northeastern University, Boston, MA 02115, USA}
\vskip2mm
\it{${}^2$ Yau Mathematical Sciences Center, Tsinghua University, Beijing 100084, China}
\vskip2mm
\it{${}^3$ Yanqi Lake Beijing Institute of Mathematical Sciences and Applications (BIMSA), Huairou District, Beijing 101408, China}
\vskip2mm
\it{${}^4$ Department of Physics, Cornell University, Ithaca, New York, USA}
\vskip2mm

\vskip5mm

\end{center}

\vspace{4mm}

\begin{abstract}
\noindent

By analyzing non-perturbative RG flows that explicitly preserve given topological symmetries, we show that each of their coarse-graining steps can be expressed as a quantum path integral of the SymTFT in one higher dimension. When the symmetries involved include the Virasoro defect lines, such as in the case of $T\bar{T}$ deformations, this RG kernel is the 3D quantum gravitational path integral. For 2D CFTs whose structure constants are under control, we identify the corresponding ground state of the SymTFT, from which the Wheeler-DeWitt equation emerges as the non-perturbative no-flux constraint: the gravitational path integral acts on this state as the projector that renders symmetry-preserving coarse-graining exact. These observations are summarized in the slogan: $\textbf{SymQRG = QG}$. The exact discrete formulation of Liouville theory in \cite{Chen:2024unp} allows us to identify a universal SymQRG kernel, constructed from quantum $6j$ symbols of $U_q(SL(2,\mathbb{R}))$: it provides a discrete realization of two copies of the Virasoro TQFT on $\Sigma\times I$, and manifests as an exact and analytic 3D background-independent MERA-type holographic tensor network. Many aspects of the AdS/CFT correspondence, including the factorization puzzle, admit a natural interpretation within this framework. We propose that the non-perturbative AdS$_3$/CFT$_2$ correspondence is a \textit{maximal} form of topological holography, in which the physical boundary is fixed by kinematics alone.

 \end{abstract}

\pagebreak
\pagestyle{plain}

\setcounter{tocdepth}{3}
{}
\vfill
\tableofcontents

 \newpage

\date{}

\section{Introduction}
\label{sec: intro}
The quest for a complete non-perturbative theory of quantum gravity remains one of the most profound challenges in modern theoretical physics. Among the most promising frameworks is the AdS/CFT correspondence \cite{Maldacena:1997re, Gubser:1998bc, Witten:1998qj}, which establishes a duality between a $D$-dimensional conformal field theory (CFT) and a gravitational theory in $(D+1)$-dimensional asymptotically Anti-de Sitter (AdS) spacetime. This duality not only provides a powerful tool for studying strongly coupled quantum field theories but also offers deep insights into the nature of spacetime itself.

A very important aspect of the AdS/CFT correspondence is that global symmetries of the field theory exhibit themselves as gauge symmetries in the bulk\cite{Witten:1998qj}. In particular, conformal transformations of the field theory are connected to a subset of diffeomorphisms in the asymptotically AdS$_{D+1}$ spacetime. Other global symmetries in the field theory also manifest as gauge symmetries in the holographic dual. This structure bears striking resemblance to the ``topological holographic principle'', proposed in the study of generalized symmetries\cite{Gaiotto:2014kfa},\footnote{Throughout this paper, by ``topological symmetries'' we mean symmetries implemented by topological defect operators, including non-invertible ones. Every global symmetry is topological in this sense; the adjective emphasizes that the basic objects entering our construction are the topological defects themselves.} where the path integral of a quantum field theory (QFT) with symmetries on a $D$-manifold $\Sigma_D$ is produced from the path integral of a topological field theory (often called SymTFT) in $(D+1)$ dimensions on a manifold of the form $\Sigma_D \times I$, with $I$ being an interval. This is often called ``sandwich construction''\cite{Gaiotto:2020iye, Ji:2019jhk, *Kong:2020cie, *Kong:2020jne, *Chatterjee:2022kxb, Freed:2022qnc, Apruzzi:2021nmk, Moradi:2022lqp}. The SymTFT is chosen so that a given global symmetry of the QFT is gauged in the TQFT \cite{cat2,Ji:2019jhk}. This strongly suggests that the two holographic principles are closely related. 

We stress at the outset, however, an important distinction. The sandwich always provides an exact rewriting of the QFT partition function; but in general, the SymTFT together with its topological boundary abstracts only the symmetry data of the theory, while the dynamics resides in the non-topological boundary condition \cite{Freed:2022qnc}. In this generic situation, the construction separates kinematics from dynamics in the spirit of representation theory, rather than constituting a holographic duality in the AdS/CFT sense, where the bulk is expected to capture the dynamics of the boundary theory. A sharp question is therefore \textit{when, and in what sense, the bulk and its topological boundary come to carry the full dynamical content}. We will see that for the 2D CFT examples we study, this happens precisely when the symmetry made explicit in the bulk is \textit{maximal}.

On the other hand, the renormalization group (RG) flow also plays an important role in the AdS/CFT correspondence. It has been proposed that the flow of field theory couplings is encoded in the AdS bulk \cite{Susskind:1998dq, deBoer:1999tgo, Skenderis:2002wp}. If the topological holographic principle is indeed related to the AdS/CFT correspondence, then it is important to understand where the RG data of the 2D theory resides in the SymTFT description. We stress that the SymTFT itself, being topological, does not flow: as we will see, the RG data is carried by the \textit{non-topological boundary condition} of the sandwich, while the SymTFT path integral supplies the \textit{kernel} relating boundary conditions at different scales.

Can the combination of topological symmetries and RG, the two most important non-perturbative aspects of strongly interacting QFTs, reveal new insights into quantum gravity? In this paper, we show that a ``topological symmetry-preserving quantum RG'' process, referred to as ``SymQRG'', is implemented by the path integral of the SymTFT in one higher dimension. The defining feature of SymQRG is that it is a \textit{projected} coarse-graining: an ordinary block-spin transformation does not, by itself, preserve a prescribed collection of topological defect lines, and each SymQRG step completes it with the projection back onto the symmetry-organized sector. This projection is not an ad hoc truncation: it is a slab of the SymTFT path integral. In particular in 2D, RG processes preserving a continuum of topological lines related to Virasoro representations\footnote{We use ``irrational'' in a relative sense: a theory may be rational with respect to an extended chiral algebra and yet possess an infinite, continuously labeled collection of topological lines commuting with the Virasoro algebra alone. It is the latter collection that enters our construction; in particular, ``irrational'' does not simply mean ``infinite''. See Sec.~\ref{dynamicskinematics}.} produce a precise 3D {\it quantum} gravitational path integral that is given by two copies of the Virasoro/Teichmüller TQFT\cite{Verlinde:1989ua, EllegaardAndersen:2011vps, Collier:2023fwi, Chen:2024unp}: when the symmetry concerned contains such a collection of topological line operators\cite{Gaiotto:2014kfa, Verlinde:1988sn}, the SymTFT is a long-range entangled topological order\cite{Turaev:1992hq, Levin:2004mi, Chen:2010gda, Kitaev:2005dm, *PhysRevLett.96.110405} that contains 3D quantum gravity.

We summarize this principle as
\[
\textbf{SymQRG(Virasoro lines) = QG}~,
\]
whose precise meaning is the following: \textit{the projector that renders Virasoro-preserving coarse-graining exact and truncation-free is the three-dimensional quantum gravitational path integral.} The scope of this statement is laid out in Sec.~\ref{summaryofresults}.

A class of RG flows driven by $T\bar T$ deformation\cite{Zamolodchikov:2004ce, McGough:2016lol} is proposed to be related to the bulk gravitational path integral. We show that in 2D this is indeed a special case of symmetry-preserving RG flows that can be expressed as a quantum gravitational path integral. The $T\bar T$ example also illustrates an important point: the framework is tied to the preservation of topological line data, \textit{not} to conformality of the boundary theory. The $T\bar T$-deformed theories are no longer at the fixed-point, yet they preserve every Virasoro defect line, since topological lines commute with the stress tensor, cf.~\eqref{commutewithstress}. Holography for non-fixed-point theories is therefore not outside this framework: along a symmetry-preserving trajectory, the bulk kernel stays fixed while only the physical boundary state changes.

The second example we consider arises from a discrete formalism: the exact tensor network state sum representation of 2D CFTs, in particular the Liouville theory\cite{Chen:2022wvy, Cheng:2023kxh, Chen:2024unp, Hung:2024gma}. The state sum construction in fact represents a \textit{continuous family} of symmetry-preserving continuum theories, labeled by a regulator scale, with the CFT recovered at its fixed point. This method produces for us a well-defined gravitational quantum path integral, {\it non-perturbative in $G_N$}, from a quantum RG flow. While we considered the Liouville theory explicitly in \cite{Chen:2024unp}, the symmetry-preserving RG operator $\hat{U}_{\text{RG}}$ extracted from coarse-graining the triangulation of the CFT path integral is universal: after all, it follows only from the crossing kernels of correlation functions between Virasoro representations, and is therefore shared by all theories preserving the Virasoro lines. We note that such a discrete yet exact formulation of the quantum path integral is naturally a holographic tensor network. Many well established tensor network methods could then be applied to continuum field theories. 

\subsection{Summary of results}\label{summaryofresults}

By incorporating the AdS/CFT correspondence, the best-established example of holography, into this framework and understanding it in light of the SymTFT, we gain new insights into several aspects of the correspondence. Since the claims of this paper carry different logical status, we list here the main results, together with the precise sense in which each is established.

\begin{enumerate}[label=\arabic*)]

\item \textbf{The CFT partition function is a SymTFT overlap.} When the bulk path integral includes a topological boundary condition $\mathcal{B}$, it prepares a state $|\Psi(\mathcal{B})\rangle_{\Sigma_D}$ on the surface $\Sigma_D$. The non-topological boundary condition at the other end of the sandwich can then be specified by a bra state $\langle \Omega|$ on $\Sigma_D$, which can be viewed as a basis vector in one dimension higher. The $D$ dimensional QFT \textit{partition function} is a ($D+1$) dimensional \textit{wavefunction}, which can be written as an overlap
\be \label{overlapinintro}
Z_{\text{QFT}}(\Sigma_D) = \langle \Omega | \Psi (\mathcal{B})\rangle_{\Sigma_D}~.
\ee
For 2D CFTs the construction is algebraically explicit in terms of
Moore--Seiberg and BCFT sewing data. For theories rational with respect to
a chiral algebra $V$, the representations of $V$ form a modular tensor
category $\mathcal{C}=\mathrm{Rep}(V)$
\cite{Moore:1988qv,*Moore:1989vd}, from which the SymTFT is built via the
Turaev--Viro/Levin--Wen string-net construction
\cite{Turaev:1992hq,Levin:2004mi}. In the Virasoro non-rational setting,
the analogous fusion and braiding data are continuously labeled and lie
beyond the finite semisimple framework. On the string-net Hilbert space,
$\langle\Omega|$ is constructed from Virasoro conformal blocks, and
$|\Psi(\mathcal{B})\rangle$ from BCFT structure constants. This
representation is exact, with all of its ingredients solved, for rational
CFTs and for Liouville theory
\cite{Chen:2022wvy,Cheng:2023kxh,Chen:2024unp,Hung:2024gma}
(Sec.~\ref{tensor network}).

\item \textbf{SymQRG is a projected coarse-graining, and its kernel is the SymTFT slab.} An ordinary block-spin contraction of the physical boundary state is completed by the projection back onto the symmetry-organized sector. At coincident scales, the slab is the idempotent ground-state projector $\Pi_{\Lambda}$ of the SymTFT; across scales, it is the corresponding topological cobordism kernel $\hat{U}^{\Lambda,\Lambda'}_{\text{RG}}$ (Sec.~\ref{symqrgparagraph}). The insertion of the projector involves \textit{no truncation}: it is exact precisely because the dynamical state satisfies the no-flux condition $\Pi_{\Lambda}|\Psi(\mathcal{B})\rangle=|\Psi(\mathcal{B})\rangle$. This is also what distinguishes SymQRG from the quantum RG of \cite{Lee:2013dln, Lee:2016uqc}, where the projection onto single-trace operators is a choice justified at large $N$: here the projector is \textit{canonical}, determined by the preserved topological lines, and exact at finite central charge.

\item \textbf{For Virasoro lines, the kernel is the 3D quantum gravity path integral.} When the preserved lines are those commuting with the Virasoro algebra,\footnote{Topological defect lines that commute with extended chiral algebras form a subset of those that commute with the Virasoro algebra.} the kernel is built from quantum $6j$ symbols of $U_q(SL(2,\mathbb{R}))$, and provides a discrete realization of the $\Sigma\times I$ path integral of two copies of the Virasoro TQFT (Sec.~\ref{TQFTSECTION}). This identification is conditional on the proposal of \cite{Collier:2023fwi} that the latter is the non-perturbative form of AdS$_3$ quantum gravity; its semi-classical limit is verified explicitly, with the $6j$ symbols reducing to the Einstein-Hilbert action of hyperbolic tetrahedra at large central charge. For a generic large $N$ holographic CFT, whose large number of Virasoro defect lines correspond to heavy operators, the same gravitational kernel emerges in a coarse-grained sense (Sec.~\ref{ETH}).

\item \textbf{No-flux $=$ crossing symmetry $=$ Wheeler-DeWitt.} The Wheeler-DeWitt equation enforces diffeomorphism invariance along the radial direction of AdS, and its deep connection to the Ward identities of CFT$_2$ was proposed in the beautiful works of \cite{Verlinde:1989ua, Freidel:2008sh, McGough:2016lol}. Within the sandwich construction, it is the ``vanishing flux condition'' satisfied by $|\Psi(\mathcal{B})\rangle$,\footnote{To be more precise, the physical states of the Turaev-Viro TQFT, which is the SymTFT for 2D CFTs, are equivalent to ground states of the Levin-Wen string-net model as we explain in Sec.~\ref{briefreviewtqft}, and we will use these terms interchangeably throughout the paper.} and the Callan-Symanzik equation along the RG trajectory aligns directly with it, realizing the UV/IR correspondence (Sec.~\ref{wdwstates}). On the lattice, the vanishing flux condition is related to crossing symmetry, which underpins the symmetry-preserving block-spin transformation and yields an exact, analytic multiscale entanglement renormalization ansatz (MERA)-like \cite{Vidal:2008zz} tensor network capturing the field theory across scales (Sec.~\ref{tensor network}). This elevates the original analogy between AdS/CFT and MERA \cite{Swingle:2009bg} to a precise statement.

\item \textbf{The topological boundary fixes the bulk quantum measure, and factorization follows.} While different CFTs serve as fixed points of the quantum RG process preserving the same collection of symmetries, they exhibit {\it distinct} modular invariant spectra: this information is encoded in the state $|\Psi(\mathcal{B})\rangle$ prepared by the topological boundary, which carries the CFT-specific sewing rules or operator product expansion (OPE) data\cite{Fuchs:2002cm, *Fuchs:2003id, *Fuchs:2003id1, *Fuchs:2004xi, *Fjelstad:2005ua, *Frohlich:2006ch, *Fjelstad:2006aw} and, in the bulk interpretation, the measure, contour, and projection data of the state-sum, often obscured in the semi-classical limit. The semi-classical phrase ``sum over geometries'' is incomplete until this data is specified, and different CFTs sharing the same kernel define different non-perturbative bulk measures. For fixed $|\Psi(\mathcal{B})\rangle$, products of partition functions factorize identically within the sandwich construction (Sec.~\ref{algebraicimplications}); the semi-classical factorization puzzle is thereby recast as identifying which quantum measure is approximated by the usual sum over geometries.

\end{enumerate}

Many questions about the intrinsic non-perturbative properties of quantum gravity find natural answers within our framework: for example, the factorization puzzle related to wormhole contributions, and the quantum chaos associated with scattering in black hole backgrounds. In particular, the slab kernel, determined by topological data, is universal
and acts identically whether the topological boundary encodes an individual
CFT or an ensemble average: the ensemble-versus-individual question at the
heart of the wormhole discussions is thereby pushed one layer deeper into the
sandwich, residing entirely in the state $|\Psi(\mathcal{B})\rangle$.

Meanwhile, it is well known that the SymTFT is a topological theory, and that in familiar examples the dynamical data of the QFT is carried by $\langle \Omega|$, the non-topological boundary condition of the sandwich: for the Ising CFT with its $\mathbb{Z}_2$ SymTFT, the toric code, distinct conformal primaries carrying the same $\mathbb{Z}_2$ charge are packaged together inside $\langle\Omega|$.\footnote{The same theory also illustrates the maximal case: taking the SymTFT to be the Turaev-Viro TQFT based on the full Ising fusion category, so that the non-invertible Kramers-Wannier duality line is made explicit alongside the $\mathbb{Z}_2$ line, renders the physical boundary purely kinematical, built from the $c=1/2$ conformal blocks alone; see Sec.~\ref{dynamicskinematics}.} At first glance, this appears to contradict the AdS/CFT correspondence, where the bulk is expected to capture the dynamical information of the boundary CFT. However, explicit constructions of the sandwich for 2D rational CFTs and Liouville theory \cite{Chen:2022wvy, Cheng:2023kxh, Chen:2024unp} reveal that the division of data between the two boundaries depends crucially on the amount of symmetry encoded in the bulk. As the bulk encodes a larger share of boundary theory symmetries, the complexity of $\langle \Omega|$ decreases. When the bulk fully captures all the topological symmetries of the CFT, the boundary condition descends into a collection of conformal blocks that are purely \textit{kinematical}, determined by representation theory; \textit{every} theory-specific datum (the modular invariant spectrum and the OPE coefficients) then resides in $|\Psi(\mathcal{B})\rangle$, attached to the bulk. This is the quantitative sense, anticipated above, in which the kinematical separation of topological holography upgrades to a duality in the maximal case; we spell it out in Sec.~\ref{dynamicskinematics}.

In the examples we study explicitly in this paper, ranging from rational CFTs to the Liouville theory, the corresponding holographic duals preserving the symmetries are topological. In this context it is worth recalling the distinction, drawn in the last section of \cite{Witten:1988ze}, between two kinds of metric-independent theories: those in which no metric is ever introduced, and those in which metrics are integrated over (quantum gravity). Our construction realizes a concrete interface between them: in 3D gravity, the elementary SymQRG kernel admits a metric-free topological state-sum description, while its gravitational presentation, a sum over metrics, emerges in the Virasoro case, together with the quantum measure dictated by $|\Psi(\mathcal{B})\rangle$, cf.~Sec.~\ref{TQFTSECTION}. The gravitational interpretation goes beyond either case in isolation: $|\Psi(\mathcal{B})\rangle$ selects a specific collection of quantum geometries to be summed over, so that the resulting path integral is not a sum over all metrics but over precisely those specified by the CFT data. In this sense, the bulk is simultaneously a topological field theory and a theory of quantum gravity, with the choice of topological boundary resolving the apparent distinction between the two cases in \cite{Witten:1988ze}.

The examples studied in this paper suggest a precise organizing principle uniting the non-perturbative AdS/CFT correspondence with ``topological'' holography: {\bf In AdS$_3$/CFT$_2$, the holographic bulk arises from the {\it maximal} SymTFT: the one that makes explicit {\it all} the topological lines commuting with the Virasoro algebra, non-invertible and otherwise. When the symmetry extracted is maximal, the physical boundary is fixed by kinematics alone, and the sandwich becomes a genuine 2D/3D equivalence.} The prospects for extending this principle beyond AdS$_3$/CFT$_2$ are discussed in Sec.~\ref{discussionsection}.

\paragraph{Relation to previous work.}
Several ingredients used below are established results.
The $T\bar{T}$ flow equation, finite-cutoff holography, and the Wheeler-DeWitt relation build on \cite{Zamolodchikov:2004ce, McGough:2016lol, Freidel:2008sh, Verlinde:1989ua}.
The tensor-network state-sum construction uses the Moore-Seiberg/FRS sewing data \cite{Moore:1988qv, Fuchs:2002cm} and the Turaev-Viro/Levin-Wen realization of the associated symmetry theory \cite{Turaev:1992hq, Levin:2004mi}.
The Virasoro/Teichm\"{u}ller interpretation of three-dimensional gravity follows the proposal of \cite{Collier:2023fwi, Teschner:2005bz}.
Our aim is not to claim these ingredients as new in isolation.
Rather, we reorganize them into a single SymQRG framework: the physical boundary state changes along a symmetry-preserving RG trajectory, the CFT-specific topological state satisfies the no-flux/Wheeler-DeWitt constraint, and the SymTFT slab supplies the constraint kernel.
In the Virasoro case this slab is the three-dimensional gravitational path integral.

\subsection{Outline of the paper}

The structure of the paper is as follows: In Sec.~\ref{RGsection}, we review the Wilsonian and Kadanoff RG frameworks, emphasizing their distinctions from holographic RG. To reconcile these differences, we introduce a framework termed ``quantum RG'' \cite{Lee:2013dln, Lee:2016uqc}, which we further extend into SymQRG. This section explains in detail the underlying concepts and procedures of SymQRG.

To concretize this framework, we present two examples of SymQRG:
\begin{itemize}

\item{In Sec.~\ref{TTbar}, we explore the $T\bar{T}$ deformation for continuum field theories \cite{Zamolodchikov:2004ce, McGough:2016lol}, demonstrating its natural fit within the SymQRG framework. Specifically, we show that $T\bar{T}$-deformed theories are connected through an integral kernel corresponding to the 3D gravitational path integral. We then reformulate this in operator formalism, verifying that it satisfies the Wheeler-DeWitt equation. Written in operator formalism, it is clear that there is an underlying 3D long-range entangled ground state that fulfills the no-flux condition, common to 2D CFTs.}

\item{In Sec.~\ref{tensor network}, we establish the discrete formalism for SymQRG, extending the exact tensor network state sum representation of 2D CFTs introduced in \cite{Chen:2022wvy, Cheng:2023kxh, Chen:2024unp, Hung:2024gma}. After a brief review of the Turaev-Viro TQFT and the Levin-Wen string-net model, the two quantum states in \eqref{overlapinintro} are constructed from conformal blocks and BCFT structure coefficients (``BCFT Legos'' that simultaneously build up the CFT and the bulk AdS spacetime\cite{Hung:2024gma}), and crossing symmetry is shown to yield an analytic MERA-like SymQRG network, including deformations away from the CFT fixed point. }

\end{itemize}

The formalism above applies to CFTs with any topological symmetries. In Sec.~\ref{TQFTSECTION}, we investigate the conditions under which the SymTFT is related to 3D Einstein gravity. We first focus on the case of 3D pure Einstein gravity. We elaborate on the connection between the Virasoro TQFT\cite{Collier:2023fwi} and the discrete RG operator discovered during the discretization of the Liouville theory\cite{Chen:2024unp}, constructed from representations of $U_q(SL(2,\mathbb{R}))$. We also briefly comment on the connection to quantum Teichmüller theory and the lattice Liouville theory\cite{EllegaardAndersen:2011vps, Faddeev:1985gy, *Faddeev:2000if, *Faddeev:2002ms, kashaev2008}. Geometric notions like geodesic lengths and volume in 3D arise when considering symmetries associated with these Virasoro lines, and the anyons in these theories are related to conical defects or BTZ black holes \cite{Banados:1992wn, Collier:2023fwi}. Finally, we illustrate how 3D gravity universally emerges in the semi-classical limit from any microscopic large $N$ holographic 2D CFTs, with a discrete and dense spectrum, through coarse-graining, while emphasizing its distinction from the non-perturbatively exact holographic correspondence.

In Sec.~\ref{algebra}, we explore how the algebraic consistency conditions, or the bootstrap constraints of 2D CFTs are encoded in the topological boundary condition attached to the 3D bulk, or the state $\ket{\Psi(\mathcal{B})}$, and their fundamental role in the emergence of a 3D bulk. These algebraic properties are intrinsically tied to the topological symmetries encoded in the bulk SymTFT. These conditions fall into two categories: 
\begin{itemize}

\item{\textbf{The no-flux condition and crossing symmetry:} These conditions enable the SymQRG process, rendering the block-spin step exact and ensuring consistency across scales. They manifest as the Wheeler-DeWitt equation in gravity.}

\item{\textbf{Gauging/anyon condensation and modular invariance:} These conditions guarantee a modular invariant spectrum for the 2D theory, allowing the 3D bulk to collapse into an intrinsically defined 2D theory. This can be understood as specifying a precise quantum measure for the 3D theory, a critical component in achieving a genuine 2D/3D correspondence.}
\end{itemize}

Next, we elucidate how the decomposition of CFT information into $\langle \Omega |$ and $|\Psi (\mathcal{B})\rangle$ is related to the symmetry we extract into the SymTFT. We then explore the implications of these findings for quantum gravity. Specifically, we analyze the role of null states and wormholes, and demonstrate how the topological boundary helps address the factorization puzzle. Furthermore, we explain how the kinematical mechanism underlying the OTOC
diagnostic of quantum chaos arises from symmetry data. The Moore--Seiberg
data determine the second-sheet continuation of the conformal blocks,
while the spectrum and OPE coefficients encoded in
$|\Psi(\mathcal{B})\rangle$ determine how those blocks are assembled into
the full Lorentzian correlator.

We conclude our paper with some discussions on future directions.

This paper includes six appendices. Appendix \ref{reviewtvtqft} provides details on the Turaev-Viro TQFT. Appendices \ref{statesum cft} and \ref{tensornetwork SymQRG} offer a detailed explanation of the tensor network construction of 2D CFTs and the explicit procedure for the associated SymQRG. Appendix \ref{wavefunction tensornetwork} discusses the exact tensor network construction of CFT wavefunctions from the Euclidean path integral, and the connection to perfect tensors and random tensors. Appendix \ref{reviewteichmuller} provides a brief review of quantum Teichmüller theory and the Teichmüller TQFT, their connection to the $U_q(SL(2,\mathbb{R}))$ $6j$ symbols, and the lattice Liouville theory. Appendix \ref{ttbardetails} collects the intermediate steps of the formal manipulations of Sec.~\ref{TTbar}.

\section{From Wilson-Kadanoff RG to SymQRG}\label{RGsection}
\begin{figure}
	\centering
	\includegraphics[width=0.78\linewidth]{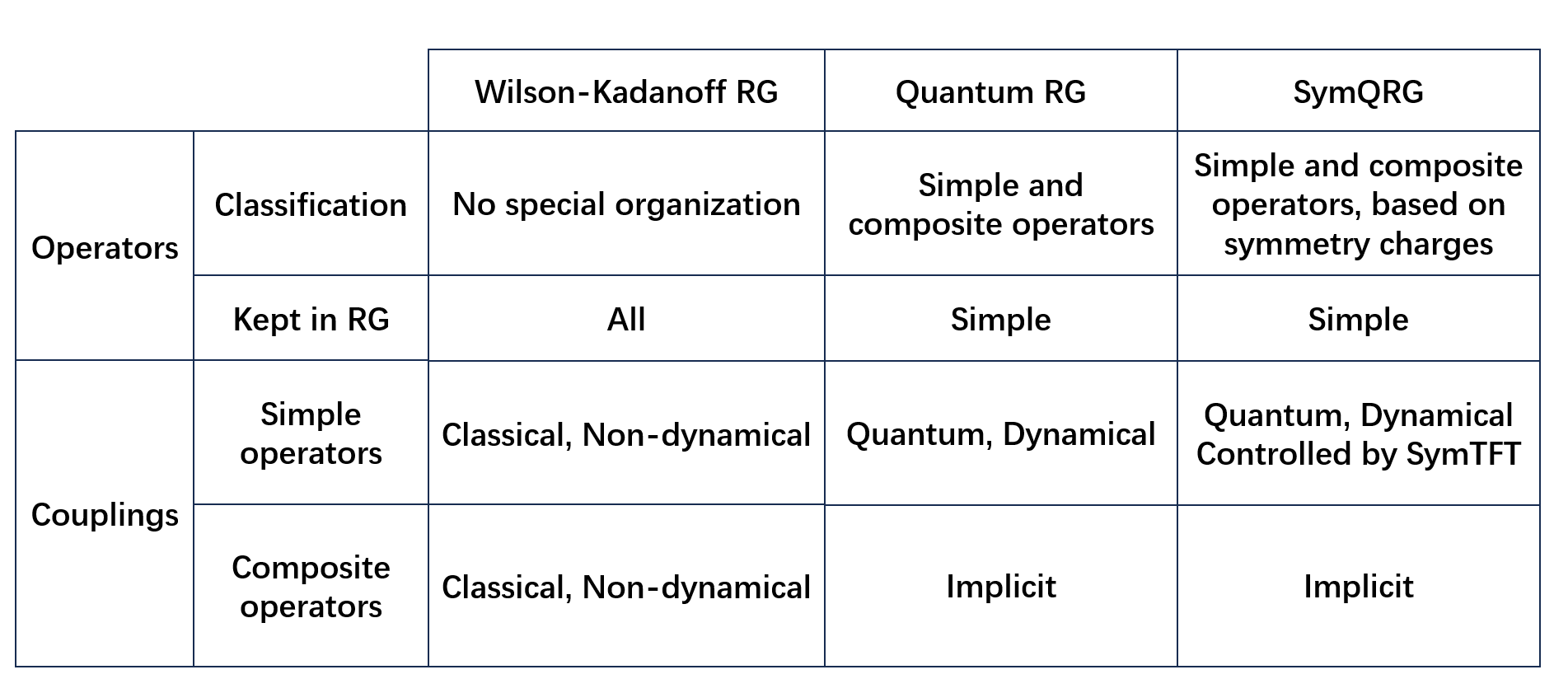}
	\caption{Comparison between three different RG procedures. Wilson-Kadanoff RG keeps all the couplings during each RG step, and the couplings are classical and non-dynamical. Quantum RG projects onto a subset of simple operators during each RG step, which makes the corresponding couplings quantum and dynamical in the emergent bulk. SymQRG projects onto the sector organized by the preserved topological defect lines, and the corresponding SymTFT supplies the no-flux projector/slab kernel implementing this projection.}
	\label{rgcomparison}
\end{figure}

The AdS/CFT correspondence has been famously argued to be related to RG flows from its very inception \cite{Maldacena:1997re, Gubser:1998bc, Witten:1998qj, Susskind:1998dq}. However, there are notable differences between holographic RG flows and the conventional Wilson-Kadanoff RG flow. In this section, we begin by briefly reviewing the Wilson-Kadanoff RG flow and then demonstrate the need to upgrade it to ``Quantum RG''\cite{Lee:2013dln, Lee:2016uqc}. We further extend this framework to SymQRG, where the RG step is completed by a fixed symmetry-theory kernel: the SymTFT slab projects the boundary data back onto the no-flux sector associated with the preserved topological defect lines. The comparison between these concepts is summarized in Fig.~\ref{rgcomparison}.

\subsection{Wilson-Kadanoff RG}

Let us briefly recall the Wilson-Kadanoff RG flow\cite{RevModPhys.55.583, PhysicsPhysiqueFizika.2.263}. 
Starting from a theory at momentum cutoff $\Lambda_{\text{RG}}$, the procedure has two steps. 
The first is \textit{coarse-graining}: integrating out modes in a momentum shell 
$\Lambda_{\text{RG}}'<|k|<\Lambda_{\text{RG}}$ produces an effective action at a lower 
cutoff $\Lambda_{\text{RG}}'$, without altering the partition function.\footnote{We always 
take the exact RG perspective in this paper; in practice, coarse-graining usually involves 
a truncation in couplings.} The second is \textit{rescaling}: momenta and fields are 
rescaled so that the cutoff is restored to $\Lambda_{\text{RG}}$ and the kinetic term 
remains canonical. Together, the two steps generate a flow in the space of couplings. 
Theories along the same RG trajectory are related by the Callan-Symanzik equation, 
which ensures that a deformation can be compensated by a rescaling.

\subsection{Quantum RG}\label{quantumrgsection}

A well-known tension between Wilsonian RG and holography is that 
RG couplings are classical, whereas their bulk duals are quantum 
and dynamical\cite{Heemskerk:2010hk, *Faulkner:2010jy, Lee:2012xba}; 
most importantly, the boundary metric $h_{ij}$ is a fixed source 
for the stress tensor, while the bulk metric $g_{\mu\nu}$ fluctuates. 
Lee's ``Quantum RG''\cite{Lee:2013dln, Lee:2016uqc} resolves this 
by projecting, at each RG step, onto a basis of single-trace operators: 
double-trace operators generated by coarse-graining are re-expressed 
in terms of single-traces and discarded, converting the sources into 
dynamical bulk fields. We adopt this perspective for the $T\bar{T}$ 
deformation in Sec.~\ref{TTbar}, where the procedure explicitly 
promotes $h_{ij}$ to the dynamical metric $g_{ij}$.

The single-trace projection in Quantum RG is, however, a choice 
justified at large $N$: it depends on the specific operator content 
of the theory and has no canonical definition at finite $N$. As detailed in Sec.~\ref{symqrgparagraph}, we replace it with a projection that is \textit{canonically} determined by the topological defect lines one chooses to preserve.

\subsection{SymQRG: Topological Symmetry-Preserving Quantum RG}\label{symqrgparagraph}
\subsubsection{General framework}
In this paper, we take a step further to explain the meaning of this projection into simple operators from a symmetry principle that extends beyond the large $N$ limit, highlighting its crucial role in the context of the AdS$_3$/CFT$_2$ correspondence. The key insight is that holography can be understood as an RG transformation that \textit{explicitly} preserves certain \textit{topological symmetries} of the 2D theories. This collection of symmetries, generically including non-invertible symmetries, follows from the (non-rational generalization of) representations of the chiral algebra inherent in any 2D CFTs, as first discovered by Moore and Seiberg\cite{Moore:1988uz, Moore:1988qv, *Moore:1989vd}. These topological symmetries manifest themselves as topological line operators in the CFT\cite{Verlinde:1988sn, Petkova:2000ip, Bhardwaj:2017xup, Chang:2018iay, Thorngren:2019iar}. 
The connection between this symmetry and the complete set of CFT gluing rules was elucidated in a series of seminal works collectively referred to as the FRS construction in the literature, named after Fuchs, Runkel and Schweigert, as well as Fröhlich and Fjelstad \cite{Fuchs:2002cm, *Fuchs:2003id, *Fuchs:2003id1, *Fuchs:2004xi, *Fjelstad:2005ua, *Frohlich:2006ch, *Fjelstad:2006aw}. The emergence of a 3D hyperbolic geometry from 2D CFTs is precisely captured by this ``Virasoro modular geometry'' for irrational CFTs\cite{Verlinde:1989ua, Jackson:2014nla, Teschner:2005bz, Friedan:1986ua, Dijkgraaf:1987vp}. More details on the emergence of the quantum group/non-commutative geometry $U_q(SL(2,\mathbb{R}))$ and its connection to 3D gravity are the focus of Sec.~\ref{TQFTSECTION}.

In general, a given set of symmetries in a 2D theory is associated with a 3D (topological) theory that follows from gauging the set of symmetries. For conventional discrete group symmetries with symmetry group $G$, the 3D theory corresponds to a topological gauge theory with gauge group $G$. More generally, for a rational collection of non-invertible  symmetries, the ``gauged'' theory is a 3D TQFT\cite{Bhardwaj:2017xup, Lin:2022dhv}, which is the SymTFT. 
The Hilbert space of the SymTFT defined on a given 2D manifold $\Sigma$ provides a symmetry-preserving basis for the construction of symmetry-preserving 2D theories. In this light, the 3D quantum gravitational path integral itself functions as the ``SymTFT'' for irrational 2D CFTs. This SymTFT can be further used to carry out an RG flow that explicitly preserves topological symmetries\cite{Chen:2022wvy, Cheng:2023kxh, Hung:2024gma, Chen:2024unp}.  We refer to this framework as ``topological symmetry-preserving quantum RG'' or ``SymQRG''.

Let us briefly summarize the idea and procedure behind SymQRG. Two concrete realizations of this framework are presented  in Sec.~\ref{TTbar} for the continuum $T\bar{T}$ deformation, and in Sec.~\ref{tensor network} for the exact discrete real-space tensor network representation of 2D field theories. 
\begin{figure}
	\centering
	\includegraphics[width=0.6\linewidth]{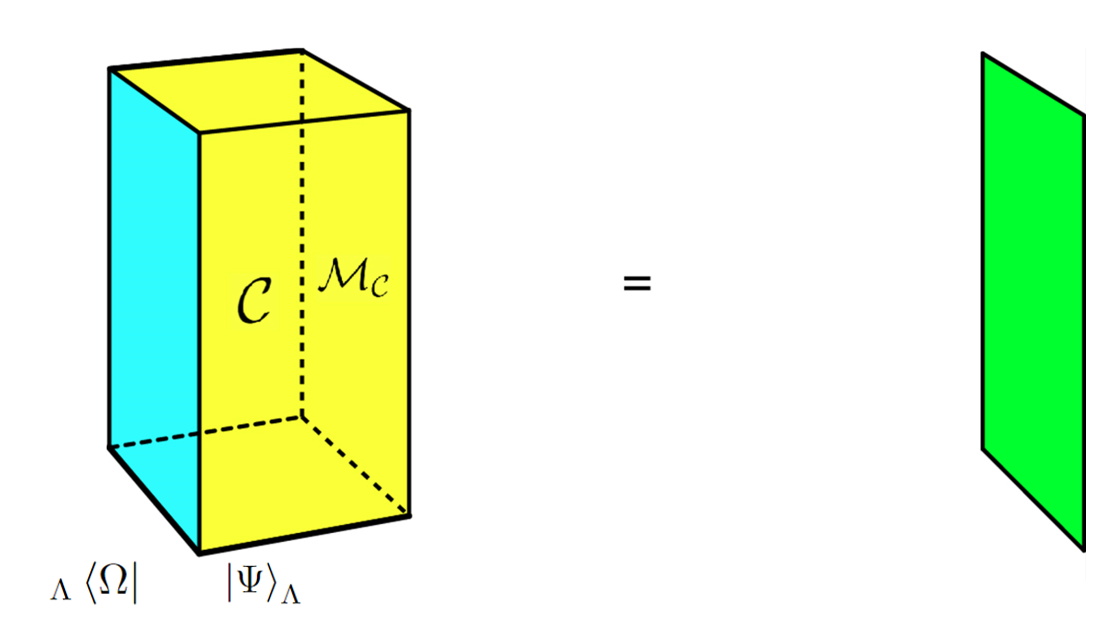}
	\caption{An illustration of the sandwich construction, where any $D$-dimensional field theory is equivalent to collapsing the two boundaries of a $D+1$-dimensional SymTFT, which governs the topological symmetries of the original theory. A topological boundary condition and a physical boundary condition are specified on the two boundaries. The precise meaning of $\mathcal{C}$ and $\mathcal{M_C}$ in the context of 2D CFTs will be explained in \ref{roleofpsi module category}.}
	\label{sandwich}
\end{figure}

The idea is that we can always express the 2D QFT partition functions as an overlap between two 3D quantum states,
\be \label{overlap}
Z={}_\Lambda \bra{\Omega^\lambda(\gamma_{ij})} \Psi \rangle_{\Lambda}~,
\ee
where the subscript $\Lambda$ denotes some reference scale (which we will make precise in explicit constructions), $\gamma_{ij}$ is the metric on which the theory is defined and $\lambda$ labels the dimensionful couplings of the theory. In Sec.~\ref{TTbar}, $\lambda$ refers to the coupling for the $T\bar{T}$ deformation, which has mass dimension $-2$. This is in turn related to the size of the holes we introduce on the lattice in Sec.~\ref{tensor network}. In the language of generalized symmetries\cite{Gaiotto:2014kfa}, this overlap realizes the ``sandwich construction'' \cite{Gaiotto:2020iye, Ji:2019jhk, Freed:2022qnc, Apruzzi:2021nmk, Moradi:2022lqp}, as shown in Fig.~\ref{sandwich}, where ${}_\Lambda\bra{\Omega^\lambda(\gamma_{ij})}$ corresponds to a \textit{physical} boundary condition that captures geometrical notions such as distances and positions in a 2D QFT. For instance, in our $T \bar{T}$ example in Sec.~\ref{wdwstates}, this state can be interpreted as the finite cutoff ``position space'' basis states \cite{PhysRev.160.1113, *PhysRevD.28.2960} in 3D quantum gravity; in Sec.~\ref{tensor network}, we use the Hilbert space of the Levin-Wen string-net model\cite{Levin:2004mi} underlying the tensor network construction to write down a continuous family of states, labeled by the size of the holes. In both of these cases, $\lambda=0$ corresponds to an RG fixed point (CFT). We also want to emphasize that these states describe the \textit{kinematical} aspects of the 2D theory. \textit{Dynamics} are  encoded in the ket state $\ket{\Psi}_{\Lambda}$, which corresponds to the state obtained from a 3D path integral associated with a \textit{topological} boundary condition\footnote{This combination is referred to as a ``quiche'' in \cite{Freed:2022qnc}.}. The state contains the algebraic information of 2D CFT, such as its sewing rules, i.e.~\textit{all} the OPE coefficients \cite{Fuchs:2002cm, *Fuchs:2003id, *Fuchs:2003id1, *Fuchs:2004xi, *Fjelstad:2005ua, *Fjelstad:2006aw}. 
The holographic principle thus equates this 3D wavefunction with 2D partition functions, and we will elaborate more on this in Sec.~\ref{algebra}. 

In terms of the 3D bulk theory that captures the non-invertible symmetries of the 2D CFT, the state $|\Psi \rangle_{\Lambda}$ is a topological ground state satisfying the no-flux constraint
\be \label{noflux}
\hat{\mathcal{H}} |\Psi \rangle_{\Lambda}=0~.
\ee
For the $T\bar{T}$ deformation, this constraint equation manifests itself as the Wheeler-DeWitt equation. In the example of the real-space tensor network, this condition implies that this state is a ground state of the underlying topological order, describing the long-range entanglement in 2D CFTs from the consistency conditions of the sewing/crossing relation. We emphasize that it is this state $\ket{\Psi}_{\Lambda}$ that underlies the emergence of the third radial direction, and the \textit{non-perturbatively exact} nature of the $3D/2D$ correspondence. 

SymQRG is the statement that there exist different physical boundary conditions, corresponding to different states $_{\Lambda}\bra{\Omega}$, whose overlap with a fixed topological ground state $\ket{\Psi}_{\Lambda}$ gives different theories along an RG trajectory. These theories are interconnected through the inclusion of an integral kernel corresponding to additional path integrals in the bulk SymTFT. The presence of a no-flux ground state guarantees that the Callan-Symanzik equation naturally implies the Wheeler-DeWitt equation. We can find the $_{\Lambda}\bra{\Omega}$ states using the following procedure inspired by our tensor network construction on the lattice\cite{Chen:2022wvy, Cheng:2023kxh, Chen:2024unp, Hung:2024gma}.

We first pass to a different representation of the original overlap:
\be \label{rgoperator}
{}_\Lambda \bra{\Omega^\lambda(\gamma_{ij})} \Psi \rangle_{\Lambda}={}_\Lambda \bra{\Omega^\lambda(\gamma_{ij})} \hat{U}^{\Lambda,\Lambda'}_{\text{RG}} |\Psi \rangle_{\Lambda'}~,
\ee
where $|\Psi \rangle_{\Lambda'}$ is a realization of the topological ground state on a different probe scale. Since this is a topological ground state, the RG operator $\hat{U}^{\Lambda,\Lambda'}_{\text{RG}}$ that changes the representation at different scales for $\ket{\Psi}$ in this context is precisely given by the 3D bulk TQFT path integral\footnote{In any explicit realization, the ground state wave function depends on extra structures on the boundary surface not intrinsic to the theory, e.g.~a boundary metric, or a triangulation in a discrete model as discussed in Sec.~\ref{tensor network}; this is what ``representation at a scale'' refers to.}, which in this case is the 3D quantum gravity path integral.

We now define the new bra state as:
\be \label{change_scale}
{}_{\Lambda'} \bra{\Omega'^\lambda(\gamma_{ij})} \equiv {}_\Lambda \bra{\Omega^\lambda(\gamma_{ij})} \hat{U}^{\Lambda,\Lambda'}_{\text{RG}}~.
\ee
This process precisely matches the notion of coarse-graining/spin-blocking, since we get two different representations of the same partition function at two different scales,
\be
{}_\Lambda \bra{\Omega^\lambda(\gamma_{ij})} \Psi \rangle_{\Lambda}={}_{\Lambda'} \bra{\Omega'^\lambda(\gamma_{ij})} \Psi \rangle_{\Lambda'}~.
\ee

Let us pause to state the structure of this operation precisely, since it is the defining feature of SymQRG. The operator $\hat{U}^{\Lambda,\Lambda'}_{\text{RG}}$ is the SymTFT path integral on a slab $\Sigma \times I$, whose two boundary components carry the representation data of the scales $\Lambda$ and $\Lambda'$. When $\Lambda = \Lambda'$ (at coincident scales), the slab is an \textit{idempotent projector}
\be \label{slabprojector}
\Pi^{\mathcal{S}}_{\Lambda}= Z_{\text{SymTFT}_{\mathcal{S}}}(\Sigma_{\Lambda} \times I)~, \qquad \big(\Pi^{\mathcal{S}}_{\Lambda}\big)^2=\Pi^{\mathcal{S}}_{\Lambda}~,
\ee
onto the subspace satisfying the no-flux condition \eqref{noflux}.\footnote{Here $\mathcal{S}$ denotes the preserved collection of topological defect lines. Since a single collection is in play throughout, we suppress this label in what follows.} On the lattice, this is the ground-state projector of the Levin-Wen model, built from the $\omega$-loops reviewed in Appendix~\ref{reviewtvtqft}. The scale-changing slab $\hat{U}^{\Lambda,\Lambda'}_{\text{RG}}$ is the corresponding topological cobordism kernel: it composes by gluing,
\be \label{slabcomposition}
\hat{U}^{\Lambda,\Lambda'}_{\text{RG}}\, \hat{U}^{\Lambda',\Lambda''}_{\text{RG}}=\hat{U}^{\Lambda,\Lambda''}_{\text{RG}}~,
\ee
reduces to $\Pi_{\Lambda}$ at coincident scales, and acts as a partial isometry between the no-flux sectors at the two scales. A single SymQRG step \eqref{change_scale} is therefore an ordinary block-spin contraction of the physical boundary state, \textit{completed by the projection back onto the symmetry-organized sector}. The projection involves no truncation: its insertion into the overlap \eqref{overlap} is exact precisely because the dynamical state satisfies $\Pi_{\Lambda} \ket{\Psi}_{\Lambda}=\ket{\Psi}_{\Lambda}$, which is nothing but the no-flux condition \eqref{noflux}. This is also the precise sense in which SymQRG upgrades the quantum RG of Sec.~\ref{quantumrgsection}: there, the projection onto single-trace operators is a choice justified at large $N$; here, the projector is \textit{canonical}: it is the ground-state projector of the SymTFT, determined by the preserved topological lines; it is exact at finite central charge, and it possesses a path-integral representation of its own. When the preserved lines are the Virasoro lines, that path integral is the 3D quantum gravity path integral, and $\Pi_{\Lambda}\ket{\Psi}_{\Lambda}=\ket{\Psi}_{\Lambda}$ is the Wheeler-DeWitt equation.

After rescaling, we get a deformed state,
\be \label{relabel0}
{}_{\Lambda'} \bra{\Omega'^\lambda(\gamma_{ij})} \to {}_{\Lambda} \bra{\Omega'^\lambda(\gamma_{ij})}~.
\ee
From the Callan-Symanzik equation, we know that the effect of this deformation is the same as directly deforming the coupling, so we can equate
\be \label{newstate}
{}_{\Lambda} \bra{\Omega^{\lambda(\frac{\Lambda}{\Lambda'})^2} (\gamma_{ij})}={}_{\Lambda} \bra{\Omega'^\lambda(\gamma_{ij})}~.
\ee

Thus, the deformed partition function with a different coupling can be written in terms of the original partition function as
\be
\begin{aligned} \label{kernelequation1}
\mathcal{Z}^{\lambda(\frac{\Lambda}{\Lambda'})^2}({\gamma_{ij}})&={}_{\Lambda} \bra{\Omega^{\lambda(\frac{\Lambda}{\Lambda'})^2} (\gamma_{ij})}\Psi \rangle_{\Lambda}\\
&=\int \mathcal{D}h_{ij} {}_{\Lambda} \langle{\Omega^{\lambda(\frac{\Lambda}{\Lambda'})^2} (\gamma_{ij})} |\Omega^{\lambda} (h_{ij})\rangle_{\Lambda}\,_{\Lambda} \langle{\Omega^{\lambda} (h_{ij})}|\Psi \rangle_{\Lambda}\\
&=\int \mathcal{D}h_{ij} {}_{\Lambda} \langle{\Omega^{\lambda} (\gamma_{ij})}| e^{-\int_{\lambda}^{\lambda (\frac{\Lambda}{\Lambda'})^2} \hat{H}_{\text{grav}}} |\Omega^{\lambda} (h_{ij})\rangle_{\Lambda}\,_{\Lambda} \langle{\Omega^{\lambda} (h_{ij})}|\Psi\rangle_{\Lambda}\\
&=\int \mathcal{D}h_{ij} K_{\text{grav}}(\gamma_{ij},h_{ij},\lambda (\frac{\Lambda}{\Lambda'})^2,\lambda) \mathcal{Z}^{\lambda}(h_{ij})~,
\end{aligned}
\ee
where in the second line, we insert a complete basis labeled by the metric, and the kernel $K_{\text{grav}}(\gamma_{ij},h_{ij},\lambda (\frac{\Lambda}{\Lambda'})^2,\lambda)$ relating different deformed partition functions is a gravitational path integral. 
The coupling $\lambda$ plays the role of a ``radial time'' (see \eqref{lambdaradial}), so that \eqref{kernelequation1} is the exact analogue of the Heisenberg-picture path integral $\Psi^t(x)=\int dx'\, K(x,x',t,t_0)\,\Psi^{t_0}(x')$ in quantum mechanics\footnote{Recall that basis states have time dependence in the Heisenberg picture, because they are eigenstates of time-dependent operators.}. The solution to the $T\bar{T}$ deformation takes precisely this form; the deformed partition function is the Wheeler-DeWitt wave functional in the Heisenberg picture. This comparison makes manifest the existence of an underlying quantum state $\ket{\Psi}_{\Lambda}$, which we will show to be a long-range entangled ground state of the topological order in the bulk, associated to the Moore-Seiberg data. The states ${}_\Lambda \bra{\Omega^\lambda(\gamma_{ij})}$ correspond to finite cutoff Dirichlet boundary conditions in AdS/CFT\footnote{Alternatively, these states can be understood as the implementation of mixed boundary conditions at asymptotic infinity.}.

We also note that in quantum mechanics the ``bulk'' connecting time slices is generated by {\it one} theory, whereas in holography each radial slice carries a \textit{different} 2D theory\footnote{We should emphasize that the Hamiltonians of the lower dimensional theories resulting from different boundary conditions on a radial slice of the bulk should not be confused with the radial Hamiltonian of the bulk quantum gravity.}: the Hilbert space of the 3D gravitational bulk on a 2D slice is the \textit{space of 2D theories with symmetries} \cite{Lee:2016uqc, Ma:2020pmt}, with different 2D theories labeled by different boundary conditions ${}_{\Lambda} \bra{\Omega^\lambda(h_{ij})}$ and $\ket{\Psi}_{\Lambda}$.
Note that this is not the complete theory space for 2D QFTs. Rather, the 3D gravitational bulk restricts the space of theories to those preserving the same topological symmetries (with at most spontaneous symmetry breaking). In a non-perturbative setting, the symmetry provides an exact basis labeling different quantum theories. 

This entire procedure is deeply inspired by the formalism introduced by Lee and formalized in \cite{Lee:2016uqc}, with the crucial addition discussed above: since 3D bulk gravity is essentially a topological order with long-range entanglement, quantum RG should incorporate generalized symmetries, realized through the long-range entangled topological ground state $\ket{\Psi}_{\Lambda}$ in the sandwich construction.
Finally, the topological boundary in $\ket{\Psi}_{\Lambda}$ implies that the bulk path integral spans a sandwich that can be compressed into an intrinsically 2D theory, encapsulating the fact that the constructed 2D theory is indeed anomaly free. 
Thus, $\ket{\Psi}_{\Lambda}$ carries the necessary ingredients of a 3D/2D duality (i.e.~{\it equivalence}).

A simple torus example illustrating the overlap decomposition for rational CFTs is given in Appendix~\ref{torusexample}.

\section{$T\bar{T}$ deformation for 2D CFTs as SymQRG} \label{TTbar}

The purpose of this section is not to rederive the theory of $T\bar{T}$ deformations \cite{Zamolodchikov:2004ce, Smirnov:2016lqw} from scratch.
The flow equation, its finite-cutoff holographic interpretation \cite{McGough:2016lol}, and its relation to the Wheeler-DeWitt equation \cite{Freidel:2008sh} have been developed in a series of works \cite{Lee:2013dln, Mazenc:2019cfg, Tolley:2019nmm, Belin:2020oib, Kim:2021qbi, Araujo-Regado:2022gvw}.
We use this well-understood example as a controlled continuum test of the SymQRG framework introduced in Sec.~\ref{symqrgparagraph}.
The new point is the reorganization of these known facts into the sandwich language: the iterative $T\bar{T}$ deformation builds the 3D gravitational path integral as the SymTFT slab, while the Heisenberg-picture state $\ket{Z_{\text{CFT}}}$ satisfies the Wheeler-DeWitt equation, which is precisely the no-flux condition \eqref{noflux} of the SymTFT\footnote{In \cite{Mazenc:2019cfg}, the $T\bar{T}$-deformed partition functions were proposed to be viewed as wavefunctions in 3D gravity in the Schr\"odinger picture; it is the Heisenberg picture that fits more naturally into our framework.}.
In Sec.~\ref{qrg for ttbar}, we derive the integral transform relating $\mathcal{Z}^\mu(\gamma_{ij})$ at different couplings and identify its kernel with the 3D gravitational path integral.
In Sec.~\ref{ttbar from 3d}, we show the converse: the 3D bulk gravitational path integral naturally gives rise to the $T\bar{T}$ deformation.
In Sec.~\ref{wdwstates}, we rewrite these results in the operator formalism, making the connection between the Wheeler-DeWitt equation and the no-flux condition manifest.

\subsection{3D phase space gravitational path integral from iterative $T \bar{T}$ deformation} \label{qrg for ttbar}

In this subsection, we first focus on the simpler case where the deformation parameter $\lambda$ is constant. The generalization to spatially varying deformations $\lambda(x)$ is discussed in \cite{Lee:2013dln, Belin:2020oib}.

We start with the CFT partition function defined on a 2D Euclidean metric $\gamma_{ij}$, denoted as $\mathcal{Z}_{\text{CFT}}[\gamma_{ij}]$. The $T \bar{T}$ deformation on 2D curved spacetimes generates a family of deformed theories parameterized by $\lambda$, which satisfy the flow equation \cite{McGough:2016lol, Belin:2020oib, Mazenc:2019cfg, Tolley:2019nmm}\footnote{We use a different convention from \cite{McGough:2016lol, Belin:2020oib}, where $\gamma_{ij}^{\text{ours}} = \frac{1}{8} \gamma_{ij}^{\text{theirs}}$.},
\be \label{flowequation}
\begin{aligned}
\frac{d}{d \lambda} \mathcal{Z}^{\lambda}[\gamma_{ij}]=-\int d^2 x\left[ \frac{c}{48 \pi \lambda} \sqrt{\gamma} R(\gamma_{ij}) +\frac{12 \pi}{c \sqrt{\gamma}}\left(\frac{1}{2}(\gamma_{ik}\gamma_{jl}+\gamma_{il}\gamma_{jk})-\gamma_{ij}\gamma_{kl} \right) \frac{\delta}{\delta \gamma_{ij}} \frac{\delta}{\delta \gamma_{kl}}\right]\mathcal{Z}^{\lambda}[\gamma_{ij}]~.
\end{aligned}
\ee
Introducing $\mu$ as a reference scale with mass dimension $-2$, the interpretation of which in the context of AdS$_3$ and CFT$_2$ will be clarified below, we define a dimensionless parameter $z$ as: 
\be \label{logtrans} 
\lambda(z) = \mu e^{z}~. 
\ee 
The flow equation can then be solved for small $\delta z$ as follows:
\begin{equation} \label{deformation}
\mathcal{Z}^{\lambda(\delta z)}[\gamma_{ij}] =e^{ \delta S(\mu, \delta z, -i \frac{\delta}{\delta \gamma_{ij}},\gamma_{ij})}\mathcal{Z}^{\lambda(0)}[\gamma_{ij}]~,
\end{equation}
where the deformation action functional is:
\be
\delta S(\mu, \delta z, -i \frac{\delta}{\delta \gamma_{ij}},\gamma_{ij})=- \mu \delta z \int d^2x\left[ \frac{c}{48 \pi \mu} \sqrt{\gamma} R(\gamma_{ij}) +\frac{12 \pi}{c \sqrt{\gamma}}\left(\frac{1}{2}(\gamma_{ik}\gamma_{jl}+\gamma_{il}\gamma_{jk})-\gamma_{ij}\gamma_{kl} \right) \frac{\delta}{\delta \gamma_{ij}} \frac{\delta}{\delta \gamma_{kl}}\right]~.
\ee
Expanding \eqref{deformation} to leading order in $\delta z$ reproduces the flow equation \eqref{flowequation}. The first term in $\delta S$ arises from the change in Casimir energy (related to the Wess-Zumino consistency condition and diffeomorphism invariance in the emergent bulk\cite{Lee:2012xba, Shyam:2016zuk, Belin:2020oib}), while the second is the double-trace $T\bar{T}$ operator.

Following \cite{Lee:2013dln}, we reformulate the double-trace deformation in terms of single-trace operators by introducing an auxiliary pair of 2D fields. Writing $\delta S'(\lambda, \delta \lambda, \cdot)$ for the deformation action in terms of $\lambda$ directly (i.e.\ replacing $\mu \delta z$ with $\delta\lambda$ in $\delta S$), we have:
\begin{equation}\label{integral deform}
\mathcal{Z}^{\lambda+\delta \lambda}[\gamma_{ij}] = \int \mathcal{D}h^{(1)}_{ij} \mathcal{D}\pi^{(1),ij}\;  e^{\int d^2x\, i \pi^{(1),ij}(\gamma_{ij}-h^{(1)}_{ij})+\delta S'(\lambda, \delta \lambda, \pi^{(1),ij},\gamma_{ij})}\mathcal{Z}^{\lambda}[h^{(1)}_{ij}]~.
\end{equation}
The auxiliary fields serve as Lagrange multipliers encoding that the metric sources the stress tensor; integrating out $\pi^{(1),ij}$ recovers the Hubbard-Stratonovich solution of \eqref{flowequation} (Appendix~\ref{ttbardetails}).

Introducing $\lambda=\mu e^{-z}$ with $z=m\delta z$ counting the iterative RG steps, and iterating \eqref{integral deform} in the limit $\delta z \to 0$, the 2D auxiliary fields are promoted to 3D fields $h_{ij}(x,\lambda)$ and $\pi^{ij}(x,\lambda)$, transforming the boundary metric into a dynamical 3D field \cite{Lee:2013dln}. The Lagrange-multiplier terms assemble into a canonical kinetic term $-i\pi^{ij}\partial_\lambda h_{ij}$ (Appendix~\ref{ttbardetails}). With boundary conditions $h_{ij}(\mu)=\gamma_{ij}$ and $h_{ij}(\epsilon)=h^{\epsilon}_{ij}$, we obtain:
\begin{equation} \label{deformed5}
 \begin{aligned}
    \mathcal{Z}^{\mu}[\gamma_{ij}] =& \int \mathcal{D} h^\epsilon_{ij}  \int_{\substack{h_{ij}(\mu)=\gamma_{ij} \\ h_{ij}(\epsilon)=h^\epsilon_{ij}}} \mathcal{D} h_{ij} \mathcal{D}\pi^{ij} \exp \left[ \int_{\mu}^\epsilon d\lambda \int d^2x \left(-i \pi^{ij}\partial_\lambda h_{ij} - \frac{1}{2\lambda} \left(\frac{24\pi \lambda}{c \sqrt{h}}(\pi^{ij}\pi_{ij}-\pi^2) \right. \right. \right. \\
    & \left. \left. \left. - \frac{c \sqrt{h}}{24 \pi } R(h_{ij})\right) \right) \right] \mathcal{Z}^{\epsilon}[h^{\epsilon}_{ij}]~.
 \end{aligned}
\end{equation}

This is a 3D phase-space path integral closely resembling 3D gravity; we show next that, after addressing counterterm subtleties, the kernel is indeed the 3D gravitational path integral.

\subsection{$T\bar{T}$ deformation from 3D gravitational path integral}\label{ttbar from 3d}

It was further proposed in \cite{McGough:2016lol}, building upon earlier results in \cite{Freidel:2008sh}, that a Wheeler-DeWitt wavefunction $\Psi(g)$ in 3D quantum gravity can be obtained from the deformed partition functions $Z^{(\mu)}(\gamma)$ via\footnote{We choose the same sign as in \cite{McGough:2016lol, Hartman:2018tkw} for the $T\bar{T}$ deformation and counter term, which corresponds to the opposite sign leading to the Nambu-Goto string action\cite{Cavaglia:2016oda}. This choice of sign gives the correct conformal anomaly equation in the Euclidean signature as in \eqref{anomaly}.}:
\be
\label{wdw} \Psi(\gamma/\mu) = e^{\frac{1}{8\pi G_N \mu} \int d^2x \sqrt{\gamma}} \mathcal{Z}^{(\mu)}[\gamma_{ij}]~, 
\ee
where the induced metric on a constant $\mu$ cutoff surface is related to the field theory metric by $g_{ij}=\frac{\gamma_{ij}}{\mu}$, mirroring the standard relation in holography as $\mu \to 0$. The additional term in the exponent represents a 2D cosmological constant and is associated with the holographic counterterm. In Sec.~\ref{wdwstates}, we will show that this manipulation turns the flow equation \eqref{flowequation} into the Wheeler-DeWitt equation\cite{McGough:2016lol}.

We now demonstrate that equation \eqref{wdw} is precisely equivalent to deforming the 2D CFT partition function using the 3D gravitational path integral, expressed as follows\footnote{Subtleties related to the flat measure in the phase-space path integral of gravity are analyzed in detail in \cite{Han:2009bb, Belin:2020oib}.}:
\begin{equation}
\label{bulk}
 \begin{aligned}
    \Psi(\gamma/\mu) =& \int \mathcal{D} h^\epsilon_{ij} \int_{\substack{g_{ij}(\mu)=\frac{\gamma_{ij}}{\mu} \\ g_{ij}(\epsilon)=\frac{h^\epsilon_{ij}}{\epsilon}}} \mathcal{D} g_{ij} \mathcal{D}\Pi^{ij} \exp \left[ \int_{\mu}^\epsilon d\lambda \int d^2x \left(-i \Pi^{ij}\partial_\lambda g_{ij} -\frac{1}{2\lambda} \left( \frac{16\pi G_N}{\sqrt{g}}(\Pi^{ij}\Pi_{ij}-\Pi^2) \right. \right. \right. \\
    & \left. \left. \left. - \frac{\sqrt{g}}{16\pi G_N} (R(g_{ij})+2) \right) \right) \right] e^{\frac{1}{8\pi G_N \epsilon}\int d^2x \sqrt{h^\epsilon}}\mathcal{Z}^{\epsilon}[h^{\epsilon}_{ij}]~.
 \end{aligned}
\end{equation}
Here, the kernel is the 3D bulk gravitational path integral in radial gauge, $ds^2=\frac{d\lambda^2}{4\lambda^2}+\frac{h_{ij}}{\lambda}dx^i dx^j$ (lapse $N=\frac{1}{2\lambda}$, shift $N^i=0$; generalizations to non-uniform $\lambda(x)$ and nonzero shift are discussed in \cite{Lee:2013dln, Belin:2020oib}). Under the identification
\be \label{lambdaradial}
\lambda=\frac{2 \pi G_N}{r^2}~,
\ee
this is the conventional Fefferman-Graham form\cite{McGough:2016lol}, with $r$ the AdS radial direction and $e^{\frac{1}{8\pi G_N \epsilon}\int d^2x \sqrt{h^\epsilon}}\mathcal{Z}^{\epsilon}[h^{\epsilon}_{ij}]$ the bare partition function after holographic renormalization\cite{Balasubramanian:1999re, Belin:2020oib, Witten:2022xxp}.

To relate \eqref{bulk} to the 2D RG framework, we rescale
\be \label{rescaling}
\Pi^{ij}=\lambda \pi^{ij}~, \quad g_{ij}=\frac{h_{ij}}{\lambda}~,
\ee
and shift $\pi^{ij} \to \pi^{ij}-\frac{i}{16\pi G_N \lambda} \sqrt{h}h^{ij}$. The change of variables and momentum shift reorganize the gravitational phase-space path integral into the iterative $T\bar{T}$ kernel; the boundary terms cancel the cosmological counterterm at $\epsilon$ and generate the boundary term at $\mu$ in \eqref{wdw}. The intermediate steps are collected in Appendix~\ref{ttbardetails}; the result is
\begin{equation} \label{finalanswer}
 \begin{aligned}
   \Psi(\gamma/\mu)=& e^{\frac{1}{8\pi G_N \mu}\int d^2x \sqrt{\gamma}}\int \mathcal{D} h^\epsilon_{ij}  \int_{\substack{h_{ij}(\mu)=\gamma_{ij} \\ h_{ij}(\epsilon)=h^\epsilon_{ij}}} \mathcal{D} h_{ij} \mathcal{D}\pi^{ij} \exp \left[ \int_{\mu}^\epsilon d\lambda \int d^2x \left(- i \pi^{ij}\partial_\lambda h_{ij} \right. \right.  \\
   &\left. \left. - \frac{1}{2\lambda} \left(\frac{16\pi G_N\lambda}{\sqrt{h}}(\pi^{ij}\pi_{ij}-\pi^2) - \frac{\sqrt{h}}{16\pi G_N \lambda} R(\frac{h_{ij}}{\lambda})\right) \right) \right] \mathcal{Z}^{\epsilon}[h^{\epsilon}_{ij}]~.
 \end{aligned}
\end{equation}
With $R(h_{ij}/\lambda)=\lambda R(h_{ij})$ and the Brown-Henneaux relation\cite{Brown:1986nw} ($\ell_{\text{AdS}}=1$),
\be \label{brown}
c=\frac{3}{2 G_N}~,
\ee
\eqref{finalanswer} matches \eqref{wdw}, establishing the equivalence between the $T\bar{T}$ deformation and the 3D gravitational path integral\cite{Hartman:2018tkw, Araujo-Regado:2022gvw}.

\subsection{Wheeler-DeWitt equation from $T\bar{T}$ deformation}\label{wdwstates}
We now recast the gravitational path integral \eqref{bulk} in operator formalism, making the no-flux condition and the underlying quantum state explicit.

We introduce metric eigenstates $\bra{g_{ij} = \frac{\gamma_{ij}}{\lambda}, \lambda}$ in the Heisenberg picture \cite{PhysRev.160.1113, *PhysRevD.28.2960}, playing the role of ${}_{\Lambda}\bra{\Omega^\lambda(\gamma_{ij})}$, and define the state $\ket{Z_{\text{CFT}}}$ via
\be \label{overlap1}
\bra{g_{ij},\lambda} Z_{\text{CFT}}\rangle=e^{\frac{1}{8\pi G_N }\int d^2x \sqrt{g}}{\mathcal{Z}}^{\lambda}[\lambda g_{ij}]~.
\ee
Inserting a complete set of metric eigenstates and integrating out the conjugate momenta $\Pi^{ij}$ in \eqref{bulk} (with integration limits reversed to describe evolution into the bulk), the Wheeler-DeWitt wavefunction becomes
\be
 \begin{split}
     \label{eq:RGevolve}
   \Psi(\gamma/\mu) & \equiv \bra{g_{ij}=\frac{\gamma_{ij}}{\mu},\mu} Z_{\text{CFT}}\rangle\\
   &=\int \mathcal{D} h^\epsilon_{ij} \bra{g_{ij}=\frac{\gamma_{ij}}{\mu},\mu} g_{ij}=\frac{h_{ij}^\epsilon}{\epsilon},\epsilon \rangle \bra{g_{ij}=\frac{h_{ij}^\epsilon}{\epsilon},\epsilon} Z_{\text{CFT}}\rangle~.
 \end{split}
\ee
Introducing the Hamiltonian density
\be
\hat{\mathcal{H}}=-\frac{16\pi G_N}{\sqrt{\hat{g}}}(\hat{\Pi}^{ij}\hat{\Pi}_{ij}-\hat{\Pi}^2)+ \frac{\sqrt{\hat{g}}}{16\pi G_N} (R(\hat{g}_{ij})+2)~,
\ee
then \eqref{eq:RGevolve} can be written as
\begin{equation} \label{rgoperatorinttbar}
    \begin{split}
        \Psi(\gamma/\mu)&=\int \mathcal{D} h^\epsilon_{ij} \bra{g_{ij}=\frac{\gamma_{ij}}{\mu},\epsilon} e^{ -\int_\epsilon^\mu \frac{d\lambda}{2\lambda} \int \text{d}^2x\hat{\mathcal{H}} }\ket{g_{ij}=\frac{h_{ij}^\epsilon}{\epsilon},\epsilon} \bra{g_{ij}=\frac{h_{ij}^\epsilon}{\epsilon},\epsilon} Z_{\text{CFT}}\rangle \\
        &= \bra{g_{ij} = \frac{\gamma_{ij}}{\mu},\epsilon} \hat{U}_{\text{3D gravity}} |Z_{\text{CFT}}\rangle~,
    \end{split}
\end{equation}
where $\hat{U}_{\text{3D gravity}}$ is the radial evolution operator, and $\ket{Z_{\text{CFT}}}$ is exactly the $\ket{\Psi}_{\Lambda}$ state underlying 2D CFTs.
In the metric basis, $\hat{\mathcal{H}}$ becomes a differential operator $\mathcal{H}(\lambda,h_{ij},\frac{\delta}{\delta h_{ij}})$ (Appendix~\ref{ttbardetails}).

Thus, we have the Wheeler-DeWitt equation,
\be \label{operatordiff}
\begin{aligned}
\mathcal{H}(\mu,\gamma_{ij},\frac{\delta}{\delta \gamma_{ij}}) \Psi(\gamma/\mu)&=\mathcal{H}(\mu,\gamma_{ij},\frac{\delta}{\delta \gamma_{ij}}) \bra{g_{ij}=\frac{\gamma_{ij}}{\mu},\epsilon} \hat{U}_{\text{3D gravity}}|Z_{\text{CFT}}\rangle\\
&= \bra{g_{ij}=\frac{\gamma_{ij}}{\mu},\epsilon} \hat{\mathcal{H}} \hat{U}_{\text{3D gravity}}|Z_{\text{CFT}}\rangle\\
&= \bra{g_{ij}=\frac{\gamma_{ij}}{\mu},\epsilon}  \hat{U}_{\text{3D gravity}} \hat{\mathcal{H}}|Z_{\text{CFT}}\rangle~.
\end{aligned}
\ee
In the second equation, we apply the Baker-Campbell-Hausdorff formula and use the fact that the terms arising from the commutator of the Hamiltonian constraints vanish due to the momentum constraint (2D diffeomorphism) \cite{Araujo-Regado:2022gvw, Lee:2013dln}.

Now we show that this is equal to zero, which comes from:
\be \label{wdwcft constraint}
\hat{\mathcal{H}}|Z_{\text{CFT}}\rangle=0~.
\ee
Evaluating $\bra{g_{ij},\epsilon}\hat{\mathcal{H}}\ket{Z_{\text{CFT}}}$ explicitly in the metric basis (Appendix~\ref{ttbardetails}), the bulk cosmological constant is cancelled once the Hamiltonian passes through the boundary cosmological counterterm, and a linear term absent in the Hamiltonian is generated; the residual double-derivative term vanishes as $\epsilon \to 0$, reflecting the irrelevance of the $T\bar{T}$ deformation under RG flow. Together, these lead to the 2D conformal anomaly equation:
\be \label{anomaly}
\left(-2 h^\epsilon_{ij} \frac{\delta}{\delta h^\epsilon_{ij}}+ \frac{\sqrt{h^\epsilon}}{16\pi G_N \epsilon} R(\frac{h^\epsilon_{ij}}{\epsilon}) \right)\mathcal{Z}^{\epsilon}[h^{\epsilon}_{ij}]=\left(-2 h^\epsilon_{ij} \frac{\delta}{\delta h^\epsilon_{ij}}+ \frac{\sqrt{h^\epsilon}}{16\pi G_N } R(h^\epsilon_{ij}) \right)\mathcal{Z}^{\epsilon}[h^{\epsilon}_{ij}]=0~.
\ee
The WdW/anomaly connection was highlighted by Freidel \cite{Freidel:2008sh} building on Verlinde \cite{Verlinde:1989ua}; in the $T\bar{T}$ context \cite{McGough:2016lol}, cf.~\eqref{wdw}, the finite cutoff sits at $r_c=\sqrt{2\pi G_N/\mu}$.
The basis states $\bra{g_{ij}=\gamma_{ij}/\mu,\mu}$ are Dirichlet boundary conditions on finite-cutoff surfaces \cite{Witten:1988hc, Guica:2019nzm, Guica:2021pzy, *Guica:2022gts}, preserving the Virasoro algebra. The WdW equation is basis-independent.

Using techniques similar to \eqref{operatordiff}, one shows \cite{Araujo-Regado:2022gvw} that $\mathcal{Z}^\mu(\gamma_{ij})$ in \eqref{deformed5} satisfies the Callan-Symanzik equation,
\be \label{callen}
\left(\int d^2x \pi^i_i-i \mu \frac{\partial}{\partial \mu} \right)\mathcal{Z}^\mu(\gamma_{ij})=0~.
\ee
This implies that the $T \bar{T}$ deformed theories lie on a single RG trajectory.

In fact, it also means that the overlap in \eqref{overlap1} and the Wheeler-DeWitt wavefunction are ``time-independent'', as
\be
\bra{g_{ij},\lambda} Z_{\text{CFT}}\rangle=e^{\frac{1}{8\pi G_N }\int d^2x \sqrt{g}}{\mathcal{Z}}^{\lambda}[\lambda g_{ij}]=e^{\frac{1}{8\pi G_N }\int d^2x \sqrt{g}}{\mathcal{Z}}^{\lambda'}[\lambda' g_{ij}]=\bra{g_{ij},\lambda'} Z_{\text{CFT}}\rangle
\ee
or
\be \label{Uactstrivially}
\hat{U}_{\text{3D gravity}} |Z_{\text{CFT}}\rangle=|Z_{\text{CFT}}\rangle~,
\ee
being consistent with \eqref{wdwcft constraint}.

This example perfectly illustrates the SymQRG paradigm introduced in Sec.~\ref{symqrgparagraph}, rooted in the fact that the original CFT defines a 3D quantum state, satisfying \eqref{wdwcft constraint}. This no-flux condition indicates that this state corresponds to a ground state of a 3D TQFT. At first glance, the connection to topological symmetries may seem obscure. The quick answer lies in the following observation: the $T\bar{T}$ deformation is a deformation that preserves all the topological symmetries, as the topological defect lines $\hat{X_i}$ commute with the stress tensor at any point on the 2D surface by definition,
\be \label{commutewithstress}
[\hat{T}(z),\hat{X_i}]=[\hat{\bar{T}}(\bar{z}),\hat{X_i}]=0~.
\ee
In the next section, we introduce a discrete framework for SymQRG, utilizing the exact tensor network state sum representation of 2D CFTs, which will make the connection to topological symmetries more explicit. We will return to the question of when this bulk SymTFT describes 3D gravity in Sec.~\ref{TQFTSECTION}.

\section{SymQRG for exact tensor network state sum representation of 2D CFTs}\label{tensor network}

In this section, we briefly recall the exact tensor network state sum representation of 2D QFTs and show how it furnishes a discrete, exactly solvable realization of the SymQRG framework. We demonstrate how the $\ket{\Psi}_{\Lambda}$ state is constructed from the BCFT structure coefficients, while the $_{\Lambda}\bra{\Omega}$ state arises from conformal blocks, discretizing the CFT. The interplay between these states, through the crossing symmetry of the CFT, enables SymQRG and facilitates the construction of an exact MERA-like tensor network \cite{Vidal:2008zz} with an emergent RG direction in the bulk. 

In Sec.~\ref{briefreviewtqft}, we begin with a brief review of the discrete Turaev-Viro TQFT\cite{Turaev:1992hq}, which is the SymTFT of the 2D CFTs, and then explain its connection to the Levin-Wen string-net model\cite{Levin:2004mi}. Following this, in Sec.~\ref{SymQRGTN}, we revisit the tensor network state sum representation of the 2D QFTs, and show how it is related to SymQRG. Finally, in Sec.~\ref{Tensornetwork}, we summarize the pivotal role tensor networks play in describing both 2D field theories and their corresponding 3D bulk theories. 

Additional details on the Turaev-Viro TQFT can be found in appendix \ref{reviewtvtqft}. For more information on the tensor network construction of 2D CFTs and the explicit procedure for SymQRG, see appendix \ref{statesum cft} and appendix \ref{tensornetwork SymQRG}. Readers interested in the exact tensor network construction of CFT wavefunctions from the Euclidean path integral can refer to appendix \ref{wavefunction tensornetwork}.

\subsection{Turaev-Viro TQFT and Levin-Wen string-net model}\label{briefreviewtqft}

The SymTFT for a 2D CFT whose chiral modules form a modular tensor category $\mathcal{C}$ \cite{Moore:1988qv} is the Turaev-Viro TQFT \cite{Turaev:1992hq} constructed from $\mathcal{C}$, equivalent in the Hamiltonian picture to the Levin-Wen string-net model \cite{Levin:2004mi, Kirillov:2011mk}. We will use three standard facts about this theory (details are reviewed in Appendix~\ref{reviewtvtqft}):
\begin{enumerate}
\item[(i)] Its Hilbert space on a 2D surface $\Sigma$ is the string-net Hilbert space: a tensor product $\bigotimes_e \mathcal{H}_e$ over edges of a trivalent graph dual to a triangulation of $\Sigma$, with each $\mathcal{H}_e$ spanned by simple objects in $\mathcal{C}$.
\item[(ii)] The Hamiltonian consists of vertex projectors (Gauss's law) and plaquette projectors (no-flux condition \eqref{Omegaflux}). Its ground states are the physical states of the Turaev-Viro TQFT, obtained by the 3D path integral on a manifold $\mathcal{M}$ with $\partial \mathcal{M} \supset \Sigma$.
\item[(iii)] By the Pentagon relation \eqref{pentagon1}, ground-state wavefunctions can be freely re-triangulated: they are fixed-point wavefunctions under RG with zero correlation length \cite{Levin:2004mi, PhysRevB.79.195123, *KOENIG20102707}, and exhibit long-range entanglement enforced by the Gauss's law and no-flux constraints. One main result of this paper is that this long-range entanglement is also the crux of the Wheeler-DeWitt equation in 3D gravity.
\end{enumerate}

\subsection{SymQRG for tensor network state sum representation of 2D CFT path integrals}\label{SymQRGTN}

\noindent\textit{Status and purpose.} 
The tensor network state sum
representation of 2D CFT path integrals was developed in
\cite{Chen:2022wvy, Cheng:2023kxh, Hung:2024gma, Chen:2024unp} (see also
\cite{Brehm:2021wev,*Brehm:2024zun}), building on \cite{Vanhove:2018wlb,
Aasen:2016dop, Aasen:2020jwb} and the foundational FRS construction
\cite{Fuchs:2002cm, *Fuchs:2003id, *Fuchs:2003id1, *Fuchs:2004xi,
*Fjelstad:2005ua, *Fjelstad:2006aw}. The construction itself is thus not new. We emphasize, however, that the generalized-symmetry/SymTFT perspective was an organizing principle behind these constructions from the start: it dictates the string-net Hilbert space on which the two states are defined, and it explains why the cloaking boundary condition on the holes is the right choice, being the one transparent to all the topological lines, which is what renders the regularization symmetry-preserving. What is new in this paper is making this role explicit: the tensor network is interpreted as a discrete, exactly solvable realization of the SymQRG framework, where the SymTFT is the Turaev-Viro TQFT and the coarse-graining is implemented by the recoupling of $6j$ symbols. We also want to stress that the construction is more general than the examples in which every ingredient is solved explicitly (rational CFTs and the Liouville theory): theories sharing the same SymTFT have the same SymQRG kernel, as we will see explicitly below. We briefly recall the essential ingredients below.


The result can be explicitly written as an overlap \eqref{overlap} in the Hilbert space of the Levin-Wen string-net models.

\begin{figure}
	\centering
	\includegraphics[width=0.3\linewidth]{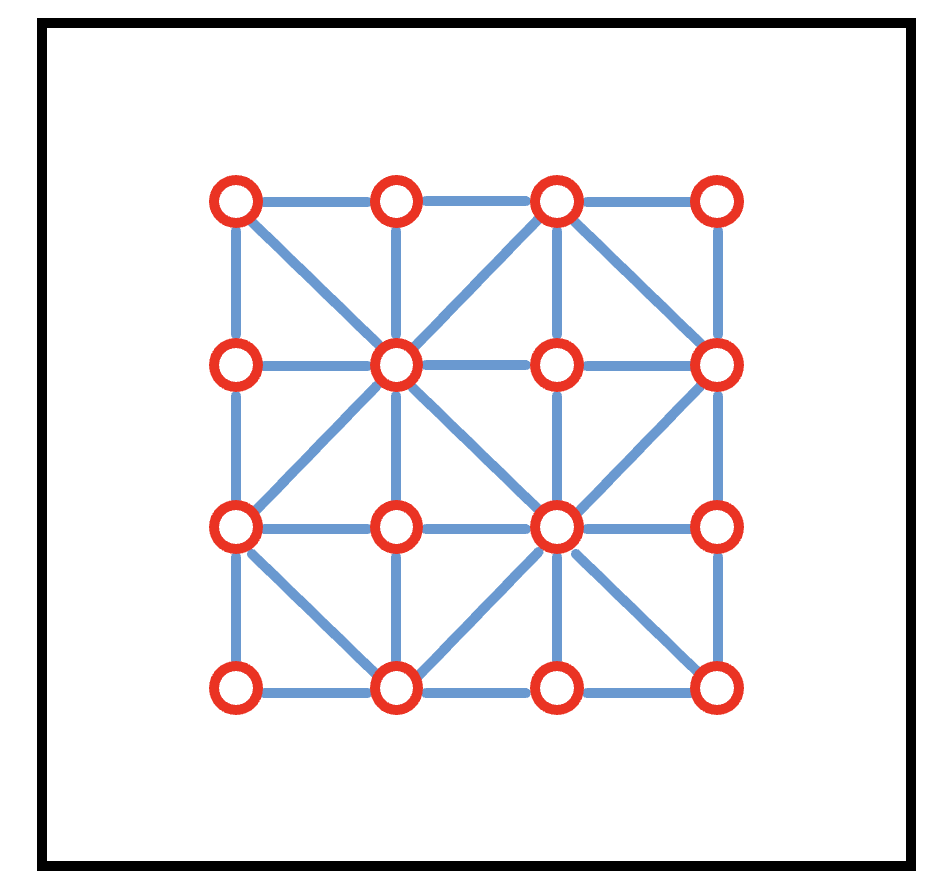}
	\caption{A tiling of 2D manifold using chipped triangles. We specify boundary conditions on the holes colored red.}
	\label{tiling}
\end{figure}
We cut open the 2D manifold into chipped triangles, as illustrated in Fig.~{\ref{tiling}}. The tiny holes in the figure have radius $R$, and are introduced in \cite{Chen:2022wvy, Cheng:2023kxh, Hung:2024gma, Chen:2024unp} for regularization purposes. They  are taken to zero at the end of the computation to recover the original CFT. However, in this paper, we demonstrate that these holes serve a more significant role: by allowing $R$ to vary, they introduce a continuous family of topological symmetry-preserving deformations away from the original CFT. These deformations are analogous to the $T\bar{T}$ deformations discussed in Sec.~\ref{TTbar}, and similarly arise from gluing 3D gravitational path integrals (represented here by quantum $6j$ symbols) onto the 2D theories. 

We first fix some conformal boundary conditions $\sigma_a$ on these tiny circles, and each chipped triangle is an open pair of pants, with states labeled by the primaries $\alpha_i$ and descendants $I_i$ running on the edges. The CFT path integral on each chipped triangle gives,
\be
\mathcal{T}^{\sigma_1\sigma_2\sigma_3}_{(\alpha_1, I_1) (\alpha_2, I_2)(\alpha_3, I_3)}(\triangle, R) = \tilde{C}^{\sigma_3\sigma_1\sigma_2}_{\alpha_1 \alpha_2\alpha_3} \tilde{\gamma}^{\alpha_1 \alpha_2\alpha_3}_{I_1I_2I_3}(R)~.
\ee
The tilde indicates a special choice of normalization explained in \eqref{blockgauge}. In the diagrammatic notation used below, $\tilde{\gamma}^{\alpha_1 \alpha_2 \alpha_3}_{I_1 I_2 I_3}(R)$ is drawn as black dashed lines and $\tilde{C}^{\sigma_3 \sigma_1 \sigma_2}_{\alpha_1 \alpha_2 \alpha_3}$ as blue solid lines with red circles. We suppress the dependence on the background metric for simplicity; see Appendix \ref{statesum cft}.

There is a remarkable formula relating the BCFT OPE coefficients to the (generalized) $6j$ symbols, first discovered by Runkel for diagonal CFTs\cite{Runkel:1998he, *Runkel:1999dz} and generalized to non-diagonal CFTs in the FRS framework\cite{Fuchs:2002cm}, which reads
\be \label{rationalC1}
\tilde{C}^{\sigma_3\sigma_1\sigma_2}_{\alpha_1 \alpha_2\alpha_3} = \left(d_{\alpha_1} d_{\alpha_2} d_{\alpha_3}\right)^{1/4} \begin{Bmatrix}
    \alpha_1 & \alpha_2 & \alpha_3  \\
    \sigma_3 & \sigma_1 & \sigma_2
    \end{Bmatrix},
\ee
where $d_{\alpha_i}$ is the quantum dimension associated with $\alpha_i$. In general, the label sets for the primaries and boundary conditions are different, and the curly brackets denote the generalized $6j$ symbols with mixed indices. These generalized symbols can also be expressed in terms of the standard $6j$ symbols (with all indices in the same label set of primaries) multiplied by additional factors corresponding to a topological boundary condition in the 3D Turaev-Viro TQFT\cite{Fuchs:2002cm, Lootens:2020mso}. This is the local form of the separation: the conformal-block factor $\tilde{\gamma}$ contributes to the physical boundary state, while the $6j$-symbol/OPE data $\tilde{C}$ contributes to the topological state.

Summing over all intermediate primaries and descendants\footnote{In a theory with a continuous family of primaries such as Liouville theory, this sum becomes an integral\cite{Chen:2024unp, Hung:2024gma}.}, we get the partition function for fixed boundary conditions $\{\sigma_{a}\}$:
\be \label{Zproduct0}
\mathcal{Z}(\{\sigma_a\},R)=\sum_{\{\alpha_i\}}  \sum_{\{I_i\}}     \prod_{\triangle} \mathcal{T}^{\sigma_a\sigma_b\sigma_c}_{(\alpha_i, I_i) (\alpha_j, I_j)(\alpha_k, I_k)}(\triangle, R)~,
\ee
where the sum over descendants corresponds to conformal blocks.

With fixed boundary conditions, inserting holes turns on a relevant deformation that does not recover the CFT in the IR \cite{Brehm:2021wev}. The resolution \cite{Chen:2022wvy, Cheng:2023kxh, Hung:2024gma, Chen:2024unp, Hung:2019bnq, Brehm:2021wev} is to sum over boundary conditions with weight $\omega_{\sigma_a} \propto d_{\sigma_a}$ (quantum dimension) for each vertex $a$:
\be \label{Zproduct01}
\mathcal{Z}(R)= \sum_{\{\sigma_a\}} \prod_v \omega_{\sigma_a} \sum_{\{\alpha_{i}\}}  \sum_{\{I_{i}\}}   \prod_{\triangle} \mathcal{T}^{\sigma_{a}\sigma_{b}\sigma_c}_{(\alpha_{i}, I_{i}) (\alpha_{j}, I_{j})(\alpha_{k}, I_{k})}(\triangle, R)~.
\ee
This ``cloaking'' or ``shrinkable'' boundary condition\cite{Hung:2019bnq, Brehm:2021wev, Chen:2022wvy, Cheng:2023kxh, Hung:2024gma, Chen:2024unp} has two key effects. In the dual closed channel, the weighted sum of boundary states produces the vacuum Ishibashi state,
\be \label{vacIshibashi}
\ket{0}\rangle=\ket{0}+\frac{2}{c} L_{-2} \bar{L}_{-2} \ket{0}+\ldots
\ee
The leading deformation $\frac{2}{c} L_{-2} \bar{L}_{-2} \ket{0}$ corresponds to the $T\bar{T}$ operator \cite{Brehm:2021wev}, giving at small $R$:
\be \label{TTbarvev}
R^4 \frac{2}{c} \sum_{v} \langle T \bar{T}(v) \rangle+\ldots
\ee
Since $T\bar{T}$ is irrelevant, the theory flows back to the original CFT as $R \to 0$. Moreover, all topological defect lines can freely traverse the holes equipped with this ``cloaking boundary condition'' \cite{Brehm:2021wev, Chen:2024unp, Vanhove:2018wlb, Aasen:2016dop, Aasen:2020jwb}, ensuring that the deformation preserves all topological symmetries at any $R$.

The weighted sum also connects to the SymTFT. Using the Levin-Wen Hilbert space $\mathcal{H}_\Lambda=\text{Span} \big\{\ket{\{\alpha_{i}\}}_{\Lambda}\big\}$ introduced in Sec.~\ref{briefreviewtqft}\footnote{We assume multiplicity one throughout; see \cite{Levin:2004mi} for the general case. We thank Greg Moore and Sahand Seifnashri for discussions on this point.}, we rewrite the partition function \eqref{Zproduct01} as an overlap:
\be \label{strange_correlator}
\mathcal{Z}(R)={}_\Lambda \langle \Omega^R| \Psi \rangle_{\Lambda}~.
\ee  
where
\be \label{psistate}
\ket{\Psi}_{\Lambda}=\sum_{\{\sigma_{a}\}}  \prod_v \omega_{\sigma_{a}} \sum_{\{\alpha_{i}\}}  \prod_{\triangle} \tilde{C}^{\sigma_{c}\sigma_{a}\sigma_{b}}_{\alpha_{i} \alpha_{j}\alpha_{k}}  \ket{\{\alpha_{i}\}}_{\Lambda}~,
\ee
and
\be \label{omegacft}
{}_\Lambda \langle \Omega^R| =  \sum_{\{\alpha_{i}\}}{}_{\Lambda}\bra{\{\alpha_{i} \}} \sum_{\{I_{i}\}}     \prod_{\triangle}  \tilde{\gamma}^{\alpha_{i} \alpha_{j}\alpha_{k}}_{I_{i} I_{j} I_{k}}(R)~. 
\ee
The state ${}_\Lambda \langle \Omega^R|$ encodes all position-dependent contributions, with the wavefunctions being chiral conformal blocks, thus embodying the local Virasoro algebra $V$. The primary labels $\alpha_{i}$ correspond to the objects in the representation category $\mathcal{C}$ of $V$, where $\mathcal{C} = \text{Rep}(V)$.

On the other hand, the state $\ket{\Psi}_{\Lambda}$ captures all BCFT OPE coefficients. Remarkably, this state is precisely a ground state of the Levin-Wen string-net model, or equivalently, the Turaev-Viro theory constructed from $\mathcal{C}$. In particular, it satisfies the no-flux constraint \eqref{noflux}.

\begin{figure}
	\centering
	\includegraphics[width=0.45\linewidth]{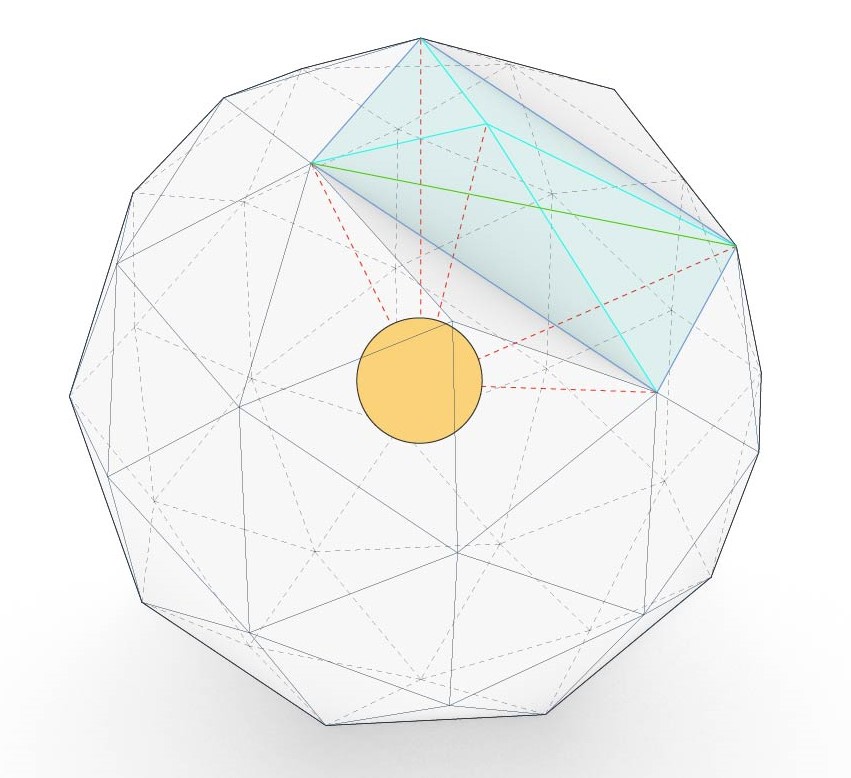}
	\caption{The ground state $\ket{\Psi}_{\Lambda}$ on $S^2$ from a Turaev-Viro path integral in a 3D ball. The orange blob is the topological boundary (Sec.~\ref{roleofpsi module category}). Two representations on cyan/green edges are connected by an additional layer of tetrahedra (grey), leading to \eqref{eq:URG} (Fig.~\ref{removingbubble}).}
	\label{azure}
\end{figure}
Both states are PEPS-type tensor-network states \cite{PhysRevA.70.060302, *Verstraete:2004cf}; crucially, ${}_{\Lambda} \bra{\Omega^R}$ has \textit{infinite} bond dimension from the infinite descendant tower, enabling an exact representation of the \textit{continuum} CFT and the entire family of deformed theories $Z(R)$. The construction generalizes straightforwardly to correlation functions \cite{Chen:2024unp}.

To recover the CFT partition function, we take
\be \label{disentangle}
\mathcal{Z}_{\text{CFT}}=\lim_{R \to 0} (\prod_{v} e^{-\frac{c}{6} \ln(R)}) {}_\Lambda \langle \Omega^R| \Psi \rangle_{\Lambda}~.
\ee
The meaning of this factor will be explained after the SymQRG procedure below.

This ``strange correlator'' formulation \cite{Vanhove:2018wlb, PhysRevLett.112.247202} naturally provides a real-space tensor network RG algorithm, exploiting the topological fixed-point property of $\ket{\Psi}_{\Lambda}$ \cite{PhysRevLett.99.120601, *PhysRevLett.115.180405, *PhysRevLett.118.110504, *PhysRevLett.118.250602, PhysRevB.79.195123, *KOENIG20102707}. The procedure parallels the $T\bar{T}$ deformation of Sec.~\ref{TTbar}, now realized with lattice TQFT techniques (appendix \ref{reviewtvtqft}).

The RG map (Fig.~\ref{removingbubble}; details in appendix \ref{tensornetwork SymQRG}) follows from crossing symmetry, giving the transformation rule at scales $\Lambda$ and $\Lambda'=\sqrt{2} \Lambda$:
\be \label{eq:URG}
\ket{\Psi}_{\Lambda}=\hat{U}^{\Lambda, \Lambda'}_{\text{RG}} \ket{\Psi}_{\Lambda'}~.
\ee

\begin{figure}
	\centering
	\includegraphics[width=0.85\linewidth]{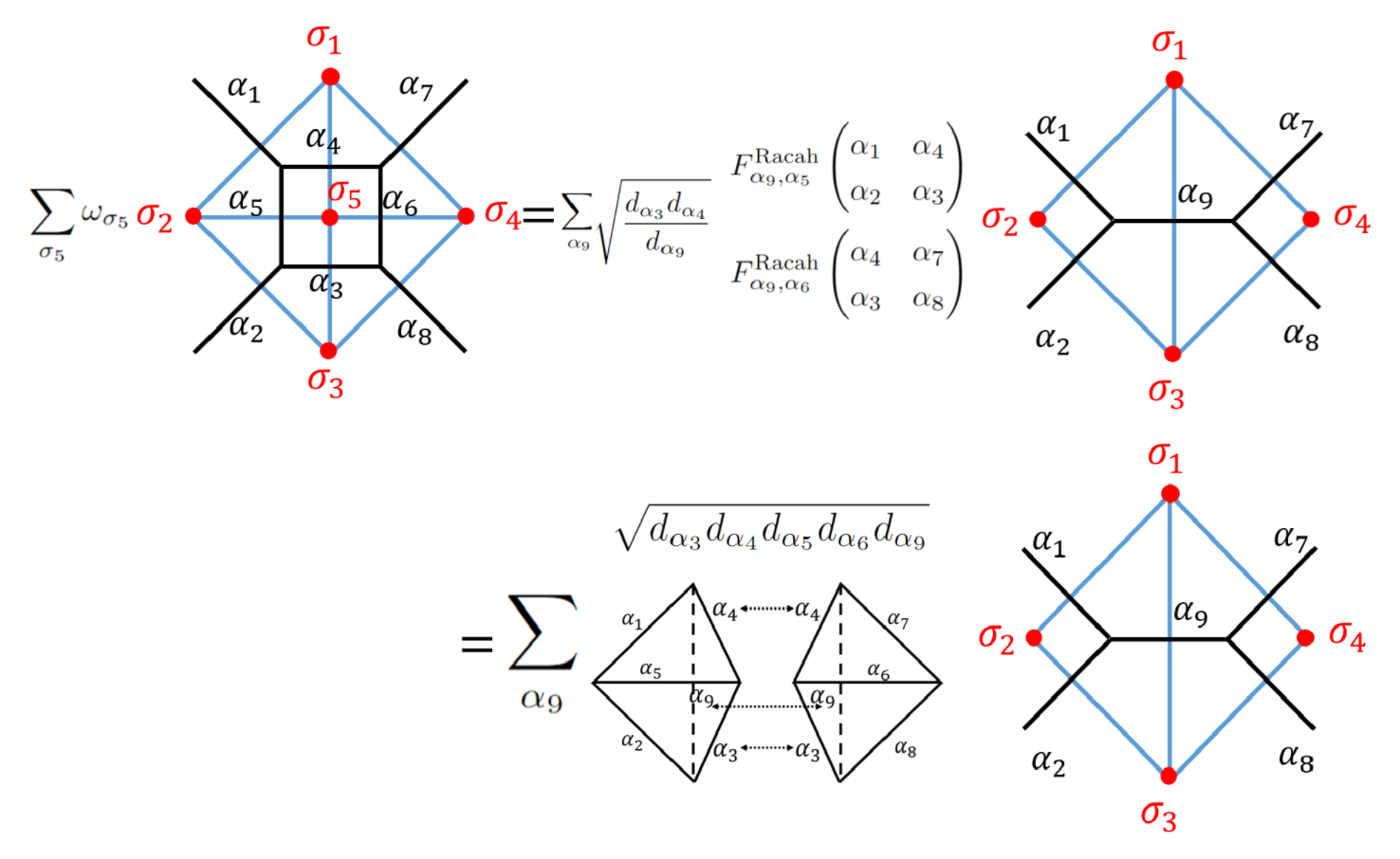}
    \caption{Coarse-graining equation of the Levin-Wen ground state $\ket{\Psi}_{\Lambda}$.}
	\label{removingbubble}
\end{figure}

Under this transformation, four BCFT structure coefficients reduce to two by gluing a layer of tetrahedra (Turaev-Viro path integral), as in Fig.~\ref{azure}. The Levin-Wen ground states are exact fixed points of this block-spin map\cite{PhysRevB.79.195123}. As outlined in Sec.~\ref{symqrgparagraph}, the same procedure defines the coarse-graining map for ${}_\Lambda \langle \Omega^R|$:
\be \label{RG_lattice}
{}_{\Lambda'} \langle \Omega^{'R}|={}_{\Lambda} \langle \Omega^R| \hat{U}^{\Lambda, \Lambda'}_{\text{RG}}~.
\ee

This ensures that the overlap remains invariant across different scales. Now contracted with the ${}_{\Lambda} \langle \Omega^R|$ state, the extra layers of tetrahedra accumulate with each RG step, assembling the bulk SymTFT path integral and realizing the UV/IR correspondence in the AdS/CFT correspondence\cite{Susskind:1998dq}. This is the lattice realization of the projected coarse-graining defined in Sec.~\ref{symqrgparagraph}: the same-scale version of the gluing is precisely the Levin-Wen no-flux projector $\Pi_{\Lambda}$ of \eqref{slabprojector}, and the ordinary block-spin contraction becomes symmetry-preserving exactly because it is completed by this projection.

The RG operator depends only on $6j$ symbols with indices $\alpha_i \in \mathcal{C}$ (specifying the SymTFT), not on the theory-specific boundary conditions $\sigma_a$; it is therefore universal for all theories sharing the same topological symmetries.

Repeated application yields a multilayer quantum circuit resembling MERA \cite{Vidal:2008zz}, relating theories at different scales along the same symmetry-preserving RG trajectory and building the CFT path integral and the bulk quantum spacetime simultaneously from BCFT building blocks \cite{Hung:2024gma, VanRaamsdonk:2018zws}. At this stage, the procedure establishes a ``pregeometry'' \cite{Luo:2016leh}; the connection to 3D gravity and emergent metric structures is the focus of the next section.

\begin{figure}
	\centering
	\includegraphics[width=0.85\linewidth]{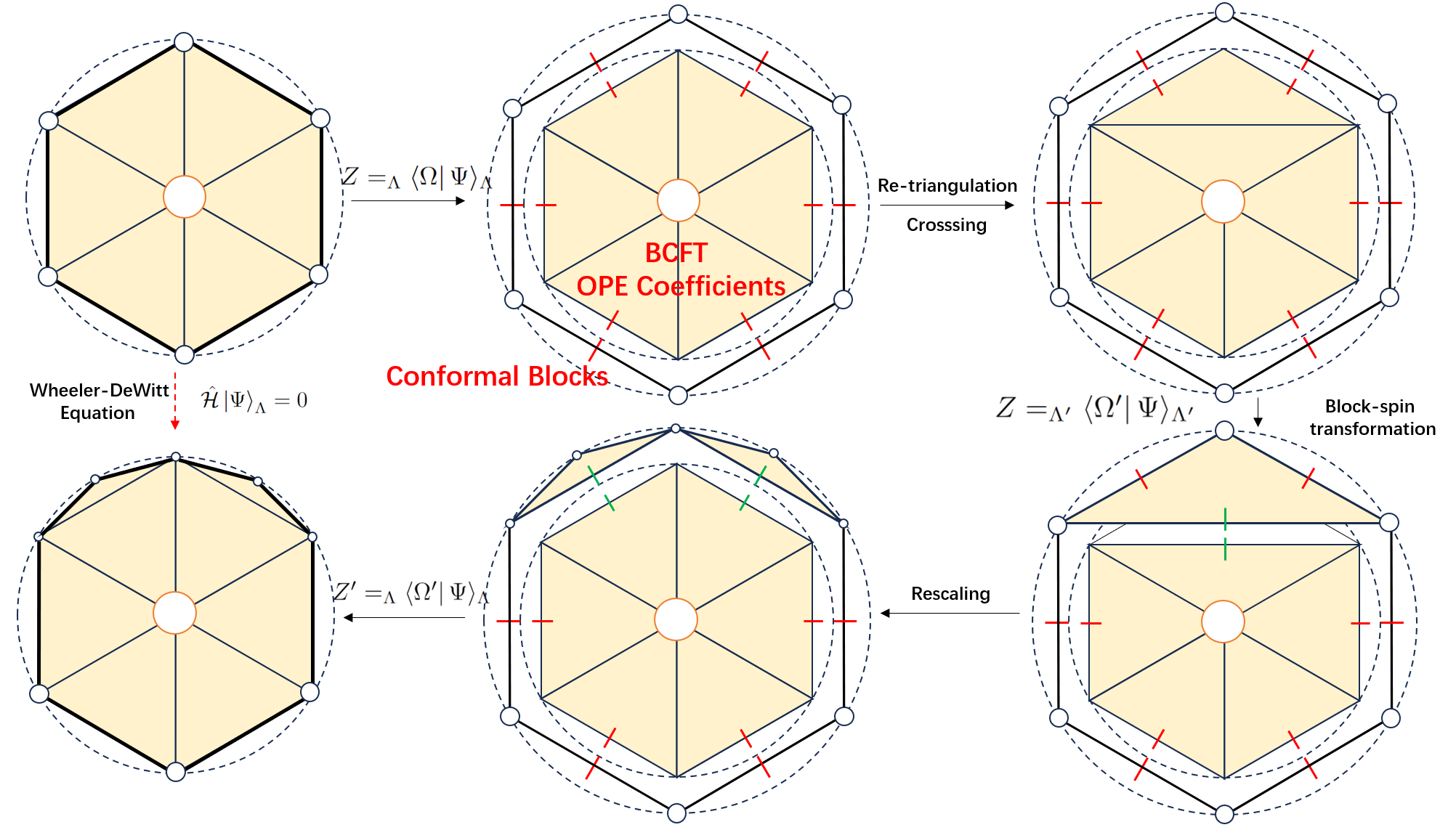}
	\caption{The SymQRG coarse-graining procedure. The second step uses topological invariance (2D CFT crossing) to re-triangulate $\ket{\Psi}_{\Lambda}$. The third step contracts $_{\Lambda}\bra{\Omega}$ on red edges into $_{\Lambda'}\bra{\Omega'}$ on green edges, realizing a symmetry-preserving block-spin transformation. Before rescaling (bottom right): coarse-grained lattice. After rescaling: the overlap gives a deformed path integral with smaller holes, corresponding to the IR direction in the bulk.}
	\label{coarsegrain}
\end{figure}

Incorporating the rescaling step, the entire procedure is illustrated in Fig.~\ref{coarsegrain}\footnote{These 2D diagrams correspond precisely to tensor networks related to 2D JT gravity and will be presented in \cite{jtgravity}.}.

As demonstrated in \cite{Chen:2022wvy, Cheng:2023kxh, Hung:2024gma, Chen:2024unp}, the state ${}_{\Lambda} \langle \Omega^R|$ constructed from conformal blocks is a fixed point of this RG procedure in the limit $R \to 0$. For $R>0$, using the properties of conformal blocks and crossing kernels (see Appendix~\ref{tensornetwork SymQRG} for the general procedure and explicit formulae), the coarse-grained wavefunction takes the form:

\be	\includegraphics[width=0.6\linewidth]{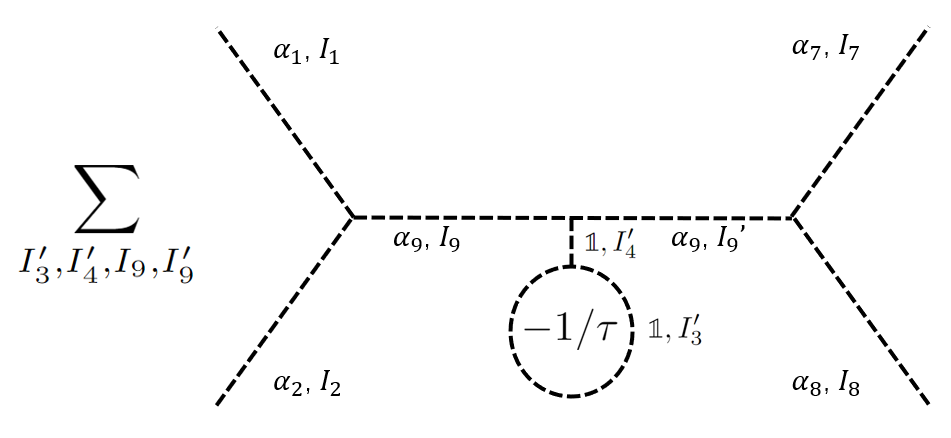}
    \label{crossinggamma3}
\ee

In the limit $R \to 0$, the tiny bubble factorizes, contributing the universal Casimir energy $e^{c \ln(R)/6}$, while the rest reduces to conformal blocks on a coarser triangulation, confirming that the construction is a fixed point. After rescaling, the procedure yields smaller holes (RG towards the IR); reversing it produces larger holes, implementing the $T\bar{T}$ deformation \eqref{TTbarvev}.

\subsection{Wheeler-DeWitt equation and Callan-Symanzik equation from tensor networks}

The tensor network enables a discrete realization of the Wheeler-DeWitt equation. When the deviation of the boundary state $\langle \Omega |$ from the CFT fixed point corresponds to $T\bar T$ deformation, we expect it should recover flows described by (\ref{operatordiff}) and thus (\ref{flowequation}). 

The line of logic has been outlined in Secs.~\ref{symqrgparagraph} and \ref{wdwstates} for the continuous formulation, and we will make various ingredients explicit in the discrete formulation here too. 

The key ingredient is that in the partition function expressed as in (\ref{strange_correlator}), the wave-function $|\Psi \rangle_\Lambda$ as shown in (\ref{psistate}) is a ground state of the Levin-Wen model satisfying
\be \label{LWmodel}
\hat{\mathcal{H}}_{\Lambda} |\Psi\rangle_\Lambda =0,
\ee
where $\hat{\mathcal{H}}_{\Lambda} $ is the Hamiltonian of the Levin-Wen model defined on the same lattice $\Lambda$ \cite{Levin:2004mi}, and it imposes the analogue of the no-flux condition as in the continuous Chern-Simons theory\footnote{It is convenient for us to shift the ground state energy to 0, which can always be arranged.}. Equation \eqref{LWmodel} is the explicit discrete version of \eqref{wdwcft constraint}, anticipated in \eqref{noflux} which is essentially the Wheeler-DeWitt equation.

To make the comparison with Sec.~\ref{TTbar} explicit, and to recover the familiar Callan-Symanzik equation, we again consider a theory deformed perturbatively away from a CFT fixed point, which was shown above to be given by the $T\bar T$ deformation. As explained in \eqref{vacIshibashi} and \eqref{TTbarvev}, the small radius $R$ of each hole is equivalent to introducing a finite coupling to $T\bar T$ in the hole, at least to the first leading contribution from the small $R$ expansion. Comparing with (\ref{flowequation}), we have, to linear order in the $T\bar T$ perturbation coupling $\lambda$
\be \label{Rvslambda}
\frac{2R^4}{c} \approx \frac{12\pi R^2 \lambda} {c} \implies \lambda \approx \frac{R^2}{6\pi},
\ee
where we have approximated the deformation inserted in the holes $\int d^2 x \sqrt{\gamma} T\bar T $ by
\be \label{convert1}
\int d^2 x \, \sqrt{\gamma}\,T\bar T \approx  R^2 \sum_v T\bar T(v) .
\ee

As expected, the hole size $R$ is related to the $T\bar T $ coupling $\lambda$. The precise constant of proportionality is not important. 
We can thus use ${}_{\Lambda}\langle \Omega^R|$ interchangeably with ${}_{\Lambda}\langle \Omega^\lambda|$ at least for small $R$ given the conversion (\ref{Rvslambda}).
The scale $\Lambda$ that appeared in (\ref{overlap}) now takes a precise meaning of the lattice scale chosen in the surface triangulation of the path-integral. Since this triangulation scale is arbitrarily chosen in representing the continuum field theory, one can consider it as a renormalization scale or a metric scale. 
Now as explained in Sec.~\ref{RGsection}, the bulk SymTFT produces a symmetry-preserving RG operator $U_{\textrm{RG}}^{\Lambda, \Lambda'}$ that changes the scale $\Lambda$. This operator is constructed explicitly using quantum $6j$ symbols in (\ref{eq:URG}).

Equation (\ref{LWmodel}) ensures the relation (\ref{eq:URG}) relating the ground state at different scales. 
This in turn implies the equality. 
\be
{}_\Lambda \bra{\Omega^{\lambda}}  \Psi \rangle_{\Lambda}={}_{\Lambda} \bra{\Omega^{\lambda}}  \hat{U}^{\Lambda, \Lambda'}_{\textrm{RG}}| \Psi \rangle_{\Lambda'}. 
\ee

Now a lot of the discussion in Sec.~\ref{symqrgparagraph} takes definite shape in the discrete formulation. 
As in (\ref{change_scale}) the RG operator here can be absorbed in ${}_{\Lambda}\langle \Omega^\lambda|$ as in (\ref{RG_lattice}) to produce 
\be
{}_{\Lambda'}\langle \Omega'^\lambda| \equiv {}_{\Lambda} \bra{\Omega^{\lambda}}  \hat{U}^{\Lambda, \Lambda'}_{\textrm{RG}}.
\ee
Now we want to define a {\it new} state by relabeling the lattice scale back to $\Lambda$ as advocated in (\ref{relabel0}) and (\ref{newstate}). i.e.
\be
{}_{\Lambda}\langle \Omega'| \equiv {}_{\Lambda'}\langle \Omega'^\lambda| \mathcal{D}^{\Lambda,\Lambda'} = {}_{\Lambda}\langle \Omega^{\lambda'}|, \qquad \lambda' \approx \lambda (\Lambda/\Lambda')^2 ,
\ee
where $\mathcal{D}^{\Lambda,\Lambda'}$ is simply a relabeling (dilatation) operator; we reserve the symbol $\Pi$ for the no-flux projector \eqref{slabprojector}. To leading order in $R$, the new state is nothing but the original boundary condition with the holes rescaled after an RG step that combines four triangles into two, i.e. $R' = R \Lambda/\Lambda'$, explaining the approximate equality above, confirming equation (\ref{newstate}). $R$ needs to be small because the finite size of the hole contains more perturbation than the $T\bar T$ deformation alone, and the RG process would generally generate flow in couplings of higher powers of the stress tensor too. To leading order in $R$ however the RG process is only shifting the $T\bar T$ coupling.
The combined operation of RG and rescaling the lattice size is equivalent to the conformal transformation $h_{ij} \partial_{h_{ij}}$. We show that such a transformation is equivalent to shifting the $T\bar T$ coupling and now we are ready to write down the Callan-Symanzik equation.  
We can arrange the re-triangulation that defines the RG operator such that $\Lambda'$ is arbitrarily close to $\Lambda$. 
Therefore we can obtain an infinitesimal form of the flow equation
\be
{}_\Lambda \langle \Omega^{\lambda'} | \Psi\rangle_\Lambda - {}_\Lambda \langle \Omega^\lambda | \Psi\rangle_\Lambda \approx \delta \lambda \frac{d{}_\Lambda \langle \Omega^\lambda | \Psi\rangle_\Lambda}{d\lambda} \approx 
\delta \lambda \left( \sum_v(\frac{c}{12 \lambda}  )  +  \frac{12\pi}{c} \int d^2 x \sqrt{\gamma }\,T\bar T\right) {}_\Lambda \langle \Omega^\lambda | \Psi\rangle_\Lambda,
\ee

where we have made use of the conversion in (\ref{Rvslambda}) and (\ref{convert1}). The term proportional to $c$ follows from the fact that there is a universal factor in the small $R$ limit at every hole proportional to $\exp(c/6 \ln R)$, as explained below (\ref{crossinggamma3}). 

The expression ${}_\Lambda \langle \Omega^\lambda | \Psi\rangle_\Lambda$ diverges as $R\to 0 $, and regularization was introduced in (\ref{disentangle}). 

Therefore we could consider the flow equation of the regularized path-integral. Instead of including a factor for each vertex as in (\ref{disentangle}), we distribute these factors to individual triangles, as befits a tensor network representation built from local triangle data. For an equilateral triangulation, where each vertex is shared by 6 triangles, the natural assignment is a factor of $\exp(-c \ln R/36)$ per vertex of each triangle, i.e.~we have
\be
{}_\Lambda \langle \Omega^\lambda | \Psi\rangle_{\Lambda,\textrm{Reg}} \equiv e^{-\frac{c \ln R}{36}\sum_v\chi(v)}{}_\Lambda \langle \Omega^\lambda | \Psi\rangle_\Lambda,
\ee
where $\chi(v)$ is the number of triangles sharing this vertex $v$.
Then the flow equation of the regulated partition function is given by
\be
\lambda \frac{d{}_\Lambda \langle \Omega^\lambda | \Psi\rangle_{\Lambda,\textrm{Reg}}}{d\lambda}
\approx \left( \frac{c}{48\pi }  \int_M d^2x \sqrt{\gamma}\textrm{Ri} +  \frac{12\pi \lambda}{c} \int d^2 x \sqrt{\gamma }\,T\bar T\right) {}_\Lambda \langle \Omega^\lambda | \Psi\rangle_{\Lambda,\textrm{Reg}},
\ee

where we have made use of
\be
\sum_{v \in G} (6-\chi(v)) = 6\chi_M = \frac{6}{4\pi}\int_M d^2x \sqrt{\gamma} \textrm{Ri},
\ee
for some 2D manifold $M$ triangulated by a graph $G$, and $\chi_M$ the Euler characteristic of the manifold $M$. Ri is used to denote the Ricci scalar to avoid confusing it with the hole radius $R$. 
Contrasting with the flow equation (\ref{flowequation}), the right-hand side carries an overall minus sign, consistent with the $T\bar{T}$ coupling having opposite sign in \eqref{TTbarvev} compared to Sec.~\ref{TTbar}. This sign convention agrees with other lattice constructions of $T\bar T$ deformation \cite{Jiang:2023rxa}; constructing lattice models that reproduce the finite-$r$ cutoff sign of \cite{McGough:2016lol} remains an open problem.

$T\bar T$ deformation is only one very specific choice of symmetry preserving deformations away from the CFT fixed point.
One can describe more general quantum RG flows with more generic ${}_\Lambda\langle \Omega |$ that introduces other deformations away from the CFT fixed point. In a general quantum RG flow, the discrete form of the Wheeler-DeWitt equation generically follows from
\be \label{rgoperator1}
{}_\Lambda \bra{\Omega^{\lambda'}}  \hat{\mathcal{H}}_{\Lambda} \hat{U}^{\lambda' , \lambda}_{\textrm{RG}}| \Psi \rangle_{\Lambda}={}_\Lambda \bra{\Omega^{\lambda' }} \hat{U}^{\lambda' , \lambda}_{\textrm{RG}}  \hat{\mathcal{H}}_{\Lambda} | \Psi \rangle_{\Lambda}=0~,
\ee
where we have defined 
\be
\hat{U}^{\lambda' , \lambda}_{\textrm{RG}} \equiv \hat{U}^{\Lambda , \Lambda'}_{\textrm{RG}} \mathcal{D}^{\Lambda,\Lambda'}~.
\ee

As discussed above, this property enables us to perform the SymQRG, gluing additional layers of the SymTFT path integral onto the boundary state. This shows the connection between crossing/re-triangulation invariance and the Wheeler-DeWitt equation, which is responsible for the emergence of a 3D bulk.

In the next section, we apply this formalism when the bulk SymTFT contains 3D gravity. In gravitational language, the no-flux condition states that torsion vanishes and curvature is constant; this parallels the flat-holonomy Hamiltonian constraint in spin foam models \cite{Bonzom:2009zd, *Bonzom:2011hm}, enforced by the Pentagon identity.

\subsection{Roles of tensor networks}\label{Tensornetwork}

The tensor network plays three essential roles in this construction.

\textit{Locality.} At its core, the tensor network is the non-perturbative manifestation of \textit{locality}: the path integral over an extended region is built by recursively gluing local pieces, with the gluing captured by tensor contraction.
The FRS theorem provides the local tensor network for $\ket{\Psi}_{\Lambda}$ (topological/algebraic sewing rules) \cite{Fuchs:2002cm, *Fuchs:2003id, *Fuchs:2003id1, *Fuchs:2004xi, *Fjelstad:2005ua, *Fjelstad:2006aw}, while ${}_{\Lambda} \bra{\Omega}$ \cite{Chen:2022wvy, Cheng:2023kxh, Hung:2024gma, Chen:2024unp} provides the geometrical/complex-analytical half.
This locality also makes the boundary-to-bulk map explicit, offering a non-perturbative realization of the Ryu-Takayanagi formula \cite{Ryu:2006bv} and bulk reconstruction \cite{Hamilton:2005ju, *Hamilton:2006az, Czech:2012bh, Jafferis:2015del, Almheiri:2014lwa, *Dong:2016eik}. Although our tensor network is graph-independent, a MERA-like triangulation simplifies entanglement computations, with the ``quasi-perfect'' property of the CFT tensors \cite{Chen:2024unp, Pastawski:2015qua} ensuring that the RT surface precisely determines the entanglement spectrum (Appendix~\ref{wavefunction tensornetwork}).

\textit{Symmetry and error correction.} The tensor indices are labeled by the topological symmetries of the CFT: the Hilbert space is that of the Levin-Wen string-net model, where topological symmetries manifest as ribbon operators. The SymQRG projection, governed by quantum $6j$ symbols, mirrors the single-trace projection of Quantum RG but is now canonically fixed by the preserved symmetry. This naturally induces \textit{gauge symmetry} in the bulk and \textit{global symmetry} on the boundary\footnote{Tensor network toy models for holography involving gauge symmetry in the bulk have been explored in \cite{Donnelly:2016qqt, *Qi:2022lbd, *Dong:2023kyr, *Akers:2024ixq, Akers:2024wab}.}, as in the WZW/Chern-Simons correspondence \cite{Singh:2013sda, *Singh:2012np}. Moreover, since our construction involves a topological order, the holographic code is a \textit{fault-tolerant topological} error-correcting code\cite{Kitaev:1997wr, Verlinde:2012cy, Almheiri:2014lwa, Pastawski:2015qua}, with quantum information stored through the long-range entanglement of the gravitational theory\cite{Verlinde:2013vja}.

\textit{Background independence.} This framework provides an exact 2D/3D equivalence at any finite $N$, without assuming pre-existing geometries \cite{Bao:2024hwy}. Quantum geometries emerge naturally from the CFT data, and the resulting bulk involves a sum over them; bulk operators are inherently defined within the CFT \cite{Verlinde:2015qfa, *Goto:2016wme, *Lewkowycz:2016ukf, *Goto:2017olq}. The bulk/boundary lattice has nothing to do with target-space geometries\cite{Chen:2024unp, Hung:2024gma, Akers:2024wab}: the emergence of geometry arises from the algebraic data in the CFT admitting geometrical interpretations. The large $N$ semi-classical results are recovered at saddle points, as we elaborate in Sec.~\ref{algebra}.

\section{Virasoro SymQRG kernel and three-dimensional gravity}\label{TQFTSECTION}

In Sec.~\ref{TTbar}, we discussed the connection between an RG flow driven by $T\bar{T}$ deformations and the gravitational path integral. Specifically, we showed that coarse-graining the $T\bar{T}$-deformed CFT is equivalent to treating the background metric as a dynamical variable and evolving it according to Einstein's gravitational evolution. This highlights one of the central themes of the current paper, the continuum version of the slogan stated in the introduction: the projector/slab kernel that renders Virasoro-preserving quantum RG exact is the three-dimensional gravitational path integral. In Sec.~\ref{tensor network}, we have illustrated how to construct the SymQRG evolution on a lattice that explicitly preserves topological symmetries. 

However, it is perhaps not yet obvious how the RG flow driven by $T\bar T$ deformation and the SymQRG are related, and what makes the emergent bulk theory {\it Einsteinian}? We have alluded to the fact that the choice of $T\bar T$ deformation is special because it preserves topological symmetries, and an RG process that preserves topological symmetries should only change couplings of symmetry-preserving operators. In the case of 2D CFTs, this 3D bulk being Einsteinian is by no means an accident. In this section, we identify the topological defect lines associated with 3D gravity. We begin with the case of 3D pure gravity, and then explore the connection to general irrational large $N$ holographic CFTs, where 3D Einstein gravity serves as the universal approximate SymTFT.

When an RG process preserves a continuously labeled family of inequivalent
simple topological lines, the SymTFT encoding them is necessarily
non-rational and lies beyond the finite semisimple framework familiar from
rational examples. Classical AdS$_3$ gravity provides a natural starting
point: its first-order action can be written as that of a Chern-Simons
theory with gauge algebra
$\mathfrak{sl}(2,\mathbb{R})\oplus\mathfrak{sl}(2,\mathbb{R})$
\cite{Achucarro:1986uwr,Witten:1988hc}. This action-level relation alone,
however, neither fixes the admissible quantum line spectrum nor establishes
that the quantum gravitational path integral realizes this SymTFT.
Combining the Virasoro TQFT proposal reviewed below with our SymQRG
construction, we argue that the three-dimensional quantum gravitational
path integral realizes the non-rational SymTFT associated with the
preserved Virasoro lines. On a slab, the gravitational path integral is
precisely the Virasoro-preserving SymQRG kernel. The discrete Liouville
constructions of Refs.~\cite{Chen:2024unp,Hung:2024gma} then furnish an
explicit state-sum realization of this slab kernel in terms of the left-
and right-moving Virasoro TQFT factors; at large central charge, the
quantum $6j$ symbols reproduce the Einstein-Hilbert action of hyperbolic
tetrahedra. Remarkably, Virasoro TQFT has continuously labeled anyonic
sectors represented by Wilson lines; in the gravitational description,
insertions of these lines correspond to BTZ black holes \cite{Banados:1992wn}.

We first review the Virasoro TQFT proposal, and then construct the SymQRG kernel from the
discrete $U_q(SL(2,\mathbb{R}))$ data. We discuss its relation to quantum
Teichm\"uller theory and lattice Liouville theory in
Appendix~\ref{reviewteichmuller}, and conclude by explaining how the
pure-gravity construction extends, in a coarse-grained sense, to generic
large-$N$ holographic CFTs, distinguishing the exact finite-$N$ 2D/3D
construction from the universal semi-classical Einstein sector.



\subsection{Virasoro TQFT}
The equality of the classical gravity and Chern-Simons actions does not
imply an equivalence of the corresponding quantum theories
\cite{Witten:2007kt,Collier:2023fwi}. Directly quantizing the full moduli
space of flat
$PSL(2,\mathbb{R})\times PSL(2,\mathbb{R})$
connections includes configurations that do not define non-degenerate
metrics. In the Lorentzian canonical description of pure AdS$_3$ gravity
without conical defects, the non-degeneracy condition selects the
Teichm\"uller component in each chiral sector before quantization, so that
\[
\mathcal{M}_{\mathrm{grav}}(\Sigma)
=
\mathcal{T}(\Sigma)\times{\mathcal{T}}(\Sigma).
\]
The two factors correspond to the left- and right-moving sectors. A chiral
Virasoro TQFT is built by quantizing one such component
\cite{Verlinde:1989ua,EllegaardAndersen:2011vps,Collier:2023fwi}; its
Hilbert space is the space of chiral Virasoro conformal blocks, and its
Gauss-law constraint is equivalent to the Virasoro Ward identity
\cite{Verlinde:1989ua}. Virasoro TQFT should therefore not be understood
as a direct quantization of the full Chern-Simons theory, but as a chiral
building block of the proposed gravitational prescription.


A key observation is that the Virasoro TQFT on $\Sigma \times I$ gives \cite{Collier:2023fwi}:
\be \label{virasoro_liouville}
Z_{\textrm{Vir}}(\Sigma \times I) = \int d^{3g-3 + n}\vec{P} \rho^C_{g,n}(\vec P) |\mathcal{F}^C_{g,n}(\vec P) \rangle \otimes \langle \mathcal{F}^C_{g,n}(\vec P) |~,
\ee
where $|\mathcal{F}^C_{g,n}(\vec P) \rangle$ are chiral Virasoro conformal blocks on $\Sigma_{g,n}$ and $\rho^C_{g,n}(\vec P)$ is the Liouville OPE density. For a fixed-topology hyperbolic manifold $M$ in the class considered in
Ref.~\cite{Collier:2023fwi}, the proposed gravitational prescription
combines the left- and right-moving Virasoro TQFT factors with the
appropriate mapping-class-group sum:
\be
Z_{\text{grav}}(M) = \frac{1}{|\textrm{Map}_0(M, \partial M)|}\sum_{\gamma \in \textrm{Map}(\partial M)/\textrm{Map}(M, \partial M)} |Z_{\text{Vir}}(M^{\gamma})|^2~,
\ee
where the sum over mapping class group images ensures modular invariance (an alternative route via anyon condensation is discussed in Sec.~\ref{roleofpsi module category}; see also \cite{Benini:2022hzx}).


Within this proposal, the RG operator \eqref{rgoperatorinttbar} should be
expressible in terms of the left- and right-moving Virasoro TQFT factors.
Since Virasoro descendants do not appear explicitly in the present
formulation, we seek an alternative discrete realization, which we identify
in the next subsection using quantum $6j$ symbols of
$U_q(SL(2,\mathbb{R}))$.

\subsection{Discrete realization of the Virasoro RG operator}

To find an alternative quantum formulation of the evolution operator in (\ref{rgoperatorinttbar}), we seek inspiration from the recent work of deriving the holographic bulk dual of the Liouville theory \cite{Chen:2024unp, Hung:2024gma}. The final answer we get will be universal for theories sharing the same topological symmetries, not restricted to the Liouville theory we start with.

In \cite{Chen:2024unp, Hung:2024gma}, it is observed that the Liouville CFT admits a discretization as described in the previous section, leading to (\ref{disentangle}), despite the fact that the Liouville CFT is an irrational theory.

For the Liouville theory, the BCFT structure coefficients are known exactly \cite{Fateev:2000ik, Teschner:2000md, Ponsot:2001ng}. With $Q \equiv b + b^{-1}$, $c = 1 + 6Q^2$, and Virasoro primaries parametrized as $\alpha = Q/2 + iP_\alpha$, the conformal dimensions are
\be \label{conformaldim}
h_\alpha=\frac{c-1}{24}+P_\alpha^2~,\quad P_{\alpha}>0.
\ee
In a normalization detailed in \cite{Chen:2024unp}\footnote{This normalization, in which operators are invariant under $P\to -P$, makes the BCFT structure constants and crossing kernels directly proportional to $U_q(SL(2,\mathbb{R}))$ $6j$ symbols, taking exactly the same form as in rational CFTs \eqref{rationalC1}. The relevant representations are those of the ``modular double'' \cite{Faddeev:1999fe, Ponsot:1999uf, *Ponsot:2000mt}, a non-rational analogue of the Kazhdan-Lusztig equivalence \cite{kazhdan}.}, the structure coefficients are given by $b$-$6j$ symbols of $U_q(SL(2,\mathbb{R}))$ with $q = e^{i\pi b^2}$:
\bea \label{bcft coefficients}
C^{\sigma_3,\sigma_1,\sigma_2}_{\alpha_1,\alpha_2,\alpha_3}=\left(\mu(P_{\alpha_1}) \mu(P_{\alpha_2}) \mu(P_{\alpha_3})\right)^{1/4} \begin{Bmatrix}
\alpha_1 & \alpha_2 & \alpha_3\\
\sigma_3 & \sigma_1 &  \sigma_2
\end{Bmatrix}_b~.
\eea
The Virasoro crossing kernel in this normalization is
\be
F^{\text{Racah}}_{\alpha_6,\alpha_5} \begin{pmatrix}
\alpha_1 & \alpha_4
\\
\alpha_2 & \alpha_3
\end{pmatrix} =\frac{1}{\sqrt{2}}\sqrt{\mu(P_{\alpha_5}) \mu(P_{\alpha_6})}
\begin{Bmatrix}
\alpha_1 & \alpha_4 & \alpha_5\\
\alpha_3 & \alpha_2 & \alpha_6
\end{Bmatrix}_b~.
\ee

The parallel of these formulas with rational theories allows the Liouville CFT path integral $Z_{\text{Liouville}}(\Sigma)$ on a Riemann surface $\Sigma$ to also be expressed as a strange correlator analogous to (\ref{strange_correlator}),
i.e.~
\be \label{Liouvillestrange}
\mathcal{Z}_{\text{Liouville}}(R) = {}_\Lambda\langle \Omega^R|  \Psi_{\text{Liouville}} \rangle_\Lambda~,
\ee
where
$ |\Psi_{\text{Liouville}} \rangle_\Lambda$ is given by
\begin{align}\label{Psistateliouville}
&|\Psi_{\text{Liouville}} \rangle_\Lambda=  \prod_v \int_0^\infty dP_{\sigma_a}   
  \mu(P_{\sigma_a}) \prod_e \int_0^\infty dP_{\alpha_i}  \sqrt{\mu(P_{\alpha_i})}  \prod_{\triangle} 
\begin{Bmatrix}
    \alpha_i & \alpha_j & \alpha_k  \\
    \sigma_c & \sigma_a & \sigma_b
    \end{Bmatrix}_b|\{\alpha_i\}\rangle_\Lambda~,
\end{align}
where the Plancherel measure $\mu(P_\alpha)$ is given by
\be \label{plancherel}
\mu(P_{\alpha})=S_{0 \alpha}=4\sqrt{2} \sinh(2\pi P_\alpha b)\sinh(\frac{2\pi P_\alpha}{b})~,
\ee 
and $R$ parametrizes the hole size as explained in Sec.~\ref{SymQRGTN}.
Similarly, $_\Lambda\langle \Omega^R|$ is again given by
\be\label{Omegastate}
_\Lambda\langle \Omega^R|=  \prod_i \int_0^\infty dP_{\alpha_i}  {}_{\Lambda}\langle\{\alpha_i \}|\sum_{\{I_i\}}\prod_{\triangle}\left( \tilde\gamma^{\alpha_i \alpha_j \alpha_k}_{I_i I_j I_k}(R)\right)~,
\ee 
which are constructed from Virasoro conformal blocks as discussed in the previous section. 
Note that, similar to the discussion in the previous section, the vector space in which $|\Psi\rangle_\Lambda$ and $_\Lambda \bra{\Omega}$ reside is spanned by the tensor products of the edge Hilbert space, in the chosen triangulation of the Riemann surface $\Sigma$, and each edge is colored by the Virasoro primary labels $\alpha$.

These quantum $6j$ symbols of $U_q(SL(2,\mathbb{R}))$ also satisfy a Pentagon relation given by
\begin{equation} \label{pentagon}
\begin{aligned}
\frac{1}{\sqrt{2}}\int_0^\infty d P_{\delta_1} \mu(P_{\delta_1}) \begin{Bmatrix}
\alpha_1 & \alpha_2 & \beta_1\\
\alpha_3 & \beta_2 & \delta_1
\end{Bmatrix}_b \begin{Bmatrix}
\alpha_1 & \delta_1 & \beta_2\\
\alpha_4 & \alpha_5 & \gamma_2
\end{Bmatrix}_b \begin{Bmatrix}
\alpha_2 & \alpha_3 & \delta_1\\
\alpha_4 & \gamma_2 & \gamma_1
\end{Bmatrix}_b = \begin{Bmatrix}
\beta_1 & \alpha_3 & \beta_2\\
\alpha_4 & \alpha_5 & \gamma_1
\end{Bmatrix}_b \begin{Bmatrix}
\alpha_1 & \alpha_2 & \beta_1\\
\gamma_1 & \alpha_5 & \gamma_2
\end{Bmatrix}_b~.
\end{aligned}
\end{equation}

Therefore, like in (\ref{eq:URG}) and Fig.~\ref{removingbubble}, one can construct a QRG operator $\hat{U}^{\Lambda, \Lambda'}_{\text{RG}} $, which amounts to gluing tetrahedra, now using these quantum $6j$ symbols of $U_q(SL(2,\mathbb{R}))$. Let us denote that by $\hat{U}^{\Lambda, \Lambda'}_{\text{RG}}(U_q(SL(2,\mathbb{R})))$. When we use this operator to perform the QRG, it manifestly preserves the topological defect lines associated to \eqref{conformaldim}. Now we would like to argue that this is the discrete and non-perturbative version of the gravitational evolution effecting the RG flow operator $\hat{U}_{\text{3D gravity}}$ obtained in (\ref{rgoperatorinttbar}), i.e.
\be
\hat{U}_{\text{3D gravity}}{\underset{\text{Continuum}}{=}}\hat{U}_{\text{Virasoro TQFT}^2}{\underset{\text{Discrete}}{=}}\hat{U}^{\Lambda, \Lambda'}_{\text{RG}}(U_q(SL(2,\mathbb{R})))~.
\ee
In the language of Sec.~\ref{symqrgparagraph}, this identifies the Virasoro projector: the same-scale slab built from the $U_q(SL(2,\mathbb{R}))$ $6j$ symbols is the constraint projector $\Pi_{\Lambda}$ whose path-integral representation is the 3D quantum gravity path integral on $\Sigma\times I$. Initial evidence lies in the large $c$ semi-classical limit of the quantum $6j$ symbols of $U_q(SL(2,\mathbb{R}))$.
As shown in \cite{Chen:2024unp}, in the limit $b\to 0$, 
\begin{align}  \label{eq:classlim2}
\begin{Bmatrix}
\frac{Q}{2} + i \frac{l_4}{2\pi b}& \frac{Q}{2} + i \frac{l_5}{2\pi b} &\frac{Q}{2}+ i \frac{l_6}{2\pi b} \\
\frac{Q}{2} + i \frac{l_1}{2\pi b} &\frac{Q}{2} + i \frac{l_2}{2\pi b} & \frac{Q}{2}+ i \frac{l_3}{2\pi b}
    \end{Bmatrix}_b     \underset{b\to 0}{\to}    \exp{ \left(- \frac{ \textrm{Vol}(T\{l_i\}) + \sum_i l_i \theta_i/2   }{\pi b^2}  \right)}~, 
\end{align}
where  Vol$(T\{l_i\})$ is the volume of a hyperbolic tetrahedron with edges labeled $\{i\}$ and geodesic lengths given by $\{l_i\}$\footnote{The usual connection between $6j$ symbols and hyperbolic volume is in terms of the dihedral angles. We have used an identity to turn this into a relation in terms of geodesic lengths; see \cite{Chen:2024unp}.}, and $\theta_i$ is the dihedral angle sandwiched between the faces in the tetrahedron sharing the edge $i$. Thus, geometrical notions such as geodesic length naturally emerge in the large $N$ limit of $U_q(SL(2,\mathbb{R}))$, assigning a clear bulk geometrical interpretation to the primary label $P$.

The Plancherel measure that shows up at every internal label that is integrated over, contributes in the small $b$ limit,
\be \label{eq:geod}
\ln \mu(P(l)) \approx \frac{l}{4G_N}~,  \qquad P =  \frac{l}{2\pi b}~.
\ee
As the generalization of quantum ``dimension'', this is indeed related to the Cardy formula for the density of states \cite{Cardy:1986ie} and the entropy of the BTZ black hole \cite{McGough:2013gka}, a connection on which we will elaborate shortly.

Collecting the quantum $6j$ symbols and the Plancherel measure in $\hat{U}^{\Lambda, \Lambda'}_{\text{RG}}$, it gives in the $b\to 0$ limit
\be
\exp(-S_{EH} + \frac{\sum_e (2\pi - \Theta_e) l_e}{8 \pi G_N} )~,
\ee
where 
\be
S_{EH} =- \frac{1}{16\pi G_N}\int_{H} d^3 x \sqrt{g} (R +2)~,
\ee
and we plugged in $R=-6$. $\Theta_e = \sum_i \theta_i$ is the sum over dihedral angles of tetrahedra sharing this internal edge $e$, and can be identified with the Gibbons-Hawking-York term in the presence of codimension-two corner singularities, which is also called the Hayward term\cite{PhysRevD.47.3275, Takayanagi:2019tvn}. They do not contribute when one considers saddles with smooth geometries.  
The gravitational constant is related to the central charge by the standard Brown-Henneaux relation \eqref{brown}, $c  \approx 6/{b^2} = {3} /{2 G_N}$.

This means that $\hat{U}^{\Lambda, \Lambda'}_{\text{RG}}(U_q(SL(2,\mathbb{R})))$ reduces to the semi-classical gravitational integral in the AdS spacetime at the limit $b\to 0$ or $c\to \infty$. At finite $c$, the theory involves a summation over all possible \textit{quantum geometries} corresponding to each RG step, which are constructed from the complete set of quantized geodesic configurations building up the quantum spacetimes.

\begin{figure}
	\centering
	\includegraphics[width=0.8\linewidth]{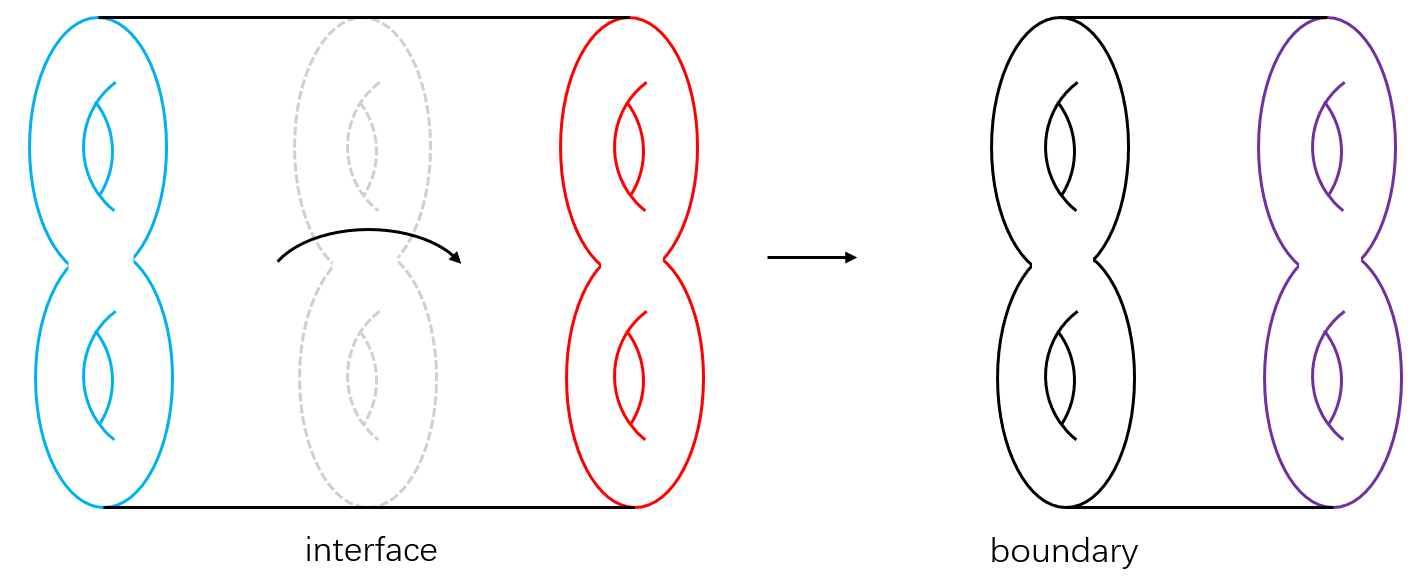}
	\caption{Left: the Virasoro TQFT computation on $\Sigma \times I$. Right: folding along the middle interface produces the state $|\Psi_{\text{Liouville}}\rangle$ from two copies of the Virasoro TQFT.}
	\label{wormholefolding}
\end{figure}

To show that $\hat{U}^{\Lambda, \Lambda'}_{\text{RG}}(U_q(SL(2,\mathbb{R})))$ is also the desired flow operator non-perturbatively in $b$, we wish to show that this is the correct discrete version of two copies of the Virasoro TQFT evaluated on the same manifold of the form $\Sigma \times I$. 
To do so, we use (\ref{virasoro_liouville}). Consider folding the path integral of the Virasoro TQFT on $\Sigma \times I$ in the middle of $I$. This produces a state $|\Psi\rangle$
\be \label{Liouvilegauging}
|\Psi\rangle = \int d\vec P \,\rho^C_{g,n}(\vec P) |\mathcal{F}^C_{g,n}(\vec P)\rangle  \otimes| \overline{\mathcal{F}^C_{g,n}}(\vec P) \rangle~,
\ee
where the ``folded bra'' becomes $| \overline{\mathcal{F}^C_{g,n}}(\vec P) \rangle$ which now denotes the anti-holomorphic basis for conformal blocks. 
The state $|\Psi\rangle$ is thus a sum over basis holomorphic and antiholomorphic conformal blocks in exactly the linear combination that coincides with the Liouville theory. Our state $|\Psi_{\text{Liouville}}\rangle$ is indeed the same state by construction (despite being expressed in a basis that breaks up the closed conformal blocks into open ones tiling the same surface). Therefore, our discrete representation of $|\Psi_{\text{Liouville}}\rangle$ should be equivalent to the folded Virasoro TQFT evaluated on $\Sigma \times I$. 
In particular, the RG operator $\hat{U}^{\Lambda, \Lambda'}_{\text{RG}}(U_q(SL(2,\mathbb{R})))$, which essentially grows the $I$ direction of $\Sigma \times I$ in $|\Psi_{\text{Liouville}}\rangle$, using tetrahedra associated to the $U_q(SL(2,\mathbb{R}))$ $6j$ symbols, should also be a discrete realization of two copies of the Virasoro TQFT on $\Sigma \times I$, as illustrated in Fig.~\ref{wormholefolding}.  In other words, the quantum $6j$ symbols indeed reproduce two copies of the Virasoro TQFT, as expected of the quantized theory of gravity. This relation is reminiscent of the connection between rational Turaev-Viro TQFT and two copies of Chern-Simons theories we reviewed in Appendix \ref{reviewtvtqft}\footnote{The CFT in that case is the non-chiral Wess-Zumino-Witten model\cite{Witten:1988hf, Elitzur:1989nr}.}.

This provides a rare example of a discrete, Turaev-Viro-type state-sum for a non-rational Virasoro/Teichm\"{u}ller TQFT, extending the Ponzano-Regge model \cite{Regge:1961px, *ponzano} to spacetimes with negative cosmological constant. The $6j$ symbols are diffeomorphism-invariant; boundary triangulations can be freely changed via the 3-2 Pachner move (Pentagon relation), while the 4-1 move diverges \cite{Collier:2023fwi, Hung:2024gma}. The SymQRG procedure avoids this by using only the 3-2 move for block-spin transformations, with a canonical bulk triangulation (Fig.~\ref{azure}) that keeps the procedure finite. We stress that incorporating the asymptotic boundary $\bra{\Omega}$ is essential for a well-defined answer \cite{Maldacena:1997re, Gubser:1998bc, Witten:1998qj}.

The bulk theory contains a continuous spectrum of topological line operators \cite{Chen:2024unp}, the counterparts of Wilson lines in Chern-Simons theory. They correspond to Verlinde lines in the Liouville CFT, and the anyons are related to BTZ black holes \cite{Banados:1992wn, McGough:2013gka}: the conformal dimensions \eqref{conformaldim} correspond to hyperbolic $SL(2,\mathbb{R})$ holonomies, while the Plancherel measure $\mu(P_\alpha)$ \eqref{plancherel} reproduces the Cardy density of states \cite{Cardy:1986ie} and, semi-classically via \eqref{eq:geod}, the BTZ entropy. The $\omega$-loop projector \eqref{Omegaflux} becomes $\hat{\omega} = \int d\alpha\, \mu(P_\alpha) \mathcal{P}^\alpha$, whose eigenvalues give the ``quantum area operator'' for generalized entropies \cite{Mertens:2022ujr, *Wong:2022eiu, Chua:2023ios}\footnote{Similar defect operators appear in JT gravity \cite{Jafferis:2019wkd} and topological string theory \cite{Donnelly:2020teo, *Jiang:2020cqo}. The interpretation of these operators as wormholes and black holes is discussed in \cite{Chua:2023ios}.}.

The $T\bar T$ deformation preserves these Virasoro topological lines. Therefore, $T\bar T$-driven flows are symmetry-preserving RG flows admitting a gravitational interpretation.

In the class of examples controlled by the Virasoro/Teichm\"{u}ller data, the corresponding SymQRG operator is expressed in terms of
$\hat{U}^{\Lambda, \Lambda'}_{\text{RG}}(U_q(SL(2,\mathbb{R})))$. Although we obtained this operator by studying $|\Psi_{\text{Liouville}}\rangle$, it goes far beyond the Liouville theory: all the operator does is make the symmetries preserved in the RG process explicit, with their representation theory determining its form, so the Liouville theory merely serves as a vehicle for deriving it. In other words, $|\Psi_{\text{Liouville}}\rangle$ is one solution to (\ref{eq:URG}), but {\it every} modular invariant CFT (which we do not yet know other than Liouville theory) preserving at least the same collection of symmetries would supply an independent solution to (\ref{eq:URG}). 
The take-home message is: symmetries help us identify a {\it universal} QRG operator $\hat{U}^{\Lambda, \Lambda'}_{\text{RG}}(U_q(SL(2,\mathbb{R})))$ that admits an interpretation as a {\it quantum gravitational path integral}. In other words, 
\be
\textrm{SymQRG(Virasoro lines)} = \textrm{QG}
\ee
which is the central result of the current paper.

\paragraph{Relation to the lattice Liouville theory.} There is a known lattice integrable model, built from the operators of quantum Teichmüller theory, that recovers the Liouville CFT in the thermodynamic limit \cite{Faddeev:1985gy}. Its transfer matrix is assembled from the same tetrahedral building blocks discussed above, and it supplies another seed state ${}_{\Lambda}\bra{\Omega}$ for the SymQRG flow, away from the RG fixed point. We review this construction and its relation to our framework in Appendix~\ref{latticeliouvilleappendix}.

\subsection{Exact 2D/3D equivalence at finite $N$ and universal semi-classical gravity dual at large $N$}\label{ETH}

\begin{figure}
	\centering
	\includegraphics[width=0.7\linewidth]{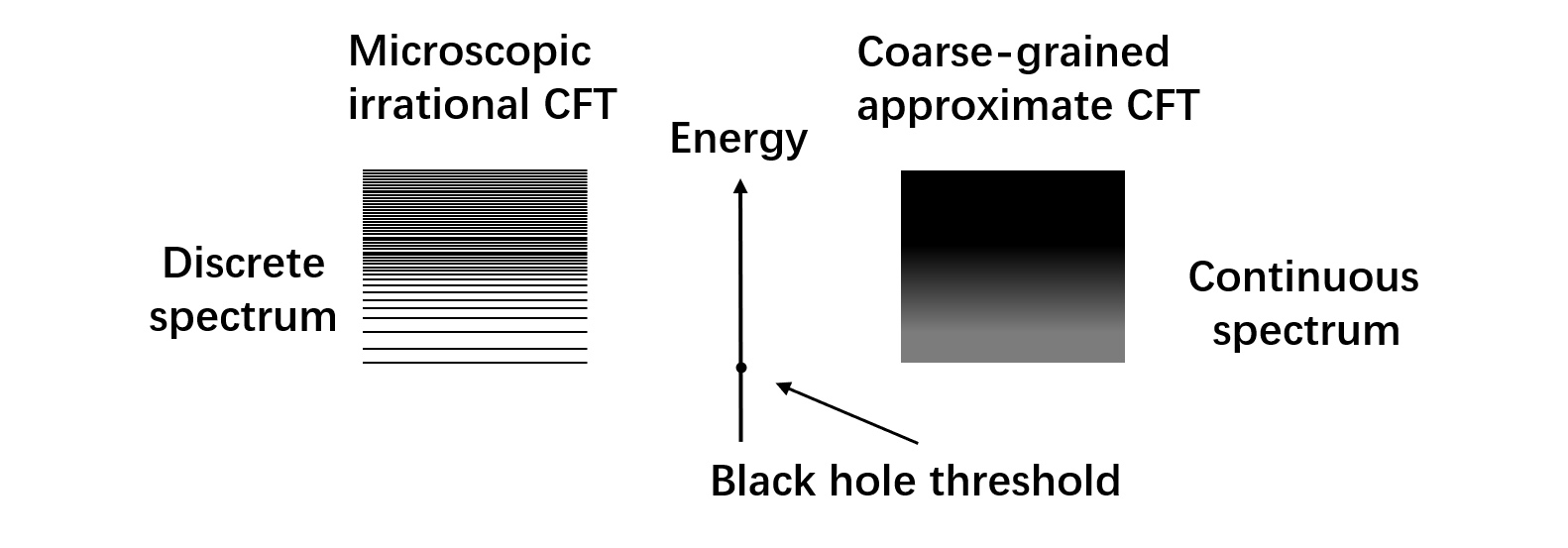}
	\caption{Comparison of the spectra of individual irrational large $N$ holographic CFTs (left, discrete but dense) and of the coarse-grained theory (right, continuous).}
	\label{spectrumaverage}
\end{figure}

In this section, we focus on the continuous spectrum \eqref{conformaldim}, which describes theories related to pure gravity, for example the Liouville theory\cite{Chen:2024unp, Hung:2024gma}. What is the connection with general holographic CFTs? The answer to this question was first proposed in \cite{Jackson:2014nla}, and recently also emphasized in the discussion for the role played by ensemble averages in holography in terms of coarse-graining and the eigenstate thermalization hypothesis (ETH)\cite{Srednicki:1994mfb, *PhysRevA.43.2046, Belin:2020hea, Saad:2019pqd, *Stanford:2020wkf, *Pollack:2020gfa, Belin:2023efa, Jafferis:2024jkb, Chandra:2022bqq}.

Before addressing this question, let us first revisit the concept of ``Quantum Gravity''. Quantum gravity is defined as a theory that is non-perturbatively complete and reduces to Einstein gravity in the semi-classical limit.

General irrational large $N$ holographic CFTs are expected to have a discrete spectrum instead of a continuous one that we studied in this section, and similarly for the associated topological defect lines. However, the general large $N$ holographic CFTs have a large gap and a dense spectrum for heavy states above the black hole threshold that is approximately continuous; see Fig.~\ref{spectrumaverage}. The idea is that for individual 2D holographic CFTs, the heavy state dynamics of these theories can be universally approximated by the results outlined in this section. The coarse-grained result has a continuous spectrum \eqref{conformaldim} with Cardy density of states\cite{Cardy:1986ie}, and the heavy-state matrix elements are given by random matrices, from a generalized version of the ETH \cite{Srednicki:1994mfb, *PhysRevA.43.2046, Jackson:2014nla, Chandra:2022bqq}. These coarse-grained matrix elements involving heavy states can be obtained by studying the bootstrap equations involving identity in certain dual channels\cite{Collier:2019weq, Kusuki:2021gpt, *Kusuki:2022wns, Numasawa:2022cni}, and then applying an argument similar to \cite{Hartman:2014oaa} to the OPE coefficients, to extend the regime of validity in the large $N$ limit. The final result is universally captured by the quantum group $U_q(SL(2,\mathbb{R}))$, where quantum Teichmüller theory and geometries emerge. More details on the average over open structure coefficients in a CFT and its connection to the emergence of gravity will be presented in a forthcoming publication \cite{ourRTN}. 

In other words, even with the presence of additional fields and fine structures in a 2D CFT, the universality of heavy-state statistics at large $N$ ensures the universal emergence of a thermodynamic description with a 3D Einsteinian gravity sector in the bulk. States below the threshold will contribute to extra fields coupled to the gravity sector. For example, we can have worldline actions coupled to the Einstein-Hilbert action as in \cite{Chandra:2022bqq}. In this sense, gravity serves as a universal long-range entangled phase of matter, when there are infinitely many degrees of freedom.\footnote{In fact, since we propose that the bulk emerges from the Virasoro fusion kernels, including the light states related by analytic continuation, the entire CFT may be understood as arising from anyon condensation within a larger TQFT, where the Virasoro TQFT appears as a subsector.} Building on this discussion, we emphasize the two distinct roles that Liouville theory plays. On one hand, it serves as an exact dual to pure gravity in a precise sense \cite{Collier:2023fwi, Coussaert:1995zp, *Henneaux:2019sjx, Chua:2023ios, Chen:2024unp}. On the other hand, it functions as a universal approximation for generic 2D holographic theories when viewed through a coarse-grained lens \cite{Jackson:2014nla, Collier:2019weq, Chandra:2022bqq, Belin:2023efa, Jafferis:2024jkb}.

The above line of logic on coarse-graining explains why the bulk dual theory universally behaves as semi-classical gravity at large $N$. In terms of the RG operator and 3D quantum states, this is the statement that\footnote{This should be compared with an intrinsic average of CFTs (or equivalently Lagrangian algebras, as we will explain), which correspond to a state of the form $\ket{\Psi}=\sum_{i} p_i \ket{\Psi}_i$, where $p_i$ represents the chosen weights for the ensemble average. This class of ket states is interpreted as having a bulk dual consisting only of smooth geometries, as studied in \cite{Maloney:2020nni, Afkhami-Jeddi:2020ezh, Aharony:2023zit}, and more recently in connection with SymTFT in \cite{Barbar:2023ncl, Dymarsky:2024frx}.}:
\be
\begin{aligned}
\hat{U}_{\text{RG}}&\overset{\text{Large N}}{\underset{\text{Coarse-graining}}{\approx}} \hat{U}_{\text{RG}, \text{3D gravity}}\\
\ket{\Psi}  &\overset{\text{Large N}}{\underset{\text{Coarse-graining}}{\approx}}\ket{\Psi}_{\text{3D gravity}}
\end{aligned}
\ee
However, equipped with our machinery, especially the exact tensor network construction of the boundary CFT and bulk dual, we always have an \textit{exact, non-perturbative} 2D/3D equivalence at finite $N$: the sandwich construction holds on symmetry grounds, the RG map relies only on exact crossing symmetry, and therefore any self-consistent 2D CFT yields an exact three-dimensional state-sum representation. With a discrete spectrum, the bulk deviates from a purely geometrical Einstein theory: light matter fields and stringy fields appear, with couplings fine-tuned to satisfy the non-perturbative bootstrap equations\cite{Jackson:2014nla}, and it approximates the geometrical bulk described in this section only in the semi-classical limit. The AdS/CFT correspondence is thus not merely an approximate correspondence valid after coarse-graining at the large $N$ semi-classical level: its true power lies in its unique ability to define non-perturbative completions of Einstein gravity through CFTs.

\section{Quantum Gravity from Algebra}\label{algebra}
\subsection{Topological holographic principle}\label{topologicalholography}

The topological holographic principle was proposed in \cite{Gaiotto:2020iye, Ji:2019jhk,*Kong:2020cie,*Kong:2020jne,*Chatterjee:2022kxb, Freed:2022qnc, Apruzzi:2021nmk, Freed:2018cec, Kong:2019byq, Moradi:2022lqp, Bhardwaj:2017xup, Albert:2021vts, Vanhove:2018wlb, Aasen:2020jwb}. As reviewed in Sec.~\ref{symqrgparagraph}, it states that the path integral of a $D$ dimensional theory with symmetry on a manifold $\Sigma_D$ can be expressed in terms of the path integral of a $D+1$ dimensional TQFT over the ``sandwich'' $\Sigma_D \times I$, with a physical boundary condition (our state $_{\Lambda}\bra{\Omega}$) on one side, and a topological boundary condition (our SymTFT ground state $\ket{\Psi}_{\Lambda}$, combined with the bulk) on the other.

In the context of 2D CFTs, the connection with a 3D TQFT is well-established. The 3D Chern-Simons theory path integral evaluated on a 3D manifold $M_3$ with boundary $\partial M_3$ famously reduces to the WZW model on the 2D boundary \cite{Witten:1988hf, Elitzur:1989nr}. It is also well known that the 2D boundary theory is anomalous and can only exist as boundaries of a 3D theory. The anomaly is canceled by the 3D bulk via the anomaly inflow mechanism. In the sandwich construction, the resulting 2D theory is anomaly-free and thus exists as a consistent stand-alone theory. The anomaly-free condition is encoded in the topological boundary of the sandwich. 

In the previous sections, we have focused primarily on the physical boundary, represented by the state $_{\Lambda}\bra{\Omega}$, and its connection to the RG flow via the bulk SymTFT. In doing so, we have assumed the existence of a CFT, corresponding to a well-defined state $\ket{\Psi}_{\Lambda}$ that encodes the topological boundary. One of the key observations in this paper is that each individual 2D CFT corresponds to a specific state $\ket{\Psi}_{\Lambda}$. In this section, we reverse the logic and focus on constructing the topological boundary condition. The key questions to consider are the following. How can we determine a topological boundary condition for a given SymTFT? What are the algebraic conditions, and how do they relate to CFT consistency relations?
To begin, we provide a brief overview of the FRS construction
\cite{Fuchs:2002cm}. For a CFT rational with respect to a chiral algebra
$V$, the Moore--Seiberg data of the representations of $V$ form a modular
tensor category
$\mathcal{C}=\mathrm{Rep}(V)$
\cite{Moore:1988qv,*Moore:1989vd}. The chiral algebra $V$ is encoded in the $\bra{\Omega}$ part of the overlap, appearing as conformal blocks. Using the Turaev-Viro construction \cite{Turaev:1992hq}, a 3D TQFT can be constructed using $\mathcal{C}$, which serves as the SymTFT associated with the symmetries of the theory. The Virasoro non-rational case considered elsewhere in this paper is
treated as a continuously labeled extension of this structure beyond the
finite semisimple framework. To obtain an intrinsic 2D CFT, it suffices to determine a suitable topological boundary condition, enabling the sandwich construction to function. In the language of the overlap \eqref{overlap}, this corresponds to identifying \textit{which} topological ground state matches each individual CFT.

FRS provided an answer to this question \cite{Fuchs:2002cm}. In fact, we have already encountered this answer when working backwards from a CFT. These answers are encoded in the BCFT structure coefficients used to construct the state $\ket{\Psi}_{\Lambda}$. In Sec.\ref{roleofpsi module category}, we briefly outline the algebraic conditions that the topological boundary condition must satisfy and explain how these conditions relate to the CFT consistency requirements. From these algebraic properties, quantum gravity in the bulk emerges naturally. In Sec.\ref{algebraicimplications}, we discuss the implications of these algebraic properties for quantum gravity. Specifically, we address the factorization puzzle concerning wormhole contributions in the gravitational path integral and explain how the out-of-time-ordered correlator is inherently encoded within the symmetry framework.

\subsection{Looking for $\ket{\Psi}_{\Lambda}$: From crossing symmetry to anyon condensation/gauging}

\label{roleofpsi module category}
\subsubsection{Crossing symmetry and emergence of 3D bulk}

The CFT path integral was found to be related to a wavefunction $\ket{\Psi}_{\Lambda}$ introduced in (\ref{psistate}). 
This is supposed to be \textit{a} ground state of the Turaev-Viro (or Levin-Wen) model. 

On a general 2D manifold $\Sigma$ with noncontractible cycles, the ground-state subspace is generically degenerate. For example, on a torus, the number of degenerate ground states of the Turaev-Viro model with input fusion category $\mathcal{C}$ is equal to the number of simple objects in its Drinfeld center $D(\mathcal{C})$. 
The basis of ground states on the torus can be constructed by inserting Wilson lines/ribbon operators along the non-contractible cycle in the path integral over a solid torus. 

Which of the ground states is selected in (\ref{psistate})? 
The information lies in the labels $\sigma$ in (\ref{psistate}). In the CFT, $\sigma$ labels the conformal boundary conditions, at least one of which preserves one copy of the chiral algebra.

The set of labels corresponds to objects of a \textit{module category} $\mathcal{M}_{\mathcal{C}}$ of ${\mathcal{C}}$, each defining a topological boundary condition of the Turaev-Viro TQFT\cite{Fuchs:2002cm, 2012CMaPh.313..351K, Fuchs:2012dt, Bhardwaj:2017xup}. Bulk anyons $c\in\mathcal{C}$ act on boundary excitations $m\in\mathcal{M}_{\mathcal{C}}$ via fusion maps $W_{cm}^{m'}$\cite{Kitaev:2011dxc, Kong:2014qka, Kong:2017hcw}, and the associativity of these maps\cite{etingof2003,Fuchs:2002cm} introduces coefficients $\tilde C^{n m m'}_{c_1 c_2 c_3}$ that play the role of boundary-bulk $6j$ symbols with indices in both $\mathcal{C}$ and $\mathcal{M}_{\mathcal{C}}$. These satisfy a pentagon equation:
\be \label{4ptcrossing1}
\tilde{C}^{m_4 m_1 m_2}_{c_1 c_5 c_4}\tilde{C}^{m_2 m_3 m_4}_{c_3 c_5 c_2} =\sum_{c_6}\tilde{C}^{m_3 m_1 m_2}_{c_1 c_2 c_6}\tilde{C}^{m_1 m_3 m_4}_{c_3 c_4 c_6} F^{\text{Racah}}_{c_6,c_5} \begin{pmatrix}
c_1 & c_4 
\\
c_2 & c_3 
\end{pmatrix}.
\ee
This is the crossing invariance of BCFT four-point functions (\ref{4ptcrossing}), which ensures re-triangulation invariance and enables the symmetry-preserving RG procedure; in the bulk, it corresponds to gluing additional tetrahedra via the SymTFT $6j$ symbol. 

The ground states of the Levin-Wen and Turaev-Viro models satisfy the no-flux constraint, and it has been shown in \cite{Levin:2004mi, PhysRevB.79.195123} that this constraint ensures crossing symmetry and re-triangulation invariance, as expressed in \eqref{4ptcrossing1}. Conversely, wavefunctions derived from \eqref{4ptcrossing1} inherently satisfy the no-flux condition, identifying them as ground states. Intuitively, this connection arises because the plaquette projection operator in the Levin-Wen model is constructed by effectively gluing a tetrahedron onto the wavefunctions. The work of FRS \cite{Fuchs:2002cm, *Fuchs:2003id, *Fuchs:2003id1, *Fuchs:2004xi, *Fjelstad:2005ua, *Frohlich:2006ch, *Fjelstad:2006aw} demonstrates an equivalence between the associativity coefficients $\tilde C^{n m m'}_{c_1 c_2 c_3}$ and the BCFT OPE coefficients, translating them into the crossing relation in the CFT. This identification is a direct manifestation of locality in the boundary-to-bulk map. 

This is precisely the statement that \textit{crossing symmetry} of the 2D CFT is responsible for the emergence of a 3D bulk. This was first proposed in the large $N$ limit in higher dimensions\cite{Heemskerk:2009pn}. Here we see an exact quantum statement in the context of the AdS$_3$/CFT$_2$ correspondence. Under SymQRG, these crossing constraints manifest themselves as the Wheeler-DeWitt equation.

Four-point crossing symmetry alone is insufficient to define self-consistently the CFT defined on arbitrary 2D manifolds. The additional constraints on the topological boundary are related to \textit{modular invariance} in the CFT. More specifically, as already mentioned above, the CFT boundary conditions are also classified by the objects of $\mathcal{M}_{\mathcal{C}}$. This piece of data is closely related to the modular invariant \textit{spectrum} of the full CFT. 

\subsubsection{Modular invariance and anyon condensation/gauging}

As shown by FRS \cite{Fuchs:2002cm, *Fuchs:2003id, *Fuchs:2003id1, *Fuchs:2004xi, *Fjelstad:2005ua, *Frohlich:2006ch, *Fjelstad:2006aw}, a self-consistent, modular-invariant CFT on arbitrary Riemann surfaces requires a symmetric special Frobenius algebra $\mathcal{A} \subset \mathcal{C}$, whose module category $\mathcal{M}(\mathcal{A})$ can be identified with the boundary-condition category $\mathcal{M}_{\mathcal{C}}$ (see \cite{Lou:2020gfq} for a concise review). 

\begin{figure}
	\centering
	\includegraphics[width=0.8\linewidth]{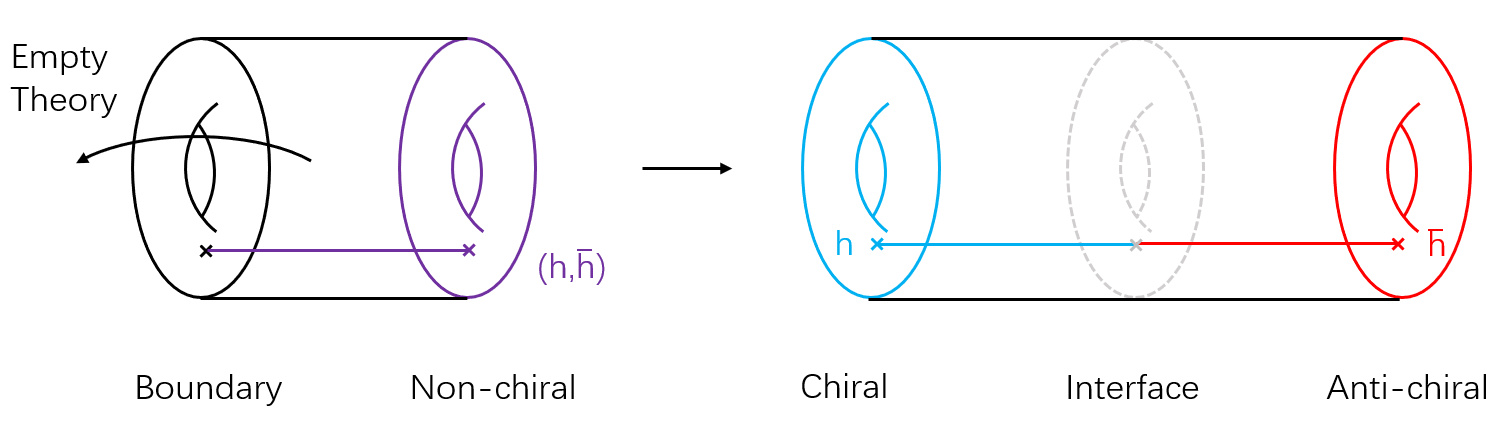}
	\caption{We can unfold the boundary to be an interface or a surface operator. This operator pairs the left movers and right movers into a modular invariant theory.}
	\label{foldingsandwich}
\end{figure}

\begin{figure}
    \centering
    \begin{minipage}[b]{0.45\linewidth}
        \centering
        \begin{overpic}[width=\linewidth]{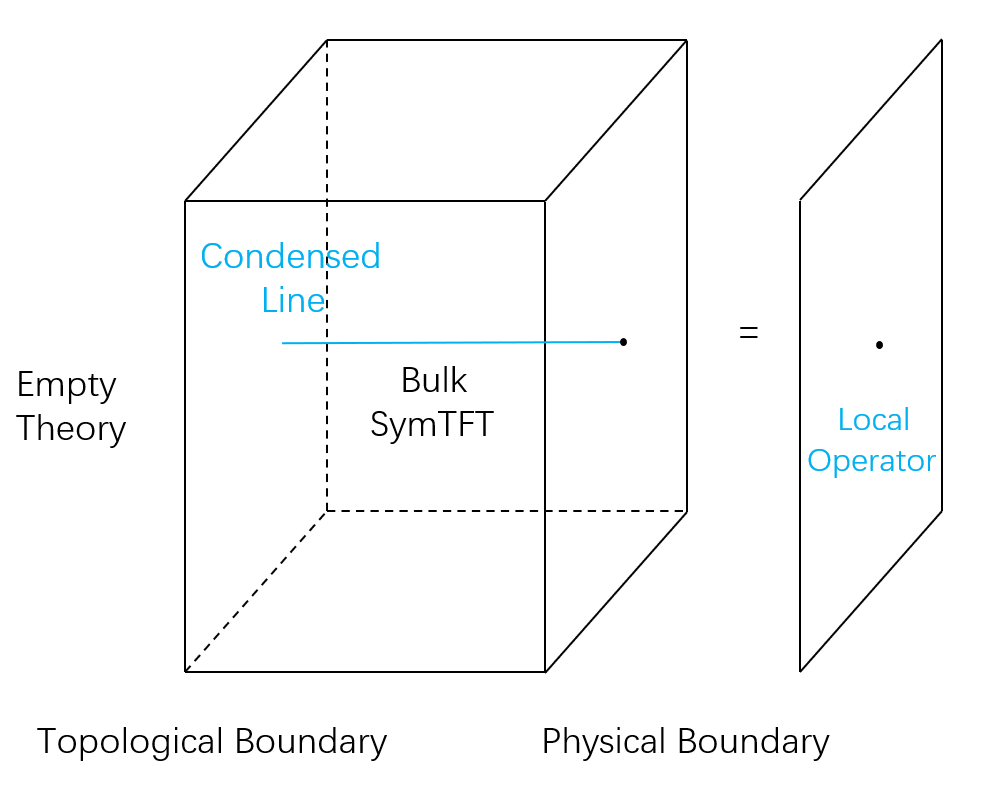}
        \end{overpic}
    \end{minipage}
    \hfill
    \begin{minipage}[b]{0.45\linewidth}
        \centering
        \begin{overpic}[width=\linewidth]{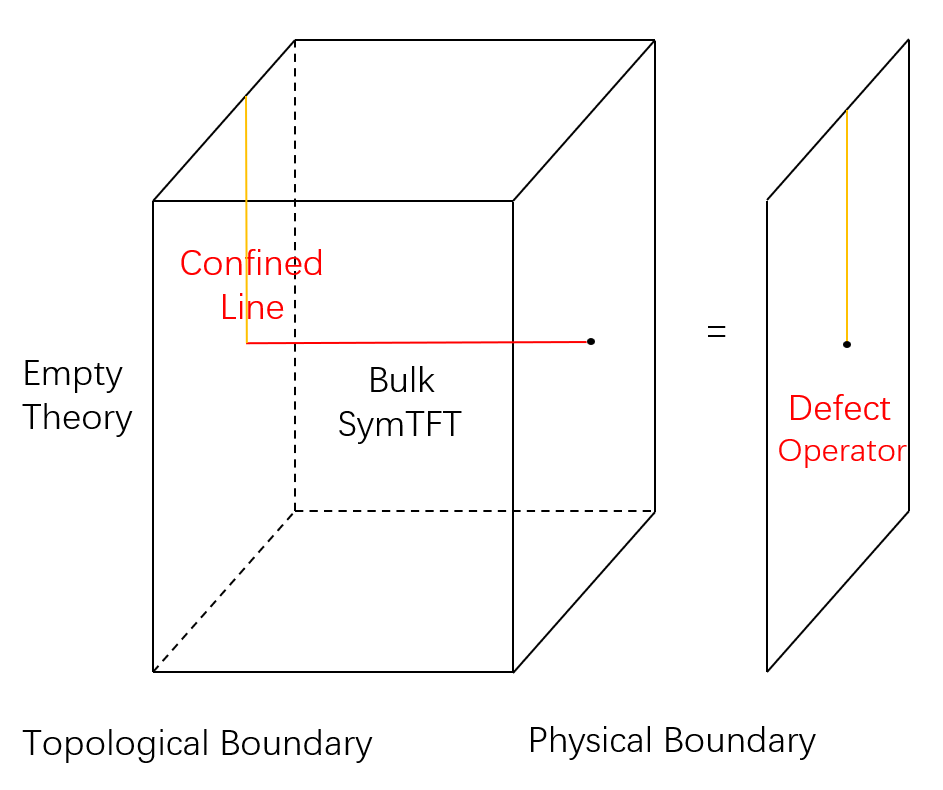}
        \end{overpic}
    \end{minipage}
    \caption{Left: local operators in 2D come from condensed line operators in the 3D bulk; Right: defect operators in 2D come from confined line operators in the 3D bulk. In both pictures, there is only an empty theory to the left of the topological boundary after anyon condensation. }\label{sandwich operators}
\end{figure}

In the left-right CFT with SymTFT $\mathcal{C} \boxtimes  \overline{\mathcal{C}}$, $\mathcal{A}$ defines a topological boundary condition $\mathcal{B}_\mathcal{A}$\cite{Fuchs:2002cm}, across which ``anyon condensation''\cite{Bais:2002pb, Bais:2008ni, Bais:2008xf, Kong:2013aya} converts the topological theory to the trivial phase. The condensed anyons form a Lagrangian algebra $\mathcal{L}_\mathcal{A} \subset \mathcal{C}\boxtimes \overline{\mathcal{C}}$. Bulk lines that do not condense are ``confined'' to the boundary, where they become the topological defect lines of the 2D theory\cite{Komargodski:2020mxz, Lin:2022dhv} (see Fig.~\ref{sandwich operators}). The Frobenius algebra conditions ensure that all bulk lines are either condensed or confined, leaving an empty theory beyond the topological boundary. 

The Lagrangian algebra $\mathcal{L}_\mathcal{A}$ is in one-to-one correspondence with modular invariants\cite{Kong:2009inh}; each topological boundary $\mathcal{B}_\mathcal{A}$ thus ensures a modular-invariant spectrum. Equivalently, the condensation can be understood as gauging higher-form symmetries\cite{Benini:2022hzx}\footnote{This idea dates back to \cite{Maldacena:2001ss, Gukov:2004id}, and we thank Anatoly Dymarsky for bringing it to our attention.}: before gauging, each TQFT ground state produces a non-modular-invariant combination of conformal blocks\cite{Witten:1988hf}; after gauging, a unique ground state is singled out, collapsing the 3D bulk and yielding an anomaly-free 2D theory. The pairing of left and right movers can also be visualized by unfolding the doubled theory into a surface operator\cite{Kapustin:2010if, Fuchs:2012dt, 2012CMaPh.313..351K} (Fig.~\ref{foldingsandwich}), which lies at the heart of the FRS construction\cite{Fuchs:2002cm}. 

For example, diagonal RCFT corresponds to condensing diagonal pairs $i\otimes \bar i$, with $\mathcal{M}_{\mathcal{C}} = \mathcal{C}$ and $\mathcal{A} = 1$. Liouville theory corresponds to this diagonal gauging for two copies of the Virasoro TQFT\cite{Chen:2024unp, Collier:2023fwi}, as in \eqref{Liouvilegauging}. Other choices of $\mathcal{A}$ yield theories with non-zero spinning states\footnote{A nontrivial Frobenius algebra $\mathcal{A}$ will provide a theory other than Liouville theory, whose bulk SymTFT is two copies of the Virasoro TQFT, potentially with non-zero spinning states.}; for general irrational CFTs, finding non-trivial $\mathcal{A}$ remains an important open problem\footnote{Strictly speaking, the continuously labeled Virasoro setting has
infinitely many inequivalent simple sectors and therefore lies beyond the
standard finite semisimple notion of a fusion category. However, the success in Liouville theory and the existence of the Virasoro TQFT\cite{Collier:2023fwi} suggest the existence of a generalized mathematical framework that extends to the irrational case.}.

The bottom line is that each such Frobenius algebra $\mathcal{A}$ determines a \textit{modular invariant} full CFT associated with the Moore-Seiberg data $\mathcal{C}$, and at the same time, the $\mathcal{M}(\mathcal{A})$ for given $\mathcal{A}$ can be identified with $\mathcal{M}_{\mathcal{C}}$\cite{Fuchs:2002cm, *Fuchs:2003id, *Fuchs:2003id1, *Fuchs:2004xi, *Fjelstad:2005ua, *Frohlich:2006ch, *Fjelstad:2006aw}. In other words, the Frobenius algebra $\mathcal{A}$ determines both the full modular invariant CFT spectrum, and also its boundary conditions and boundary changing operators. 

We are now ready to address the question of the precise state $\ket{\Psi}_{\Lambda}$ selected by a given CFT. This state arises directly from the bulk path integral combined with the topological boundary. Specifically, the state resides on one of the boundaries of $\Sigma_2 \times I$, while the other boundary is labeled by $\mathcal{M}_{\mathcal{C}}$, as depicted in Fig.~\ref{sandwich}\footnote{There is an alternative way to understand this state. Instead of directly using the generalized $6j$ symbols as in (\ref{psistate}), one can also consider coloring the topological boundary by the coefficients $\mu$ defining the product morphism of the corresponding Frobenius algebra $\mathcal{A}$, as illustrated in \cite{Fuchs:2002cm, Chen:2022wvy}.}. The topological holographic principle implies that the topological boundary conditions of the Turaev-Viro TQFT associated with $\mathcal{M}_{\mathcal{C}}$, or equivalently $\mathcal{M}(\mathcal{A})$, provide the complete data required to determine the full CFT. We can explicitly express the corresponding state for any 2D CFT on the string-net Hilbert space using the BCFT OPE coefficients, as shown in \eqref{psistate}\footnote{This state for which the labels $\sigma \in \mathcal{M}_{\mathcal{C}}$ has appeared and been discussed in the tensor network literature exploring different Levin-Wen ground states\cite{Fuchs:2002cm, Lootens:2020mso}. In some cases, different $\mathcal{M}_{\mathcal{C}}$ can provide different representations of the same ground state.}.

\subsection{Dynamics vs. kinematics: where is the information encoded?}\label{dynamicskinematics}

We have discussed in detail the roles played by the state $\ket{\Psi}$, which encodes information about the topological boundary of the sandwich; and the state $\bra{\Omega}$, which governs the physical boundary. In \cite{Gaiotto:2020iye}, it is observed that the physical boundary contains the dynamics of the theory, while the SymTFT and the topological boundary are responsible solely for the symmetries, when the symmetries form a discrete group. This result appears to contradict our discussion and the AdS/CFT correspondence, where $\bra{\Omega}$ seems to be merely kinematical, constructed from conformal blocks, while the state $\ket{\Psi}$, which is associated with a topological boundary, carries the structure coefficients of the CFT, which are typically associated with the theory's dynamics.

In fact, for each individual theory, there can be multiple ways to realize it as a sandwich, and, as anticipated in the Introduction, the more of the theory's symmetry is made explicit in the SymTFT, the simpler the physical boundary state: in the maximal case it becomes purely kinematical, and the bulk together with the topological boundary determines the theory's dynamics. In 2D CFT, the corresponding maximal bulk theory arises from imposing all lines commuting with the Virasoro symmetry. If the theory possesses an extended chiral algebra, the bulk theory constructed from it will generally explicitly impose fewer symmetries, as the requirement that the lines commute with the full chiral algebra imposes additional constraints dictated by the extended algebra.\footnote{We thank Conghuan Luo, Shu-Heng Shao and John McGreevy for the comments and discussions on this point.} The collection of such topological lines is generally infinite, even in rational theories with an extended chiral algebra. In particular, when focusing only on the Virasoro algebra, the structure of many rational CFTs resembles that of irrational CFTs. In this light, we believe that many more theories may admit a non-trivial holographic bulk.

The Ising CFT discussed in the Introduction illustrates both ends of this dichotomy: with only the $\mathbb{Z}_2$ symmetry made explicit, the SymTFT is the toric code and $\bra{\Omega}$ retains the OPE data; extracting the complete set of symmetries, including the non-invertible Kramers-Wannier line, upgrades the SymTFT to the Turaev-Viro TQFT/Levin-Wen model based on the Ising category, and, as in other rational theories, the physical boundary becomes purely kinematical. We propose that the AdS/CFT correspondence can be understood as the \textit{maximal} SymTFT sandwich reconstruction of the CFT.

The sandwich construction thus cleanly separates the kinematics of the lower-dimensional QFT into a boundary condition, with the dynamics captured by the bulk together with the topological boundary; this is precisely the division of labor realized in the AdS$_3$/CFT$_2$ examples studied here.
\begin{figure}
	\centering
	\includegraphics[width=0.75\linewidth]{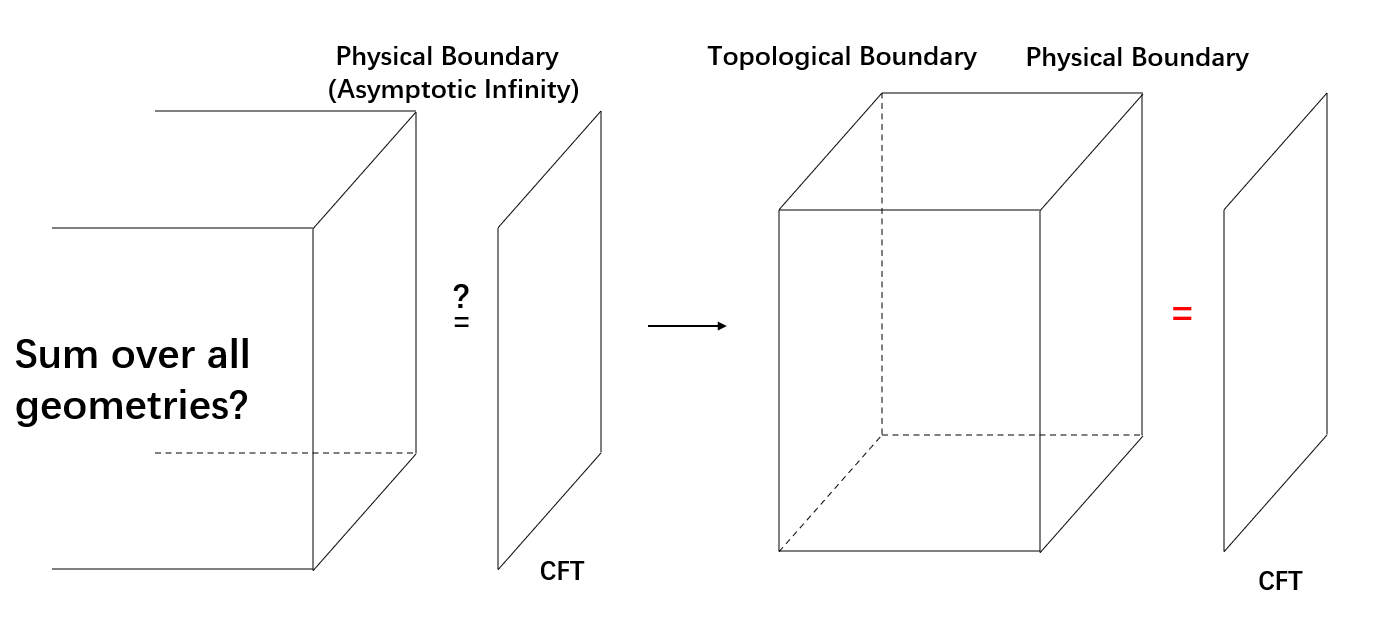}
	\caption{On the left, we have the conventional AdS/CFT dictionary, where the bulk path integral involves summing over geometries compatible with a given boundary condition. In general, this sum does not match the partition function of a CFT. On the right, the sandwich construction introduces a topological boundary that can be interpreted as a choice of measure for the bulk path integral. This choice ensures that the bulk path integral precisely reproduces the partition function of a CFT.}
	\label{adscftandsandwich}
\end{figure}

We emphasize that in the conventional study of the AdS/CFT correspondence, we only specify the asymptotic boundary condition. This precisely corresponds to our $_\Lambda\bra{\Omega}$ state, which is constructed from conformal blocks and represents the conformal symmetry at the boundary. The conventional approach then involves summing over all the smooth bulk saddles compatible with these boundary conditions. For example, with a torus boundary, we include a sum over $SL(2,\mathbb{Z})$ modular images, corresponding to different contractible cycles in the solid torus, and this ensures the modular invariance of a potential 2D dual \cite{Maldacena:1998bw, Maloney:2007ud}. However, this approach generally does not yield a {\it bona fide} unitary 2D quantum mechanical theory. In 2D CFTs, modular invariance typically arises from a carefully organized spectrum of states that transform into each other under $SL(2,\mathbb{Z})$, rather than from summing over modular images of individual sectors. This latter perspective allows us to interpret the topological boundary condition together with the 3D bulk, or $\ket{\Psi}_{\Lambda}$ as determining the quantum measure of the 3D gravity path integral, effectively selecting which bulk geometries (and also non-geometrical solutions) to include in order to construct a dual to 2D CFTs by design. Notice that even in the semi-classical limit, the choice of quantum measure determines which saddles are included in the saddle point approximation when the integration contour is deformed\footnote{A simple example is the Stokes phenomenon; see for example \cite{Witten:2010cx, Harlow:2011ny}.}. Different choices of $\ket{\Psi}_{\Lambda}$ would give us different UV completions for the bulk quantum gravity theory, and the dictionary is not complete before we specify this information\footnote{Unless there are some other reasons for there to be a unique choice of $\ket{\Psi}_{\Lambda}$\cite{McNamara:2020uza}, so we can leave this choice implicit.}. See Fig.~\ref{adscftandsandwich} for an illustration of these two dictionaries. We will argue below that this gives an exact factorization mechanism once the CFT-specific topological boundary is fixed, with the semi-classical factorization puzzle recast as a question about the bulk quantum measure. If we view the CFT as providing a self-consistent UV completion of AdS quantum gravity, the question is not whether one should sum over ``all'' geometries in an unrestricted sense, but which quantum measure, contour, and projection data define the non-perturbative bulk theory associated with a given CFT; this question is inherently answered by the structure of the CFT.\\

We note that our perspective here differs from the gauging-based discussion of \cite{Benini:2022hzx}, where the corresponding bulk is viewed as trivial (the sandwich squeezed flat, leaving only the empty theory beyond the topological boundary); we return to this comparison in Sec.~\ref{algebraicimplications}.

Finally, let us comment on the observation that flows driven by $T\bar T$ in {\it any} CFT appear to always produce an emergent gravity bulk, referred to as a ``fake bulk'' in \cite{Belin:2020oib}. In light of the discussion in this paper, it becomes clear that the bulk described in \cite{Belin:2020oib} corresponds to $\hat{U}^{\Lambda, \Lambda'}_{\text{RG}}$ in (\ref{eq:URG}). As we have demonstrated, the $T\bar T$ deformation preserves all topological symmetries, and the RG flow it generates can be expressed through an RG kernel, effectively represented by a gravitational path integral over quantum geometries. However, the complete bulk path integral is governed by the state $\ket{\Psi}_{\Lambda}$, with the complete quantum measure determined by the infrared topological boundary condition. This boundary condition specifies which quantum geometries are included in the path integral. In the context of \cite{Belin:2020oib}, this crucial aspect is implicitly embedded within the initial CFT path integral. 

We have demonstrated that the exact correspondence between a 2D boundary and a 3D bulk theory arises from the full set of CFT self-consistency relations, which can be succinctly summarized by \textit{crossing symmetry} and \textit{modular invariance} \cite{Moore:1988qv, Moore:1988uz, Friedan:1986ua, Fuchs:2002cm}. These self-consistency relations of 2D CFTs can also be elegantly encapsulated into algebraic equations connected to the 3D bulk SymTFT.

In summary, the \textit{crossing symmetry} ensures the existence of a topological ground state ${\ket{\Psi}}$ that satisfies the no-flux condition. This property allows the SymTFT slab to relate physical boundary states at different cutoff scales, turning the RG Callan-Symanzik equation into the Wheeler-DeWitt/no-flux constraint. On the other hand, \textit{modular invariance} provides the framework to end this procedure by specifying the quantum measure in 3D. This measure is encoded in the choice of a specific ground state ${\ket{\Psi}}$, which allows the termination and collapse of the 3D bulk into an intrinsically well-defined 2D theory.

\subsection{Implications for Quantum Gravity}\label{algebraicimplications}

We have explored the algebraic constraints on the topological boundary and their physical significance. In this subsection, we will examine the implications of these algebraic properties within the symmetry framework for Quantum Gravity. Specifically, we will first demonstrate how the topological boundary addresses the factorization puzzle. Following this, we will show that the second-sheet analytic continuation of the conformal blocks entering the Lorentzian out-of-time-ordered correlator is exactly determined by the Moore-Seiberg data of 2D CFTs, revealing Chaos from (topological) Order.

\subsubsection{Wormholes, null states, and factorization}

The inclusion of spacetime wormhole saddles in the gravitational path integral has led to important insights, from reproducing the Page curve\cite{Almheiri:2019qdq, Penington:2019kki} to the factorization puzzle\cite{Maldacena:2004rf, *Yin:2007at, *Arkani-Hamed:2007cpn}: should a given bulk describe a unique CFT, or an ensemble average\cite{Saad:2019lba, McNamara:2020uza}? A non-perturbative resolution requires understanding the null states and gauge equivalences in quantum gravity\cite{Marolf:2020xie, Jafferis:2017tiu}\footnote{See also earlier works pointing out that the number of connected components of space cannot serve as a non-perturbative diffeomorphism-invariant operator in quantum gravity \cite{Marolf:2012xe, *Bao:2015nca, *Berenstein:2016pcx, *Berenstein:2017abm}.}.

In our framework, there are two types of constraints in the CFT and its dual gravity, as explained in Sec.~\ref{roleofpsi module category}. The first is the Wheeler-DeWitt equation, which ensures that $\ket{\Psi}_{\Lambda}$ is a ground state of the Turaev-Viro model and satisfies the no-flux condition, allowing us to re-triangulate the 2D radial slice. In the language of the CFT, this corresponds to the crossing invariance condition \eqref{4ptcrossing}. The second constraint is the gauging constraint in the bulk SymTFT, associated with modular invariance on the boundary. This constraint seems to be the one encoding the gauge equivalence of different topologies, a perspective also suggested in \cite{Benini:2022hzx}.

\begin{figure}
	\centering
	\includegraphics[width=0.6\linewidth]{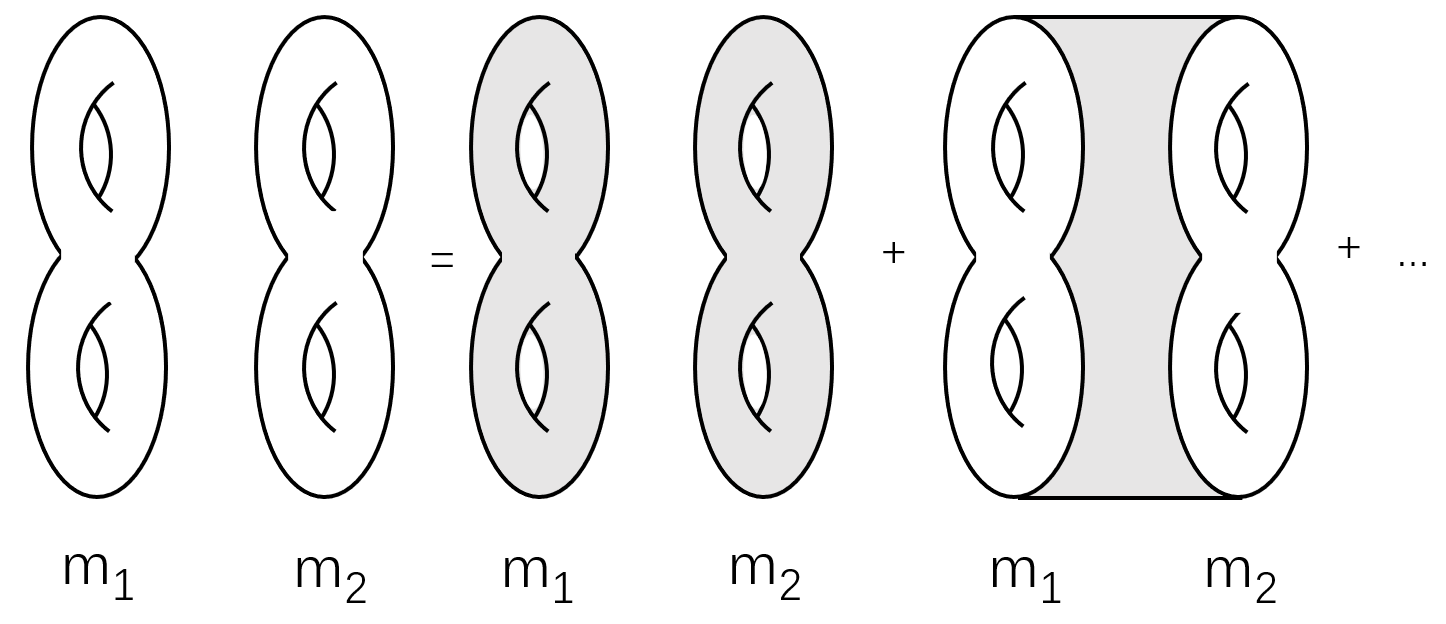}
	\caption{The computation of $Z_1(m_1) Z_2(m_2)$ in the conventional semi-classical holographic dictionary: the two leading saddles are the disconnected configuration and the connecting wormhole, and the result does not factorize.}
	\label{factorization}
\end{figure}

\begin{figure}
	\centering
	\includegraphics[width=0.75\linewidth]{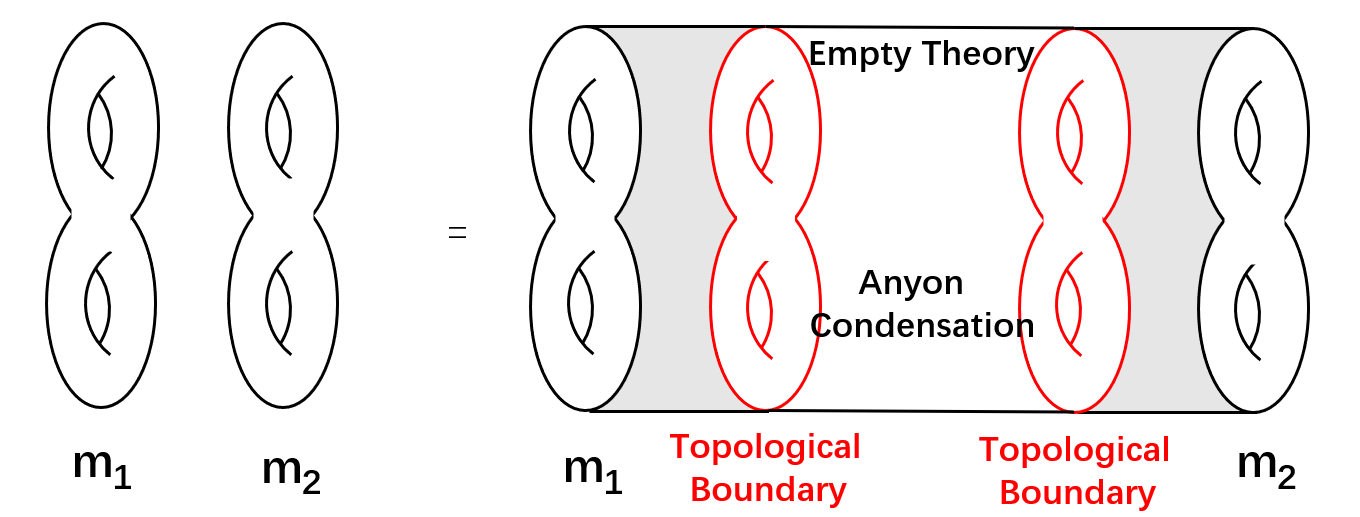}
	\caption{In the sandwich construction, the same computation factorizes by construction: the topological boundaries condense or confine all bulk anyons, so the ``connected'' bulk drawn above is in fact two ``disconnected'' pieces.}
	\label{factorization2}
\end{figure}

Let us illustrate this with the factorization puzzle for partition functions\cite{Maldacena:2004rf}. Consider computing two copies of the CFT partition functions, $Z_1(m_1) Z_2(m_2)$, where $m_1$ and $m_2$ are the moduli of the 2D CFTs. According to the conventional semi-classical holographic dictionary, this corresponds to fixing boundary conditions with two asymptotic boundaries and finding all bulk solutions compatible with these boundary conditions. The two dominant saddles in this setup are the disconnected configuration and the connected wormhole configuration, as illustrated in Fig.~\ref{factorization}\footnote{We deliberately use the genus two Riemann surfaces as our 2D boundary manifolds to get a connected wormhole saddle in 3D. In the case of torus boundaries in pure gravity, it is known that there are no such saddles. However, we can perform an exact computation for these off-shell configurations using some assumptions\cite{Saad:2019lba, Cotler:2020ugk, Jafferis:2024jkb}. However, subtleties with respect to the bulk mapping class group in 3D gravity and the connection to random matrix spectral statistics in 2D \cite{Cotler:2020ugk, DiUbaldo:2023qli} are still not completely understood. Our exact formulation might give hints on this problem.}. This leads to the factorization puzzle. However, in the sandwich construction, the way to recover the original factorized answer is straightforward: we simply insert two topological boundaries into the ``connected bulk'', effectively splitting it into two ``disconnected'' pieces separated by the empty theory. This factorization arises because anyons in the bulk are either condensed (they are gauged and disappear) or confined (they are stuck on the topological boundary), preventing them from leaking into the intermediate empty region. See Fig.~\ref{factorization2} for an illustration. The sandwich construction thus provides an exact factorization mechanism, built from the algebraic properties of the CFT and SymTFT, once the CFT-specific topological boundary is fixed. We emphasize that it is not necessary to stick to this language of topological boundary conditions to observe factorization. Instead, we can interpret it as defining the quantum measure for the bulk 3D gravity path integral as we explained above. From this perspective, the factorization of the sandwich-defined bulk theory arises from a precise definition of the measure for the bulk path integral (which might include non-geometrical contributions).

Each choice of topological boundary condition appears to correspond to a choice of an ``$\alpha$ state'' in the language of \cite{Marolf:2020xie, Coleman:1988cy, *GIDDINGS1988854}, where the partition functions factorize, as suggested by \cite{Benini:2022hzx}; in our formalism, this is encoded in the state $\ket{\Psi}_{\Lambda}$. Our perspective differs slightly from \cite{Benini:2022hzx}, however: we neither collapse the entire bulk onto the physical boundaries nor view it as entirely trivial. The bulk remains non-trivial within the sandwich, and the semi-classical factorization puzzle is recast as identifying the quantum measure for the gravitational path integral over this non-trivial bulk, encoded in the topological boundary condition, or the state $\ket{\Psi}_{\Lambda}$. We emphasize that this reorganization relies precisely on the
topological nature of the slab: the kernel is universal, and it acts
identically on the state of an individual CFT and on an ensemble mixture (cf.~Sec.~\ref{ETH}), both satisfying the
no-flux condition. Semi-classical computations that probe only the
gravitational kernel are therefore common to both readings; whether a given
bulk describes one theory or an ensemble is decided at the innermost layer of
the sandwich, by the topological-boundary state alone.

The AdS/CFT correspondence is expected to identify a given CFT with a particular quantum theory of gravity. From the SymTFT perspective, the non-perturbative completion contains UV-sensitive data such as the quantum measure that is specified by the dual CFT.  Generic choices or assumptions about such data specifying the bulk path integral most likely do not have a quantum CFT dual. We believe that translating these insights back into the semi-classical picture can guide us toward a deeper understanding of quantum gravity.

Translating Fig.~\ref{factorization2} back into the semi-classical picture of Fig.~\ref{factorization}, and exploring connections to earlier proposals in other gravitational or string-theoretic models \cite{Marolf:2020xie, Blommaert:2019wfy, *Blommaert:2021fob, Saad:2021rcu, *Saad:2021uzi, Eberhardt:2021jvj, *Eberhardt:2020bgq}, are important directions for future investigation, as is the precise relation between the Hopf algebra described in \cite{Gesteau:2024gzf}\footnote{Quantum groups are Hopf algebras by definition. In the case of rational CFTs, recovering a quantum group from the modular tensor category associated with the CFT can be done by the Tannaka duality.}, the ``baby universe category'' proposed in \cite{jacob}, and the Moore-Seiberg symmetry category $\mathcal{C}$ for 2D CFTs.

\subsubsection{Chaos from (topological) Order}

Recent studies have revealed the fascinating connection between quantum gravity and quantum chaos \cite{Shenker:2013pqa, kitaev}. It is natural to ask, where is quantum chaos in our formalism based on topological symmetries? Naively, the study of symmetry provides an organizing framework and defines the superselection sectors of a theory, positioning it in seemingly direct contrast to the disordered nature of chaos, making them, at best, tangential properties of the theory. However, we will see that in 2D CFTs the Moore--Seiberg data exactly
determine the second-sheet analytic continuation of the conformal blocks
entering the Lorentzian OTOC. The full OTOC additionally depends on the
spectrum and OPE coefficients, which in our framework are encoded in the
topological-boundary state.

For any two CFT primary operators $W$ and $V$, the OTOC is defined as the normalized four-point function
\be
C(t)=\frac{\langle W(t) V W(t) V \rangle_{\beta}}{\langle W(t) W(t) \rangle_{\beta} \langle V V \rangle_{\beta}}~,
\ee
in the thermal state with inverse temperature $\beta$. 

\begin{figure}[H]
	\centering
	\includegraphics[width=0.4\linewidth]{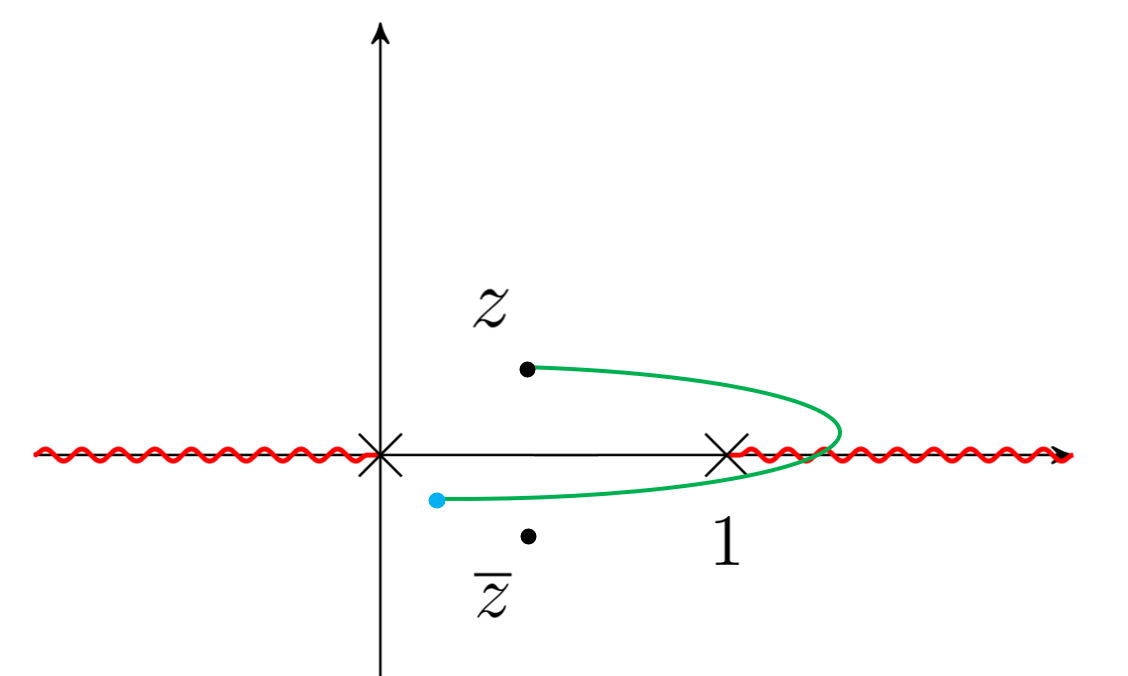}
	\caption{The kinematical regime related to the computation of the OTOC is a Lorentzian kinematics that lies on the ``second sheet'', meaning we need to traverse the branch cut on $(1, \infty)$ and pick up a monodromy for the chiral conformal blocks.}
	\label{analyticblock}
\end{figure}

Lorentzian OTOCs are obtained by analytic continuation of Euclidean four-point functions expanded in conformal blocks $\mathcal{F}^{VV}_{WW}(h_i,z)$\cite{DSDlorentzian, Cornalba:2006xk, *Cornalba:2006xm, *Cornalba:2007zb, *Cornalba:2007fs, Shenker:2013pqa}\footnote{Studies of Lorentzian CFT in the analytic conformal bootstrap program in recent years have led to fruitful new results, see \cite{Hartman:2022zik}.}. In the OTOC channel, the cross ratio $z$ traverses the branch cut at $z=1$ (Fig.~\ref{analyticblock}), picking up a monodromy that distinguishes the Lorentzian correlator from its Euclidean counterpart\cite{Hartman:2015lfa, Caron-Huot:2017vep}. In the Regge limit $z, \bar{z} \to 0$, assuming Virasoro identity block dominance, one recovers the maximal Lyapunov exponent for holographic CFTs\cite{Roberts:2014ifa, Fitzpatrick:2014vua}.

\begin{figure}
	\centering
	\includegraphics[width=0.6\linewidth]{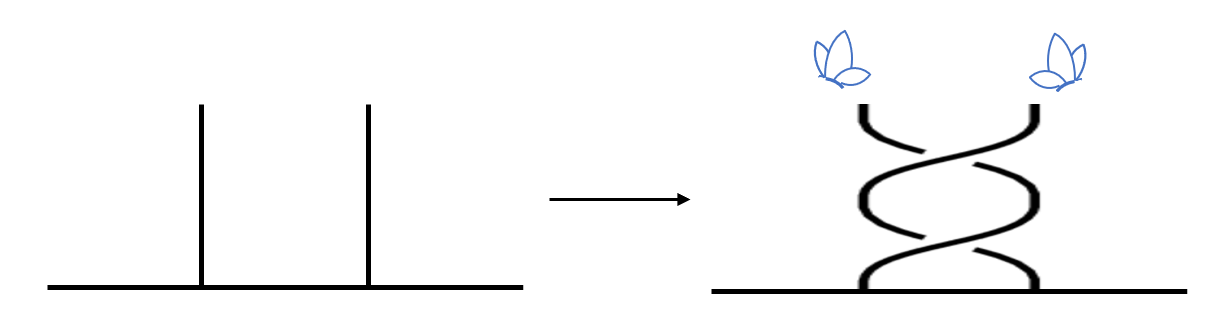}
	\caption{The monodromy of conformal blocks is determined by the braid matrix of the underlying Moore-Seiberg data. This controls the Lyapunov behavior and the butterfly effect, encapsulating the idea of ``Chaos from Order''.}
	\label{braidingblock}
\end{figure}

At the level of conformal blocks, this analytic continuation is exact and
requires no additional dynamical assumption: winding around the branch
point $z=1$ is implemented by a monodromy matrix $M$
\cite{Moore:1988qv,*Moore:1989vd}, such that
\be
\mathcal{F}^{VV}_{WW}(h_i,z) \xrightarrow{(1-z) \to e^{-2\pi i} (1-z)} \sum_{h_j}M_{ij} \mathcal{F}^{VV}_{WW}(h_j,z)~.
\ee
In addition, the monodromy matrix can be expressed in terms of the $6j$ symbols, as \footnote{We use a slightly different convention from the sections above, where now the chiral primaries are labeled by their conformal dimensions $h$ instead of the previously used $\alpha$.} 
\be
M_{ij}=\sum_{h_k} e^{-2\pi i(h_k-h_W-h_V)} F_{h_i, h_k} \begin{pmatrix}
h_V & h_V 
\\
h_W & h_W 
\end{pmatrix} F_{h_k,h_j} \begin{pmatrix}
h_V & h_W 
\\
h_V & h_W  
\end{pmatrix}~.
\ee
The fact that the second-sheet continuation of the conformal blocks can
be written in terms of the topological data is certainly not a
coincidence. The reason is that the monodromy is just a combination of braid moves, as shown in Fig.~\ref{braidingblock}\footnote{Here, we are employing a reversed logic compared to \cite{Moore:1988qv, Moore:1988uz}: Suppose we already possess the Moore-Seiberg data of the CFT. We can use this data to determine the conformal blocks on the second sheet. Notably, this data was originally derived by first analyzing the monodromies of the conformal blocks.}. From the bulk-boundary correspondence\cite{Witten:1988hf}, or unfolding the sandwich construction, we can also see the monodromy explicitly as anyons nontrivially linking themselves in the bulk, as shown in \cite{Caputa:2016tgt, *Gu:2016hoy}.

This point of view of understanding quantum chaos and OTOC from the exchange algebra was emphasized in the context of 2D CFT in \cite{Jackson:2014nla, Turiaci:2016cvo, Kusuki:2019gjs}. In the context of holography, this is the dual of the bulk exchange algebra found in \cite{Schoutens:1993hu, *Kiem:1995iy}, responsible for the exponential blueshift coming from the Dray and 't Hooft shockwave effect\cite{DRAY1985173}. As noted in \cite{Jackson:2014nla}, the goal of conformal bootstrap is to ensure that correlation functions remain single-valued in Euclidean signature. In other words, quantum statistics are invisible or ``confined'' in Euclidean, non-chiral, full CFTs. However, after we solve the bootstrap equations which are topological in nature as we explained, and manage to hide the quantum statistics, these topological data will get ``deconfined'' in Lorentzian signature, controlling the Lorentzian dynamics, causality (commutator), and information scrambling of the theory, which gives ``\textit{Chaos from Order}''. 

Finally, we want to point out an interesting fact. Rational CFTs, whose SymTFTs are governed by modular tensor categories, which are finite in nature, are \textit{not} chaotic. Quantum chaos only appears when we have irrational CFTs, or infinitely many degrees of freedom. This suggests ``infinitely more is more different''\cite{kltalk}, and we will discuss the quest for some new infinite-dimensional mathematical structure in the discussion section. 

\section{Discussion}\label{discussionsection}
In this paper, we have shown how a holographic bulk emerges naturally from considerations of a topological symmetry-preserving quantum RG flow. When the symmetry preserved is the infinite collection of Virasoro lines in a 2D theory, then a 3D gravitational bulk emerges. 
The full bulk path-integral however sensitively depends on the complete non-perturbative definition of the CFT: the latter essentially determines a 3D state $|\Psi\rangle$ that gives the precise quantum measure of the bulk path-integral, and this information is theory dependent even for theories preserving the same collection of symmetries. 

So far the explicit examples we have considered, where non-perturbative structure coefficients have been solved completely, are restricted to rational CFTs, and also the Liouville theory, which, despite being irrational, is very special. The lack of a vacuum state in the normalizable physical spectrum of Virasoro TQFT\cite{Collier:2023fwi} and thus its diagonal gauging\footnote{The normalizable physical states, forming a complete basis in the Hilbert space of the Virasoro TQFT, are given by Liouville conformal blocks corresponding to hyperbolic monodromies. This set includes all states above the black hole threshold but excludes the vacuum state \cite{Collier:2023fwi}. The vacuum state has been argued to arise from a path integral that does not produce normalizable states within the TQFT.} leads to a density of states that deviates from Cardy's universal result, making it a rather atypical example. 

While a certain choice of measure of the bulk path-integral may lead to ensemble averages of CFTs, we hope that we have shown, at least with certain examples, that in principle an appropriate measure can be chosen to recover a precise CFT.  How such a quantum measure can be constructed for more general irrational CFTs, perhaps through the search for a {\it Frobenius algebra} of the {\it ``category''} of Virasoro lines, is certainly worth further exploration. 

It would be extremely important to find examples that are more typically holographic, although we believe the evidence presented here suggests that the framework of SymQRG/QG continues to hold more generally. Nevertheless, our construction appears to suggest that a non-perturbative understanding of the CFT offers the ultimate guidance to constructing the corresponding bulk Quantum Gravity. 

We conclude this paper with a discussion of some interesting future directions.

\subsubsection*{von Neumann algebra and 2D CFT}

As discussed above, the bulk dual of the CFT is represented by a sum over quantum geometries, providing an exact description at finite $N$. In the large $N$ limit, certain quantum geometries emerge as dominant saddles in this sum, allowing for a semi-classical, low-energy geometrical description. To better understand this emergence, recent studies have proposed algebraic quantum field theory (AQFT), specifically von Neumann algebras, as a suitable framework, noting the appearance of emergent Type III von Neumann algebra structures in the large $N$ limit\cite{Leutheusser:2021qhd, *Leutheusser:2021frk}.

A natural question arises: how are these two distinct algebraic approaches to field theories and emergence of geometry related, and how are categorical structures, such as the $6j$ symbols, encoded within von Neumann algebras? In 2D CFT, there appears to be a dictionary connecting these approaches. AQFT in 2D is rigorously constructed using ``conformal nets'', and its connection to the functorial formalism is discussed in \cite{Carpi:2015fga}. A key element in this framework is the role of superselection sectors and ``subfactor theory'' \cite{Jones:1983kv, *Longo:1989tt}\footnote{We thank Keiichiro Furuya for discussions on this topic and collaboration on a related project.}. This approach might provide a new understanding of the Type II algebra structures observed in the context of higher-dimensional gravity \cite{Witten:2021unn}, which play a role in black holes and cosmology \cite{Chandrasekaran:2022eqq, *Chandrasekaran:2022cip, vanderHeijden:2024tdk}. Applied to 3D gravity, this Type II algebra is analogous to the JT gravity case studied in \cite{Penington:2023dql, *Kolchmeyer:2023gwa}. In the current situation, we even achieve its UV completion, since we start from a 2D CFT. We will report further results on these ideas in future work.

\subsubsection*{Infinitely more is more different}

New structures are known to emerge in the study of quantum many-body problems, often encapsulated by the slogan ``more is different'' \cite{doi:10.1126/science.177.4047.393}. In Sec.~\ref{tensor network}, we saw an example of this principle, where an \textit{infinite} bond dimension for the $\bra{\Omega}$ state allows us to place a gapless field theory on a lattice in an exact fashion. Here, it is probably more appropriate to say that ``infinitely more is more different'' \cite{kltalk}. This infinity stems from the infinite spacetime entanglement inherent in field theories, analogous to the Type III algebra structure in quantum field theories \cite{Witten:2018zxz}.

In the case of irrational CFTs, an additional layer of infinity appears due to the infinite number of primary fields in $\ket{\Psi}$. This infinity drives the large $N$ limit of a CFT and enables chaotic behavior, in contrast to rational CFTs. It mirrors the emergence of Type III factors in the large $N$ limit as discussed above \cite{Leutheusser:2021qhd, *Leutheusser:2021frk}. For the SymTFT description, this implies that an infinite number of anyons are required to describe the bulk theory of gravity. Consequently, the Virasoro-line data of an irrational CFT lie beyond a
fusion category in the standard sense, which has only finitely many
isomorphism classes of simple objects. This fact also limits our Turaev-Viro construction for $U_q(SL(2,\mathbb{R}))$: as explained in Sec.~\ref{TQFTSECTION}, the 4-1 Pachner move diverges, while the construction on manifolds with asymptotic boundaries, combined with boundary conformal blocks, remains convergent and consistent with the Virasoro/Teichmüller TQFT \cite{Chen:2024unp, Hung:2024gma}. This suggests the existence of a novel mathematical structure, essential for the study of \textit{quantum gravity}.

This resonates with an analogy from quantum mechanics: the commutation relation
\be
[\hat{x},\hat{p}]=i \hbar
\ee
cannot hold in a finite-dimensional Hilbert space, as taking the trace on both sides leads to a contradiction. A precise understanding of these new categorical structures, which involve infinitely many inequivalent
simple sectors, remains an open and important question.

\subsubsection*{Stringy dual}

In principle, the strategies proposed in this paper can be applied to the study of theories with stringy duals, which exhibit a more exotic spectrum of physical states, leading to non-geometric objects arising from higher-spin fields. For instance, the bulk dual of symmetric orbifolds of certain CFTs has been shown to correspond to tensionless strings in AdS$_3$ \cite{Eberhardt:2018ouy, *Eberhardt:2019ywk}. It would be interesting to examine the topological symmetries of these theories in greater detail, to understand how the bulk AdS$_3 \times$ S$^3 \times \mathbb{T}^4$ emerges from our construction and how it relates to the worldsheet perspective. The bulk dual would still be built from tetrahedra, but some of the charges might acquire the interpretation of extra compact dimensions, rather than the geodesic distances of the pure-gravity case.

\subsubsection*{Higher dimensional AdS/CFT correspondence}

Our tensor network formalism depends significantly on the unique properties of 3D gravity and 2D CFT, particularly its topological nature. Extending this line of reasoning may provide insight into constructing a non-perturbative bulk dual in higher dimensions. However, since gravitons are dynamical in higher dimensions, we want to explore several possibilities to address this tension.  One possibility, as noted in \cite{Apruzzi:2024htg}, is that the bulk theory may lose its topological character when the symmetry involves a continuum of topological lines. A second possibility is that the bulk topological invariance could be spontaneously broken, even if it initially exhibits a topological structure. Alternatively, it is possible that the bulk theory remains genuinely topological in the non-perturbative regime. The latter two possibilities might be related if we expand around fixed backgrounds.

An important lesson from the study of string theory and recent studies of generalized symmetries is that understanding the non-perturbative aspects of quantum field theories or their gravitational duals benefits greatly from a categorical approach, which involves examining structures across various dimensions and their interactions. Consequently, in higher-dimensional CFTs and holographic correspondences, the analysis of local correlation functions alone seems to be insufficient to capture all non-perturbative information, and a comprehensive understanding of the theory requires examining the roles of defects and boundaries within it. Finally, we note that the higher-dimensional analogue of the $T\bar{T}$ deformation currently only applies in the semi-classical large $N$ limit\cite{Hartman:2018tkw, Araujo-Regado:2022gvw}. Away from the large $N$ limit, does this SymTFT perspective provide a non-perturbative understanding?

\subsubsection*{Holographic complexity}
The volume/action of the bulk theory is proposed to be connected to the notion of complexity in the CFT \cite{Susskind:2014rva, *Stanford:2014jda, *Brown:2015bva, Belin:2021bga, *Belin:2022xmt, *Myers:2024vve}, although precise definitions on both sides have been tricky to identify. Our discrete formulation of the path-integral could be used to define notions of complexity in field theories that translate directly between the bulk and the boundary: each $U_q(SL(2,\mathbb{R}))$ $6j$ symbol (the volume of a hyperbolic tetrahedron) is a ``gate'' building up the path-integral, to which a definite complexity can be assigned. This seems particularly promising given that Liouville theory has been proposed to capture path integral complexity for 2D CFTs\cite{Caputa:2017urj, *Caputa:2017yrh}.

 \section*{Acknowledgement}

We thank Nathan Benjamin, Christian Ferko, Keiichiro Furuya, Hao Geng, Sarah Harrison, Daniel Jafferis, Hong Liu, Conghuan Luo, Juan Maldacena, John McGreevy, Shinsei Ryu, Sahand Seifnashri, Shu-Heng Shao, Herman Verlinde and Frank Verstraete for helpful discussions. We thank Wei Bu, Wan Zhen Chua, Christian Ferko, Hao Geng, Ziming Ji, Chen-Te Ma, Yixu Wang, Zixia Wei and Chen Yang for comments and discussions on the draft. We acknowledge the use of OpenAI's GPT 5.6 and Anthropic's Claude Fable 5 for language editing of this manuscript. We thank Lin Chen, Gong Cheng, Keiichiro Furuya, Hao Geng, Zheng-Cheng Gu, and Bing-Xin Lao for collaboration on related projects. We thank Zhenhao Zhou for plotting the 3D diagrams in this paper. YJ thanks Juan Maldacena for the invitation and hospitality during a visit to the Institute for Advanced Study, where the idea of this paper originated from discussions. YJ and NB acknowledge the support by the U.S. Department of Energy ASCR EXPRESS grant, Novel Quantum Algorithms from Fast Classical Transforms, and Northeastern University. LYH acknowledges the support of NSFC (Grant Nos. 11922502 and 11875111). ZL is supported by NSF grant PHY-2110463.

\appendix

\section{Review of Turaev-Viro TQFT and fusion category} \label{reviewtvtqft}

In this appendix, we provide a review of the Turaev-Viro TQFT\cite{Turaev:1992hq}. We start from the Turaev-Viro TQFT associated to two copies of the continuum Chern-Simons field theory, and explain its connection to the Roberts invariants\cite{ROBERTS1995771} defined using handle decomposition. For the Roberts invariant, the $\omega$ projection operator plays an important role, which we also use in the main text. Then we provide a review of fusion category theory and discuss the Turaev-Viro TQFT based on fusion categories.

\subsection{Turaev-Viro TQFT from Chern-Simons theory}

The Turaev-Viro TQFT was originally constructed to produce path integrals over 3D manifolds and recover topological invariants. While the Turaev-Viro TQFT is a discrete TQFT\cite{Turaev:1992hq}, it has been shown that a subclass of it is equivalent to two copies of Chern-Simons (CS) theory with a generic compact (non-) Abelian gauge group $G$\cite{Witten:1988hf}, which is easier to understand for field theorists. Here we give an introduction first through the lens of CS theory, before giving the general formulation based on fusion categories. 

Consider taking a pair of Chern-Simons gauge theories with the same compact gauge group $G$ (such as $SU(2)$) but opposite chirality. 
i.e. The action is given by
\be \label{pair-cs}
S = \frac{k}{4\pi}\left( \int \textrm{tr}(A_L\wedge dA_L + \frac{2}{3} A_L\wedge A_L\wedge A_L) \right) - \frac{k}{4\pi} \left( \int \textrm{tr}(A_R\wedge dA_R + \frac{2}{3}A_R\wedge A_R\wedge A_R)\right),
\ee
for independent gauge fields $A_L, A_R$, which are Lie algebra valued one forms in the same gauge group $G$. More explicitly,
\be
A_{L,R} = (A^a_{L,R})_\mu T^a dx^\mu~,
\ee
where $T^a$ are generators of the Lie algebra. 

It was proven that these classes of theories admit an equivalent discrete Turaev-Viro formulation\cite{Turaev:1992hq}. 
We note that the equations of motion for the Lagrange multipliers $A_{L,R;0}$ impose the constraint that the field strengths vanish 
\be \label{noflux1}
F_{L,R} \equiv dA_{L,R} - i [A_{L,R},A_{L,R}] =0~.
\ee

The idea of the discrete formulation of the 3D TQFT is to construct the phase space of the fields on a simplex (i.e.~a tetrahedron $\Delta_3$),  and then assign to each field configuration $\{A\}_{\Delta_3}$ defined on a simplex a value, which can be understood as $e^{-S(\Delta_3)(\{A\})}$. The path integral $Z(\mathcal{M})$ on an arbitrary manifold $\mathcal{M}$ would be given by triangulating the manifold, coloring the simplices by different field configurations, and finally obtaining 
\be
Z(\mathcal{M}) = \sum_{\{A\}} \prod_{\Delta_3} e^{-S(\Delta_3)(\{A\})}~.
\ee

In the context of Chern-Simons gauge theories, field configurations can be described by the eigenvalues of Wilson line operators. In fact, one can consider networks of open Wilson lines that meet at vertices. 
Open Wilson lines between two points take the form
\be
W_{m_1 m_2}^j(x_1,x_2) = \mathcal{P}\left(\exp(\int_{x_1}^{x_2} dx^\mu A^a_{\mu} T^a(j)) \right)_{m_1 m_2}~,
\ee
where $\mathcal{P}$ denotes path-ordering, and $T^a(j)$ are generators of the Lie algebra in representation $j$. The indices $m_1,m_2$ correspond to the indices of the representation $j$. 

To designate a field configuration to a 3-simplex, it is natural to consider inserting a Wilson line along each edge of a tetrahedron. These Wilson lines meet at a vertex. To ensure gauge invariance, the three representations carried by the three Wilson lines meeting at a vertex should be projected to the gauge-invariant sub-space. The vertex is thus a vector in the ``intertwining space''. 
In a single chiral copy of the Chern-Simons theory, the expectation value of Wilson lines forming a tetrahedron embedded in $S^3$ takes a simple form. 

\begin{figure}
	\centering
\includegraphics[width=0.35\linewidth]{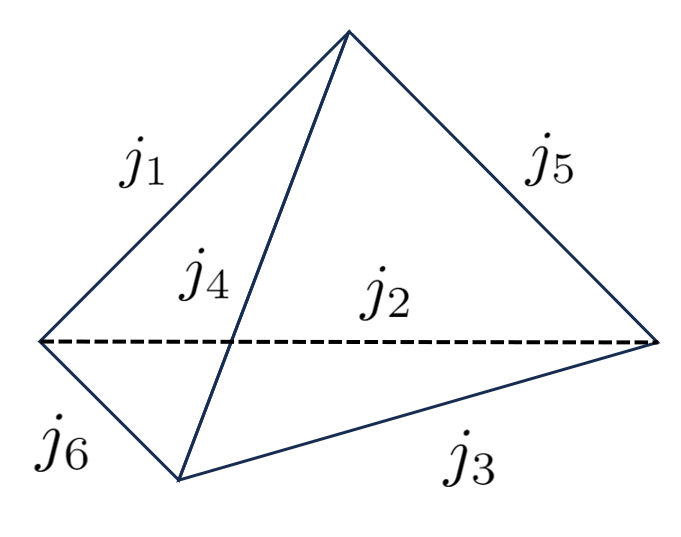}
	\caption{Wilson line network on a tetrahedron.}
	\label{wilsonlinenetwork}
\end{figure}

It is given by the quantum $6j$ symbol
\be \label{6j}
\langle \prod_{e\in \Delta_3} W^{j_e}(e) \rangle_{S^3} =  \left\{
    \begin{tabular}{ccc}
      $j_1$ & $j_2$   & $j_5$  \\
       $j_3$  & $j_4$ & $j_6$
    \end{tabular}
   \right\}_{G_k} \equiv \Gamma_3(\{j_{e\in \Delta_3}\})~,
\ee
where $j_i$ is the label of the representation a Wilson line operator took in the edge $i$, and $e$ is an edge of the tetrahedron $\Delta_3$. This configuration of Wilson lines is illustrated in Fig.~\ref{wilsonlinenetwork}. 

This result can be heuristically understood as follows in the classical limit $k\to\infty$. At every vertex $V$ where the Wilson lines meet, by gauge invariance it should produce an intertwiner, which is the usual {\it Clebsch-Gordan coefficient} for group $G$ in the $k\to \infty$ limit
\be
V^{j_1j_2j_3}_{m_1m_2m_3} = \left[
    \begin{tabular}{ccc}
      $j_1$ & $j_2$   & $j_3$  \\
       $m_1$  & $m_2$ & $m_3$
    \end{tabular}
     \right]~,
\ee
where $m_i$ corresponds to the indices in the representation $j_i$ that the Wilson line carries. 
Since there are four vertices, it would produce four such intertwiners. Using the extra fact that the local magnetic flux is constrained to be vanishing (\ref{noflux1}), each face of the tetrahedron produces delta functions connecting the indices $m_i$ from different Wilson lines. 
Finally one makes use of 
\begin{align}
\left\{
    \begin{tabular}{ccc}
      $j_1$ & $j_2$   & $j_5$  \\
       $j_3$  & $j_4$ & $j_6$
    \end{tabular}
   \right\}_G = 
   \sum_{m_i}  &\left[
    \begin{tabular}{ccc}
      $j_1$ & $j_2$   & $j_5$  \\
       $m_1$  & $m_2$ & $m_5$
    \end{tabular}
     \right] \left[
    \begin{tabular}{ccc}
      $j_2$ & $j_3$   & $j_6$  \\
       $m_2$  & $m_3$ & $m_6$
    \end{tabular}
     \right] 
     \nonumber \\
     & \left[
    \begin{tabular}{ccc}
      $j_3$ & $j_4$   & $j_5$  \\
       $m_3$  & $m_4$ & $m_5$
    \end{tabular}
     \right] 
       \left[
    \begin{tabular}{ccc}
      $j_4$ & $j_1$   & $j_6$  \\
       $m_4$  & $m_1$ & $m_6$
    \end{tabular}
     \right]
\end{align}
to recover the final result of the $6j$ symbol in (\ref{6j}) in the leading $k\to \infty$ limit. 

Equation~(\ref{6j}) is the non-perturbative result at finite $k$.
For positive integer level $k$, quantization restricts the allowed
Wilson-line sectors to the integrable highest-weight representations of
the corresponding affine Kac--Moody algebra. For $G=SU(2)$, these sectors
are labeled by 
\[
j=0,\frac12,1,\ldots,\frac{k}{2}.
\]

Each representation $j$ is associated to a ``quantum dimension'' $d_j$. In the classical case, $d_j$ is simply the actual dimension of the representation $j$. 
At finite $k$ it is an ``effective dimension'' of the representation in the following sense. Consider a number of Wilson lines ending at punctures of a 2d surface. 
In the quantized theory, one can obtain the dimension of the Hilbert space on the 2d surface with these punctures. The result is equal to the dimension of fusion space of the representations of the Wilson lines to the trivial representation. In the limit where there is a large number $N$ of the same representation $j$, the dimension of the fusion space approaches $d_j^N$, as if this $d_j$ is indeed the internal dimension of the representation $j$, which is precisely what happens in the classical case. 
This can be readily proved using the Verlinde formula that connects the modular matrix and the fusion coefficient $N_{j_1j_2}^{j_3} \in \mathbb{Z}_{\ge 0}$ that gives the number of fusion channels between $j_1 , j_2$ and $j_3$\cite{Verlinde:1988sn}. 

It can be extracted from the expectation value of a single $j$ Wilson loop\cite{Witten:1988hf}:
\be \label{qdim}
\langle W^j(C) \rangle_{S^3}  \equiv d_j =  \frac{S_{0j}}{S_{00}},
\ee
where $C$ is any closed loop embedded in $S^3$, and $S_{ij}$ is the modular matrix associated with the character $\chi_j$ of the chiral Kac-Moody representation corresponding to the Wilson line representation $j$. 
We note that the above result can also be extracted directly from $\chi_j(\tau)$,
\be
d_j = \lim_{\tau \to 0}\frac{\chi_j(\tau)}{\chi_0(\tau)}~, \qquad \chi_j(\tau) = \textrm{tr}_{j} (e^{-\tau L_0})~,
\ee
where the trace above is taken over the Kac-Moody primary $j$ and its descendant states following from quantizing the WZW model on a circle.

The quantum $6j$ symbols satisfy the Pentagon relation given by
\begin{equation} \label{pentagon1}
\begin{aligned}
\sum_{p_1} d_{p_1} \begin{Bmatrix}
j_1 & j_2 & l_1\\
j_3 & l_2 & p_1
\end{Bmatrix}_{G_k} \begin{Bmatrix}
j_1 & p_1 & l_2\\
j_4 & j_5 & h_2
\end{Bmatrix}_{G_k} \begin{Bmatrix}
j_2 & j_3 & p_1\\
j_4 & h_2 & h_1
\end{Bmatrix}_{G_k} = \begin{Bmatrix}
l_1 & j_3 & l_2\\
j_4 & j_5 & h_1
\end{Bmatrix}_{G_k} \begin{Bmatrix}
j_1 & j_2 & l_1\\
h_1 & j_5 & h_2
\end{Bmatrix}_{G_k}~.
\end{aligned}
\end{equation}

\begin{figure}
	\centering
\includegraphics[width=0.8\linewidth]{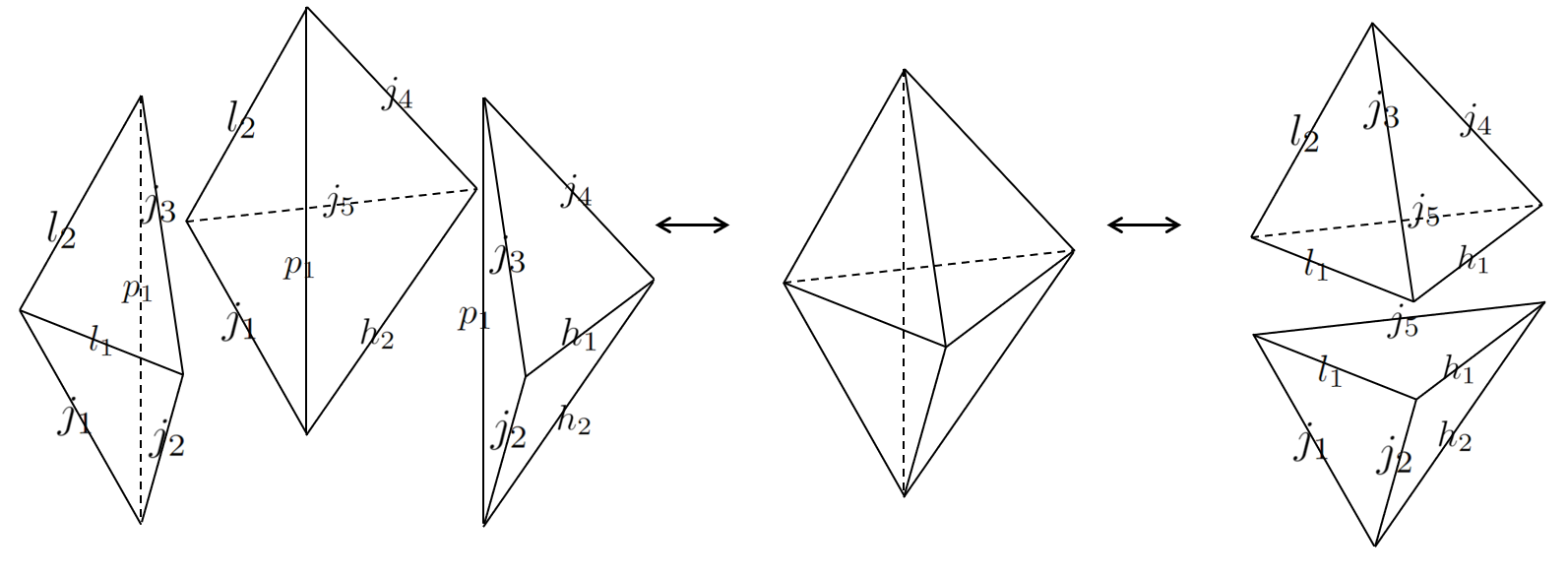}
	\caption{The Pentagon identity \eqref{pentagon1} is equivalent to the 3-2 Pachner move for tetrahedra in 3D.}
	\label{32move}
\end{figure}

The Pentagon identity can be represented graphically in 3D as Fig.~\ref{32move}. 

Now, this expectation value can be interpreted as that of the doubled theory (\ref{pair-cs}) on a tetrahedron. The intuitive way to see this is that a tetrahedron embedded in $S^3$ essentially divides it into two tetrahedra glued together at the edges. (For a lower-dimensional case, imagine drawing a triangle in $S^2$. The triangle divides $S^2$ into two triangles glued together at the edges. )
Folding the external tetrahedron into the internal one leads to the doubled theory (\ref{pair-cs}) filling up one tetrahedron. Before folding, the gauge field is taken continuous across the boundary of the tetrahedron. i.e.~
\be(A_{\textrm{inside}} - A_{\textrm{outside}})\vert_{\textrm{surface of $\Delta_3$}} = 0~.
\ee
Since the path integral outside is folded in, it corresponds to the following boundary conditions of the doubled theory
\be
(A_L + A_R)\vert_{\textrm{surface of $\Delta_3$}} = 0~,
\ee
where $A_R$ inside the tetrahedron is defined as the value of $A_{\textrm{outside}}$ through a reflection, and the interface becomes the boundary after folding. 

We can now construct the Turaev-Viro topological invariant over an arbitrary 3D manifold using the tetrahedra as building blocks\cite{Turaev:1992hq}. The process begins by triangulating the 3D manifold with tetrahedra, followed by coloring the edges with representations $j$. Then each tetrahedron is assigned a $6j$ corresponding to the six edges of the tetrahedron. Furthermore, for every edge labeled $j$ that lies in the interior of the 3D manifold and is shared by multiple tetrahedra, we include a factor of $d_j$. Finally, we sum over the coloring of all edges, 
i.e.\footnote{A square root is assigned to the quantum dimensions of the boundary edges.}\,
\be \label{TV2}
Z(\mathcal{M}) = \sum_{\{j_e\}}  \prod_{v \in \mathcal{M}}D^{-2} 
\prod_{e' \in \partial \mathcal{M}} \sqrt{d_{j_{e'}}} \prod_{e \notin \partial \mathcal{M}} d_{j_e} \prod_{\Delta_3}  \Gamma_3(\{j_{e\in \Delta_3}\})~,
\ee
where $\Gamma_3$ are $6j$ symbols as defined in (\ref{6j}). 

The normalization convention in the literature attaches a factor of $D^{-2}$ to every vertex. This normalization ensures $Z(S^2\times S^1) = 1$.

\begin{figure}
	\centering
\includegraphics[width=0.8\linewidth]{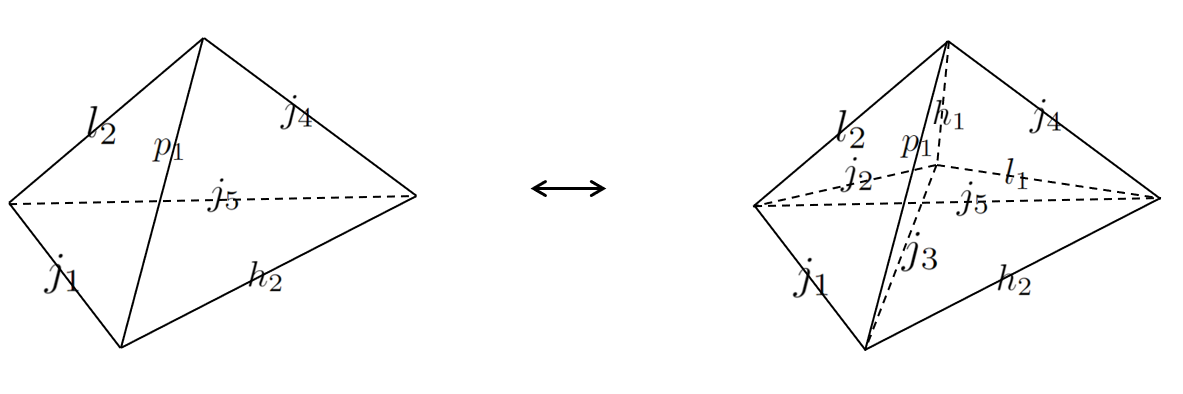}
	\caption{Equation \eqref{41move} corresponds to the Pachner 4-1 move for 3D tetrahedra.}
	\label{41movefig}
\end{figure}

Notice that a single tetrahedron can be subdivided into 4 tetrahedra. The path integral over 4 tetrahedra indeed recovers the result of 1 tetrahedron. i.e.~
\be \label{41move}
\begin{Bmatrix}
j_1 & p_1 & l_2\\
j_4 & j_5 & h_2
\end{Bmatrix}_{G_k} = \sum_{h_1,j_2,j_3,l_1} \frac{d_{h_1} d_{j_2} d_{j_3} d_{l_1}}{D^2}
 \begin{Bmatrix}
j_1 & p_1 & l_2\\
h_1 & j_2 & j_3
\end{Bmatrix}_{G_k}
  \begin{Bmatrix}
j_1 & j_3 & j_2\\
l_1 & j_5 & h_2
\end{Bmatrix}_{G_k}
 \begin{Bmatrix}
j_3 & p_1 & h_1\\
j_4 & l_1 & h_2
\end{Bmatrix}_{G_k}
 \begin{Bmatrix}
j_2 & h_1 & l_2\\
j_4 & j_5 & l_1
\end{Bmatrix}_{G_k}~.
\ee
This is depicted in Fig.~\ref{41movefig}. 

The Pentagon relation (\ref{pentagon1}) and (\ref{41move}), which correspond to the Pachner 3-2 move and the Pachner 4-1 move respectively, are the crucial identities ensuring that $Z(\mathcal{M})$ is independent of triangulation\cite{72d638c2-d595-33bc-a4e6-e007f4e4f5f5, Turaev:1992hq}. This also explains the factor $d_i$ assigned to the gluing of edges. In the context of 3D gravity, this re-triangulating invariance is closely related to 3D diffeomorphism invariance.

To make a further connection of the path integral (\ref{TV2})
with that of two copies of Chern-Simons theory, we make use of the ``Chain-Mail'' construction introduced by Roberts in \cite{ROBERTS1995771}. 

Consider a very special linear combination of the Wilson loops $\omega$ given by \cite{ROBERTS1995771}\footnote{This collection of Wilson loops is often denoted as $\Omega$ in literature. However, since $\Omega$ is already used in this paper to represent the state corresponding to the physical boundary conditions, we instead denote this loop operator by $\omega$.}:
\be \label{Omegaflux}
\omega(C) = \mathcal{N} \sum_j^{k/2} d_j W^j(C)~.
\ee

This linear combination is special in that $\omega(C)$ is essentially a projector, constraining the flux across the loop $C$ to vanish. This operator plays an important role in the definition of the Levin-Wen string-net model\cite{Levin:2004mi} reviewed in Sec.~\ref{briefreviewtqft}, and it also has a nice physical interpretation in gravity language\cite{Chua:2023ios}, as we explain in Sec.~\ref{TQFTSECTION}. Classically, the Wilson loop operator in representation $j$ on a closed loop $C$ would give the character of $j$ evaluated on the group element $g = \exp(\oint_C A)$.  
Characters satisfy the following identity:
\be
\sum_j d_j \textrm{tr}(R_j(g)) = \delta(g)~,
\ee
which indeed leads to a projection onto trivial flux through the loop $C$. 

At finite $k$, this remains a projection operator. Normalization in a rational theory (where $k$ is finite and $G$ is compact) is usually chosen as
\be
\mathcal{N}^{-2} = D^2 \equiv \sum_j d_j^2~. 
\ee

We note that $D$ is finite for finite $k$, but diverges as $k\to \infty$. This is one subtlety that forbids naive generalization of rational Turaev-Viro theories to irrational ones. 

\begin{figure}
	\centering
\includegraphics[width=0.6\linewidth]{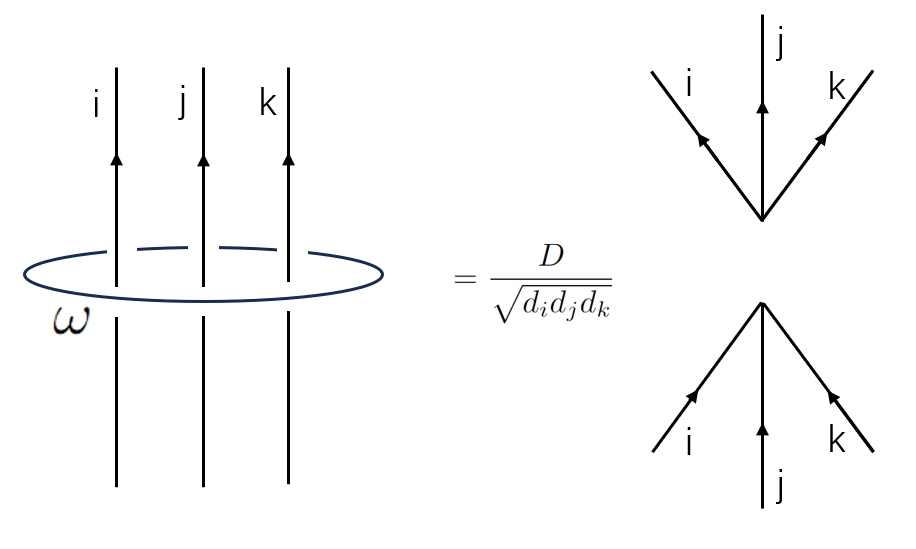}
	\caption{Property of the $\omega$ loop. This property can be used to relate the topological invariant defined by Roberts on the handle decomposition of 3D manifolds to the Turaev-Viro invariant defined using triangulation via tetrahedra.}
	\label{killing}
\end{figure}

These lines satisfy interesting properties, illustrated in Fig.~\ref{killing}. This suggests that these Wilson line configurations of $\omega$ loops can be assembled by matching the Wilson lines meeting at two vertices, and that can be interpreted as joined lines. Consequently, a continuum path integral of intertwining $\omega$ loops can be reduced to disconnected Wilson lines joined at the vertices. This is the idea behind the Chain-Mail construction. 
\begin{figure}
	\centering
\includegraphics[width=0.6\linewidth]{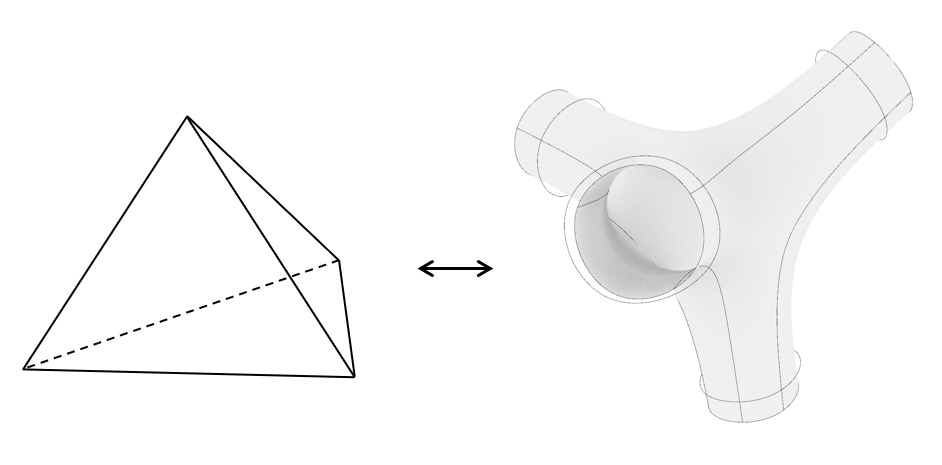}
	\caption{The tetrahedron and associated handle decomposition coming from thickening the dual. We insert the $\omega$ loops in the handle decomposition to get the Roberts invariant.}
	\label{dualomega}
\end{figure}
Starting with a triangulation $T$ of a 3D manifold, we consider the thickening of its dual complex $T^*$, which provides a handle decomposition $D$ of the original 3D manifold. The vertices of $T^*$ correspond to 0-handles, the edges of $T^*$ correspond to 1-handles, the faces of $T^*$ correspond to 2-handles, and the 3-cells correspond to 3-handles. Next, we insert $\omega$ loops on all components of the handle decomposition. This process is illustrated for each tetrahedron in Fig.~\ref{dualomega}.

By considering the dual graph corresponding to the network of Wilson lines, and recursively using the relation in Fig.~\ref{killing}, one recovers the path integral with the insertion of $\omega$ loops, which serves only to project every loop back to vanishing flux\cite{Burnell:2010mx, *2011NJPh...13f5001B}. The resultant network is called the ``Chain-Mail'' \cite{ROBERTS1995771}. The corresponding expectation value of closed Wilson loops is denoted as $C(\mathcal{M},D)$. Roberts invariants can be defined as:
\be
R(\mathcal{M})=D^{-d_3-d_0} C(\mathcal{M},D)~,
\ee
where $d_3$ and $d_0$ represent the number of 3-handles and 0-handles in $D$, respectively. As defined, this invariant is independent of the specific handle decomposition and serves as a topological invariant of $\mathcal{M}$. Using Fig.~\ref{killing}, it can also be shown that this Roberts invariant matches precisely with the Turaev-Viro invariant, $R(\mathcal{M})=Z(\mathcal{M})$\cite{ROBERTS1995771}. Finally, the expectation value of $\omega$ loops on $\mathcal{M}$ can be computed using surgery, and gives a product of two copies of the chiral Reshetikhin-Turaev-Witten invariants\cite{Witten:1988hf, Reshetikhin:1991tc, ROBERTS1995771}, consistent with our expectation.

\subsection{Turaev-Viro TQFT from fusion category}

In the above, we attempt to give an introduction to the Turaev-Viro theory to emphasize that it can be understood in field-theoretic terms as Chern-Simons theories. 
However, the Turaev-Viro TQFT can also be defined independently of the Chern-Simons theory. 

As we have seen above, the basic building block of the Turaev-Viro TQFT is given by the so-called quantum $6j$ symbol in (\ref{6j}). While it arises in the context of Chern-Simons gauge theory, the mathematical structure it belongs to can be defined independently of a field-theoretic realization. 
The mathematical structure concerned is the {\it fusion category} $\mathcal{C}$. 
There are many comprehensive reviews on the subject. (For reviews for physicists, one can consult, for example, \cite{Kitaev:2005hzj, Bhardwaj:2017xup, Komargodski:2020mxz, Aasen:2020jwb}. More mathematical treatment can be found, for example, in \cite{etingof}.)  Here, we provide a basic skeleton to make it easier for our readers to follow more sophisticated expositions. 

A fusion category $\mathcal{C}$ contains the following ingredients. 
\begin{enumerate}
   \item A finite set of isomorphism classes of {\it simple objects}
$\{a,b,c,d,\ldots\}$; general objects are finite direct sums of these.
    \item Finite-dimensional spaces of maps between objects, called
{\it morphisms}.
    \item The category $\mathcal{C}$ is equipped with a tensor product $\otimes$: 
    \be
        a\otimes b = \bigoplus_c N^c_{ab} c~,
    \ee
    where $N^c_{ab} \in \mathbb{Z}_{\ge 0}$ is the dimension of the fusion space mapping $a\otimes b$ to $c$.
\end{enumerate}
We also impose a few extra properties:
\begin{itemize}
    \item The labels $a,b,\ldots$ used below denote simple objects, for which
$\operatorname{End}(a)\cong\mathbb{C}$.
    \item There is a distinguished object $0$ that plays the role of {\it identity}, so that every object fusing with it leaves the object unchanged. 
    \item For every simple object $a$ one can also define the dual object $\bar a$ so that $0 \in a \otimes \bar a$.
\end{itemize}

\begin{figure}
	\centering
\includegraphics[width=0.6\linewidth]{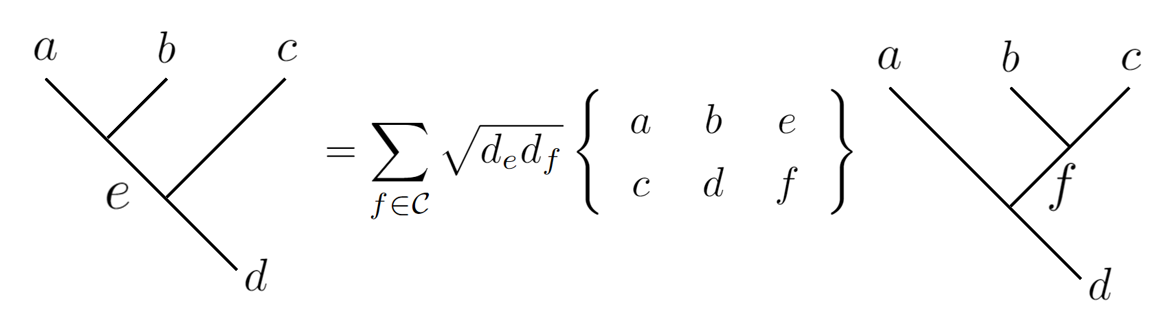}
	\caption{Associativity of fusion involves the quantum $6j$ symbol.}
	\label{crossing}
\end{figure}

The fusion product is associative in a {\it weak} sense. 
Rather than imposing that $(a\otimes b) \otimes c $ is directly equal to $a\otimes (b \otimes c) $, these two sets of fusion maps are only required to be linearly dependent. 
Denote the basis of the fusion maps from $(a\otimes b)$ to $c$ by $V_{ab}^c(\alpha)$ where $ 1\le \alpha \le N^c_{ab}\neq 0$.
For simplicity, in the following we will focus on the cases where $N^c_{ab}\leq 1$, in which case we can omit the labels $\alpha,\beta$ for each fusion channel. 
The associativity is then expressed through the $6j$ symbols as
\be \label{fusionmap}
V_{ab}^e V_{ec}^d = \sum_{f\in \mathcal{C}}\sqrt{d_e d_f}\left\{\begin{tabular}{ccc}
$a$ & $b$& $e$ \\
$c$ & $d$ & $f$\end{tabular}\right\} V_{af}^d V^f_{bc}~.
\ee
There is a very convenient diagrammatic representation of the above associativity relation, as depicted in Fig.~\ref{crossing}.

The coefficient $d_j$ is the quantum dimension of the simple object $j\in \mathcal{C}$. Note that $d_0 = 1$ and $d_a= d_{\bar a}$.
These quantum dimensions also satisfy the identity
\be
d_a d_b = \sum_c N_{ab}^c d_c~.
\ee
Quantum dimension is thus analogous to the dimension of a representation in the Chern-Simons theories discussed above. The unit object $0$ corresponds to the ``trivial representation'' in a gauge theory.

\begin{figure}
	\centering
\includegraphics[width=0.6\linewidth]{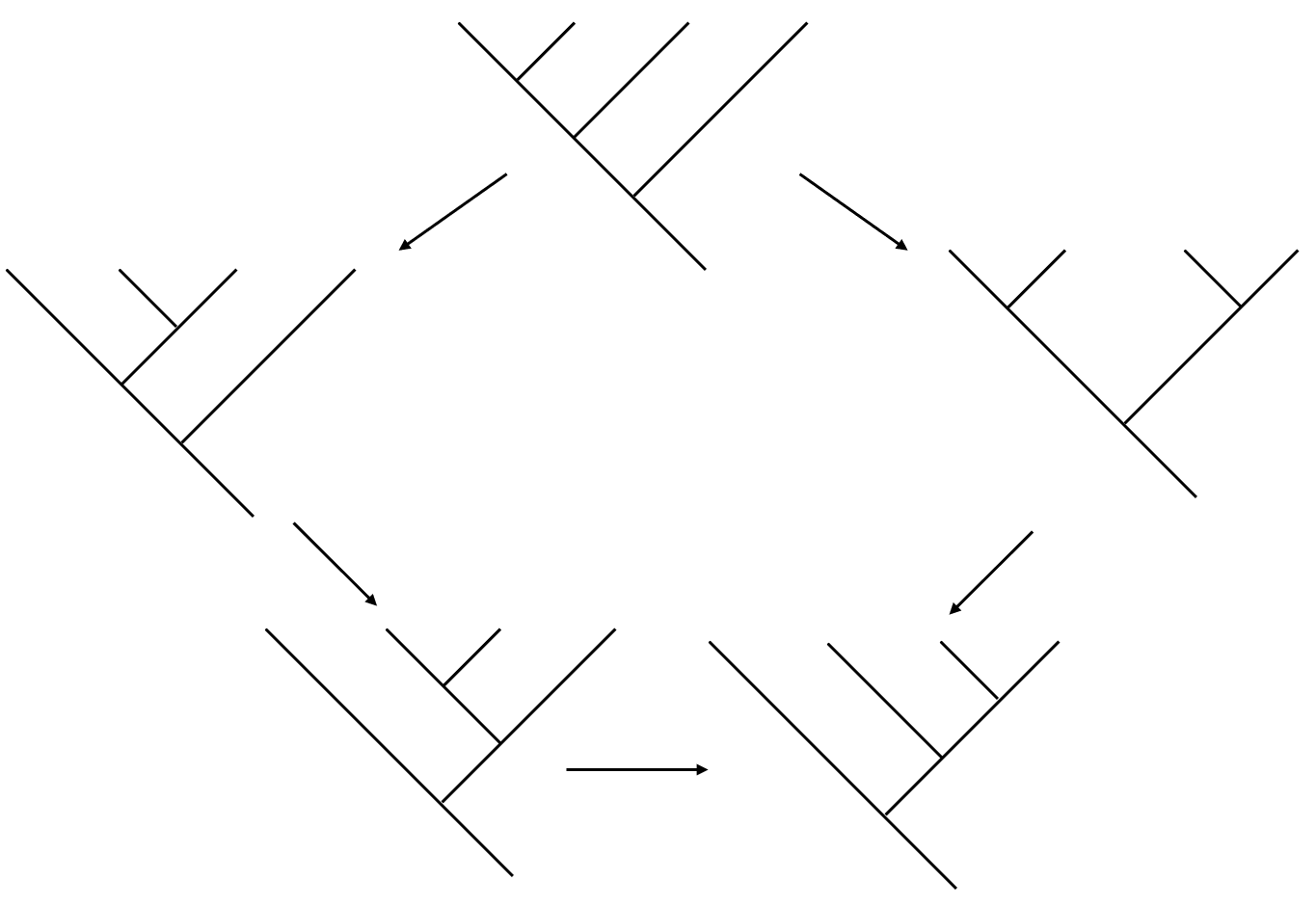}
	\caption{Pentagon identity for the quantum $6j$ symbols.}
	\label{pentagon6j}
\end{figure}

The coefficient in curly brackets is precisely the analogue of the quantum $6j$ symbol that we introduced in the context of the Chern-Simons theories. This is part of the input data that define a fusion category, and the Pentagon relation expressed in (\ref{pentagon1}) is satisfied.  The Pentagon relation expresses the consistency under different orders of fusions of 4 objects, as illustrated in Fig.~\ref{pentagon6j}. 

For example, the category of finite-dimensional complex representations
of a finite group is a fusion category. Each representation corresponds to an object in the category, and the fusion of representations corresponds to the fusion morphisms discussed above, whose associativity is indeed controlled by the $6j$ symbols.

Given a fusion category $\mathcal{C}$, it is sufficient to define a Turaev-Viro path integral on any manifold $\mathcal{M}$\cite{Turaev:1992hq} exactly as in (\ref{TV2}).

\begin{figure}
	\centering
\includegraphics[width=0.4\linewidth]{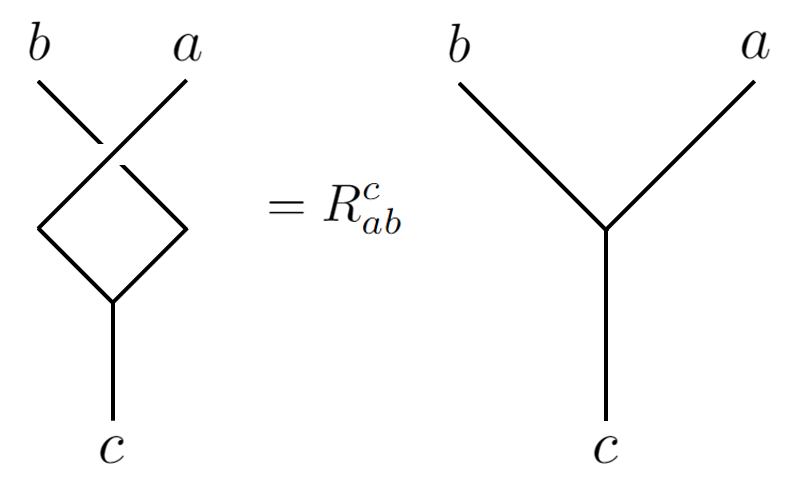}
	\caption{The braided structure is encoded in the $R$-matrix.}
	\label{rmatrix}
\end{figure}

Generically, it is possible to endow a fusion category with more structures. An important structure is the braided structure that exchanges two objects. This structure is encoded in the so-called $R$-matrix, with components $R^{c}_{ab}$ which are phases, that connects $a\otimes b$ to $b\otimes a$. 
This is illustrated in Fig.~\ref{rmatrix}. 
Note that there is a distinction between an ``over-crossing'' and ``under-crossing''. The factor that appears in an ``under-crossing'' is related to the ``over-crossing'' by a complex conjugate of $R_{ab}^c$. The $R$-matrix satisfies some consistency conditions, so that the action of braiding is consistent with the associativity relations from fusion. These consistency conditions are given by the Hexagon relations\cite{Moore:1988uz, Moore:1988qv}.

One can define an $S$-matrix for a braided tensor category. The components of the $S$-matrix are given by
\be
S_{ab} = \frac{1}{D}\sum_c N_{\bar{a}b}^c \frac{\theta_c}{\theta_a\theta_b} d_c~,
\ee
where $\theta_c$ is the ``topological spin'' of the object type $c$, and $D^2=\sum_a d_a^2$ as introduced above. It is related to the conformal dimension $h_i$ of the primary by
\be
\theta_c = e^{2\pi i h_c}~.
\ee
The square of the $R$ matrix is also related to the $h_c$ by
\be
(R_{ab}^c)^2 = e^{2\pi i (h_a + h_b - h_c) }~.
\ee

We note that when the $S$-matrix is a unitary matrix, the braided tensor category is modular. The Turaev-Viro TQFT reduces to pairs of Chern-Simons theories described previously when $\mathcal{C}$ is a {\it modular} tensor category.

\subsection{Illustrative example: torus partition function}\label{torusexample}

Before turning to the two concrete realizations in the main text, we illustrate the overlap structure with a simple example related to the torus partition functions of 2D rational CFTs. On a torus, the CFT partition function is given by
\be
Z(\tau,\bar{\tau})=\sum_{I,J} Z_{IJ} \chi_I(\tau) \bar{\chi}_J(\bar{\tau}) ~,
\ee
where $(I,J)$ label conformal primaries with chiral/anti-chiral representations $I$ and $J$, and $\chi_I(\tau)$ is the associated character with modulus $\tau$. As shown in \cite{Witten:1988hf, Elitzur:1989nr}, the Hilbert space of torus conformal blocks is isomorphic to that of a 3D TQFT on a torus, leading to a 3D quantum state
\be
\ket{\Psi}=\sum_{I,J} Z_{IJ} \ket{I,J} ~,
\ee
where $\ket{I,J}$ can be constructed from 3D solid torus path integral with Wilson line insertions.
This quantum state encodes the operators appearing in the spectrum of the 2D CFT and can alternatively be described by a topological boundary condition of the 3D TQFT (Sec.~\ref{algebra}).

The coefficients $Z_{IJ}$ represent algebraic/topological data of the CFT, ensuring a modular invariant spectrum. On more general manifolds, they are explicitly related to closed CFT OPE coefficients via the closed pair-of-pants decomposition. The observation in \cite{Fuchs:2002cm} is that these coefficients can also be constructed from BCFT open OPE coefficients, a fact leveraged to build the exact CFT tensor network. Notably, these are numbers independent of the moduli and are intrinsically topological in nature.

To recover the CFT partition function with moduli dependence, we construct a physical boundary condition in the 3D TQFT corresponding to a state $\bra{\tau,\bar{\tau}}$ with $\bra{\tau,\bar{\tau}} I,J \rangle=\chi_I(\tau)  \bar{\chi}_J(\bar{\tau})$. If the theory is deformed away from the CFT fixed point while preserving the symmetry, the bulk is preserved and a new physical boundary gives 2D partition functions depending on dimensionful couplings $\lambda$:

\be
Z^{\lambda}(\tau,\bar{\tau})
=
\sum_{I,J} Z_{IJ}\,
\mathcal{F}^{\lambda}_{IJ}(\tau,\bar{\tau})
=
\bra{\tau,\bar{\tau},\lambda}\Psi\rangle~,
\ee

where
$\mathcal{F}^{\lambda}_{IJ}(\tau,\bar{\tau})$
is the deformed sector wavefunction. Away from the conformal fixed point
it need not factorize into separately holomorphic and antiholomorphic
characters. The $T\bar T$ deformation is a notable example
\cite{Zamolodchikov:2004ce}: it leaves the topological data encoded in
$\ket{\Psi}$ unchanged, while modifying the physical-boundary
wavefunctions according to the Callan--Symanzik flow.

\section{Tensor network state sum representation of 2D CFT path integrals}\label{statesum cft}

  \begin{figure}
    \centering
    \begin{minipage}[b]{0.48\linewidth}
        \centering
        \begin{overpic}[width=\linewidth]{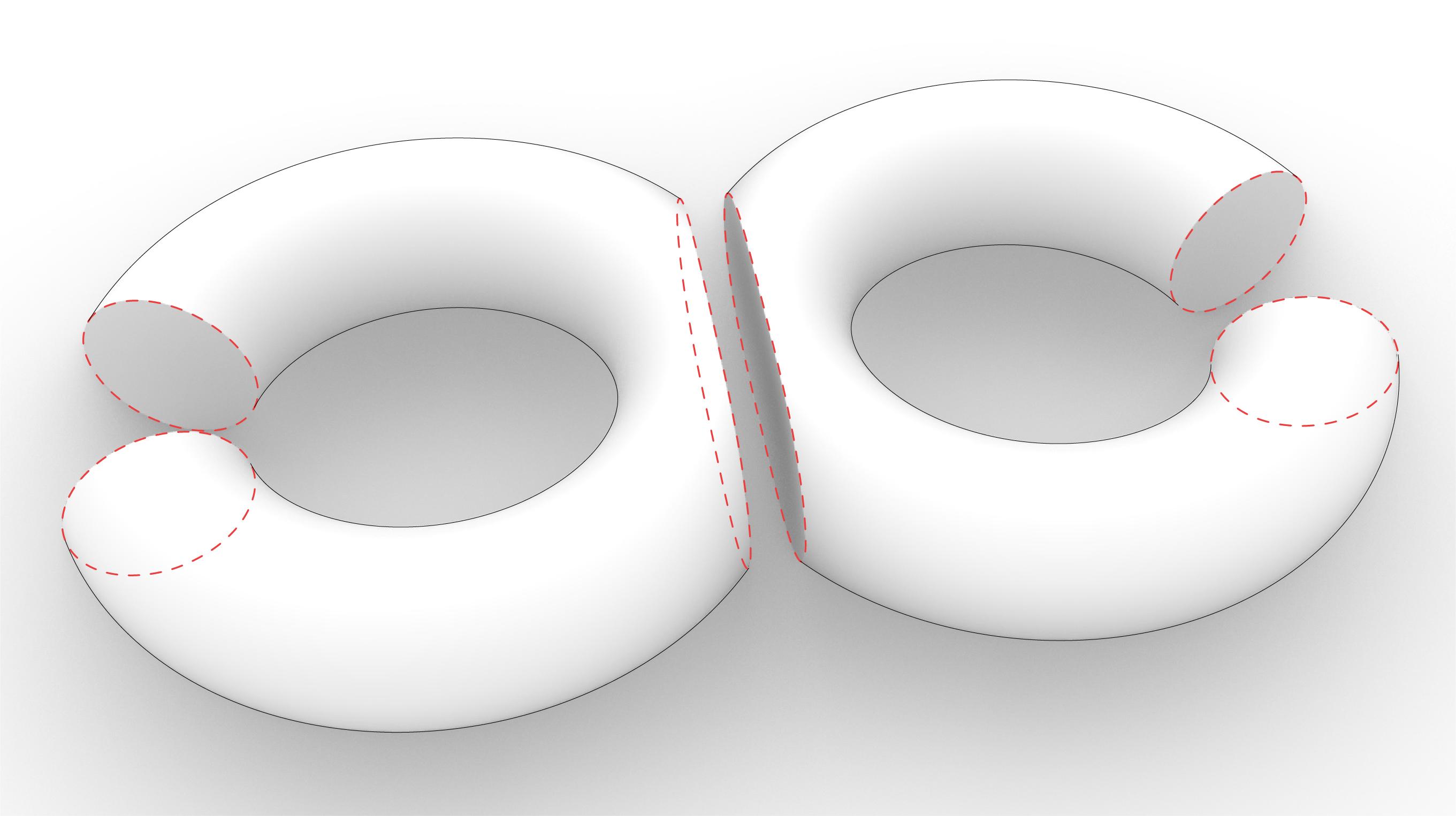}
        \end{overpic}
    \end{minipage}
    \hfill
    \begin{minipage}[b]{0.48\linewidth}
        \centering
        \begin{overpic}[width=\linewidth]{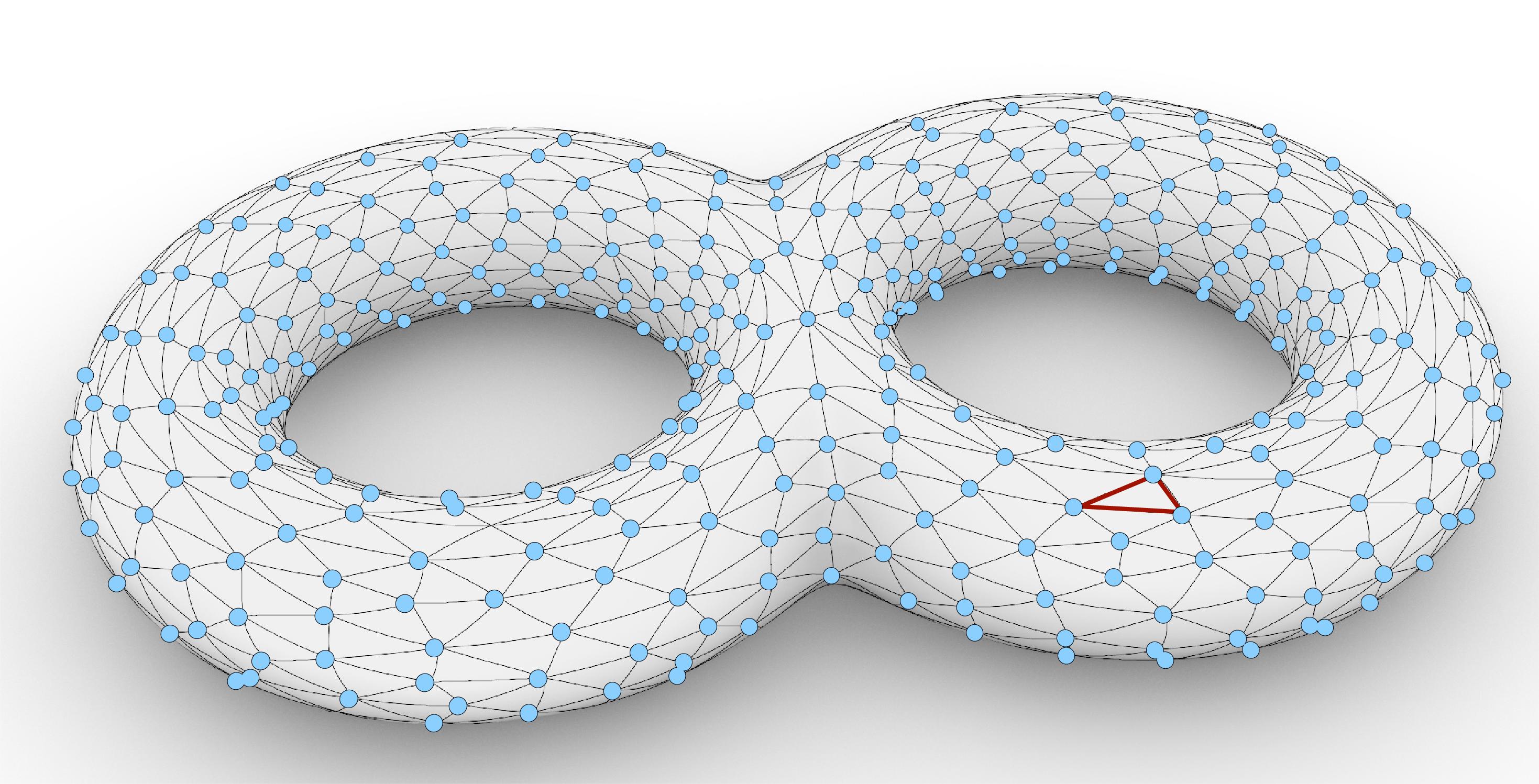}
        \end{overpic}
    \end{minipage}
    \caption{Two different cutting and gluing procedures for the computation of genus two CFT partition functions. The conventional method on the left involves the closed pair-of-pants decomposition. In contrast, we cut the manifold open into chipped triangles, which means the building blocks are open pairs of pants.}\label{cutting}
\end{figure}

In this appendix, we review in great detail the tensor network state sum representation of 2D CFT path integrals\cite{Chen:2022wvy, Cheng:2023kxh, Hung:2024gma, Chen:2024unp}.

Instead of employing the more conventional decomposition of the CFT path integral on 2D manifolds into closed pairs of pants in the closed channels defined on circles, we decompose the path integral into small triangular regions and define Hilbert spaces along the lines connecting their endpoints. For example, the distinction between these approaches in computing the genus-two partition function is illustrated in Fig.~\ref{cutting}. The tiny blue holes have radius $R$\footnote{In \cite{Chen:2024unp}, we mostly used the symbol $\epsilon$ to parameterize the size of the holes. As explained in footnote 6 of that paper, when the length of the lattice site is $\Lambda$, then $\frac{\pi}{\epsilon}=\ln (\frac{\Lambda}{R})$.}, introduced in \cite{Chen:2022wvy, Cheng:2023kxh, Hung:2024gma, Chen:2024unp} for regularization purposes. As discussed in Sec.~\ref{SymQRGTN}, taking $R \to 0$ recovers the original CFT, while allowing $R$ to vary yields a continuous family of topological symmetry-preserving, $T\bar{T}$-like deformations, obtained by gluing 3D gravitational path integrals (quantum $6j$ symbols) onto the 2D theories.

For the computation of the total CFT partition function $\mathcal{Z}_{\text{CFT}}$, the basic building block is the CFT path integral on the chipped triangles, as shown in Fig.~\ref{tiling}\footnote{We will focus on a 2D CFT defined on a flat background $\delta_{ij}$ for simplicity. However, the formalism can be easily generalized to general fixed curved backgrounds $h_{ij}$. For many 2D manifolds that do not admit a global flat metric, such a generalization must be applied.}. These represent open pairs of pants, in contrast to the more conventional closed pairs of pants commonly found in the CFT literature. The Hilbert spaces in the computation are associated with the edges of the triangles, which are lines between the two endpoints where boundary conditions must be specified.

\begin{figure}
	\centering
	\includegraphics[width=0.6\linewidth]{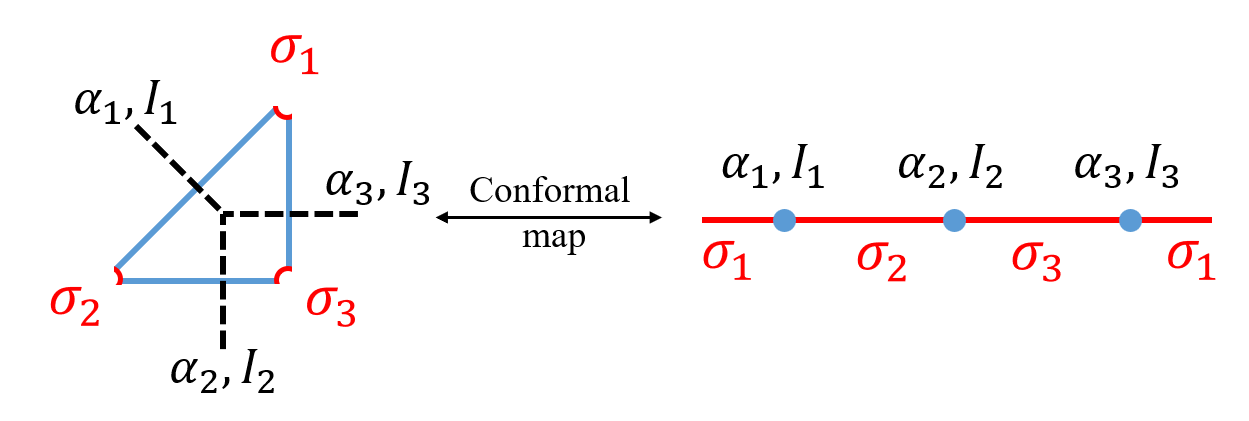}
	\caption{Conformal map from a chipped triangle (open pair of pants) to the upper half plane with three boundary-changing operators on the real line.}
	\label{UHP}
\end{figure}
We first impose and fix some conformal boundary conditions, labeled by $\sigma_i$, as illustrated in Fig.~\ref{UHP}. Once the boundary conditions are fixed, we insert a complete basis of states on the lines, compatible with the boundary conditions at the two endpoints. Through the state-operator correspondence in the open channel, these states correspond to boundary condition-changing operators such as $\Phi_{\alpha_1,I_1}^{\sigma_1 \sigma_2}$. These operators are classified by their primary label $\alpha_i$ and descendant label $I_i$ under a single copy of the Virasoro algebra. The value of the open pairs of pants involving descendant fields is entirely determined by the Virasoro symmetry from those involving only primary fields. Therefore, we will focus solely on those associated with the primary fields in the following.

\begin{figure}
	\centering
	\includegraphics[width=0.8\linewidth]{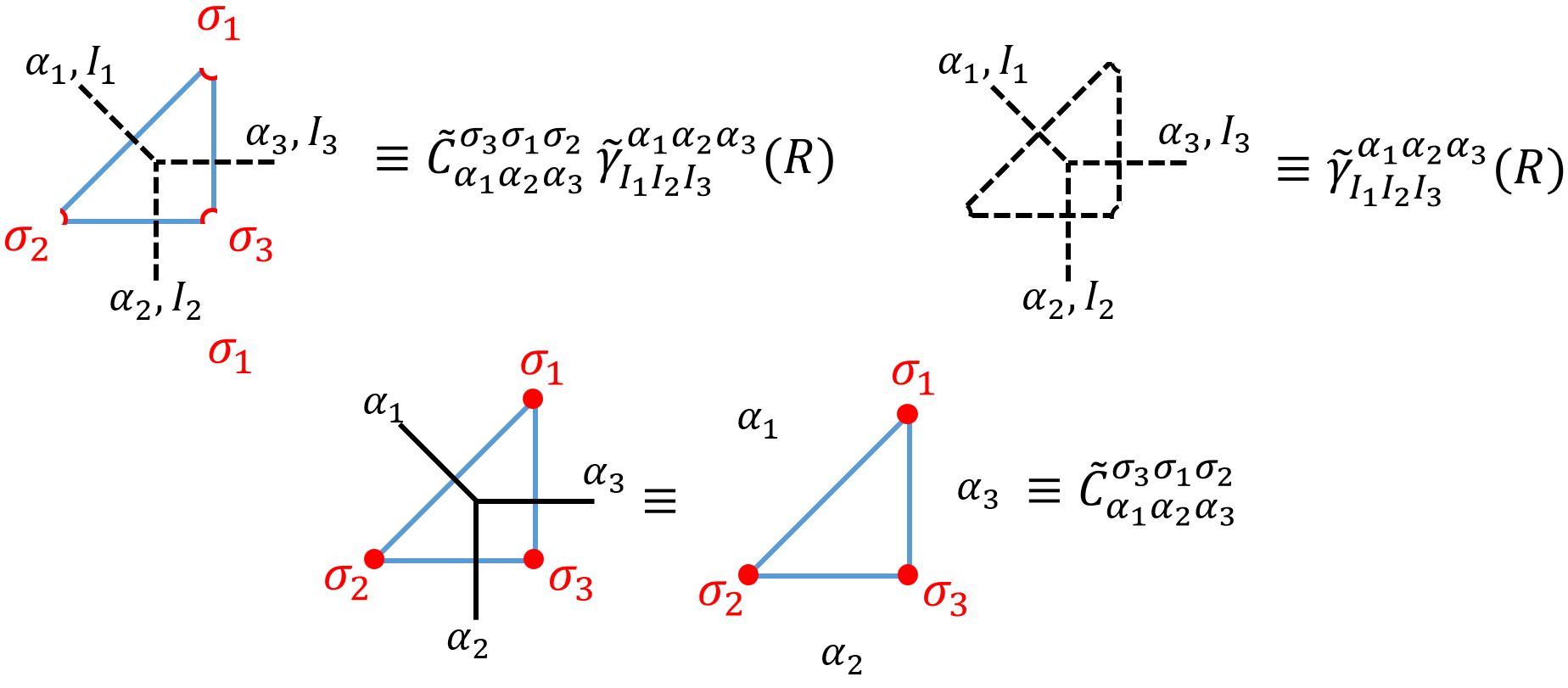}
	\caption{Diagrammatic notation for $\tilde{C}^{\sigma_3 \sigma_1 \sigma_2}_{\alpha_1 \alpha_2 \alpha_3}$ and $\tilde{\gamma}^{\alpha_1 \alpha_2 \alpha_3}_{I_1 I_2 I_3}(R)$.}
	\label{diagramthreepoint1}
\end{figure}

More concretely, we can use a conformal map to map the chipped triangles to the upper half-plane\footnote{The explicit function realizing the conformal map for the flat background $\delta_{ij}$ can be found in \cite{Brehm:2021wev, Cheng:2023kxh}. For more general background metrics, a more complicated conformal transformation will be applied.}, where the BCFT primary three-point function takes the familiar form,
\be \label{eq:3pt0}
\begin{aligned}
\langle \Phi_{\alpha_1}^{\sigma_1 \sigma_2}(x_1) \Phi_{\alpha_2}^{\sigma_2 \sigma_3}(x_2)  \Phi_{\alpha_3}^{\sigma_3 \sigma_1}(x_3)\rangle=\frac{C_{\alpha_1,\alpha_2,\alpha_3}^{\sigma_3,\sigma_1,\sigma_2}}{|x_{21}|^{\Delta_1+\Delta_2-\Delta_3} |x_{32}|^{\Delta_2+\Delta_3-\Delta_1} |x_{31}|^{\Delta_3+\Delta_1-\Delta_2}}~.
\end{aligned}
\ee
When we consider the quantities associated with descendant fields, the numerator will remain unchanged, while the denominator will acquire dependence on the descendant labels, determined by the Virasoro symmetry. When we pull the three-point function from the upper half-plane back to the original chipped triangles, we obtain,
\be
\mathcal{T}^{\sigma_1\sigma_2\sigma_3}_{(\alpha_1, I_1) (\alpha_2, I_2)(\alpha_3, I_3)}(\triangle, R) = C^{\sigma_3\sigma_1\sigma_2}_{\alpha_1 \alpha_2\alpha_3} \gamma^{\alpha_1 \alpha_2\alpha_3}_{I_1I_2I_3}(R)~.
\ee
This normalization is what we call the ``Block gauge''. For later convenience, we also introduce the ``Racah gauge'' \cite{Kojita:2016jwe, Chen:2022wvy, Cheng:2023kxh, Hung:2024gma, Chen:2024unp}, where
\be
\begin{aligned}\label{blockgauge}
\mathcal{T}^{\sigma_1\sigma_2\sigma_3}_{(\alpha_1, I_1) (\alpha_2, I_2)(\alpha_3, I_3)}(\triangle, R) &\equiv \tilde{C}^{\sigma_3\sigma_1\sigma_2}_{\alpha_1 \alpha_2\alpha_3} \tilde{\gamma}^{\alpha_1 \alpha_2\alpha_3}_{I_1I_2I_3}(R)\\
&=\left( C^{\sigma_3\sigma_1\sigma_2}_{\alpha_1 \alpha_2\alpha_3} \mathcal{N}_{\alpha_1 \alpha_2\alpha_3}\right) \frac{\gamma^{\alpha_1 \alpha_2\alpha_3}_{I_1I_2I_3}(R)}{ \mathcal{N}_{\alpha_1 \alpha_2\alpha_3}}~.
\end{aligned}
\ee
$\mathcal{N}_{\alpha_1 \alpha_2 \alpha_3}$ is a function chosen such that the crossing kernels for these rescaled conformal blocks are proportional to the Racah coefficients or the quantum $6j$ symbols, and the explicit form of these functions can be found in \cite{Kojita:2016jwe, Chen:2022wvy, Cheng:2023kxh, Hung:2024gma, Chen:2024unp}. A diagrammatic representation is introduced in Fig.~\ref{diagramthreepoint1}.

For diagonal CFTs, the label set of conformal boundary conditions coincides with the set of primary fields (this is referred to as the Cardy case \cite{Cardy:2004hm} in the literature), and the rescaled BCFT OPE coefficients are directly proportional to the quantum $6j$ symbols, as first discovered by Runkel \cite{Runkel:1998he, *Runkel:1999dz}: this is the formula quoted as \eqref{rationalC1} in the main text, with $d_{\alpha_i}$ the quantum dimension introduced above. For theories with a continuous spectrum, such as the Liouville theory, this corresponds to the Plancherel measure \cite{Chen:2024unp, Ponsot:1999uf, *Ponsot:2000mt}. In the diagonal case, the $6j$ symbol enjoys the tetrahedron symmetry, and we associate it with a tetrahedron, which serves as the fundamental building block of the Turaev-Viro TQFT, as reviewed above.

\begin{figure}[H]
	\centering
	\includegraphics[width=0.8\linewidth]{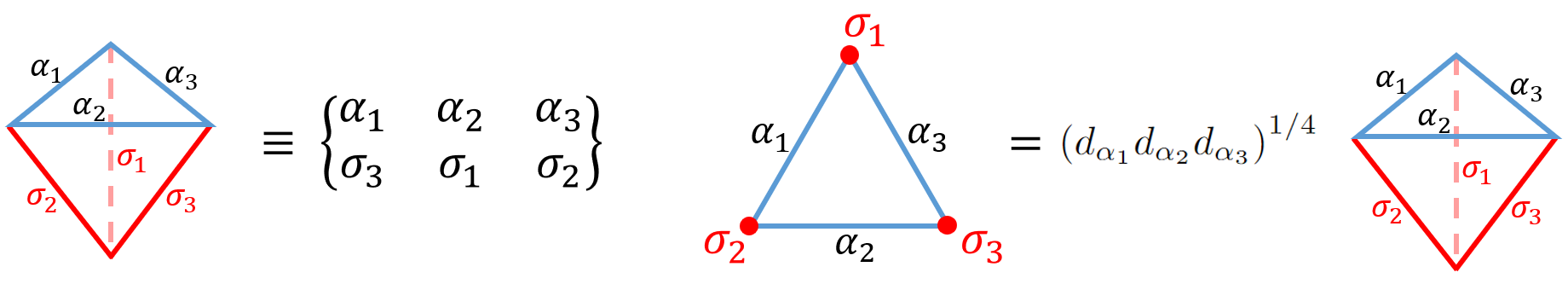}
	\label{diagonal6j}
\end{figure}

For more general non-diagonal CFTs, a formula similar to \eqref{rationalC1} holds. The boundary conditions now take values in the ``module category'', which we explain in Sec.~\ref{roleofpsi module category}. For now, it suffices to understand that this is a different label set from the primary labels, and the curly brackets denote the generalized $6j$ symbols with mixed indices. These generalized symbols can also be expressed in terms of the standard $6j$ symbols, where the standard $6j$ symbols are ``dressed'' with quantities associated with these boundary conditions \cite{Fuchs:2002cm, Lootens:2020mso}. Diagrammatically, we introduce the following representation:

\begin{figure}[H]
	\centering
	\includegraphics[width=0.8\linewidth]{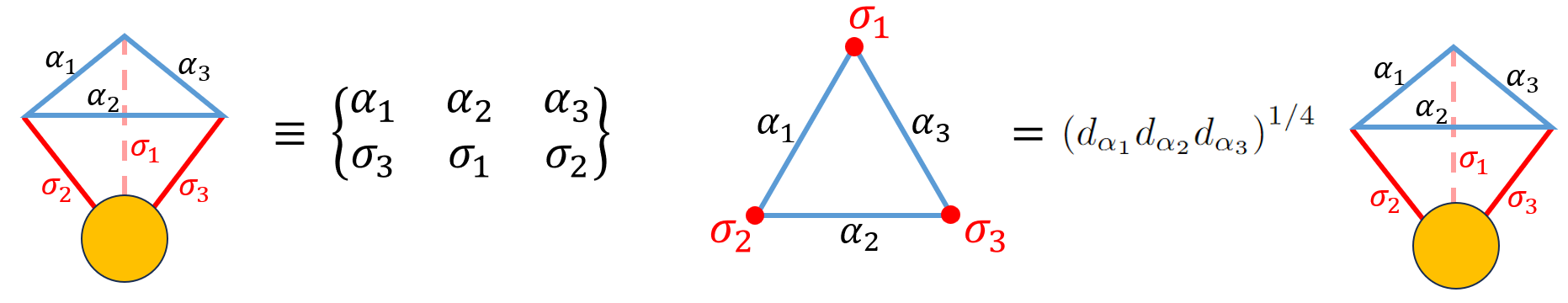}
	\label{nondiagonal6j}
\end{figure}

After summing over all intermediate primaries and descendants, we recover the partition function \eqref{Zproduct0} of the main text, for fixed boundary conditions $\{\sigma_{a}\}$.

As we explain in the main text, to ensure that we get the correct IR fixed point and reveal the connection to topological symmetries, we sum over the boundary conditions with a specific weight, proportional to the quantum dimension $\omega_{\sigma_a} \propto d_{\sigma_a}$, following \cite{Hung:2019bnq, Chen:2022wvy, Cheng:2023kxh, Hung:2024gma, Chen:2024unp, Brehm:2021wev}, leading to \eqref{Zproduct01}.

This weighted sum brings the Turaev-Viro TQFT and the string-net model into the game, and we can explicitly write the expression above as an overlap \eqref{overlap} of the PEPS tensor network states, in the Hilbert space of the string-net model, as shown in \eqref{strange_correlator}.

To recover the CFT partition function, we simply take the $R \to 0$ limit with the counterterm factor, as in \eqref{disentangle}.

\section{Tensor network SymQRG}\label{tensornetwork SymQRG}

In this appendix, we explain in more detail the tensor network SymQRG algorithm introduced in \cite{Vanhove:2018wlb, Chen:2022wvy, Cheng:2023kxh, Hung:2024gma, Chen:2024unp}.
\begin{figure}
	\centering
	\includegraphics[width=0.5\linewidth]{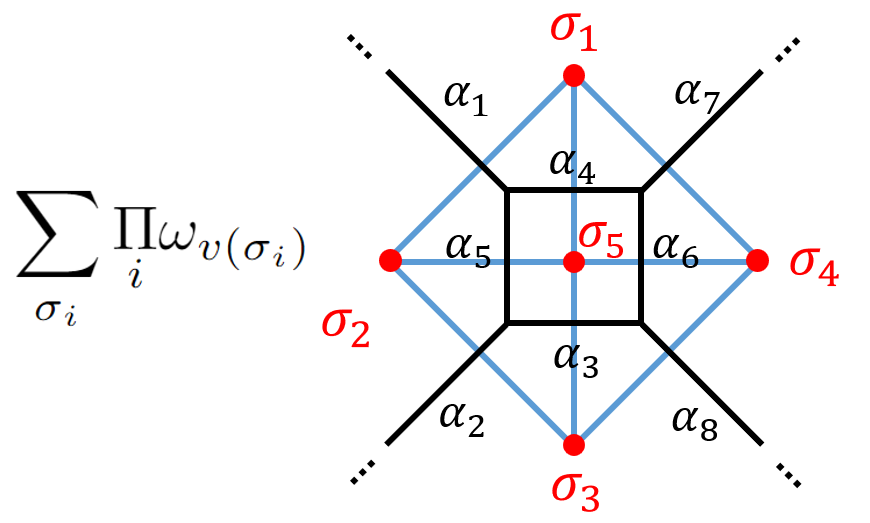}
    \caption{A local representation with five vertices and eight edges of the ground state $\ket{\Psi}_{\Lambda}$.}
	\label{localtriangle}
\end{figure}
We start with a local configuration of $\ket{\Psi}_{\Lambda}$ as in Fig.~\ref{localtriangle}. Explicitly in formula as in \eqref{psistate}, this state is locally expressed as
\be
\ket{\Psi}_{\Lambda, \text{local}}=\sum_{\sigma_{i}} \sum_{\alpha_{i}} \omega_{\sigma_1} \omega_{\sigma_2} \omega_{\sigma_3} \omega_{\sigma_4} \omega_{\sigma_5} \tilde{C}^{\sigma_5\sigma_1\sigma_2}_{\alpha_1 \alpha_5\alpha_4} \tilde{C}^{\sigma_5\sigma_2\sigma_3}_{\alpha_2 \alpha_3\alpha_5} \tilde{C}^{\sigma_5\sigma_3\sigma_4}_{\alpha_8 \alpha_6\alpha_3} \tilde{C}^{\sigma_5\sigma_4\sigma_1}_{\alpha_7 \alpha_4\alpha_6} \ket{\{\alpha_{i=1,\ldots,8} \}}_{\Lambda}~.
\ee
Alternatively, for a specific basis state labeled by the coloring $\{\alpha_{i=1,\ldots,8}\}$, the wavefunction is locally
\be
\Psi^{\Lambda}(\{\alpha_{i=1,\ldots,8} \})=\sum_{\sigma_{i}} \omega_{\sigma_1} \omega_{\sigma_2} \omega_{\sigma_3} \omega_{\sigma_4} \omega_{\sigma_5} \tilde{C}^{\sigma_5\sigma_1\sigma_2}_{\alpha_1 \alpha_5\alpha_4} \tilde{C}^{\sigma_5\sigma_2\sigma_3}_{\alpha_2 \alpha_3\alpha_5} \tilde{C}^{\sigma_5\sigma_3\sigma_4}_{\alpha_8 \alpha_6\alpha_3} \tilde{C}^{\sigma_5\sigma_4\sigma_1}_{\alpha_7 \alpha_4\alpha_6}~. 
\ee

\begin{figure}[H]
	\centering
	\includegraphics[width=0.8\linewidth]{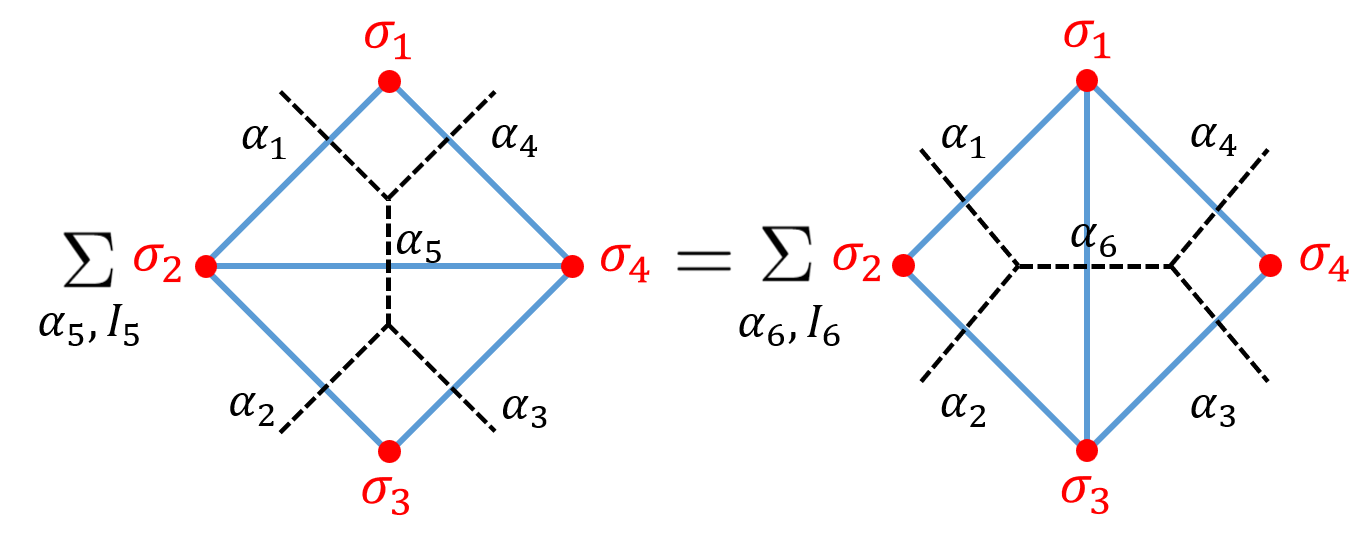}
    \caption{Crossing equation of the BCFT four point functions.}
	\label{fourpointcrossing}
\end{figure}

The BCFT OPE coefficients solve the conformal bootstrap equations. In particular, the crossing equation for the BCFT four-point functions is illustrated in Fig.~\ref{fourpointcrossing}.
As an algebraic equation, it gives
\be
\sum_{\alpha_5}\tilde{C}^{\sigma_4 \sigma_1 \sigma_2}_{\alpha_1 \alpha_5 \alpha_4}\tilde{C}^{\sigma_2 \sigma_3 \sigma_4}_{\alpha_3 \alpha_5 \alpha_2} \mathcal{\tilde{F}}^t (\alpha_5;\alpha_{i=1,4}, x_{i=1,4})=\sum_{\alpha_6}\tilde{C}^{\sigma_3 \sigma_1 \sigma_2}_{\alpha_1 \alpha_2 \alpha_6}\tilde{C}^{\sigma_1 \sigma_3 \sigma_4}_{\alpha_3 \alpha_4 \alpha_6} \mathcal{\tilde{F}}^s (\alpha_6;\alpha_{i=1,4}, x_{i=1,4})~,
\ee
where $\mathcal{\tilde{F}}^t$ and $\mathcal{\tilde{F}}^s$ are the conformal blocks in the $t$ and $s$ channel respectively in Racah gauge, and they are related via the crossing kernel:
\be
\mathcal{\tilde{F}}^t (\alpha_5;\alpha_{i=1,4}, x_{i=1,4})=\sum_{\alpha_6} F^{\text{Racah}}_{\alpha_6,\alpha_5} \begin{pmatrix}
\alpha_1 & \alpha_4 
\\
\alpha_2 & \alpha_3 
\end{pmatrix} \mathcal{\tilde{F}}^s (\alpha_6;\alpha_{i=1,4}, x_{i=1,4})~.
\ee

Thus the BCFT OPE coefficients are related by,
\be \label{4ptcrossing}
\tilde{C}^{\sigma_4 \sigma_1 \sigma_2}_{\alpha_1 \alpha_5 \alpha_4}\tilde{C}^{\sigma_2 \sigma_3 \sigma_4}_{\alpha_3 \alpha_5 \alpha_2} =\sum_{\alpha_6}\tilde{C}^{\sigma_3 \sigma_1 \sigma_2}_{\alpha_1 \alpha_2 \alpha_6}\tilde{C}^{\sigma_1 \sigma_3 \sigma_4}_{\alpha_3 \alpha_4 \alpha_6} F^{\text{Racah}}_{\alpha_6,\alpha_5} \begin{pmatrix}
\alpha_1 & \alpha_4 
\\
\alpha_2 & \alpha_3 
\end{pmatrix}~.
\ee
The crossing kernel in Racah gauge is related to the $6j$ symbols via,
\be \label{rationalkernel}
F^{\text{Racah}}_{\alpha_6,\alpha_5} \begin{pmatrix}
\alpha_1 & \alpha_4 
\\
\alpha_2 & \alpha_3 
\end{pmatrix} =\sqrt{d_{\alpha_5} d_{\alpha_6}}
\begin{Bmatrix}
\alpha_1 & \alpha_4 & \alpha_5\\
\alpha_3 & \alpha_2 & \alpha_6
\end{Bmatrix}~.
\ee

Notice that the $6j$ symbol above depends solely on the primary labels and not on the boundary conditions. Therefore, it is entirely determined by SymTFT in the bulk, rather than by the specific choice of $\ket{\Psi}_{\Lambda}$ or the topological boundary condition.

Diagrammatically, we have the following,
\begin{figure}[H]
	\centering
	\includegraphics[width=0.9\linewidth]{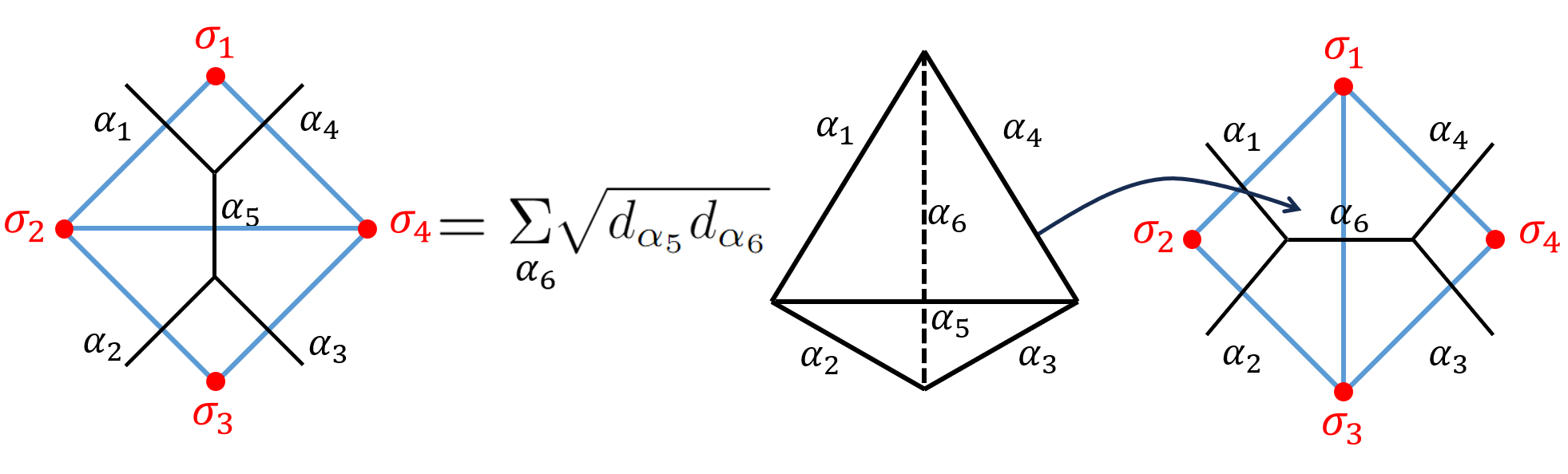}
	\label{crossingbcft}
\end{figure}

Since the wavefunctions for the string-net ground states are represented by the BCFT structure coefficients, this equation establishes a relationship between the states in two different triangulations. In the 3D perspective, this corresponds to modifying the 2D triangulation by gluing an additional layer of 3D tetrahedra, which is related to the 3-2 Pachner move.

Applying this equation twice, to both the left and right parts of Fig.~\ref{localtriangle}, we obtain:

\begin{figure}[H]
	\centering
	\includegraphics[width=1\linewidth]{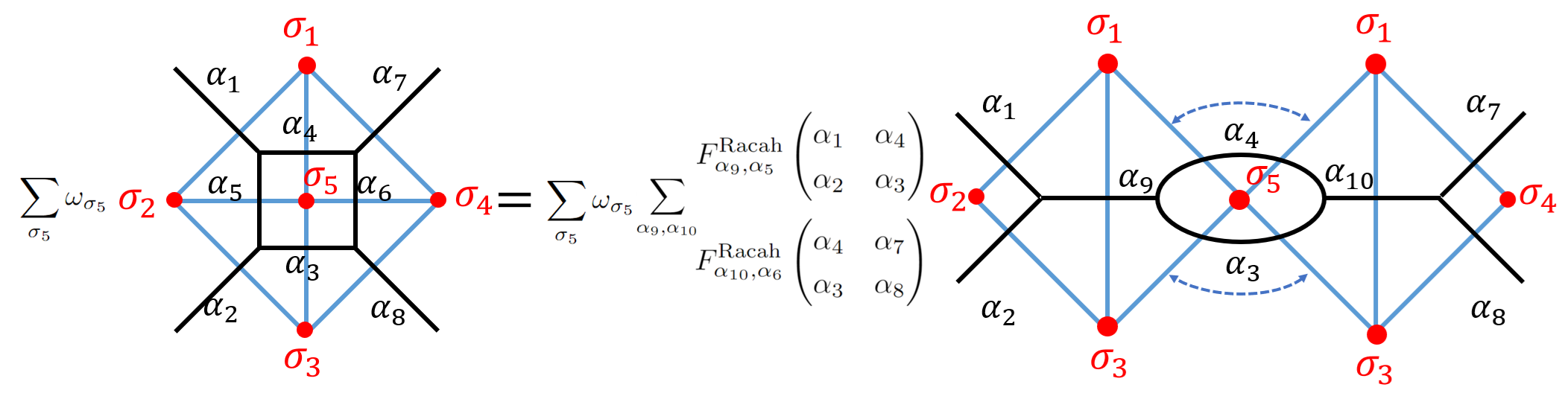}
	\label{creatingbubble}
\end{figure}

The orthogonality condition of the (generalized) $6j$ symbols is\cite{Fuchs:2002cm, *Fuchs:2004xi, Lootens:2020mso}, 

\be
\sum_{\sigma_5} d_{\sigma_5} \begin{Bmatrix}
\alpha_3 & \alpha_4 & \alpha_9\\
\sigma_1 & \sigma_3 & \sigma_5
\end{Bmatrix} \begin{Bmatrix}
\alpha_3 & \alpha_4 & \alpha_{10}\\
\sigma_1 & \sigma_3 & \sigma_5
\end{Bmatrix}=\frac{\delta_{{\alpha_9,\alpha_{10}}}}{d_{\alpha_9}} N_{\alpha_3 \alpha_4}^{\alpha_9} N_{\sigma_1 \sigma_3}^{\alpha_9}~,
\ee
where $N_{\alpha_3 \alpha_4}^{\alpha_9}$ and $N_{\sigma_1 \sigma_3}^{\alpha_9}$ take values of either $0$ or $1$, indicating whether the configuration is admissible. We can use it to get,
\begin{figure}[H]
	\centering
	\includegraphics[width=0.85\linewidth]{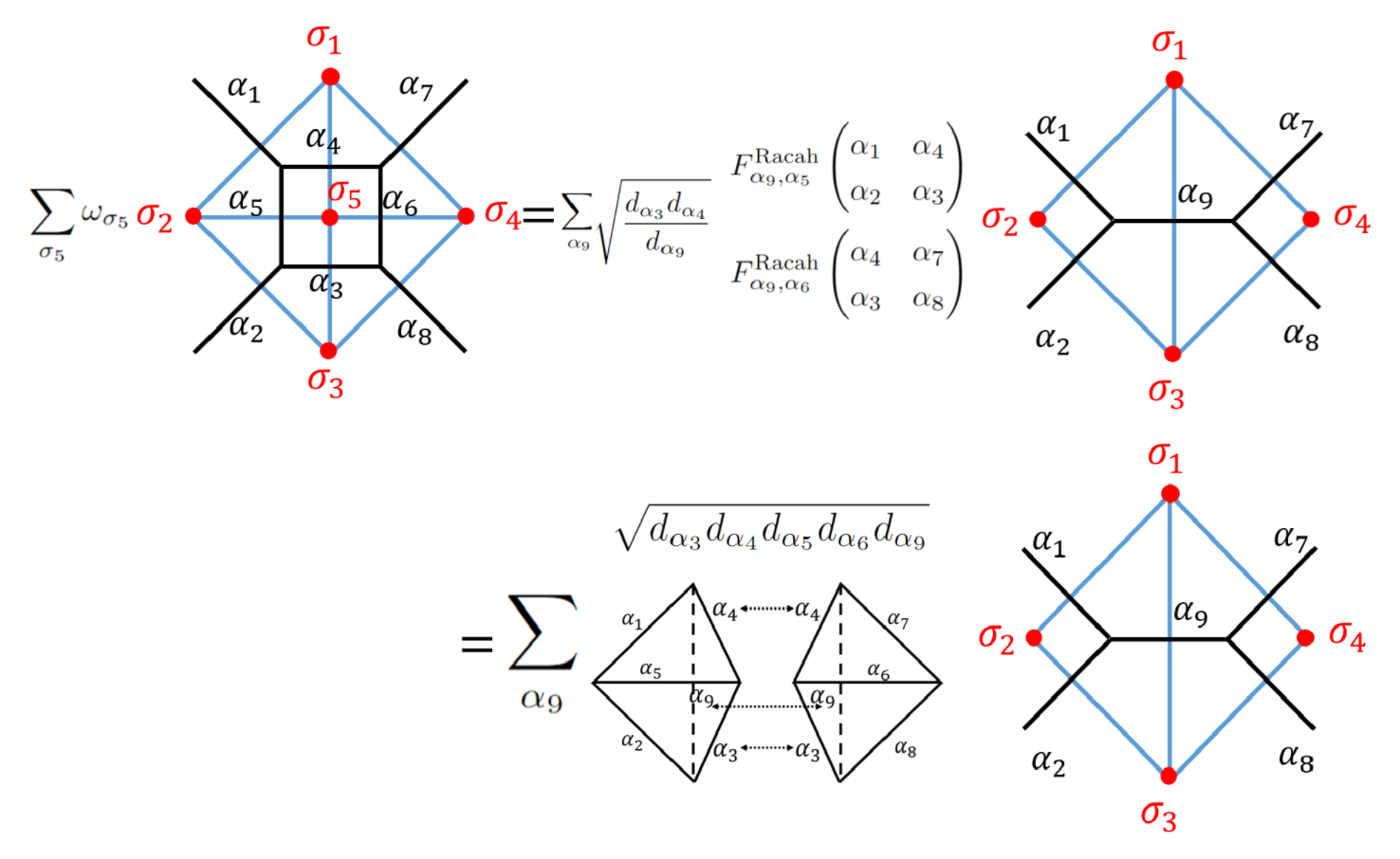}
    \caption{Coarse-graining equation of the Levin-Wen ground state $\ket{\Psi}_{\Lambda}$.}
	\label{removingbubble1}
\end{figure}

We recover the transformation rule \eqref{eq:URG} of the main text, relating the ground states at scales $\Lambda$ and $\Lambda'=\sqrt{2} \Lambda$.

As outlined in Sec.~\ref{symqrgparagraph}, the same procedure defines the coarse-graining map for ${}_\Lambda \langle \Omega^R|$, as in \eqref{RG_lattice}.

We can also incorporate the rescaling step to deform the theory from the overlap.

Now we illustrate the general procedure involving singular value decompositions for arbitrary seed states. We start with a general PEPS state ${}_{\Lambda} \langle \Omega|$. For example,  using a direct product seed state $_{\Lambda}\bra{\Omega}$ and taking its overlap with $\ket{\Psi}_{\Lambda}$ reproduces the Ising lattice model \cite{Aasen:2016dop, Vanhove:2018wlb}. By applying the RG procedure described below, the critical temperature can be determined numerically with high accuracy\cite{Chen:2022wvy}. 

\begin{figure}
	\centering
	\includegraphics[width=0.45\linewidth]{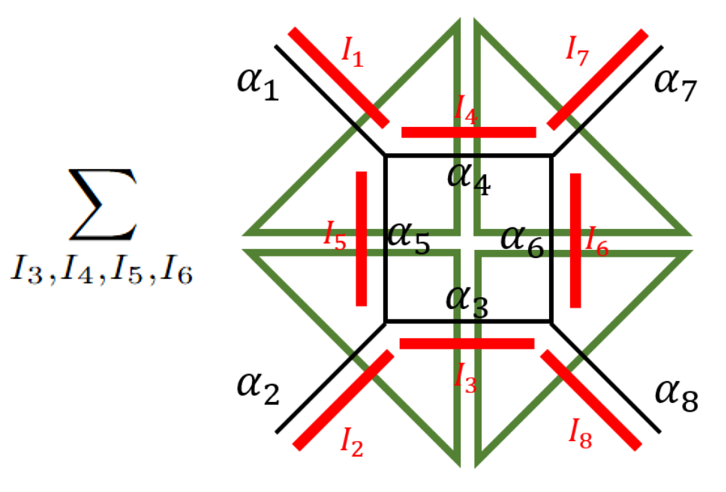}
	\caption{A local representation of a general PEPS state ${}_{\Lambda} \langle \Omega|$ to be coarse-grained.}
	\label{omegastate}
\end{figure}

A general initial PEPS state ${}_{\Lambda} \langle \Omega|$ is locally shown in Fig.~\ref{omegastate} and written as
\be
{}_{\Lambda} \langle \Omega|=\sum_{\alpha_i} {}_{\Lambda}\bra{\{\alpha_{i=1,8} \}} \sum_{I_3, I_4,I_5,I_6}    \Omega^{\alpha_1 \alpha_5\alpha_4}_{I_1 I_5 I_4} \Omega^{\alpha_2 \alpha_3\alpha_5}_{I_2 I_3 I_5} \Omega^{\alpha_8 \alpha_6\alpha_3}_{I_8 I_6 I_3} \Omega^{\alpha_7 \alpha_4\alpha_6}_{I_7 I_4 I_6}~.
\ee

The RG operator maps the state to
\be
\begin{aligned}
{}_{\Lambda'} \langle \Omega'| ={}_{\Lambda} \langle \Omega| \hat{U}^{\Lambda, \Lambda'}_{\text{RG}}=\sum_{\alpha_{i=1,2,7,8,9}} {}_{\Lambda'}\bra{\{\alpha_{i=1,2,7,8,9}\}} &\sum_{\alpha_{i=3,4,5,6}} \sqrt{\frac{d_{\alpha_3} d_{\alpha_4}}{d_{\alpha_9}}}
F^{\text{Racah}}_{\alpha_9,\alpha_5} \begin{pmatrix}
\alpha_1 & \alpha_4 
\\
\alpha_2 & \alpha_3 
\end{pmatrix}
F^{\text{Racah}}_{\alpha_{9},\alpha_6} \begin{pmatrix}
\alpha_4 & \alpha_7 
\\
\alpha_3 & \alpha_8 
\end{pmatrix}\\
& \sum_{I_3, I_4,I_5,I_6}    \Omega^{\alpha_1 \alpha_5\alpha_4}_{I_1 I_5 I_4} \Omega^{\alpha_2 \alpha_3\alpha_5}_{I_2 I_3 I_5} \Omega^{\alpha_8 \alpha_6\alpha_3}_{I_8 I_6 I_3} \Omega^{\alpha_7 \alpha_4\alpha_6}_{I_7 I_4 I_6}~.
\end{aligned}
\ee

\begin{figure}
	\centering
	\includegraphics[width=0.4\linewidth]{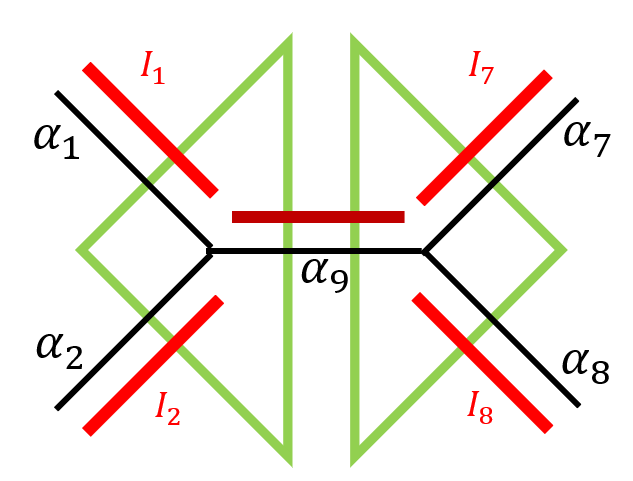}
	\caption{The singular value decomposition turns the result with four dangling physical legs into a contraction of contributions from two triangles.}
	\label{omegastate1}
\end{figure}

The next step is to express the wavefunctions as contractions of local tensors defined on the triangles as in Fig.~\ref{omegastate1}, using singular value decomposition \cite{PhysRevLett.99.120601, *PhysRevLett.115.180405, Chen:2022wvy}, as follows:
\be
\Omega^{'\Lambda'}(\alpha_{i=1,2,7,8,9})= \sum_{I_9} \Omega^{'\alpha_1 \alpha_2\alpha_9}_{I_1 I_2 I_9} \Omega^{'\alpha_8 \alpha_7\alpha_9}_{I_8 I_7 I_9}~.
\ee
In numerical computations, truncation is typically performed at this step. The bond dimensions in tensor networks are related to the amount of entanglement required to glue different pieces together. In the coarse-graining procedure towards the CFT fixed point, quantum entanglement is recursively generated until the bond dimension becomes infinite, reflecting long-range correlation which is captured by the presence of infinitely many descendant states. 

For the family of theories proposed in \eqref{strange_correlator}, the coarse-grained wavefunction is given by:
\be
\begin{aligned}
\Omega^{'\Lambda'}(\alpha_{i=1,2,7,8,9})=&\sum_{\alpha_{i=3,4,5,6}} \sqrt{\frac{d_{\alpha_3} d_{\alpha_4}}{d_{\alpha_9}}}
F^{\text{Racah}}_{\alpha_9,\alpha_5} \begin{pmatrix}
\alpha_1 & \alpha_4 
\\
\alpha_2 & \alpha_3 
\end{pmatrix}
F^{\text{Racah}}_{\alpha_{9},\alpha_6} \begin{pmatrix}
\alpha_4 & \alpha_7 
\\
\alpha_3 & \alpha_8 
\end{pmatrix}\\
& \sum_{I_3, I_4,I_5,I_6}    \tilde{\gamma}^{\alpha_1 \alpha_5\alpha_4}_{I_1 I_5 I_4} \tilde{\gamma}^{\alpha_2 \alpha_3\alpha_5}_{I_2 I_3 I_5} \tilde{\gamma}^{\alpha_8 \alpha_6\alpha_3}_{I_8 I_6 I_3} \tilde{\gamma}^{\alpha_7 \alpha_4\alpha_6}_{I_7 I_4 I_6}~. 
\end{aligned}
\ee
Diagrammatically,
\begin{figure}[H]
	\centering
	\includegraphics[width=0.8\linewidth]{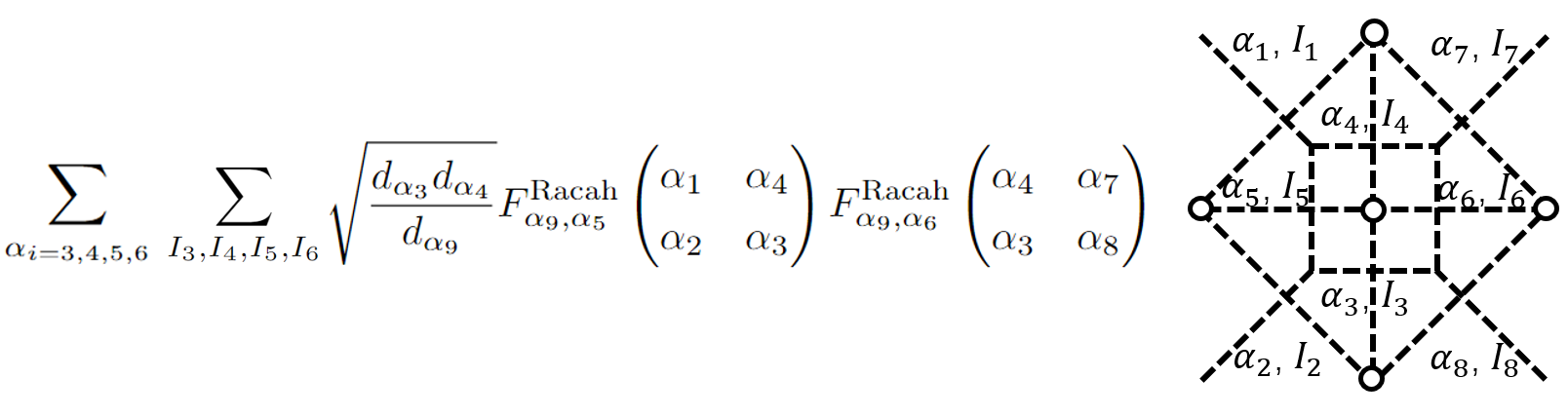}
	\label{omegagamma2}
\end{figure}
We first use the definition of crossing kernel to turn the conformal blocks on the left and right to the crossed channel.
\begin{figure}[H]
	\centering
	\includegraphics[width=0.65\linewidth]{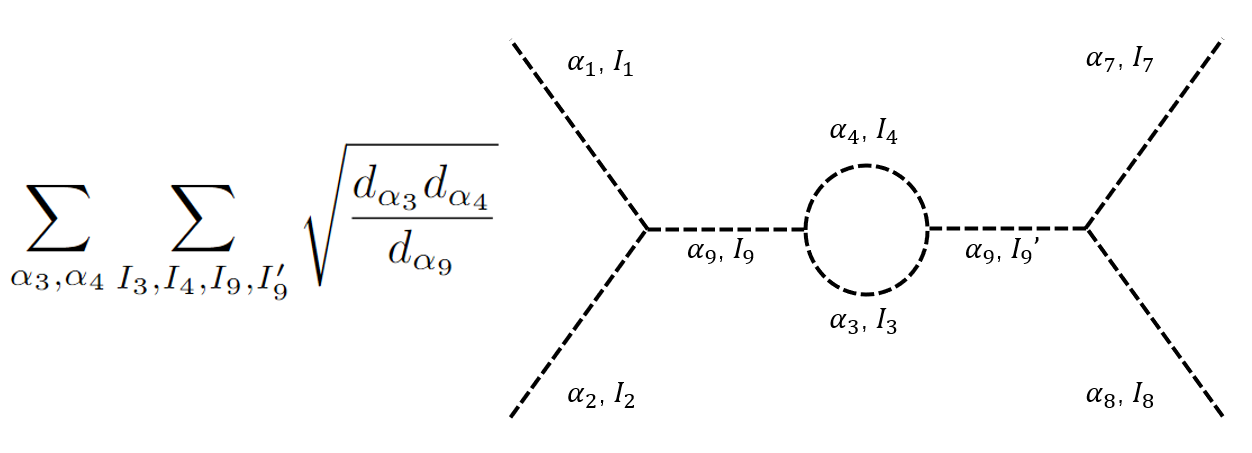}
	\label{crossinggamma1}
\end{figure}
Next, we use the identity in Racah gauge:
\be
F^{\text{Racah}}_{\mathbb{1},\alpha_4} \begin{pmatrix}
\alpha_9 & \alpha_3 
\\
\alpha_9 & \alpha_3 
\end{pmatrix}=\sqrt{\frac{d_{\alpha_4}}{d_{\alpha_3} d_{\alpha_9}}}~,
\ee
to turn this into
\begin{figure}[H]
	\centering
	\includegraphics[width=0.65\linewidth]{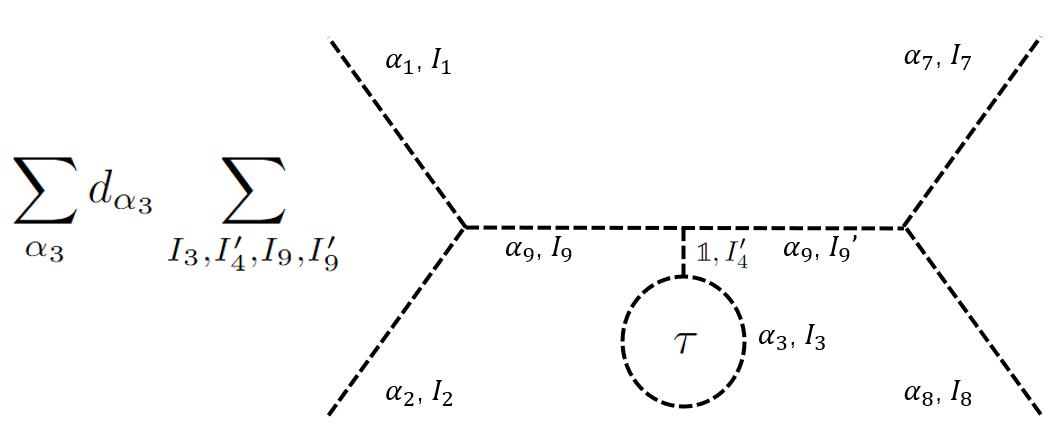}
	\label{crossinggamma2}
\end{figure}
Finally, we use the fact that $d_{\alpha_3} \propto S_{\mathbb{1} \alpha_3}$, to get,
\begin{figure}[H]
	\centering
	\includegraphics[width=0.6\linewidth]{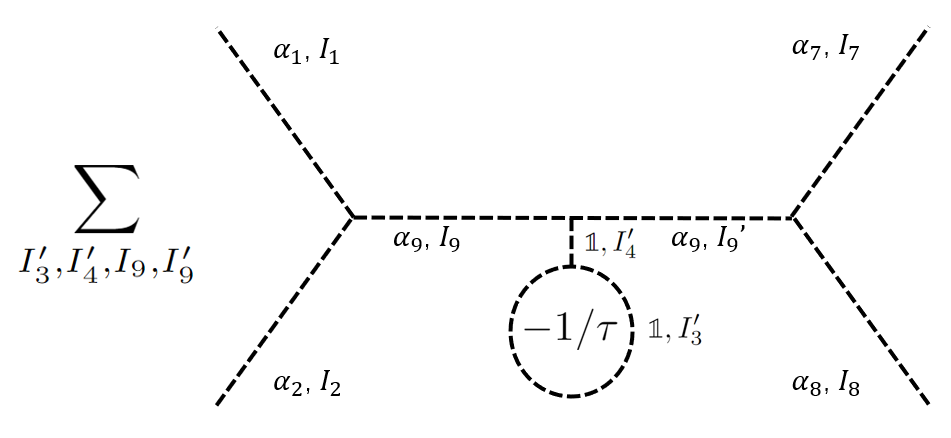}
	\label{crossinggamma4}
\end{figure}
and we turn this into the contribution of two triangles as we explain in the main text.

Away from the $R\to 0$ fixed point, singular value decomposition needs to be performed to update the tensor values assigned to the triangles after each RG step. It would be fascinating to explore in greater detail the properties of this non-perturbatively defined family of theories and their connection to continuum field theories.

\section{Tensor network for CFT wavefunctions from Euclidean path integrals, perfect tensors and random tensors}\label{wavefunction tensornetwork}

In this appendix, we explain exact tensor network construction for CFT wavefunctions from Euclidean path integrals. We also discuss the connection to perfect tensors and random tensors.

The initial motivation for connecting tensor networks to gravity did not originate from constructing a spacetime tensor network to represent the local dictionary for partition functions, as we have done above. Instead, Swingle observed that the Ryu-Takayanagi formula, used to compute the entanglement entropy of holographic CFT quantum states on codimension-one surfaces, exhibited a striking resemblance to the entanglement structure in ground states represented by tensor networks\cite{Vidal:2008zz, Swingle:2009bg}. Swingle's analogy suggested that tensor networks could encode the spatial entanglement patterns within a CFT in a manner that mirrors the holographic entanglement entropy calculation in the AdS/CFT correspondence.

\begin{figure}
	\centering
	\includegraphics[width=0.8\linewidth]{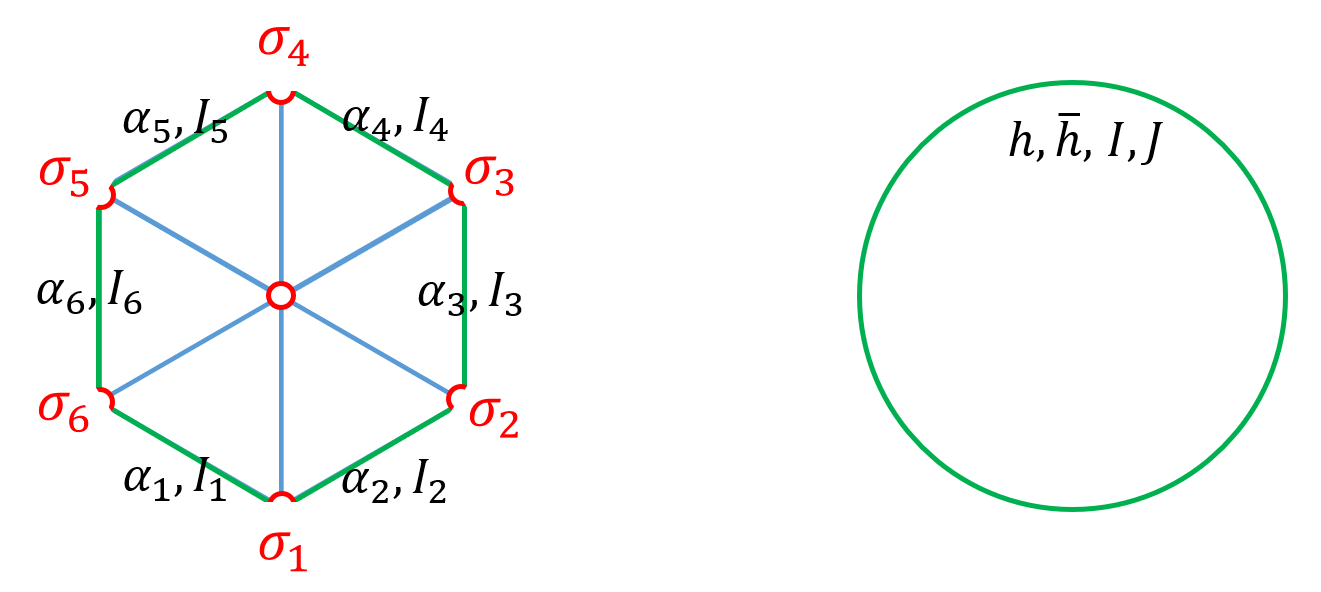}
	\caption{Two different ways of obtaining a state living on a circle by Euclidean path integral. The open CFT triangulation method provides a state with 18 labels on the left, and the closed CFT method provides a state with 4 labels on the right.}
	\label{diskpathintegral}
\end{figure}

On the other hand, as discussed in \cite{Chen:2024unp}, our approach extends the application of tensor networks by constructing a representation for codimension-one quantum states prepared through a Euclidean CFT path integral. This construction directly links to conventional studies of spatial entanglement in quantum states. In the following, we first review this construction and then demonstrate how it aligns with previously proposed ``perfect tensor network'' and ``random tensor network'' models for holography\cite{Pastawski:2015qua, Hayden:2016cfa}. Notably, our approach encodes dynamical information for both the AdS and the CFT, providing a refined exact framework for holographic tensor networks that goes beyond toy models.

\begin{figure}
	\centering
	\includegraphics[width=0.65\linewidth]{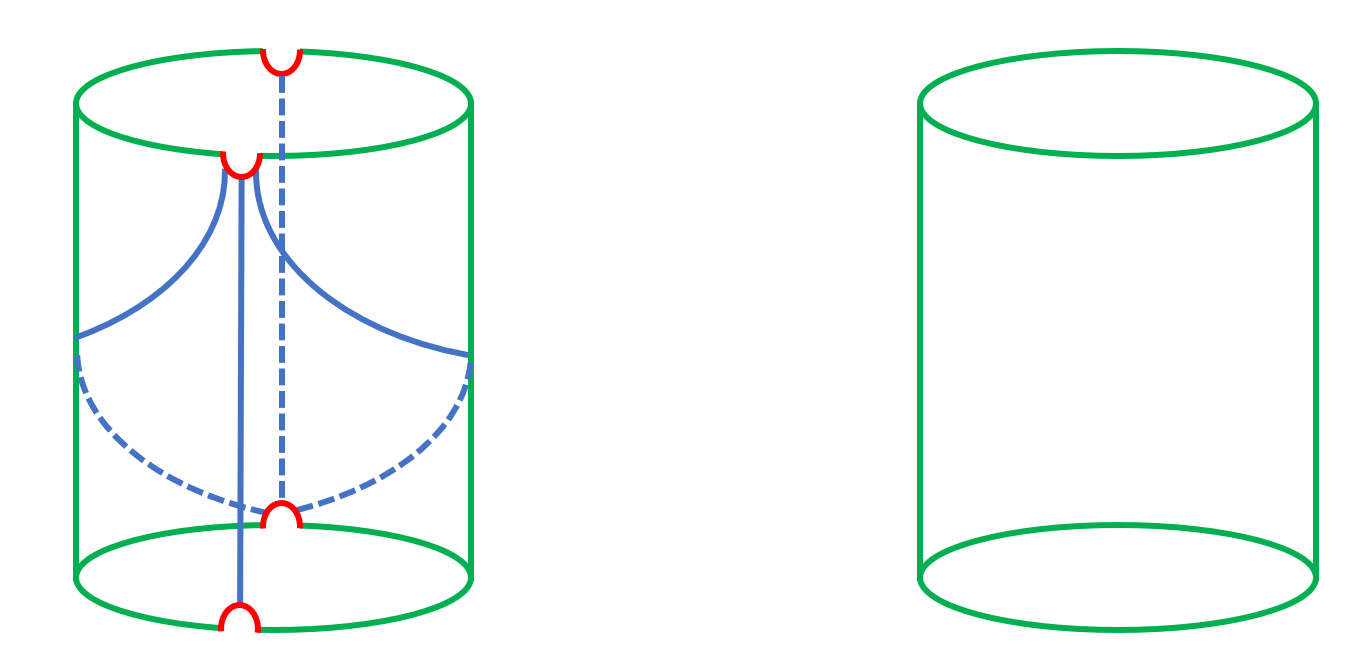}
	\caption{Two different ways of computing the CFT propagator represented by a cylinder. The closed CFT spectrum can be extracted from the open CFT computation on the left.}
	\label{cylinder}
\end{figure}

The idea here is to consider CFT path integrals defined on manifolds with boundaries rather than on closed manifolds, which yield partition functions. We will still triangulate these path integrals. For example, the CFT ground state defined on a circle can be prepared using the Euclidean path integral on a disk, as shown in Fig.~\ref{diskpathintegral}. The primary distinction from the partition function case is that some auxiliary legs, which would otherwise be contracted, now become dangling physical legs.

In Fig.~\ref{diskpathintegral}, for instance, the Hilbert space consists of 18 labels corresponding to the basis states $\{\ket{\alpha_i, I_i, \sigma_i}, i=1 \ldots 6\}$. Here, $\alpha_i$ represents the primary fields, $I_i$ the descendants, and $\sigma_i$ the boundary conditions. Typically, CFT techniques assign a Hilbert space on the circle labeled by closed CFT operators $\{\ket{h, \bar{h}, I, J}\}$, where $h$ and $\bar{h}$ are the chiral and anti-chiral conformal dimensions of the closed CFT primaries, and $I, J$ are the descendant labels. These two Hilbert spaces are, in fact, equivalent.

\begin{figure}
	\centering
	\includegraphics[width=0.65\linewidth]{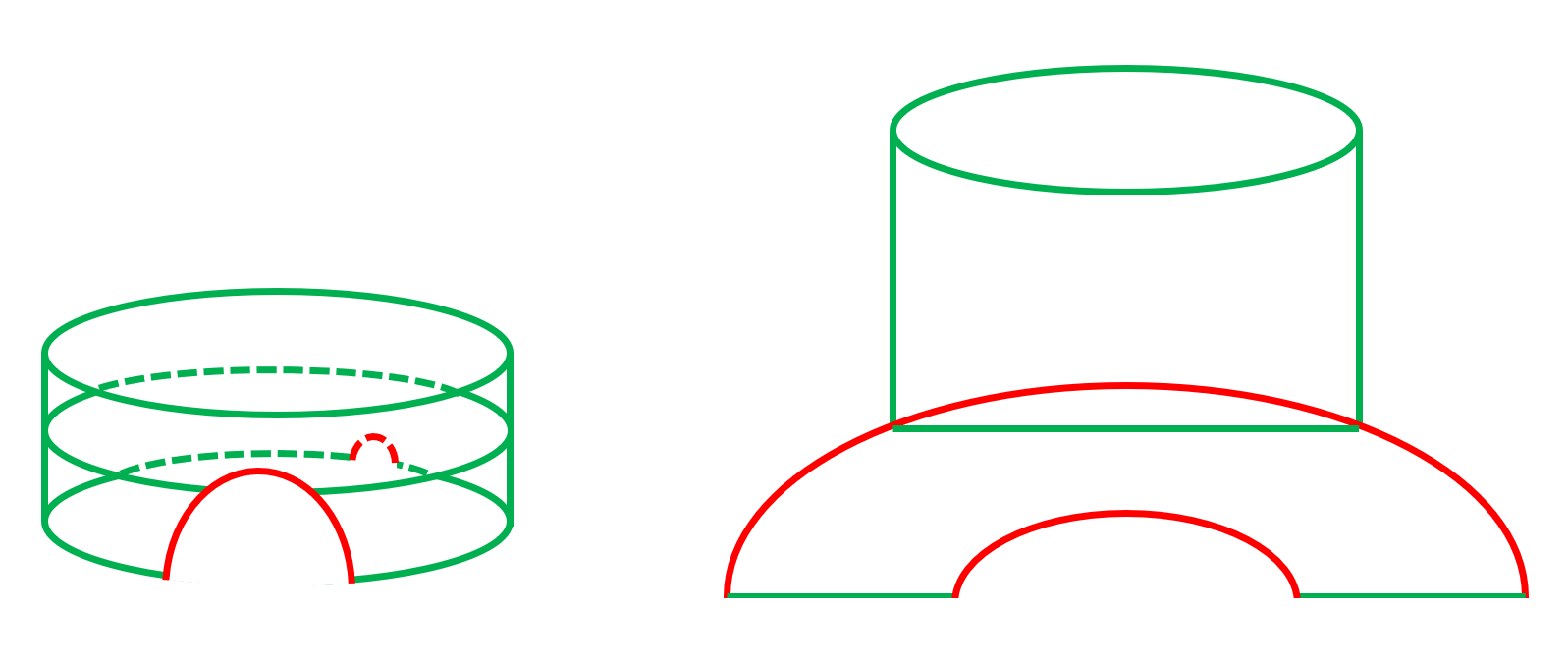}
	\caption{The map between the open and closed bases is obtained from gluing the ``whistle diagram'' (the open-closed one-point function) to the open pair of pants.}
	\label{closedandopen}
\end{figure}

To check this equivalence, consider the path integral on a cylinder, as illustrated in Fig.~\ref{cylinder}. In the closed CFT picture (right side), this computation corresponds to the matrix element of $e^{-\beta \hat{H}}$, where the indices are labeled by $\{h, \bar{h}, I, J\}$, and the eigenvalues of the transfer matrix are given by the closed CFT spectrum. On the left side, we compute the matrix element of this transfer matrix using triangulation, where the matrix indices are labeled with a different set, $\{\alpha_i, I_i, \sigma_i, i=1 \ldots 4\}$. However, upon diagonalizing this matrix, we find that the spectrum matches exactly with the closed CFT results. This equivalence has been checked numerically for several minimal models. Remarkably, it was observed that even when truncating the number of descendants to a very small number (e.g.~2 or 3) for each edge of a triangle, the resultant glued cylinder can still capture the lowest $\sim 20$ states (including closed CFT primaries and descendants) with high accuracy \cite{Cheng:2023kxh}.

The map between these two bases is shown in Fig.~\ref{closedandopen}. The equivalence of these two Hilbert spaces follows from the ability to shrink the holes in Fig.~\ref{shrinkableholes} in the $R \to 0$ limit, as detailed in appendix \ref{tensornetwork SymQRG}.

\begin{figure}
	\centering
	\includegraphics[width=0.6\linewidth]{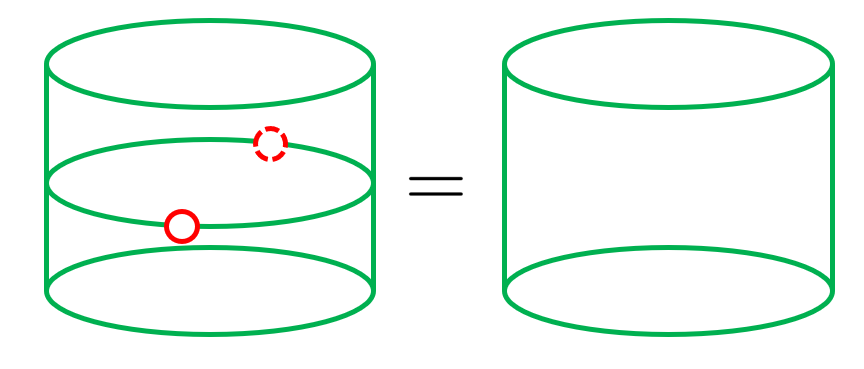}
	\caption{The equivalence of the open and closed Hilbert spaces can be seen from the fact that the tiny red holes are contractible as we showed in appendix \ref{tensornetwork SymQRG}.}
	\label{shrinkableholes}
\end{figure}

\begin{figure}
	\centering
    \includegraphics[width=0.8\linewidth]{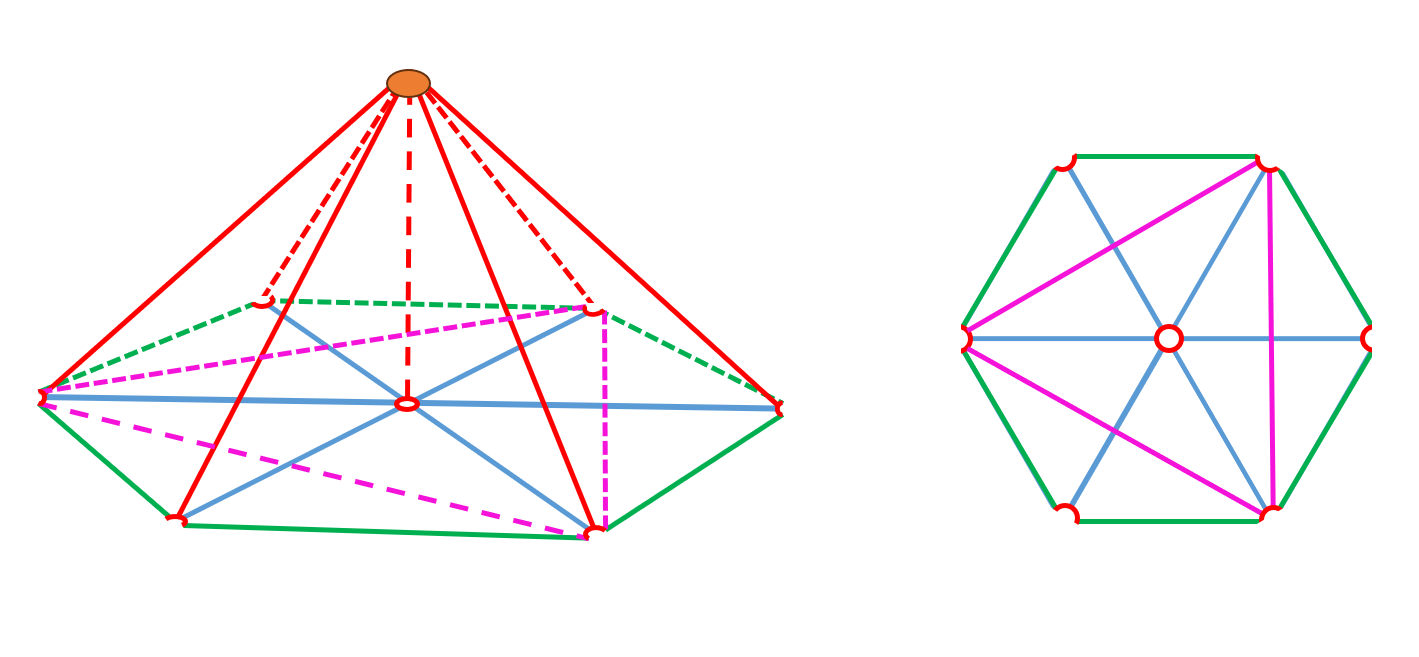}
	\caption{We can separate the OPE coefficients and the conformal blocks, as we did for the partition functions above, leading to an emergent 3D interpretation. Using the crossing move, we can also transition from the blue edges to the purple edges. The Cauchy slice in the bulk corresponds to the surface on the top with red edges.}
	\label{cauchyslice}
\end{figure}

We can also separate the BCFT OPE piece and the conformal block piece, as we did for the partition function. This decomposition is illustrated in Fig.~\ref{cauchyslice}.

The surface at the top, denoted by the red lines, represents a Cauchy slice in the bulk. Additional layers of the tensor network can be constructed using the crossing move shown in Fig.~\ref{fourpointcrossing}. For example, the next layer corresponds to the purple lines in the diagram above.

Since the tensor network reproduces the exact CFT results, it inherently incorporates the correct kinematics and dynamics of the CFT, which are both long-standing difficulties. This is achieved because one part of the tensors encodes the OPE coefficients, which captures the dynamical information of the CFT, while the other part arises from the conformal blocks, providing the correct scaling laws in the CFT\footnote{In the literature, there is another suggested method to produce a tensor network related to a Cauchy slice from QFT path integrals using ``Cauchy slice holography''\cite{Araujo-Regado:2022gvw, Soni:2024aop, soniyoutube}. It will be interesting to understand the connection with our formalism. We thank Ronak Soni for pointing this out to us.}.

Next, we will demonstrate how our wavefunction tensor network relates to and generalizes the best existing tensor network models proposed to describe holography, namely the perfect tensor network \cite{Pastawski:2015qua} and the random tensor network \cite{Hayden:2016cfa}. Our construction incorporates the intrinsic CFT data into the tensor network, which are necessary to describe both AdS and CFT, thereby establishing a connection between tensor networks and gravity that goes beyond mere analogy.

\subsubsection*{Quasi-Perfect tensor network}

The first model constructed to exactly saturate the Ryu-Takayanagi formula is the HaPPY code introduced in \cite{Pastawski:2015qua}, which uses perfect tensors for the tensor network. The model is constructed explicitly to recover wavefunctions of 2D models that live on a line. 
The holographic network that prepares the wavefunction covers a two-dimensional space in a way very similar to the MERA tensor network\cite{Vidal:2008zz}. 
The idea is to consider placing a tensor network over the dual graph of a given (regular) tiling over hyperbolic planes.
The polygons chosen in the tiling carry an even number of edges, and each host a perfect tensor with even number of legs. 
The property of a perfect tensor $U^{i_1\cdots i_{2N}}$ is such that when one treats any $N$ of the $2N$ indices as input indices and the remaining $N$ indices as output indices, the matrix remains unitary. The dangling indices in the tensor network can then be identified with the Hilbert space in which the wavefunction lives. 

Consider the computation of the entanglement entropy of a simply connected interval $A$ for the tensor network wavefunction. When the indices belonging to the complement $\bar A$ are contracted with the complex conjugate of the wavefunction, the property of a hyperbolic tiling is such that each contracted tensor living at the boundary would have more than half of its indices contracted. As a perfect tensor, this produces the identity operator, and the contraction of indices thus propagates into the bulk of the hyperbolic plane. In \cite{Pastawski:2015qua} it was shown that this process ends precisely when one reaches a minimal cut that separates $A$ and $\bar A$. The entanglement spectrum of the resultant reduced density matrix is flat, and its rank is given precisely by the number of links cut by the minimal cut across the tiling. The entanglement entropy is thus proportional to $L \ln D $, where $D$ is the bond dimension of each index and $L$ is the number of links cut by the minimal cut.
This thus reproduces the Ryu-Takayanagi formula. 
\begin{figure}
	\centering
	\includegraphics[width=0.8\linewidth]{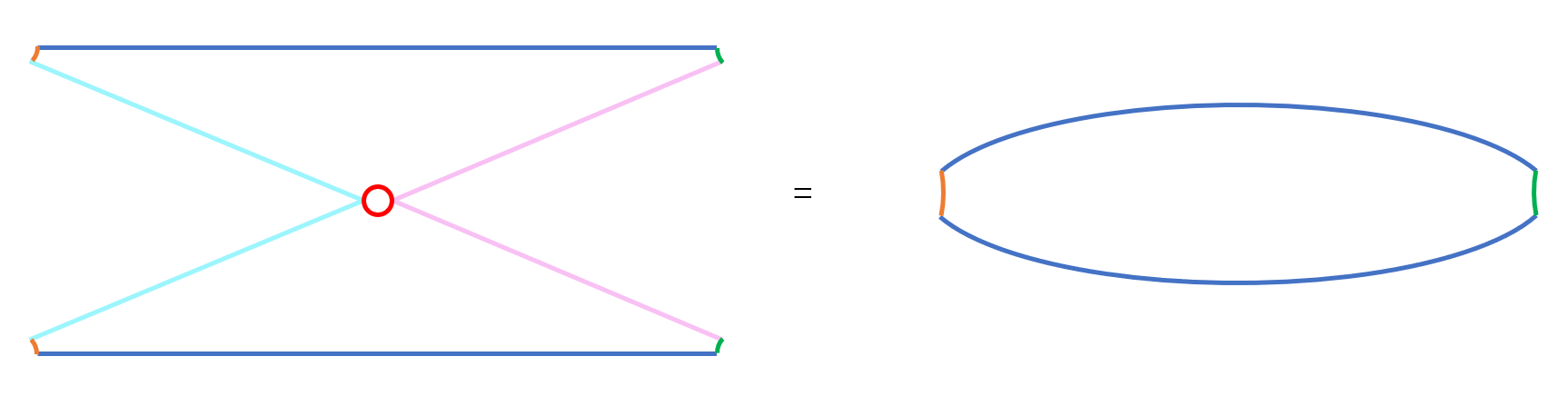}
    \caption{The ``quasi-perfect-isometric'' condition of tensor network for CFT states from Euclidean path integral.}
	\label{QPI}
\end{figure}
In our construction of the CFT wavefunction as discussed above and illustrated in Fig.~\ref{diskpathintegral}, a structure emerges that partly resembles the perfect tensor. Take any two triangles and contract any two of the edges and the corner index between the two chosen edges, resulting in the ``propagator'' of the remaining edge, as illustrated in Fig.~\ref{QPI}. This is referred to as the ``quasi-perfect-isometric'' condition in \cite{Chen:2024unp}. 

\begin{figure}
	\centering
	\includegraphics[width=0.8\linewidth]{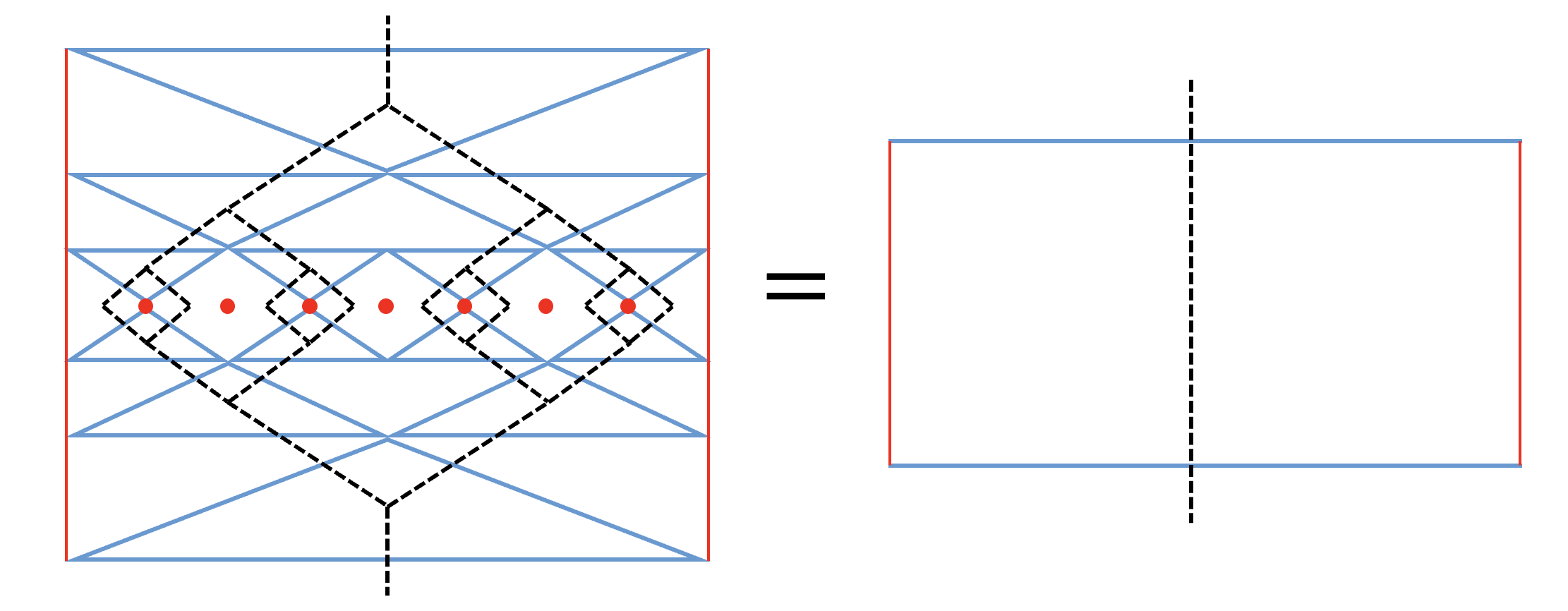}
	\caption{The ``quasi-perfect-isometric'' condition in our CFT wavefunction tensor network allows us to simplify the single-interval reduced density matrix to an open CFT propagator.}
	\label{cancel}
\end{figure}

\begin{figure}
	\centering
	\includegraphics[width=0.8\linewidth]{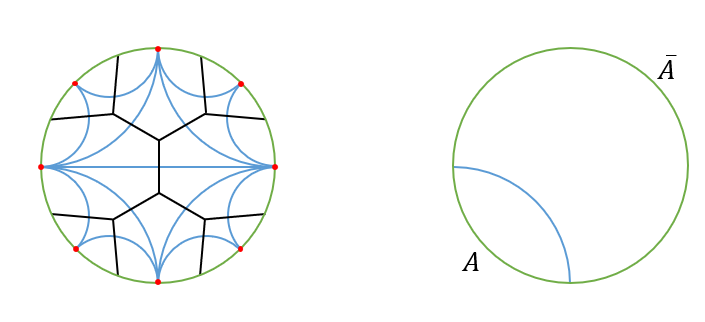}
	\caption{Left: the tensor network after repeated use of the crossing relation, in MERA-like form. Right: the cancellation in the reduced-density-matrix computation using the quasi-perfect-isometric property ends at the Ryu-Takayanagi surface.}
	\label{perfecttensor}
\end{figure}

When we construct the wavefunction with a canonical choice of triangulation as in Fig.~\ref{diskpathintegral}, it is not immediately clear how to compute the reduced density matrix. To make use of the ``quasi-perfect-isometric'' condition, we can convert the triangulation using crossing relations, as illustrated in Fig.~\ref{cauchyslice}, in which the crossing relation allows one to replace the blue lines by the purple lines. By repeated use of the crossing relation, one can convert the tensor network to take the form of a MERA-like tensor network, as illustrated in the left figure of Fig.~\ref{perfecttensor}. When we compute the reduced density matrix, the quasi-perfect isometric condition would play a similar role as the perfect condition, as illustrated in Fig.~\ref{cancel}. The cancellation ends at the Ryu-Takayanagi surface illustrated in the right figure of Fig.~\ref{perfecttensor}. 

\begin{figure}
	\centering
	\includegraphics[width=0.8\linewidth]{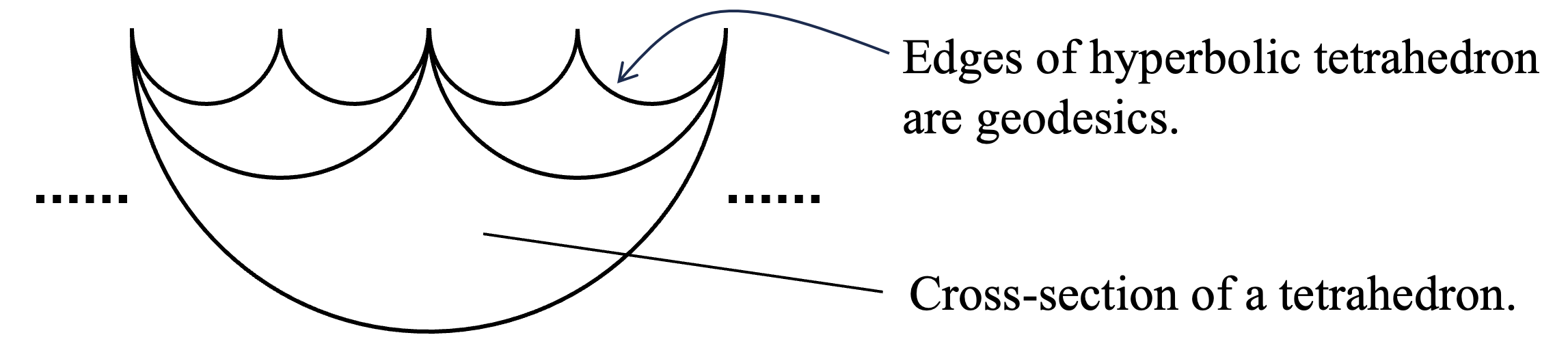}
	\caption{The edges in our tensor networks have the physical meaning of hyperbolic geodesic lengths, and this will be explained in Sec.~\ref{TQFTSECTION}.}
	\label{crosssection}
\end{figure}

The edges of the triangles correspond to geodesics in the 3D hyperbolic space, with the geodesic lengths associated with the Virasoro representation labels propagating along these edges, as illustrated in Fig.~\ref{crosssection}, turning it into a geodesic tensor network. This connection is the primary focus of Sec.~\ref{TQFTSECTION}.

As an exact reconstruction of the CFT wavefunction, the entanglement spectrum is not flat but reproduces exactly that of the CFT. 

\subsubsection*{Random tensor network}
In the discussion so far, we have been working with a given CFT, where its structure coefficients exactly solve all the consistency relations of the CFT. Alternatively, one can consider an ensemble average of CFTs by invoking the eigenstate thermalization hypothesis (ETH). (A further discussion of the connection between gravity and ETH is presented in Sec.~\ref{ETH}.) This approach involves treating the OPE coefficients as (pseudo)random variables\cite{Collier:2019weq, Belin:2020hea}. 

In the context of 3D gravity, the idea of seeing the emergence of gravity by treating closed OPE coefficients in the pair-of-pants decomposition as random tensors was introduced in \cite{Verlindetalk, *Verlindeunpublished}, and later adapted into a tensor network framework in \cite{Chandra:2023dgq}. For example, a simple MPS tensor network, constructed with random closed OPE coefficients in a 2D CFT as tensors, can be interpreted as discretizing the radial direction in the bulk dual in the semi-classical approximation, and approximately reproduces the location of the horizon\cite{Verlindetalk, *Verlindeunpublished, Chandra:2023dgq}. In our framework, which recovers the full CFT path integral, it is also natural to consider an ensemble average of the structure coefficients for boundary-changing operators.
The ensemble average of the structure coefficients for the boundary-changing operators in the large central charge limit has been studied in \cite{Kusuki:2021gpt, *Kusuki:2022wns, Numasawa:2022cni}. Applied to our formalism, this results in a random tensor network that represents the ensemble-averaged CFT path integral as outlined in \cite{Chen:2024unp}. A crucial distinction in our tensor network construction is worth emphasizing: We used the \textit{open} structure coefficients of the CFT, rather than the \textit{closed} ones, to build the CFT tensor network. From our perspective, the open structure coefficients are more directly connected to the emergence of bulk geometry and allow for discretizing both the CFT and AdS spaces arbitrarily, not only along the radial direction, and without having to come with external operator insertions.

Although the resulting random tensor network contains some ingredients similar to those in \cite{Hayden:2016cfa}, there is a key difference: Tensors at different locations in our construction are not treated as independent random variables.

\section{Quantum Teichmüller theory and its connection to $U_q(SL(2,\mathbb{R}))$ $6j$ symbols}\label{reviewteichmuller}

This appendix first provides a review of quantum Teichmüller theory and the Teichmüller TQFT, then we show how the tetrahedron operator in quantum Teichmüller theory is related to $U_q(SL(2,\mathbb{R}))$ $6j$ symbols we used in our main text.

\subsection{Reviewing quantum Teichmüller theory and Teichmüller TQFT}

The central idea of quantum Teichmüller theory and Teichmüller TQFT follows from constructing explicit representations of the mapping class group on a hyperbolic Riemann surface\cite{Teschner:2005bz}. 

In \cite{Nidaiev:2013bda}, it is shown that the quantum $6j$ symbols in $U_q(SL(2,\mathbb{R}))$ are related to the tetrahedron operator $T$ introduced by Kashaev in quantum Teichmüller theory \cite{1997q.alg.....5021K, *Kashaev:2000ku, Teschner:2005bz}. 

To appreciate the relation given in \cite{Nidaiev:2013bda}, we provide a brief review of quantum Teichmüller theory and the $T$ operator introduced in \cite{1997q.alg.....5021K, *Kashaev:2000ku}. To ensure convergence within the discrete formulation of Teichmüller TQFT, these $T$ operators were later extended to their charged counterparts, as developed in \cite{EllegaardAndersen:2011vps}.

A hyperbolic Riemann surface satisfies
\be  \label{Mhyperbolic}
M\equiv 2g + n - 2 > 0~,
\ee
where $g$ is the genus and $n$ is the number of punctures. (In general, we could consider $n$ finite boundary components instead, but in the following we will illustrate the main ideas when these boundary components have infinitesimal sizes, and thus punctures.)

Consider the space of deformations of complex structures on the surface $\Sigma$. This is called the Teichmüller space $\mathcal{T}(\Sigma)$. For a chosen complex structure on the given Riemann surface, there is a unique metric with constant negative curvature. 
The metric can be represented as $ds^2 = e^{2\phi} dz d\bar z$, where $\phi$ satisfies the Liouville equation
\be
\partial \bar\partial \phi = \frac{1}{4} e^{2\phi}~.
\ee
The Teichmüller space can be identified with the space of deformations of these constant-curvature metrics. 
The idea of Penner coordinates for the Teichmüller space is to use geodesic lengths to parameterize the space of metrics\cite{1987CMaPh.113..299P, *10.4310/jdg/1214448257}. 
Consider, thus, this specific hyperbolic conformal frame. The punctures are mapped to cusps. In the following, we would be using these two terms interchangeably.

\begin{figure}[H]
	\centering
	\includegraphics[width=0.4\linewidth]{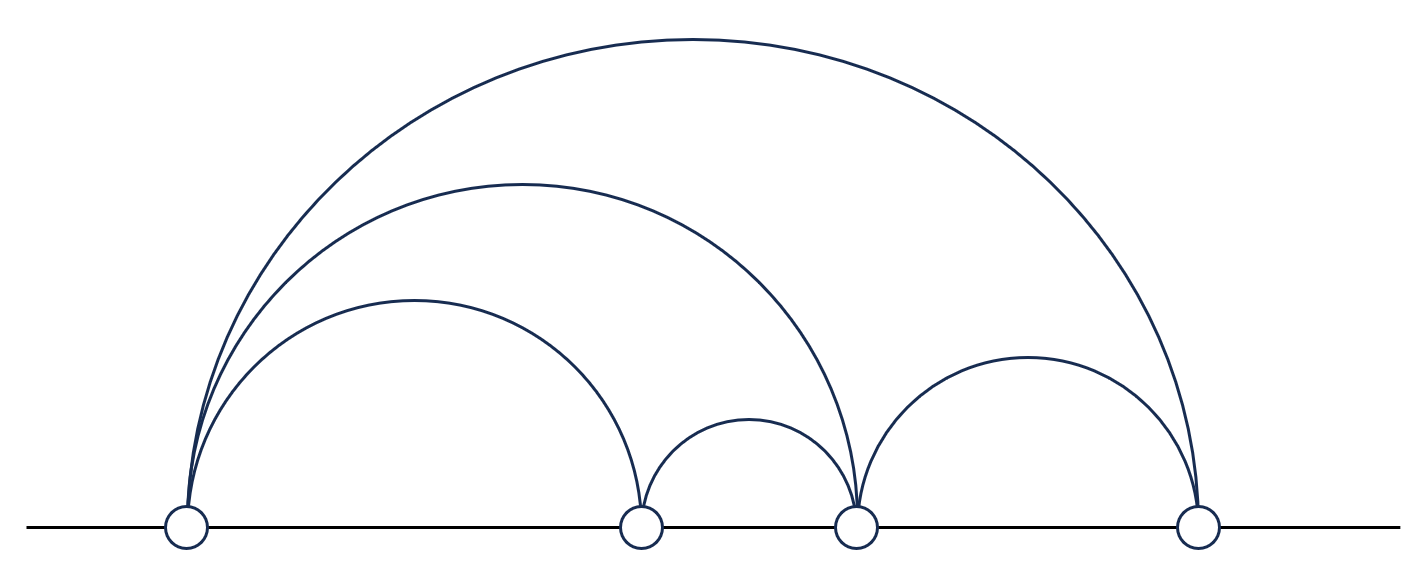}
	\caption{An ideal triangulation with four punctures.}
	\label{idealtriangle}
\end{figure}

Consider triangulating the surface. The surface of genus $g$ with cusps $n$ can be triangulated with triangles whose vertices are given by the puncture, and the edges are geodesics that run between the punctures.
This triangulation $\tau$ is called an ``ideal triangulation''. 
Since these edges connect the punctures, they have infinite geodesic lengths. 
To regulate these edge lengths, pick a closed curve of constant geodesic curvature surrounding each puncture. Each edge $e$ intersects each curve once. The regulated geodesic length $l_e$ for the edge $e$ is given by the segment between the intersections with the closed curves. See Fig.~\ref{idealtriangle} for an illustration. The regulated edge lengths $\{l_e\}$ can be taken as a coordinate of the Teichmüller space. 


The mapping class group of the surface corresponds to (orientation-preserving) diffeomorphisms not isotopic to the identity. 
i.e.~
\be
\textrm{MCG}(\Sigma) = \textrm{Diff}^+(\Sigma)/\textrm{Diff}^+_0(\Sigma)~,
\ee
where $\textrm{Diff}^+$ are orientation-preserving diffeomorphisms.

The triangulation of a surface can be changed. In particular, one can consider two neighboring triangles that share an edge. One can remove the shared edge and replace it by another edge connecting the other two vertices. Considering the dual graph of the triangles (the dual graph is called the ``fat graph'' in this context), this change in triangulation is, in fact, the crossing transformation. Under a change of triangulation, the tuples of edge lengths would change. This change can be considered as a change of coordinates of the Teichmüller space. These transformations generate the ``Ptolemy groupoid''.

\begin{figure}
	\centering
	\includegraphics[width=0.9\linewidth]{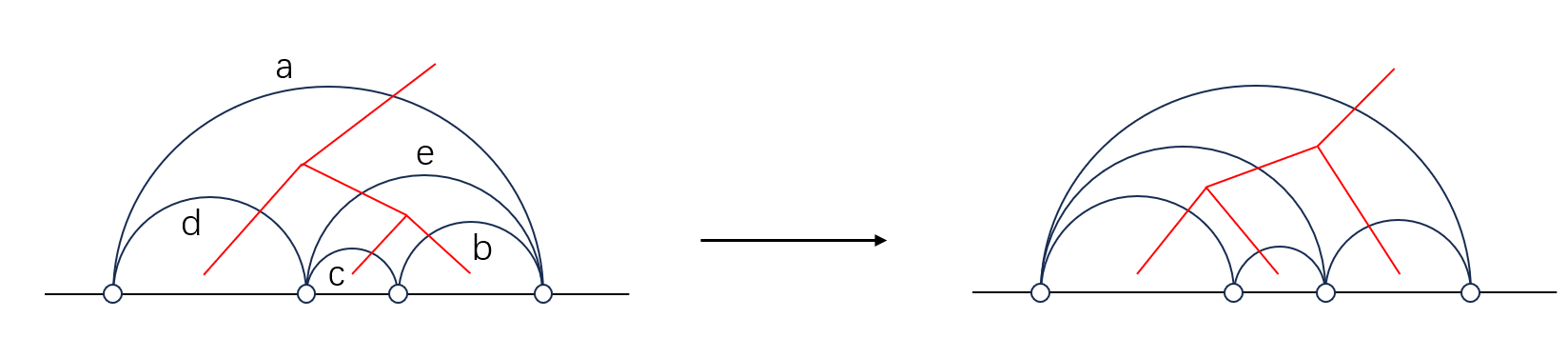}
	\caption{Change of an ideal triangulation.}
	\label{hyperboliccrossing}
\end{figure}

This transformation admits a simple expression if we introduce another set of coordinates called Fock coordinates $\{z_e\}$\cite{1997dg.ga.....2018F}, which is also attached to every edge $e$ in triangulation $\tau$. 
Each edge $e$ is shared between two triangles. Suppose that one of the triangles has two other edges $a,d$ and the other triangle has two other edges $b,c$.  This is shown in Fig.~\ref{hyperboliccrossing}. 

\begin{figure}
	\centering
	\includegraphics[width=0.6\linewidth]{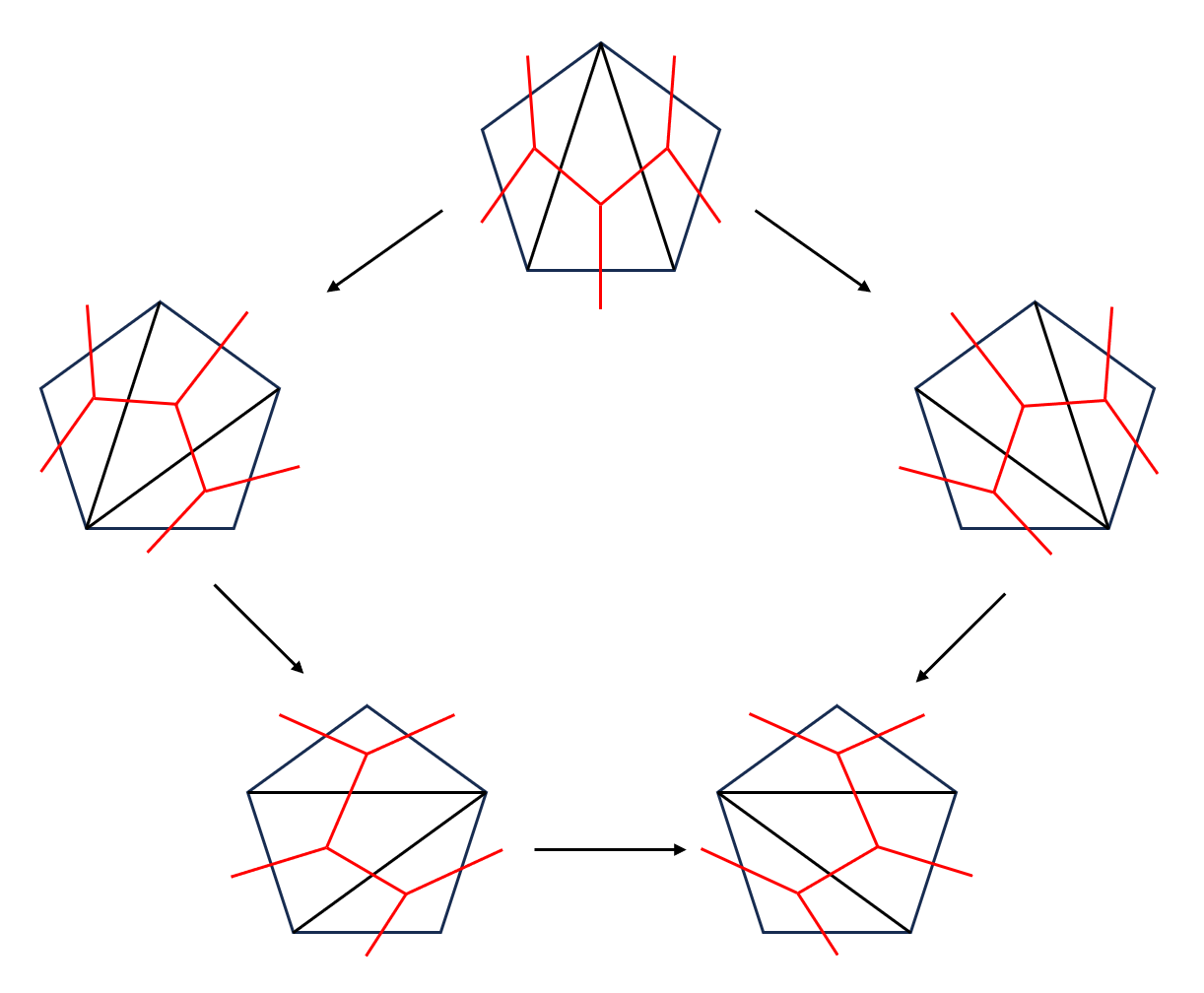}
	\caption{Consistency condition for re-triangulation.}
	\label{pentagon2}
\end{figure}

Then $z_e$ is related to the geodesic lengths of $a,b,c,d$ as
\be
z_e = l_a + l_c - l_b - l_d.
\ee
In the Fock coordinates, the ``crossing'' transformation between two triangles corresponds to the following change of coordinates
\bea  \label{flip}
z_e' = - z_e~,   &  \varphi(x) =  \ln(1+ e^x)~, \nonumber \\
z_a'= z_a - \varphi(-z_e)~, & z_b' = z_b + \varphi(z_e), \nonumber \\
z_d' = z_d + \varphi(z_e)~, & z_c' = z_c - \varphi(-z_e)~.
\eea
Since this is essentially a crossing relation from the perspective of the dual graph, these transformations should satisfy a consistency condition analogous to the Pentagon relation. 
This is depicted in Fig.~\ref{pentagon2}. 

Under the action of the mapping class group MCG($\Sigma$), a triangulation $\tau$ is mapped to another triangulation $\tau'$. Similarly, repeated application of the ``crossing transformation'' can relate one triangulation to another. This parallel suggests that representations of the mapping class group can be constructed leveraging these transformations.

The ultimate goal, however, is to obtain a quantized Teichmüller space and quantum actions of the mapping class group.

To do so, one needs to make use of the symplectic structure naturally defined in the Teichmüller space. It is called the Weil-Petersson symplectic form. In terms of the Penner coordinate with a choice of triangulation $\tau$, it is given by
\be
\omega = -\sum_{\Delta \in \tau} \left( dl_{e_1(\Delta)} \wedge  dl_{e_2(\Delta)} + dl_{e_2(\Delta)} \wedge  dl_{e_3(\Delta)} +  dl_{e_3(\Delta)} \wedge dl_{e_1(\Delta)} \right)~,
\ee
where $\Delta$ denotes triangles in $\tau$ and $e_{1,2,3}(\Delta)$ are the three edges of the triangle $\Delta$ with labels ordered counter-clockwise around the triangle. The symplectic form can, of course, also be expressed in terms of the Fock coordinates, which we shall not reproduce here. 

The quantization of the Teichmüller space is achieved by upgrading the Poisson bracket to a commutator. i.e.~
\be
[z_e, z_{e'}] = 2\pi i b^2 \{z_e, z_{e'}\}~,
\ee
where one introduces a ``Planck constant'' $b$. Ultimately, the parameter $b$ would coincide with the quantum group deformation parameter $b$ in the Liouville theory, entering the $6j$ symbols in \eqref{bcft coefficients} through $q = e^{i\pi b^2}$.

Then one needs to construct the action of the mapping class group on the quantized variables and also on the Hilbert space that follows from quantization. 
Such a transformation should preserve the Poisson bracket, while satisfying the same consistency condition of the flip, including the relation depicted in Fig.~\ref{pentagon2}, now as an operator relation. 

This is in fact possible, if the operator transformation under ``crossing'' is given precisely by (\ref{flip}), with $\varphi(x)$ upgraded to
the special function $\varphi_b(x)$ given by
\be
\varphi_b(x) = \frac{\pi b^2}{2} \int_{i 0 - \infty}^{i 0 + \infty} dw \frac{e^{- i x w}}{\sinh(\pi w) \sinh(\pi b^2 w)}~,
\ee
which reduces to $\varphi(x)$ in the limit $b\to 0$. With the explicit form of transformation, one needs to construct the operator that in fact has the correct commutation with the variables $z_e$ to generate these transformations. That would produce explicit quantum representations of the crossing relations, and subsequently the mapping class group. 

The main idea of such a construction by Kashaev in \cite{1997q.alg.....5021K, *Kashaev:2000ku} follows from expressing the Fock variables in terms of canonical variables $p,q$ that satisfy the commutation relation $[q,p] = 1 $ in the convention used here. 

\begin{figure}
	\centering
	\includegraphics[width=0.4\linewidth]{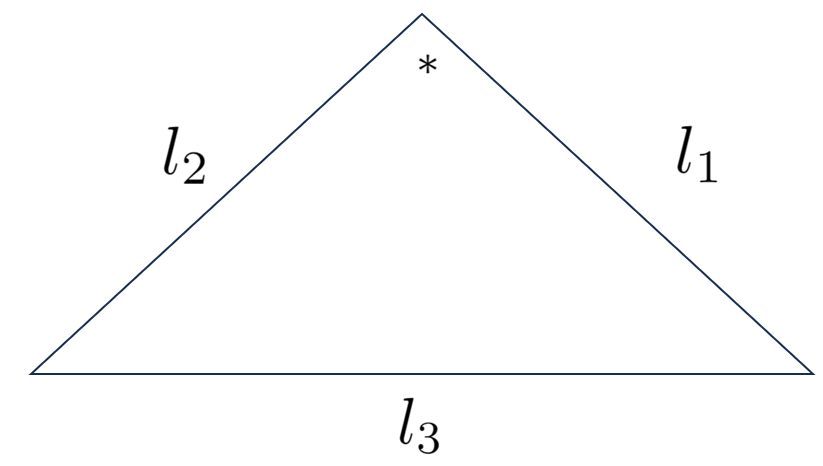}
	\caption{A decorated triangle in Kashaev's construction of quantum Teichmüller theory.}
	\label{decoratedtriangle}
\end{figure}

This relation with canonical variables is constructed as follows. For every triangle $\Delta$ in the triangulation, Kashaev assigns a canonical pair $q_\Delta, p_\Delta$. To connect them with the edge lengths, an extra structure is needed on the triangle. One particular vertex is singled out, and we will mark the vertex with a star, and this is called a decorated triangle. We label the edges of the decorated triangle $l_1, l_2, l_3$ with a particular counter-clockwise order as shown in Fig.~\ref{decoratedtriangle}. 

Then 
\be \label{Kashaev_canonical}
(q_\Delta, p_\Delta) = (l_3 - l_2, l_1 - l_2)~.
\ee
There are $3M$ edges in the triangulation, but we have introduced $4M$ variables for $2M$ triangles, more than the original set. To account for this redundancy, there is a set of constraints \cite{1997q.alg.....5021K, Teschner:2005bz}.

\begin{figure}
	\centering
	\includegraphics[width=0.8\linewidth]{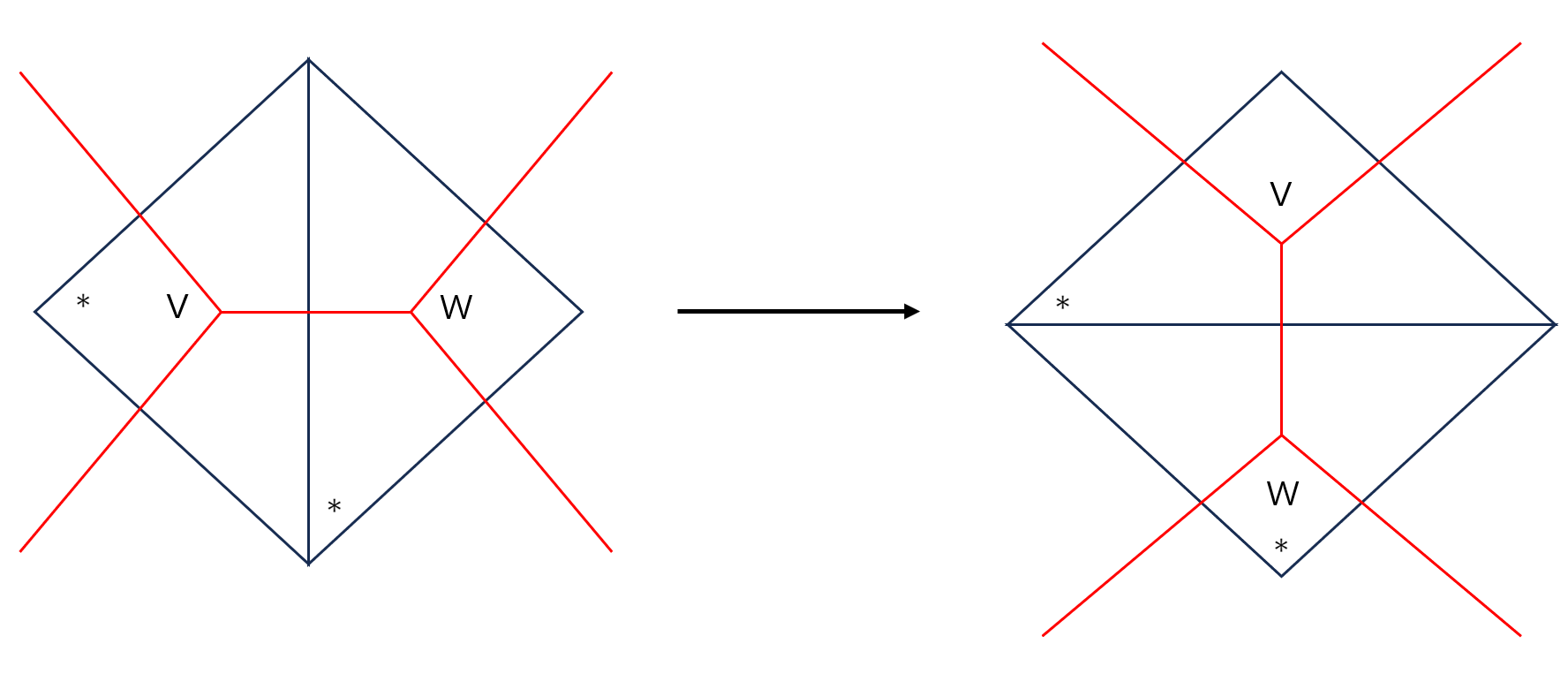}
	\caption{Crossing move in Kashaev's construction with decorations explicitly labeled.}
	\label{decoratedcrossing}
\end{figure}

The crossing relation can now be translated to a transformation $T_{vw}$ in the Kashaev variables. It is given by
\bea
&(U_v, V_v) \to (U_vU_w, U_v V_w + V_v)~, \nonumber \\
&(U_w, V_w) \to (U_w V_v(U_v V_w + V_v)^{-1}, V_w(U_v V_w + V_v)^{-1})~,
\eea
where $v,w$ label the decorated triangles, and they are sharing an edge to be transformed into the crossed triangulation. In addition, the variables in capitals are related to the Kashaev canonical variables in each triangle via 
\be
U_{v,w} = \exp(q_{v,w})~, \qquad V_{v,w} = \exp(p_{v,w})~.
\ee
The above crossing relation is depicted in Fig.~\ref{decoratedcrossing}, with the decorations shown explicitly. 

Now that the triangles are decorated, one also needs to construct transformations $A_v$ that correspond to moving the marked vertex inside a given decorated triangle $v$. 
They correspond to
\be
(q_v, p_v) \to (p_v - q_v, -q_v)~.
\ee
The important point is that $A_v$ and $T_{vw}$ are the basic building blocks in constructing the representations of the mapping class group. Moreover, Kashaev's discrete version of the Teichmüller TQFT is also built from them. 
The $T_{vw}$ operator basically takes the role of a tetrahedron. 
Explicitly in terms of the canonical variables, it is given by
\be \label{Tmatrix}
T_{vw} = e_b(q_v + p_w- q_w) e^{-2\pi i p_v q_w}~, 
\ee
where 
\be
e_b(z) = \exp\left(\frac{1}{4} \int_{i0 -\infty}^{i0 + \infty} \frac{dw}{w} \frac{e^{-2i zw}}{\sinh(b w) \sinh(w/b)}\right)~,
\ee
which is the renowned Faddeev quantum dilogarithm that played an important role in integrable models.

\begin{figure}
	\centering
	\includegraphics[width=0.5\linewidth]{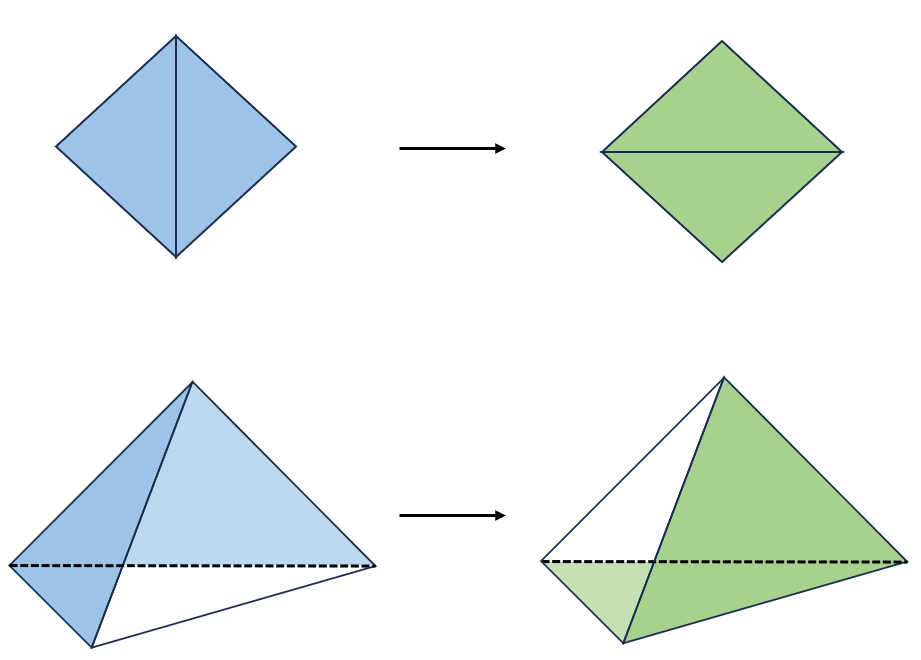}
	\caption{The change of triangulation is given by an operator $T$, and takes the role of a tetrahedron.}
	\label{tetrahedron}
\end{figure}

The TQFT on a 3-manifold is basically defined as the product of these tetrahedra that triangulate the 3-manifold, with the caveat that not all 3-manifolds can admit a finite answer and that there could be issues with divergence. Moreover, since the Hilbert space constructed from the canonical variables is infinite-dimensional and in fact continuous, to ensure convergence, the operator $T$ is further regulated by two positive variables $\alpha,\gamma$ satisfying $1/2 - \alpha - \gamma >0$. 
These are called ``charged'' tetrahedral operators
\be
T_{vw}(\alpha,\gamma) = e^{+\pi i Q^2 ((\alpha- \gamma + 1)/6)}e^{-2 \pi Q(\alpha p_w - \gamma q_w) } T_{vw} e^{2\pi  Q(\alpha p_w + \gamma q_w)}~,
\ee
which reduce to the original tetrahedral operator when these twist parameters approach 0. 

Explicit examples can be found in \cite{EllegaardAndersen:2011vps}, where the (charged) tetrahedral operator builds up the 3-manifold. 

The explicit expression for the operator $A_v$ introduced above, which moves the marked point in the triangle from one corner to its neighbor, is given by
\be \label{movemarkpt}
A_v = e^{-i \pi/3} e^{i 3\pi q_v^2} e^{i \pi (p_v+ q_v)^2}~.
\ee

\subsection{$U_q(SL(2,\mathbb{R}))$ $6j$-symbols and quantum Teichmüller theory}
Now we give an overly simplified review of the connection between the tetrahedral operator $T_{vw}$ and the $U_q(SL(2,\mathbb{R}))$  quantum $6j$ symbols \cite{Nidaiev:2013bda}. The purpose is to translate the convoluted mathematical construction into an intuitive physical picture and to appreciate the relation between the quantum $6j$ symbols and Kashaev's tetrahedra in his formulation of the Teichmüller TQFT. Readers should consult \cite{Nidaiev:2013bda,Teschner:2005bz, Teschner:2003em,EllegaardAndersen:2011vps, 1997q.alg.....5021K} for complete details. 

The connection follows from two ways of parameterizing the Riemann surface $\Sigma$. In Kashaev's discussion, one discretizes the Riemann surface into triangles. This is quite different from what field theorists, particularly string theorists, are familiar with. 
For field/string theorists, it is customary to consider cutting the Riemann surface along closed curves, and decomposing the surface into closed pairs of pants. 
There is a natural way to construct coordinates of the moduli space based on the pair-of-pants decomposition, and construct representations of the mapping class group associated to this set of coordinates. The latter construction is shown to be related to the $U_q(SL(2,\mathbb{R}))$ $6j$-symbols. 

To appreciate some more of the details,  the set of coordinates on the Teichmüller space that are natural to the closed pair-of-pants decomposition is called the Fenchel–Nielsen coordinates. 

The decomposition of the Riemann surface is achieved by a collection of cuts along non-intersecting closed curves $\{C_i\}$, which is called a cut system, where 
\be
1\le i \le 3g + n - 3 \equiv \kappa~.
\ee
For each pair of pants in which the boundary components are geodesics, one can go to the unique conformal frame with constant curvature -1 and label the surface by the geodesic lengths of the boundary components
$(L_1,L_2,L_3)$. We use capital $L_i$'s to distinguish them from the Penner coordinates. 
In addition, to specify the original Riemann surface, one has to describe how the pair of pants are glued back together along the cuts $\{C_i\}$. In general, one could twist one of the boundary components of a trinion sharing a given cut by an angle $\phi_i$ before gluing. A Dehn twist in particular would shift these twist angles by $2\pi$. 
Every cut is also assigned this twist label $\phi_i$. Together, there are $2\kappa$ variables $\{L_1,\cdots L_{\kappa}; \phi_1, \cdots \phi_{\kappa}\}$. 
The Weil-Petersson form takes the form 
\be
\omega = \sum_i^\kappa d(L_i \frac{\phi_i}{2\pi}) \wedge dL_i~.
\ee
Note that the associated Poisson bracket $\{L_i,L_j\}$ vanishes.

\begin{figure}
	\centering
	\includegraphics[width=0.4\linewidth]{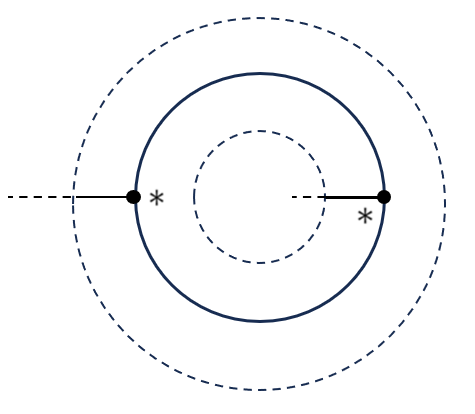}
	\caption{The connection between the Fenchel–Nielsen coordinate and the Fock coordinate can be understood by examining the triangulation of an annulus region surrounding closed geodesics.}
	\label{lolipop}
\end{figure}

There is a simple conversion between the closed geodesic lengths $L_i$ and the Fock coordinates. Recall that the Fock coordinates are defined based on an (ideal) triangulation of the Riemann surface. Consider the dual graph of the triangulation, and pick up a closed path $\gamma_i$ that is homotopic to a closed cut $C_i$ in the cut system chosen for the Fenchel–Nielsen coordinates. Then the relation between the geodesic length $L_i$ and the Fock coordinates of the ordered collection of edges $\{z_{e_m}\}_{\gamma_i}$ along the path $\gamma_i$ is given by\cite{1997dg.ga.....2018F}
\be
2 \cosh(\frac{L_i}{2}) = |\tr(X_{\gamma_i})|~, 
\ee
where $X_{\gamma_i}$ is
\be
X_{\gamma_i} = V^{\sigma_r} E(z_{e_r}) \cdots V^{\sigma_1} E(z_{e_1}), \qquad e_m \in \gamma_i~.
\ee
Here
\be
E(z) = \left(\begin{tabular}{cc}
    0 & $e^{z/2}$ \\
    $-e^{-z/2}$ & 0
\end{tabular}\right)~, \qquad V= \left( \begin{tabular}{cc}
   1  & 1 \\
   -1  & 0
\end{tabular}\right)~,
\ee
and $\sigma_m = \pm 1$, depending on whether the segment $e_m$ is turning right ($+1$) when it connects to $e_{m+1}$ or left ($-1$).
This is a classical relation. 
When one quantizes the Teichmüller space, the conversion between the Fenchel–Nielsen length operators and the Fock operators would be a nontrivial transformation. This transformation is constructed by considering an annular region which admits a canonical ideal triangulation with two triangles that would be naturally assigned two pairs of Kashaev's canonical variables. The closed geodesics at the boundary of the annulus can then be expressed in terms of these Kashaev variables. See Fig.~\ref{lolipop} for an illustration. These have been constructed in detail in \cite{Teschner:2005bz}. 
The mapping class group representations can thus alternatively be constructed with the $L_i$'s eigenstates serving as eigenbasis. 

\begin{figure}
	\centering
	\includegraphics[width=0.6\linewidth]{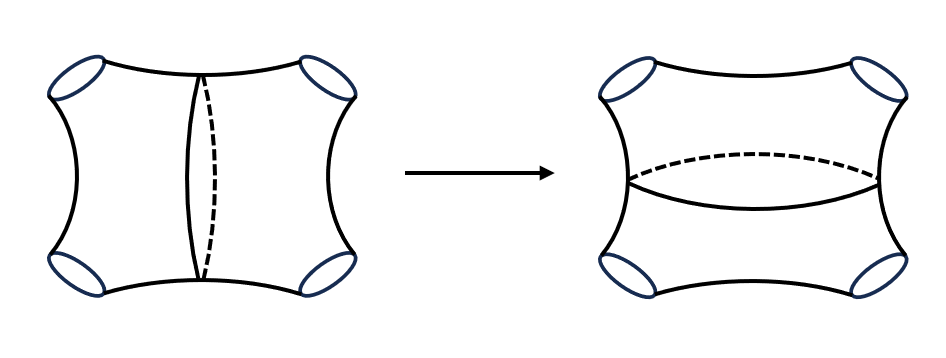}
	\caption{Crossing move for two closed pairs of pants decomposition of the four punctured sphere.}
	\label{pantcrossing}
\end{figure}

Now it comes to the emergence of the quantum $6j$ symbol. 
Consider two pairs of pants joining at one circle. This is conformally related to a sphere with four punctures. It is possible to decompose this surface differently into two other pairs of pants, each containing one boundary component from each pair of pants of the previous decomposition. This is depicted in Fig.~\ref{pantcrossing}. This is essentially the crossing transformation of the four punctures/incoming closed strings. 
The two decompositions were shown to be related precisely by the quantum $6j$ symbol of $U_q(SL(2,\mathbb{R}))$ introduced in \eqref{bcft coefficients}. Recall that each of the parameters $\alpha_i,\sigma_i$ in the $6j$ symbol is parametrized as
\be
\alpha = \frac{Q}{2} + i P_\alpha~, \qquad P_\alpha \in \mathbb{R}_{>0}~.
\ee
Here, they are related to the
geodesic length of each boundary cycle and the intermediate cycles that propagate between the two pairs of pants by
\be
P_i = \frac{L_i}{2\pi b}~.
\ee
Finally, one needs to connect the ``crossing transformation'' for the closed pairs of pants with the flipping of the triangulation in the Fock/Penner coordinates.

\begin{figure}
	\centering
	\includegraphics[width=1\linewidth]{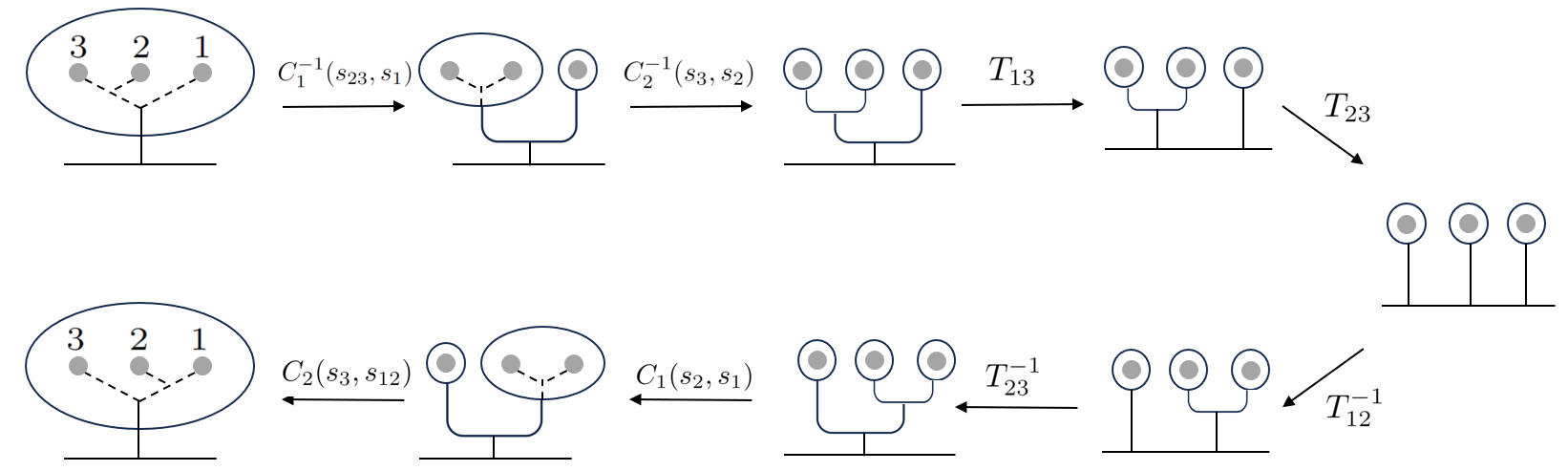}
	\caption{The computation of the $U_q(SL(2,\mathbb{R}))$ $6j$ symbol is performed by first applying the (inverse of) Clebsch-Gordan map, such that we can apply the tetrahedron flip operator $T$ on triangles, and then converting back to eigenstates of the Fenchel–Nielsen coordinates.}
	\label{holealgebra}
\end{figure}

Intuitively, in order to make explicit the intermediate propagating cycle between two pairs of pants from the Fock coordinates, the triangulation should contain paths homotopic to the intermediate cycle. It is not surprising, therefore, that to perform crossing in this picture of closed curve decomposition (i.e.~Fenchel–Nielsen coordinates), one has to change the triangulation in the Fock/Penner picture. 
These considerations produce the final relation between the quantum $6j$ symbols and Kashaev's tetrahedral operators, which is given as follows: 
\bea
&\left\{\begin{tabular}{ccc}
   $Q/2+ i P_1$  & $Q/2+ i P_2$ & $Q/2+ i P_{12}$  \\
   $ Q/2 + i P_3 $& $Q/2 + i P_4$ & $Q/2 + i P_{23}$ 
\end{tabular}\right\} \nonumber \\
&={\langle \{P_i, P_{12}\}| C_2(s_3,s_{12}) C_1(s_2,s_1) T_{23}^{-1} T_{12}^{-1} T_{23} T_{13}  C_2^{-1}(s_3,s_{2}) C_1^{-1}(s_{23},s_1)| \{P_i, P_{23}\} \rangle}~,
\eea
where $C$ is the Clebsch-Gordan map and is detailed in \cite{Nidaiev:2013bda}. The labels of the tetrahedral operator are related to a canonical ideal triangulation tiling the pairs of pants. Rather than explaining that triangulation, it is much more instructive to depict the whole procedure as in Fig.~\ref{holealgebra}.

This shows that a single quantum $6j$ symbol of $U_q(SL(2,\mathbb{R}))$ can be decomposed into four tetrahedra in Kashaev's Teichmüller TQFT (with the twist operators already sent to zero, which is acceptable since this combination is finite anyway).

\subsection{Remarks on the lattice Liouville theory from quantum Teichmüller theory}\label{latticeliouvilleappendix}

In the literature, it is known that there is a lattice integrable model that also recovers the Liouville CFT in the thermodynamic limit\cite{Faddeev:1985gy, *Faddeev:2000if, *Faddeev:2002ms, kashaev2008}. 

The transfer matrix of the lattice model is constructed using operators in the quantum Teichmüller theory of Kashaev\cite{1997q.alg.....5021K, *Kashaev:2000ku, Teschner:2005bz}. In the previous subsections of this appendix, we reviewed the construction of these operators and explained the connection to the $U_q(SL(2,\mathbb{R}))$ $6j$ symbols.

The Hilbert space of the discrete Liouville theory on a circle is constructed as follows. 
First, consider an annulus, where each side of the annulus represents a chiral/anti-chiral half of the theory. Discretize the annulus into one single row of squares, and for each square draw a diagonal say from the top left corner to the bottom right corner. This produces a simple triangulation of the annulus.
Mark a corner in each triangle.
Then attach a pair of Kashaev's canonical variable $(q_v, p_v)$ as in (\ref{Kashaev_canonical}) to each triangle labeled $\Delta_v$. 

These canonical variables are related to the discretized quantum Liouville field $r_v$ as follows \cite{kashaev2008},
\be
\rho(r_v) = \bigg\{\begin{tabular}{cc} 
$p_v+ p_{v-1}$ & if $v= 0$ mod 2, \\
$q_v + q_{v-1} $& otherwise
\end{tabular}
\ee
where $\rho$ denotes the fact that this is a reducible representation of the operator algebra satisfied by $r_v$,
where $\exp( 2\pi  b r_v )\to  \exp(-\phi(x))$ in the semi-classical limit $b \to 0$, and the continuum limit. 
\begin{figure}
	\centering
	\includegraphics[width=0.8\linewidth]{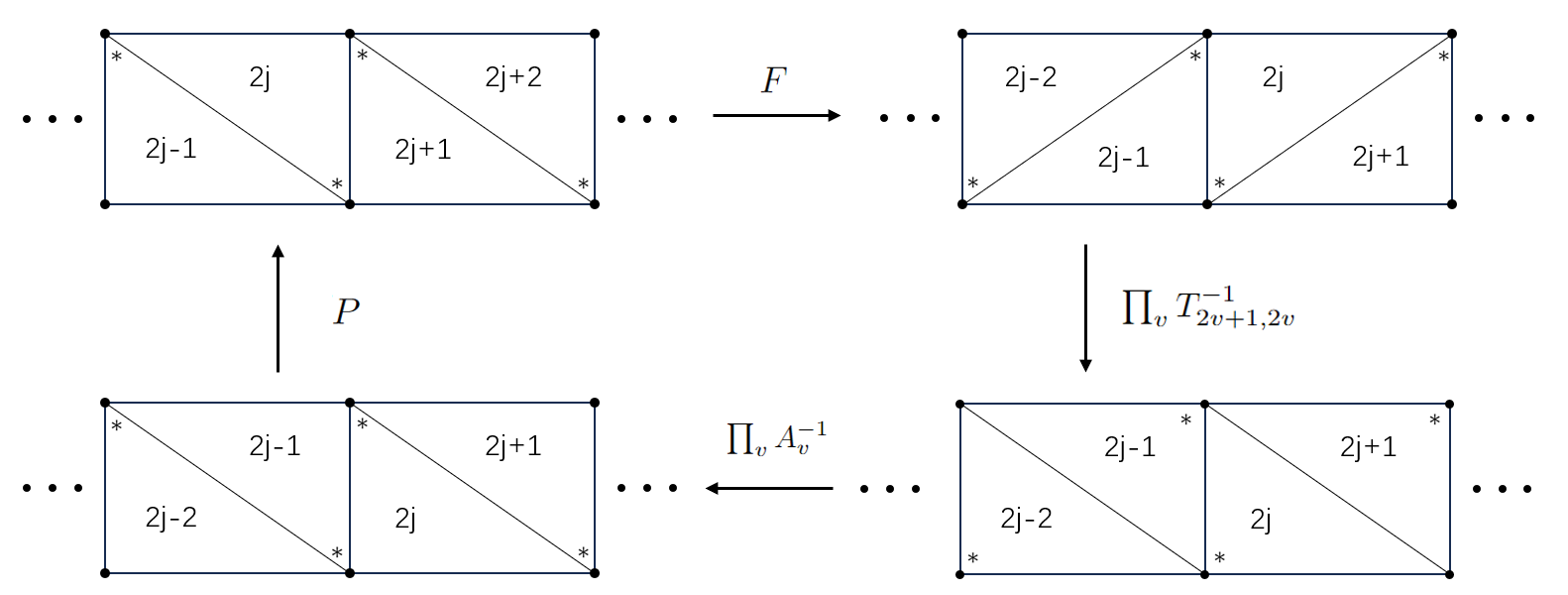}
	\caption{The transfer matrix $F$ can be constructed from the product of $T$, $A$ and $P$ operators.}
	\label{latticeliouville}
\end{figure}
The transfer matrix acts as follows. It is basically implementing the mapping class transformation on the annulus, shearing one boundary of the annulus one lattice site forward relative to the other boundary. After the shear the lattice would look different.
This is illustrated in Fig.~\ref{latticeliouville}. One can convert the sheared lattice back into its original shape using the ``crossing transformation'' as in Fig.~\ref{decoratedcrossing}, with the $T$ operator in (\ref{Tmatrix}) and also the $A$ operator introduced in (\ref{movemarkpt}) that moves the marked point. 

The transfer matrix, denoted $F$, is then given by
\be
F = \mathcal{N} P \prod_v A_v \prod_v T_{2v+1,2v}~,
\ee
for some appropriate overall normalization factor $\mathcal{N}$, and $P$ denotes the cyclic permutation of the vertex on one boundary of the annulus. 

\begin{figure}
	\centering
    \includegraphics[width=0.6\linewidth]{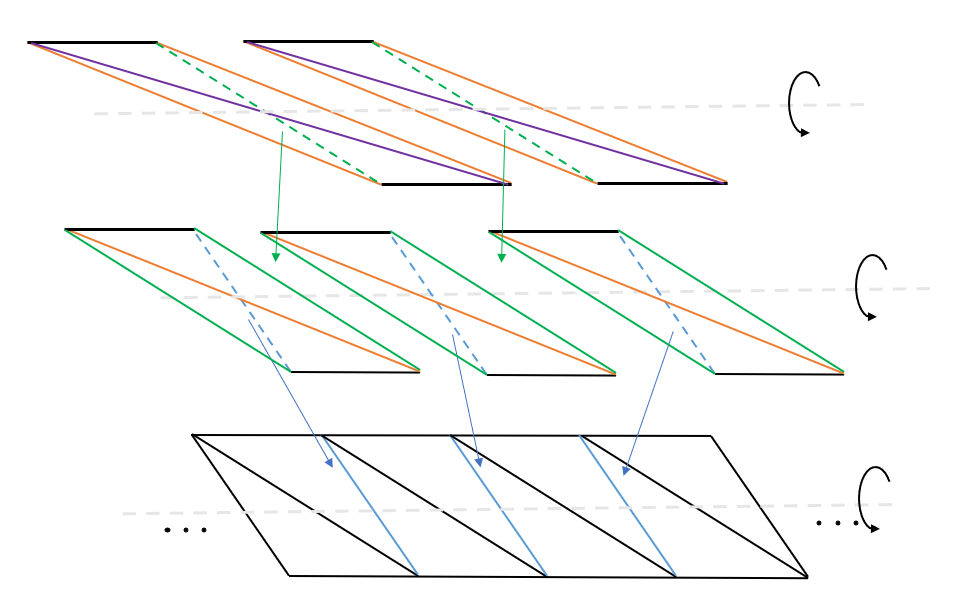}
	\caption{The path integral of the lattice Liouville theory is coming from gluing the tetrahedron operators in quantum Teichmüller theory. When we fold the diagram in the middle, this becomes the sandwich construction, with a seed state not at its RG fixed point.}
	\label{transfermatrix}
\end{figure}

The important observation is that the $T_{vw}$ operators are the building blocks of the (chiral) quantum Teichmüller theory, each being geometrically a tetrahedron. The repeated application of the transfer matrix $F$ is thus a path integral of the Teichmüller theory over a 3-dimensional manifold that takes the form of $A_2 \times I$, where $A_2$ is the annulus and $I$ is an interval corresponding to the time direction in which the transfer matrix is generating an evolution, as illustrated in Fig.~\ref{transfermatrix}.

This path integral in fact conforms with the construction of the Liouville CFT as a strange correlator \eqref{Liouvillestrange}. The slab is a chiral path integral of the Teichmüller TQFT with two boundaries, corresponding to the chiral/anti-chiral parts. It is related to the non-chiral sandwich by folding in the middle, similar to Fig.~\ref{wormholefolding}. Of course in our exact lattice construction realizing Liouville theory \eqref{Liouvillestrange}, the physical boundary condition corresponding to $\langle \Omega|$ is constructed from conformal blocks and represents an RG fixed point. Meanwhile, the discrete quantum Liouville theory construction outlined in this section supplies another seed state $\langle \Omega|$, similar to the case of the product state for the lattice Ising model\cite{Aasen:2016dop, Vanhove:2018wlb}. By applying the RG procedure outlined in Sec.~\ref{SymQRGTN} and appendix \ref{tensornetwork SymQRG} to this seed state, we expect to recover the fixed-point tensor described in \eqref{Liouvillestrange}.

\section{Details of the iterative $T\bar{T}$ derivations}\label{ttbardetails}

In this appendix, we collect the intermediate steps of the formal manipulations in Secs.~\ref{qrg for ttbar}, \ref{ttbar from 3d} and \ref{wdwstates}.

We first verify that the rewriting \eqref{integral deform} solves the flow equation \eqref{flowequation}. The intuition behind this rewriting is to introduce Lagrange multipliers to encode the fact that the metric serves as the source for the stress tensor $T$ or the momentum $\pi$ in the action. This can be explicitly verified as follows. We rewrite \eqref{integral deform} as:
\be
\mathcal{Z}^{\lambda+\delta \lambda}[\gamma_{ij}]= \int \mathcal{D}h^{(1)}_{ij} \mathcal{D}\pi^{(1),ij}  e^{\int d^2 x\left[-\delta \lambda \pi \cdot \Gamma \cdot \pi+\delta \lambda S_1 \cdot \pi-\delta \lambda R(\gamma, \lambda) \right]}\mathcal{Z}^\lambda[h^{(1)}_{ij}]~,
\ee
where
\be
\begin{aligned}
\Gamma&=-\delta(x-y) \frac{12 \pi}{c \sqrt{\gamma}} \left(\frac{1}{2}\big(\gamma_{ik}\gamma_{jl}+\gamma_{il}\gamma_{jk}\big)-\gamma_{ij}\gamma_{kl} \right)~,\\
S_1&=i\frac{\gamma_{ij}-h^{(1)}_{ij}}{\delta \lambda}~,\\
R(\gamma, \lambda)&=\frac{c}{ 48 \pi \lambda}   \sqrt{\gamma}R(\gamma_{ij})~.
\end{aligned}
\ee
By performing the Gaussian integral for $\pi$, we obtain
\be
 \int \mathcal{D}h^{(1)}_{ij}  e^{\int d^2x \left[\frac{\delta \lambda}{4} S_1 \cdot \Gamma^{-1} \cdot S_1 -\delta \lambda R(\gamma, \lambda) \right]}\mathcal{Z}^\lambda[h^{(1)}_{ij}]~,
\ee
where 
\be
\Gamma^{-1}=-c\frac{ \delta(x-y) \sqrt{\gamma}}{12\pi}\left(\frac{1}{2}(\gamma^{ik}\gamma^{jl}+\gamma^{il}\gamma^{jk})-\gamma^{ij} \gamma^{kl} \right)~.
\ee
Thus the result reads
\be
\mathcal{Z}^{\lambda+\delta \lambda}[\gamma_{ij}]= \int \mathcal{D}h^{(1)}_{ij}  e^{\int d^2 x \left[\frac{c \sqrt{\gamma}}{ 48 \pi \delta \lambda} (h^{(1)}_{ij}-\gamma_{ij})  \left(\frac{1}{2}(\gamma^{ik}\gamma^{jl}+\gamma^{il}\gamma^{jk})-\gamma^{ij} \gamma^{kl}\right) (h^{(1)}_{kl}-\gamma_{kl})-\frac{c \delta \lambda}{48\pi \lambda} \sqrt{\gamma} R(\gamma_{ij})\right] }\mathcal{Z}^\lambda[h^{(1)}_{ij}]~.
\ee

Next we perform the change of variables via 
\be
h'^{(1)}_{ij}=h^{(1)}_{ij}-\gamma_{ij}~,
\ee
then we get
\be
\mathcal{Z}^{\lambda+\delta \lambda}[\gamma_{ij}]= \int \mathcal{D}h'^{(1)}_{ij}  e^{\int d^2 x \left[\frac{c \sqrt{\gamma}}{ 48 \pi \delta \lambda}  h'^{(1)}_{ij}  \left(\frac{1}{2}(\gamma^{ik}\gamma^{jl}+\gamma^{il}\gamma^{jk})-\gamma^{ij} \gamma^{kl}\right) h'^{(1)}_{kl}-\frac{c \delta \lambda}{48\pi \lambda} \sqrt{\gamma} R(\gamma_{ij}) \right]  }\mathcal{Z}^\lambda[h'^{(1)}_{ij}+\gamma_{ij}]~,
\ee
which is precisely the solution of \eqref{flowequation} in terms of the Hubbard-Stratonovich transformation, and can be interpreted as coupling the original theory to 2D random geometry\cite{Cardy:2018sdv, McGough:2016lol}.

Next, in the iterated procedure leading to \eqref{deformed5}, the kinetic term, the initial condition and the integration limits arise as follows. By writing
\be
i \pi^{(m+1),ij}(h^{(m)}_{ij}-h^{(m+1)}_{ij})=-i (\lambda^{(m+1)}-\lambda^{(m)}) \pi^{(m+1),ij}\frac{h^{(m+1)}_{ij}-h^{(m)}_{ij}}{\lambda^{(m+1)}-\lambda^{(m)}}~,
\ee
we observe that this expression, in the limit $\delta z \to 0$, becomes a derivative term:
\be
-i d\lambda \pi(\lambda)^{ij}\partial_{\lambda} h(\lambda)_{ij}~.
\ee
The initial condition for the path integral is imposed using a delta function:
\begin{equation}
    \mathcal{Z}^{\mu}[\gamma_{ij}] = \int \mathcal{D}h^{(0)}_{ij} \mathcal{D}\pi^{(0),ij} e^{i\int d^2x \pi^{(0),ij}(\gamma_{ij}-h^{(0)}_{ij})}\mathcal{Z}^\mu[h^{(0)}_{ij}]=\int \mathcal{D}h^{(0)}_{ij} \delta(h^{(0)}_{ij}-\gamma_{ij})\mathcal{Z}^\mu[h^{(0)}_{ij}]~.
\end{equation}
The integration limits for our 3D path integral are:
\be
\begin{aligned}
\lambda_{\text{ini}}=\mu, \quad h_{ij}(\mu)=\gamma_{ij}~; \\
\lambda_{\text{fin}}=\epsilon, \quad h_{ij}(\epsilon)=h^{\epsilon}_{ij}~. 
\end{aligned}
\ee

Finally, we verify that the change of variables \eqref{rescaling} renders \eqref{bulk} manifestly equivalent to \eqref{wdw}. With this transformation, we rewrite equation \eqref{bulk} as
\begin{equation} \label{deformed}
 \begin{aligned}
  \Psi(\gamma/\mu) =&\int \mathcal{D} h^\epsilon_{ij}  \int_{\substack{h_{ij}(\mu)=\gamma_{ij} \\ h_{ij}(\epsilon)=h^\epsilon_{ij}}} \mathcal{D} h_{ij} \mathcal{D}\pi^{ij} \exp \left[ \int_{\mu}^\epsilon d\lambda \int d^2x \left(- i\lambda \pi^{ij}\partial_\lambda \left(\frac{h_{ij}}{\lambda}\right)   \right. \right. \\
    & \left. \left.  -\frac{1}{2\lambda} \left( \frac{16\pi G_N\lambda}{\sqrt{h}}(\pi^{ij}\pi_{ij}-\pi^2) - \frac{\sqrt{h}}{16\pi G_N \lambda} (R(\frac{h_{ij}}{\lambda})+2) \right) \right) \right] e^{\frac{1}{8\pi G_N \epsilon}\int d^2x \sqrt{h^\epsilon}}\mathcal{Z}^{\epsilon}[h^\epsilon_{ij}]~.
 \end{aligned}
\end{equation}
Next, we perform the change of variables:
\be \label{shift}
\pi^{ij} = \pi'^{ij}-\frac{i}{16\pi G_N \lambda} \sqrt{h}h^{ij}~,
\ee
and for simplicity, we will continue to denote the new variable $\pi'^{ij}$ as $\pi^{ij}$ to facilitate comparison, as it serves as a dummy variable. This procedure does not alter the total path integral but instead reorganizes the terms within it. It renders \eqref{bulk} manifestly equivalent to \eqref{wdw}.

First, we have the change in the first term of \eqref{bulk} as,
\be
\begin{aligned}
\delta(-i\lambda\pi^{ij}\partial_\lambda \left(\frac{h_{ij}}{\lambda}\right))&=- \frac{\sqrt{h}}{16\pi G_N } h^{ij}\partial_\lambda \left(\frac{h_{ij}}{\lambda}\right)=-\frac{1}{16\pi G_N}(\frac{\sqrt{h} h^{ij} \partial_\lambda h_{ij}}{\lambda}-\frac{2\sqrt{h}}{\lambda^2})=-\frac{1}{8\pi G_N} \partial_\lambda (\frac{\sqrt{h}}{\lambda})~.
\end{aligned}
\ee

This term is a total derivative with respect to $\lambda$, and thus the change of variables introduces two boundary terms:
\be
\int^{\epsilon}_\mu d\lambda d^2x\delta(-i\lambda\pi^{ij}\partial_\lambda \left(\frac{h_{ij}}{\lambda}\right))=\int d^2 x (\frac{1}{8\pi G_N \mu} \sqrt{\gamma}-\frac{1}{8\pi G_N \epsilon} \sqrt{h^{\epsilon}})~.
\ee

The second term precisely cancels the boundary cosmological counterterm at $\epsilon$, while the first term corresponds to the boundary term at $\mu$ in \eqref{wdw}.

Next, we consider:
\be \label{cancelbulkcc}
\begin{aligned}
\frac{16\pi G_N\lambda}{\sqrt{h}}\delta(\pi^{ij} \pi_{ij}-\pi^2)&=\frac{16\pi G_N\lambda}{\sqrt{h}}\left(\frac{i \sqrt{h} }{8\pi G_N \lambda} \pi^i_i+\frac{|h|}{128(\pi G_N \lambda)^2}\right)\\
&=2i \pi^i_i+\frac{\sqrt{h}}{8\pi G_N \lambda}~.
\end{aligned}
\ee

The second term cancels precisely with the bulk cosmological constant term in \eqref{bulk}. The first term, which is linear in momentum, represents the Weyl transformation generator. Notably, it can be combined with the term $-i \lambda \pi^{ij}\partial_\lambda \left(\frac{h_{ij}}{\lambda}\right)$ as follows:
\be \label{cancelrescaling}
\begin{aligned}
 &-i \lambda \pi^{ij}\partial_\lambda \left(\frac{h_{ij}}{\lambda}\right) - \frac{1}{2\lambda} (2i \pi^i_i)\\
=&-i\pi^{ij} \partial_\lambda h_{ij}+i\frac{\pi^i_i}{\lambda}-i\frac{\pi^i_i}{\lambda}=-i\pi^{ij} \partial_\lambda h_{ij}~.
\end{aligned}
\ee
Collecting the pieces gives \eqref{finalanswer} in the main text.

For the operator formulation of Sec.~\ref{wdwstates}: acting on wavefunctions in the metric basis for general states $\ket{\varphi}$, $\hat{\mathcal{H}}$ becomes a differential operator:
\be
\begin{aligned}
&\bra{g_{ij}=\frac{h_{ij}}{\lambda}}\hat{\mathcal{H}}\ket{\varphi}\\
=&\left[\frac{16\pi G_N \lambda}{\sqrt{h}} \left(\frac{1}{2}(h_{ik}h_{jl}+h_{il}h_{jk})-h_{ij}h_{kl} \right) \frac{\delta}{\delta h_{ij}}\frac{\delta}{\delta h_{kl}}+ \frac{\sqrt{h}}{16\pi G_N \lambda} (R(\frac{h_{ij}}{\lambda})+2) \right]\bra{g_{ij}=\frac{h_{ij}}{\lambda}}\varphi\rangle\\
=&\mathcal{H}(\lambda,h_{ij},\frac{\delta}{\delta h_{ij}})\bra{g_{ij}=\frac{h_{ij}}{\lambda}}\varphi\rangle~,
\end{aligned}
\ee
where we scaled the metric by the lapse. 
The explicit evaluation of the constraint on $\ket{Z_{\text{CFT}}}$ in this basis proceeds as follows. Explicitly, in metric basis, we have:
\begin{equation}
    \begin{split}
        &\bra{g_{ij}=\frac{\gamma_{ij}}{\mu},\epsilon}  \hat{U}_{\text{3D gravity}} \hat{\mathcal{H}}|Z_{\text{CFT}}\rangle\\
        =&\int \mathcal{D} h^\epsilon_{ij} \bra{g_{ij}=\frac{\gamma_{ij}}{\mu},\epsilon}  \hat{U}_{\text{3D gravity}} \ket{g_{ij}=\frac{h_{ij}^\epsilon}{\epsilon},\epsilon} \bra{g_{ij}=\frac{h_{ij}^\epsilon}{\epsilon},\epsilon}\hat{\mathcal{H}}| Z_{\text{CFT}}\rangle~.
    \end{split}
\end{equation}

Now we can evaluate\footnote{Some formally infinite constants are dropped, and we can think of this as a normal-ordering prescription common in $T\bar{T}$ literature.}:
\be \label{wdweqn}
\begin{aligned}
&\bra{g_{ij}=\frac{h_{ij}^\epsilon}{\epsilon},\epsilon}\hat{\mathcal{H}}| Z_{\text{CFT}}\rangle =\mathcal{H}(\epsilon,h^\epsilon_{ij},\frac{\delta}{\delta h^{\epsilon}_{ij}})\bra{g_{ij}=\frac{h^\epsilon_{ij}}{\epsilon},\epsilon}Z_{\text{CFT}}\rangle\\
&=\left[\frac{16\pi G_N \epsilon}{\sqrt{h^\epsilon}} \left(\frac{1}{2}(h^\epsilon_{ik}h^\epsilon_{jl}+h^\epsilon_{il}h^\epsilon_{jk})-h^\epsilon_{ij}h^\epsilon_{kl} \right) \frac{\delta}{\delta h^\epsilon_{ij}}\frac{\delta}{\delta h^\epsilon_{kl}}+ \frac{\sqrt{h^\epsilon}}{16\pi G_N \epsilon} \left(R(\frac{h^\epsilon_{ij}}{\epsilon})+2\right) \right]\left( e^{\frac{1}{8\pi G_N \epsilon}\int d^2x \sqrt{h^\epsilon}}\mathcal{Z}^{\epsilon}[h^{\epsilon}_{ij}]\right)\\
&=e^{\frac{1}{8\pi G_N \epsilon}\int d^2x \sqrt{h^\epsilon}} \left[\frac{16\pi G_N \epsilon}{\sqrt{h^\epsilon}} \left(\frac{1}{2}(h^\epsilon_{ik}h^\epsilon_{jl}+h^\epsilon_{il}h^\epsilon_{jk})-h^\epsilon_{ij}h^\epsilon_{kl} \right) \frac{\delta}{\delta h^\epsilon_{ij}}\frac{\delta}{\delta h^\epsilon_{kl}}-2 h^\epsilon_{ij} \frac{\delta}{\delta h^\epsilon_{ij}}+ \frac{\sqrt{h^\epsilon}}{16\pi G_N \epsilon} R(\frac{h^\epsilon_{ij}}{\epsilon}) \right]\mathcal{Z}^{\epsilon}[h^{\epsilon}_{ij}]~.
\end{aligned}
\ee
As described in Sec.~\ref{wdwstates}, the double-derivative term vanishes in the $\epsilon \to 0$ limit, and the remaining terms reproduce the conformal anomaly equation \eqref{anomaly} quoted in the main text.

{\small
\bibliographystyle{ourbst}
\bibliography{gravity}
}

\end{document}